\documentclass{emulateapj}
\usepackage{url}
\def\arcsec{$^{\prime\prime}$}
\newcommand\kms{km s$^{-1}$}

\newcommand\ha{$\rm{H}\alpha$}
\newcommand\hb{$\rm{H}\beta$}

\shorttitle{Resolved Star Formation Histories}
\shortauthors{Yoachim et al.}

\begin{document}

\title{Spatially Resolved Spectroscopic Star Formation Histories of Nearby Disks:  Hints of Stellar Migration\footnote{ This paper includes data taken at The McDonald Observatory of The University of Texas at Austin.} }
\journalinfo{Accepted for publication in ApJ}
\submitted{Accepted for publication in ApJ, \today}

\author{Peter Yoachim\altaffilmark{1}, Rok Ro{\v s}kar\altaffilmark{2}, 
Victor P. Debattista\altaffilmark{3}} \altaffiltext{1}{Department of
Astronomy, University of Washington, Box 351580, Seattle WA, 98195; {yoachim@u.washington.edu}}
\altaffiltext{2}{Institute for Theoretical Physics
University of Z\"{u}rich
Winterthurerstrasse 190
CH-8057 Z\"{u}rich
Switzerland }
\altaffiltext{3}{Jeremiah Horrocks Institute, University of Central Lancashire, Preston PR1 2HE, UK.  RCUK Fellow}

\begin{abstract}
  We use the Mitchell Spectrograph (formerly VIRUS-P) to observe 12
  nearby disk galaxies.  We successfully measure ages in the outer
  disk in six systems.  In three cases (NGC 2684, NGC 6155, and NGC
  7437), we find that a downward break in the disk surface brightness
  profile corresponds with a change in the dominant stellar population
  with the interior being dominated by active star formation and the
  exterior having older stellar populations that are best-fit with
  star formation histories that decline with time.  The observed
  increase in average stellar ages beyond a profile break is similar
  to theoretical models that predict surface brightness breaks are
  caused by stellar migration, with the outer disk being populated
  from scattered old interior stars.

  In three more cases (IC 1132, NGC 4904, and NGC 6691), we find
  no significant change in the stellar population as one crosses the
  break radius.  In these galaxies, both the inner and outer disks are
  dominated by active star formation and younger stellar populations.  

  While radial migration can contribute to the stellar populations
  beyond the break, it appears more than one mechanism is required to
  explain all of our observed stellar profile breaks.

\end{abstract}

\section{Introduction}

The stellar components of spiral galaxies are typically found embedded
in much larger HI disks \citep{Bosma81, Broeils97, Begum05,
Walter2008}.  Early work based on photographic plates found disk
galaxies to be sharply truncated \citep{vdK79,vdk81a,vdk81b}.  Recent
work by \citet{Pohlen06} based on Sloan Digital Sky Survey (SDSS)
images finds three main types of disk galaxy surface brightness
profiles.  The majority of galaxies in their sample (60\%) are best
fit with an exponential profile that transitions to a steeper
exponential (at what is commonly referred to as a break radius).
Another 30\% of galaxies have profiles which transition to shallower
scale-lengths (up-bending profiles).  Only 10\% of galaxies are well
fit by a single exponential profile.

The first step in understanding disk outskirts therefore needs to be
an explanation for the down-bending profiles.  Traditionally, it has
been assumed there is a surface density threshold below which gas will
fail to undergo star formation \citep{Kennicutt89, Schaye04}.  There
is a long history of hunting for star formation at large galactic
radii.  \citet{Ferguson98} discovered extended star formation in 3
galaxies based on \ha\ imaging.  Large samples of galaxies have been
imaged with GALEX, showing that 20\% have UV-bright emission features
beyond the traditional star formation threshold radius
\citep{Thilker05, Zaritsky07, Thilker08}.  One interpretation of the
outer disks in these systems is that, albeit at lower efficiency, the
majority of the stars formed locally \citep{Elmegreen2006a}.  However
\citet{Bigiel10} find that star formation is extremely inefficient in
the extended HI dominated envelope.  Thus the majority of stars
beyond the break in down-bending profiles probably did not form {\emph{in
situ}}.

\citet{Roskar08} present simulations that show how the combination of
a star formation threshold and secular evolution can create disks with
broken exponential profiles.  In this scenario, the region beyond the
star formation cut-off is populated by stars which migrate there from
the inner disk.  Because migration is a random process, with stars
moving both outwards and inwards, it takes stars progressively longer
to reach larger radii.  Thus this process creates an increasing mean
stellar age beyond the break radius.  Even if there is some low-level
star formation in the outer disk, the migrated population can still be
expected to dominate in most cases \citep{Mart09,Roskar2010}.
\citet{Sanchez09b} present cosmological simulations of disk galaxies
which show breaks in their exponential profiles.  While they find
significant stellar migration (60\% of stars beyond the profile break
formed in the inner disk), they attribute the profile break and
corresponding increase in stellar age beyond the break to changes in
the star formation rate.

Secular stellar migration, caused by resonant interactions with disk
structure \citep{Sellwood02,Roskar2008,Schoenrich09,Roskar2011}, has
attracted a lot of attention in recent years in part because it
violates a key assumption that has been built into decades of chemical
evolution models: that stars live out their lives where they are born
\citep{Tinsley75,Matteucci89,Carigi96,Chiappini97,Boissier99}.
Migration has many consequences; for instance it naturally explains
the flatness and large scatter in the age-metallicity relation at the
solar neighborhood \citep{Sellwood02, Roskar2008, Schoenrich09}.
Stellar migration has even been proposed to account for (a fraction
of) the Milky Way thick disk \citep{Schoenrich09,Loebman11}.

Observationally constraining how often stellar migration actually
occurs in nature is an important next step.  In external galaxies,
This is ideally done using resolved stellar populations which are
possible with the {\it Hubble Space Telescope}.  The
single-exponential galaxy NGC 300 shows no evidence of stellar
migration \citep{Gogarten10}.  However, both NGC 4244
\citep{deJong07,Roskar08} and NGC 7793 (Radburn-Smith et al. 2011,
submitted) show clear evidence of migration in the form of increasing
mean stellar age beyond the break.  Unfortunately, the number of disk
galaxies suitable for such resolved stellar population studies is not
very large.  Instead, \citet{Bakos08} stack SDSS images of galaxies
with similar profiles from the \citet{Pohlen06} sample to measure
radial color gradients.  They find that galaxies with down-bending
breaks have a $g-r$\ color minimum at the break radius.  They take
this as evidence that the stellar populations change across the break,
with the region beyond the break being redder and older.

In this paper, we aim to spectroscopically constrain the ages and
metallicities of stars in the outer disks of many nearby galaxies to
directly test if the stars are formed recently {\emph{in situ}} or
reach the outer disk through radial migration.  To this end, we use
the George and Cynthia Mitchell Spectrograph (formerly
VIRUS-P)\citep{Hill08}.  With a large field of view, and large fibers,
the Mitchell Spectrograph has proven to be an ideal instrument for
observing nearby spiral galaxies \citep{Blanc2009} and excellent at
spectroscopically observing low surface brightness regions
\citep{Murphy2011}.  In \citet{Yoachim10c} we presented the analysis
of one galaxy from our sample, NGC 6155, which showed an age upturn
beyond the break.  This paper presents our full sample.

\section{Data}

\subsection{Observations}

We obtained spatially resolved spectroscopy of our target galaxies using the Mitchell Spectrograph on the 2.7m Harlan J. Smith telescope at McDonald Observatory.  Observations were made during dark time in April, June, and August 2008 and March 2009.  Identical instrument setups were used for all observations.  The Mitchell Spectrograph with the VP-2 IFU bundle used here has a square array of 246 optical fibers which sample the 1.9\arcmin$\times$1.9\arcmin\ field-of-view with a 1/3 filling factor, with each fiber having a diameter of 4.23\arcsec.  The spectrograph images the spectrum of each fiber on a 2048$\times$2048 Fairchild Imaging CCD.  On readout, the chip was binned by a factor of two in the horizontal (spectral) dimension.  

For each galaxy, observations were taken at three unique dither positions to provide nearly complete spatial coverage.  For the more extended galaxies, we also interleaved observations of blank sky.  Individual galaxy exposures were 20 or 30 minutes with sky exposures of 4-5 minutes.  Twilight flat field frames, bias frames, and wavelength calibration Hg and Cd lamps were taken during the day.  Flux standard stars were observed in twilight.

Our targets are listed in Table~\ref{basic_properties}.  We selected galaxies to match the instrument field of view and span a range of properties, including some with downward breaking photometric profiles and some with no break.  Targets were primarily drawn from \citet{Pohlen06}, with a few additional targets selected from HyperLeda \citep{Paturel03}.

\begin{deluxetable*}{ l l l c c c c c}
\tabletypesize{\footnotesize}
\tablewidth{0pt}
\tablecaption{Galaxy Sample \label{Table_galsample}}
\tablehead{ 
\colhead{Name} &  \colhead{RA } & \colhead{Dec }  & \colhead{Inclination} &  \colhead{Exposure Time} & \colhead{v$^1$} & \colhead{v$^1_{\rm{rot}}$}& \colhead{Profile Type} \\
\colhead{} & \colhead{(2000)} & \colhead{(2000)} & \colhead{(degrees)} & \colhead{degrees E of N} & \colhead{(minutes)} & \colhead{(\kms)} & \colhead{(\kms)}
\label{basic_properties}
}
\startdata
IC 1132 & 15:40:07 & +20:40:50.1 & 8   & 140    & 4555 & 105.6 & II \\
NGC 1058 & 02:43:30 & +37:20:27.0 & 18 & 180   & 746 & 33.7 & I \\
NGC 2684 & 08:54:54 & +49:09:37.5 & 22 & 240   & 2943 &  101.4 & II \\
NGC 2942 & 09:39:08 & +34:00:22.6 & 47 & 180  & 4412 & 177.3 & II \\
NGC 3888 & 11:47:34 & +55:58:02.3 & 47 & 180  & 2496 & 203.1  & I \\
NGC 4904 & 13:00:59 & -00:01:38.8 & 54 & 180   & 1025 & 105.2  & II \\
NGC 5624 & 14:26:35 & +51:35:08.1 & 56 & 180   & 2042 & 66.9  & III \\
NGC 6060 & 16:05:52 & +21:29:06.0 & 61 & 280  & 4503 & 252.4  & II \\
NGC 6155 & 16:26:08 & +48:22:00.4 & 34 & 240  & 2582 & 109.8   & II \\
NGC 6691 & 18:39:12 & +55:38:29.5 & 38 & 180   & 6130 & 77.6  & II \\
NGC 7437 & 22:58:10 & +14:18:31.1 & 23 & 180  & 2360 & 151.9  & II \\
UGC 04713 & 09:00:20 & +52:29:39.3 & 40 & 180 & 9127 & 347.6  & II
\enddata
\tablenotetext{1}{v and v$_{\rm{rot}}$ values taken from Hyperleda}
\end{deluxetable*}

\subsection{Data Reduction}

We use the Mitchell Spectrograph software pipeline Vaccine \citep{Adams2011} to perform basic reduction tasks.  Vaccine subtracts the over-scan region and generates and subtracts a bias frame.  Bad pixels are masked.  Twilight flats and arc lamps from each night are combined.  The combined flat frame is used to trace the spatial profile of each fiber and correct for pixel-to-pixel variation.  The combined arc lamp is used to fit a wavelength solution for each fiber.  We use a 5th order polynomial and find the wavelength solutions have a typical RMS of 0.06-0.18 \AA\ ($\sim0.1$ pixel).  Vaccine then masks potential cosmic rays and collapses each fiber to a one dimensional spectrum.  The fibers are adequately spaced, and our targets are predominately low surface brightness, so we find making a fiber cross-talk correction unnecessary.

Once the Vaccine routines are completed we make the following additional reductions using standard IRAF routines and custom IDL scripts.  The reduced twilight frames are used to generate a fiber-to-fiber flux correction (similar to a longslit illumination correction).  We use the b-spline procedure described in \citet{Kelson03} to generate background frames.  For galaxy images, the bracketing sky observations are used to generate the sky frame while flux standard stars and smaller galaxies have enough empty fibers to generate a sky spectrum.  Because we bin fibers together, imperfect sky-subtraction represents a serious source of potential systematic errors.  By taking sky observations between science exposures, we are able to construct background images that are nearly perfectly matched to the science frames, mimicking the success that has been achieved with nod-and-shuffle techniques.  We have checked for over/under background subtraction by cross-correlating binned spectra with the sky spectrum.  If the sky was over (under) subtracted, the cross correlation would show a large maximum (minimum) at zero shift. We found no significant extrema in our cross-correlations.

The spectra are then rectified to a common wavelength scale, and the flux photo-spectroscopic standards are used to flux calibrate the spectra and correct for atmospheric extinction.  The spectra are then velocity shifted to the Local Standard of Rest.  Images taken at identical dither positions are averaged together with outlier rejection to eliminate any remaining cosmic rays.  Many observations were taken in non-photometric conditions, so before combining frames we scale them to a common average flux level.  

Our final reduced data for each galaxy includes 738 spectra spanning a 1.9\arcmin$\times$1.9\arcmin\ field-of-view.  The spectra have a wavelength range of $\sim$3550-5580 with 2.2 \AA\ pixels (with some fibers only reaching 5500 \AA\ due to the warping of the trace).  The FWHM of reduced arc lamp images is 5.3 \AA, giving us a resolution of $R\sim700-1000$.  The fiber-to-fiber changes in resolution are less than 10\%.  

\begin{figure*}
\plottwo{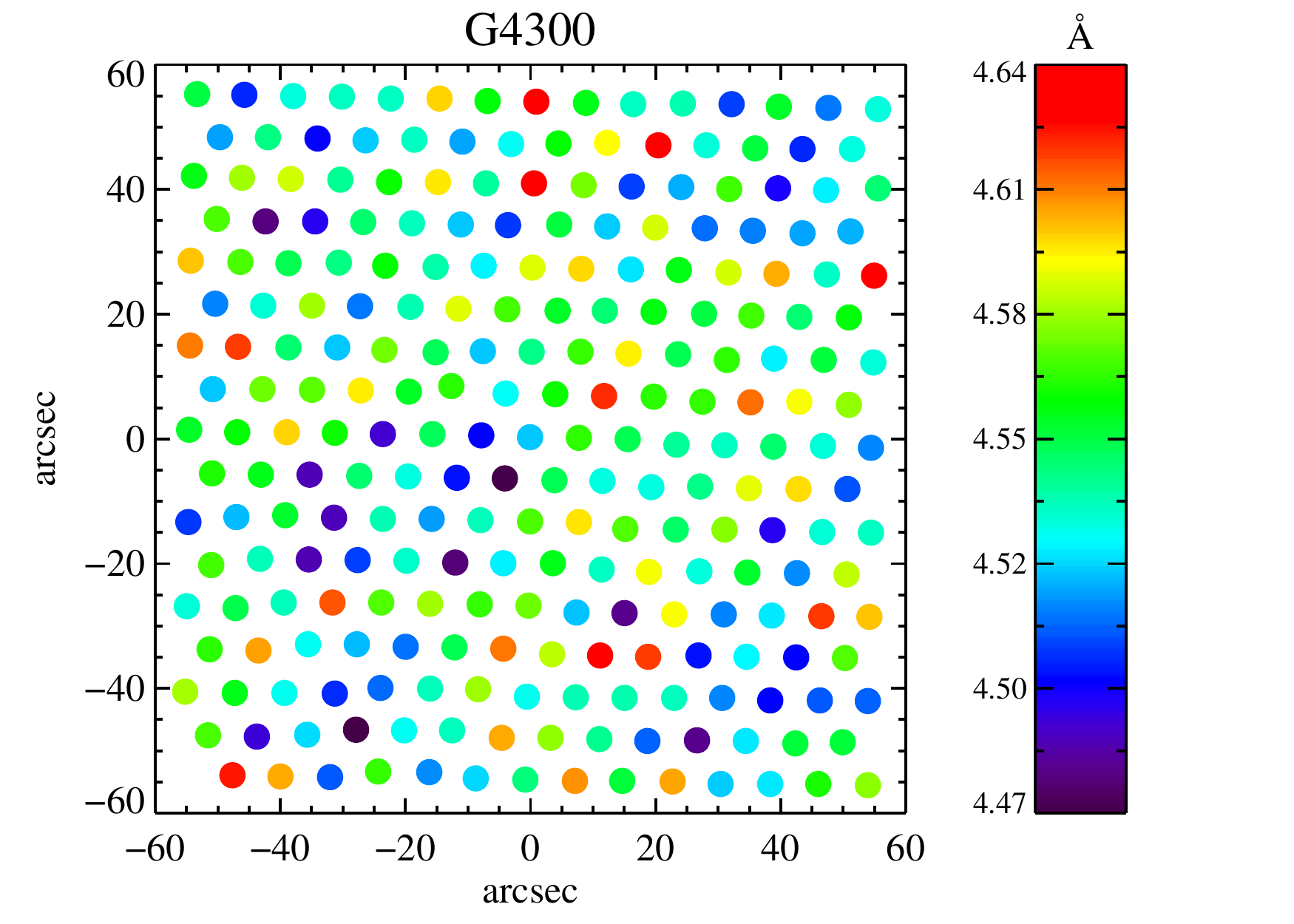}{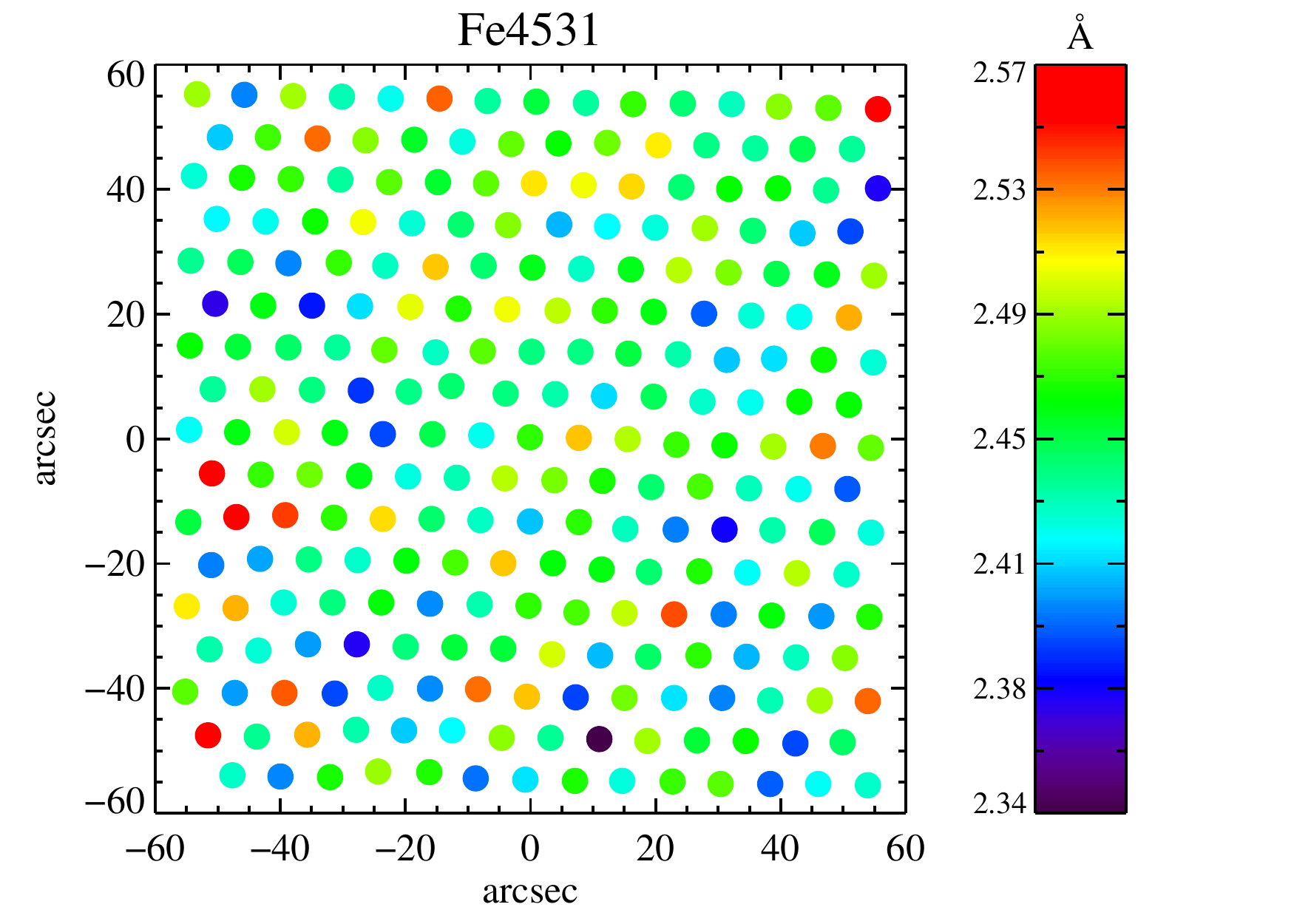} \\
\plottwo{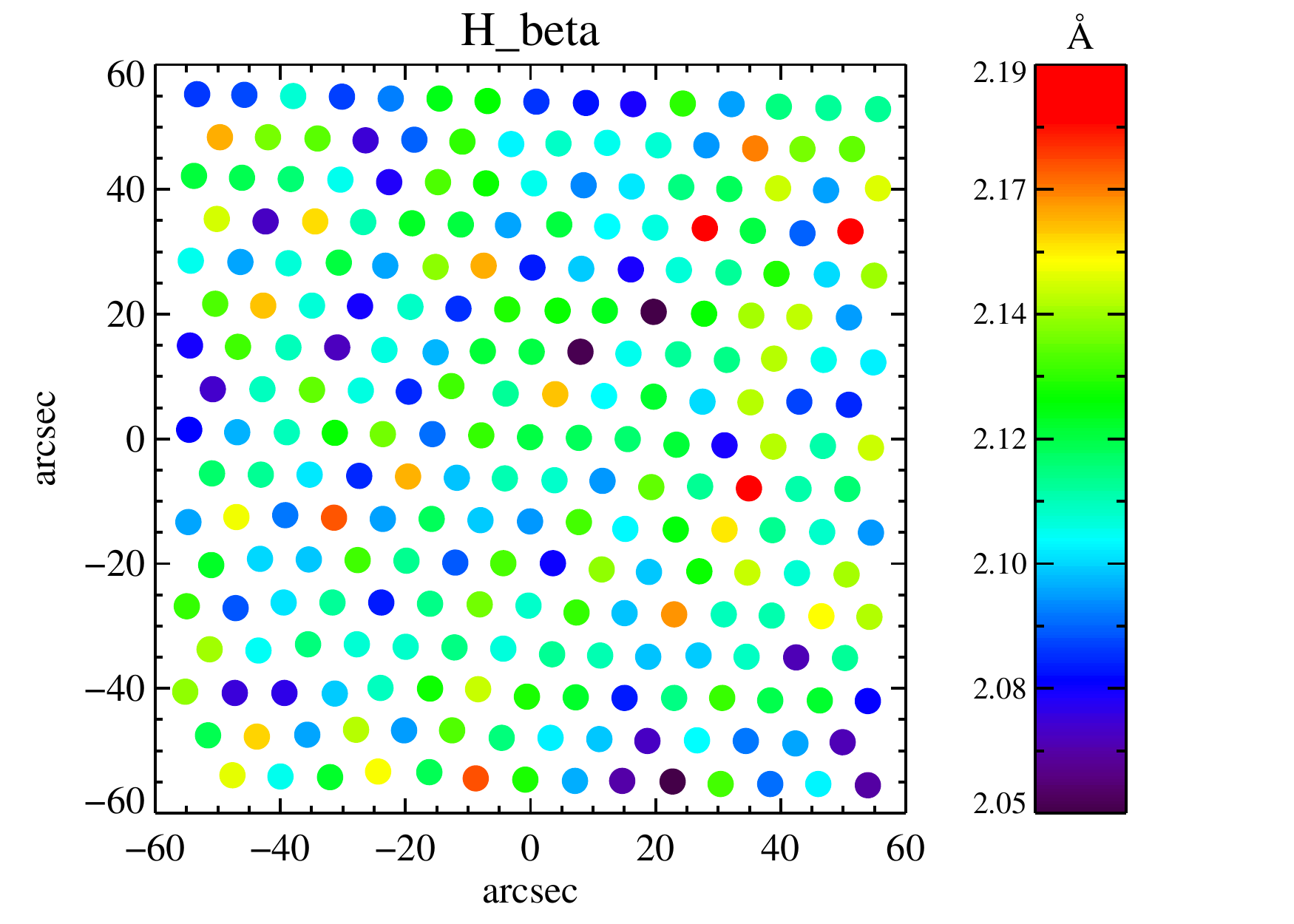}{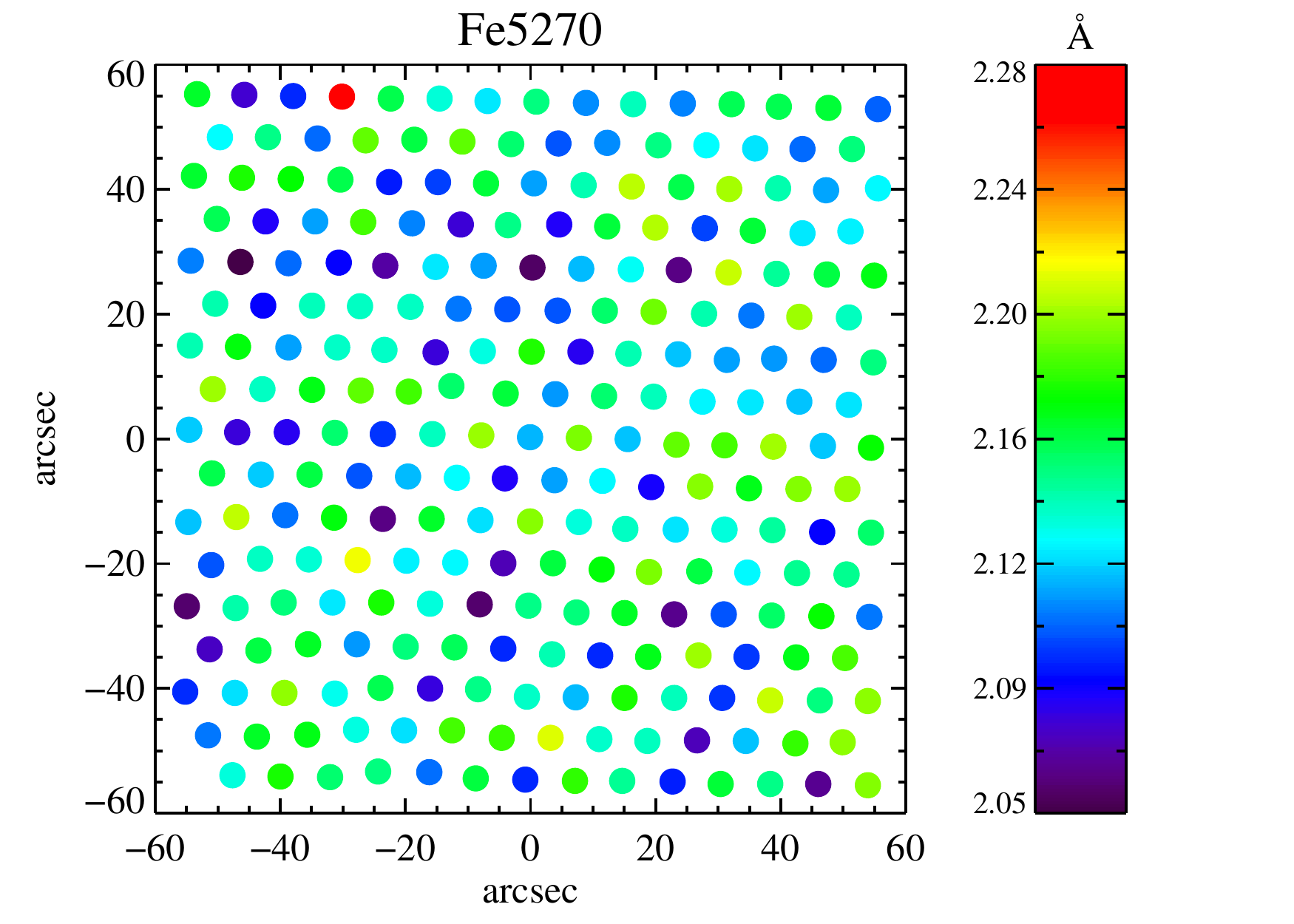}
\caption{Lick index equivalent widths measured from a twilight frame reduced identically to our science frames.  The scatter in the values is very low and shows no overall spatial structure demonstrating the extreme stability of the instrument.  Each point represents a single fiber.  \label{Lick_twi} }
\end{figure*}

The Mitchell Spectrograph utilizes a gimbal mount, such that the instrument swings freely while the telescope slews.  This mount keeps the gravity vector on the instrument constant through the night, resulting in excellent stability.  In Figure~\ref{Lick_twi}, we show Lick index equivalent width measurements made from twilight frames.  The fiber-to-fiber standard deviation of the Lick absorption equivalent widths is $\sim0.04-0.06$ \AA, or 0.7-1.8\%.  This is excellent uniformity and stability, especially compared to earlier IFU studies where variations were of order 18\% \citep[e.g., Figure~3 in][]{Ganda07}.

With our final reduced data cubes, we used Penalized Pixel-Fitting (pPXF) \citep{Cappellari04} and Gas AND Absorption Line Fitting (GANDALF) \citep{Sarzi06} on each fiber to fit stellar velocities, stellar dispersions, and emission line velocities and fluxes.  For stellar templates we use a suite of SSP spectra from \citet{Bruzual03}.  We use GANDALF to measure the [\ion{O}{2}] doublet, [\ion{O}{3}] doublet, and the higher order Balmer lines \hb, H$\gamma$, H$\delta$, H$\epsilon$ while masking the night skyline at 5577 \AA.  The fit also includes a 4th degree multiplicative polynomial to correct for any dust or flux calibration mis-match between the observations and templates.  Because our data do not extend out to \ha, we cannot use the Balmer decrement to place tight constraints on the dust extinction.  While pPXF fits the stellar velocity dispersion, our instrumental resolution is low enough ($\sim140$ \kms) that we are unable to make meaningful velocity dispersion measurements beyond the central bulges of the most massive galaxies in the sample.

The photometric flux and gaseous velocity fields were used to fit a tilted rotating disk model to each galaxy.  Once we have fixed the galaxy inclination and position angle, we median-filter the fiber surface brightness measurements and fit a broken-exponential profile (excluding the central 1-3 fibers which often have excess flux from a bulge component).  Our best-fitting profile parameters are listed in Table~\ref{Table_profilefits}.  The location of the profile break is well constrained, with uncertainties of 1-2\arcsec.  The one exception is NGC 4904 where the inclination is poorly constrained, making the break radius fit unstable.  The velocity fits were then used to shift each fiber to zero velocity.  We assign a projected radius to each fiber, based on the best-fit galaxy inclination and position angle.  Fibers with bright foreground/background objects were masked, and fibers were binned in elliptical annuli of 4 arcseconds.  We then run the binned spectra through our star formation fitting procedure (described below).

\section{Star Formation Histories}

Stellar ages and metallicities have long been estimated using prominent spectral absorption features.  The Lick indices have been used to good effect on old elliptical systems \citep[e.g.,][]{Worthey94,Trager00, Proctor02, Sansom08}.  Unfortunately, measurements of Lick indices become difficult to interpret when the stellar population cannot be approximated as a single-burst \citep{Serra07, Yoachim08b}.  When using individual equivalent widths (or broad band colors), one quickly runs into the problem of having more variables to fit than actual data points.  

Multiple authors have now attempted to model integrated galaxy spectra as the linear combination of SSP populations \citep[e.g.,][]{Fernandes05,Ocvirk06,MacArthur09,Tojeiro07b,Koleva09,Chilingarian07,Sanchez2010}. See \citet{Chen10} for a comparison of many of these codes.

A fundamental problem in fitting multiple Simple Stellar Populations (SSP) spectra to an observed galaxy is answering the question of how many spectra one needs?  At low signal-to-noise in particular, the fraction of star formation in a particular age bin can become very uncertain.  This can result in wildly different values for the average age of a population depending on the spectral range, signal-to-noise (SNR), and number of SSP models used in the decomposition.

Inspired by the star formation histories derived from HST observations of resolved galaxies, we adopt model spectra that are continuous, rather than a combination of discrete SSPs.  One important result of HST observations is that all galaxies host old stellar populations at {\emph{all radii}} \citep{Williams09,Williams08,Gogarten10,Barker06}.  This implies that any SFH decomposition that does not include old stars is probably flawed.

We continue the technique introduced in \citet{Yoachim10c}, and explore the parameter space of exponentially rising and falling star formation histories.  This technique should be expected to perform poorly in cases where the star formation histories are particularly bursty (e.g., dwarf galaxies or regions of barred galaxies, \citet{Sanchez-Blazquez2011}), but we still expect to out-perform simple SSP decomposition.  While hierarchical merging might lead to bursty star formation histories, we are primarily interested in the outskirts of spiral galaxies where observations of resolved stars have shown relatively smooth SFHs \citep{Gogarten10}.

\subsection{Spectral Templates}

\begin{figure*}
\plottwo{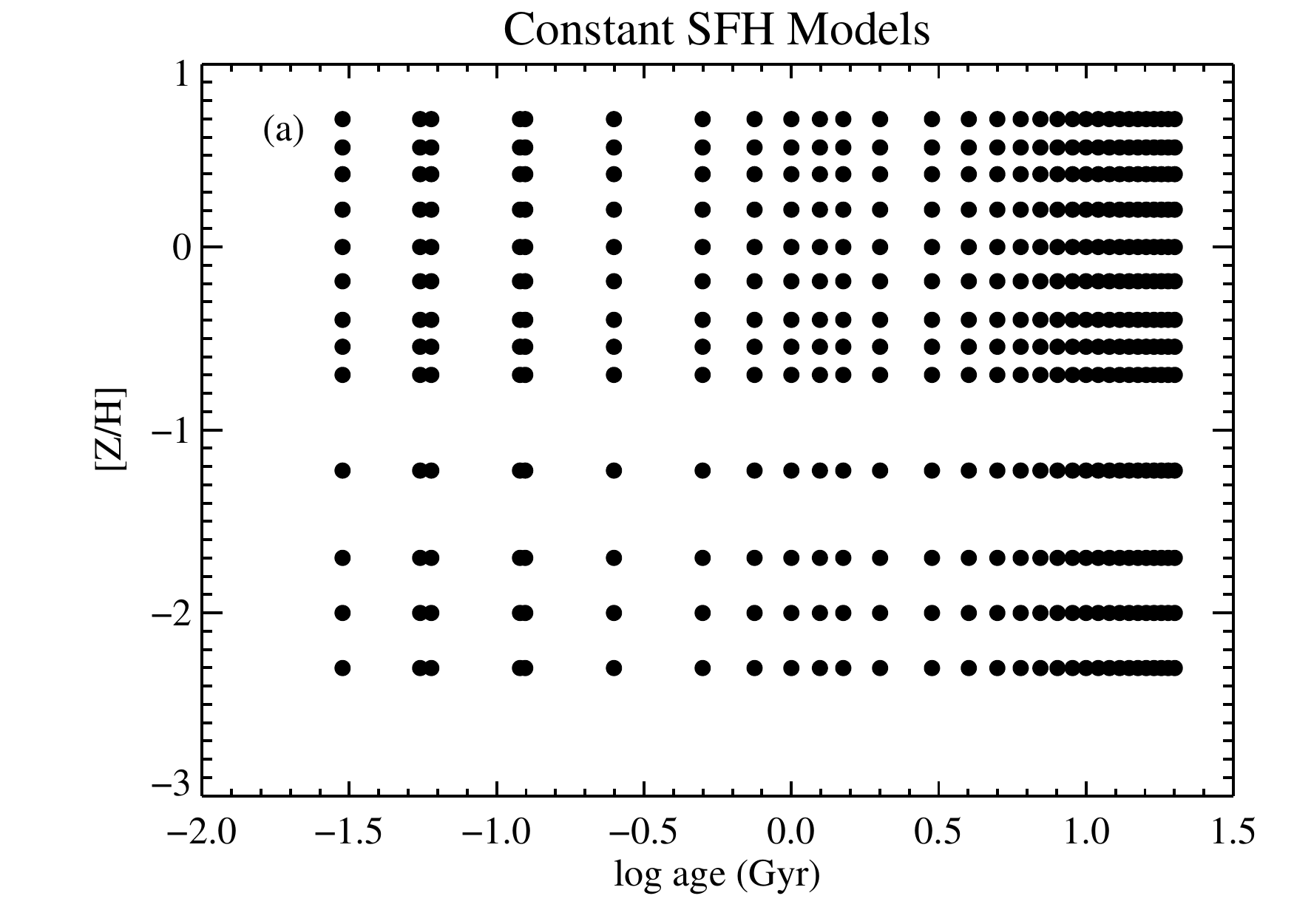}{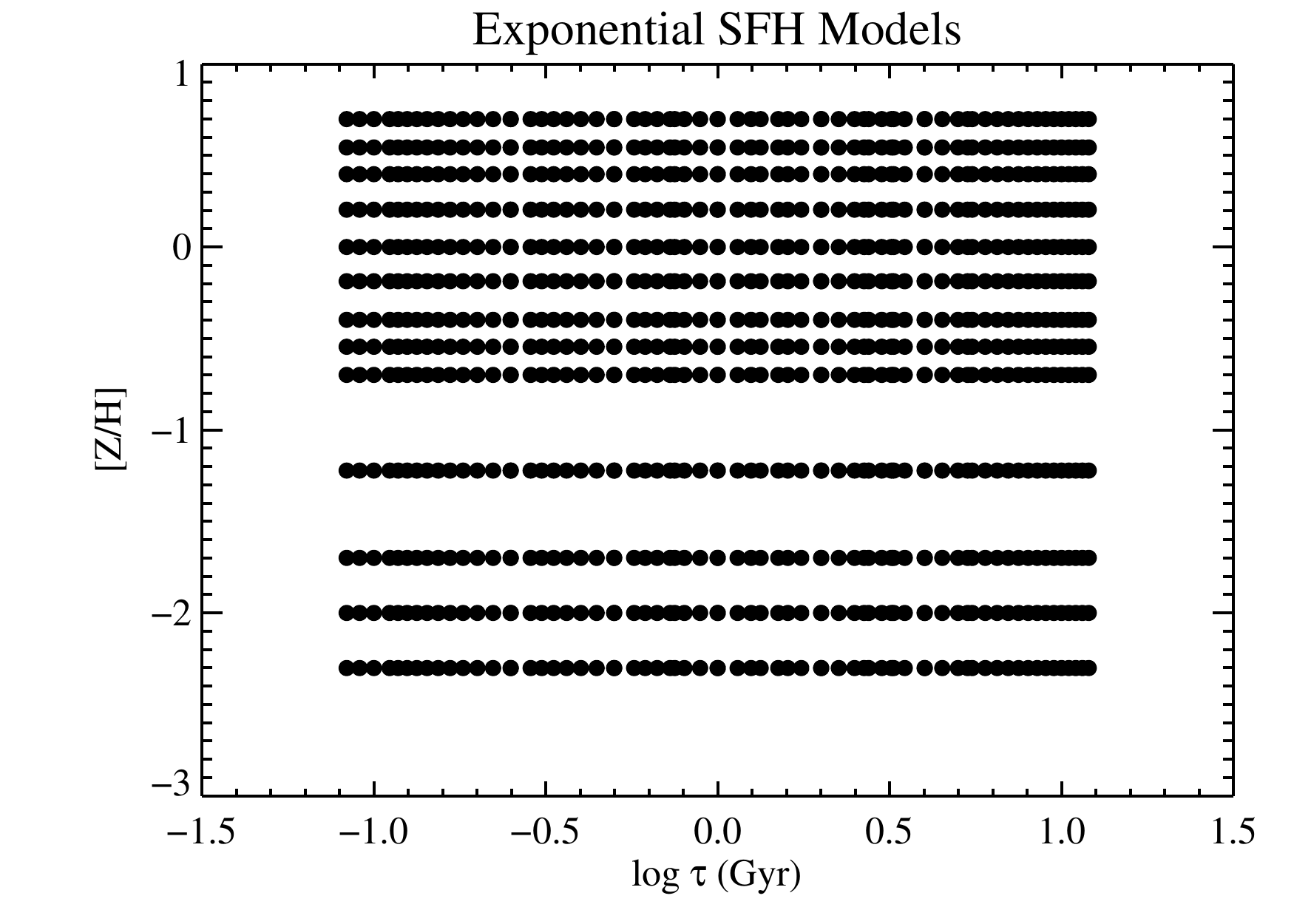}
\caption{Parameter space covered by our spectral synthesis models.  We span a large range of metallicities and include very old stellar populations as well as populations dominated by recent formation.  \label{model_params}}
\end{figure*}

We use the stellar syntheses code \citet{Bruzual07} which updates \citet{Bruzual03} to include effects of TP-AGB stars.  We have used the GALEXV package to generate a large suite of template spectra.  The parameter space of the models is plotted in Figure~\ref{model_params}.  The GALEXV code allows arbitrary SFHs, but includes only seven metallicity models.  We have generated 6 additional metallicities by interpolating between the GALEXV models.

\begin{figure*}
\epsscale{.35}
\plotone{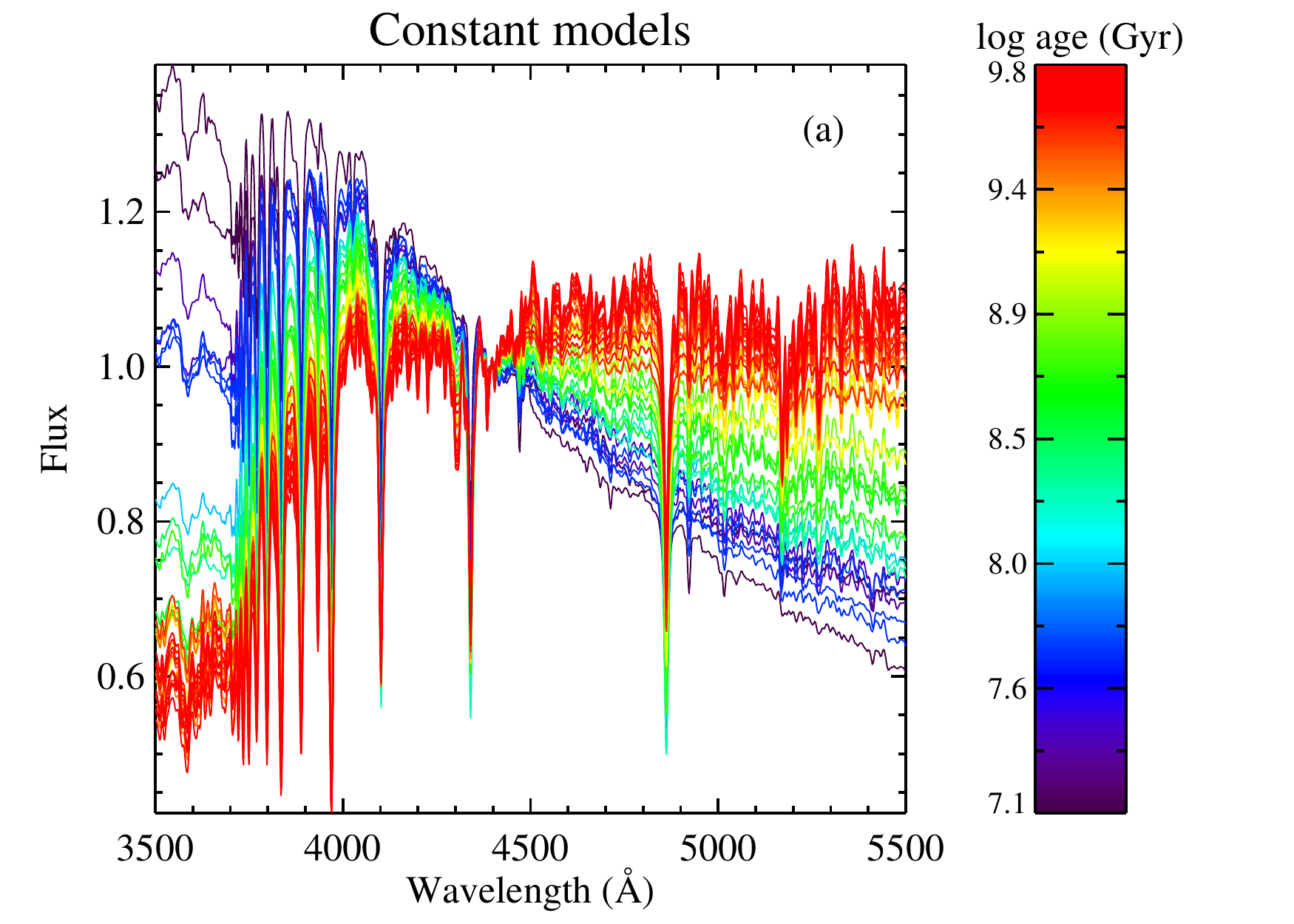}
\plotone{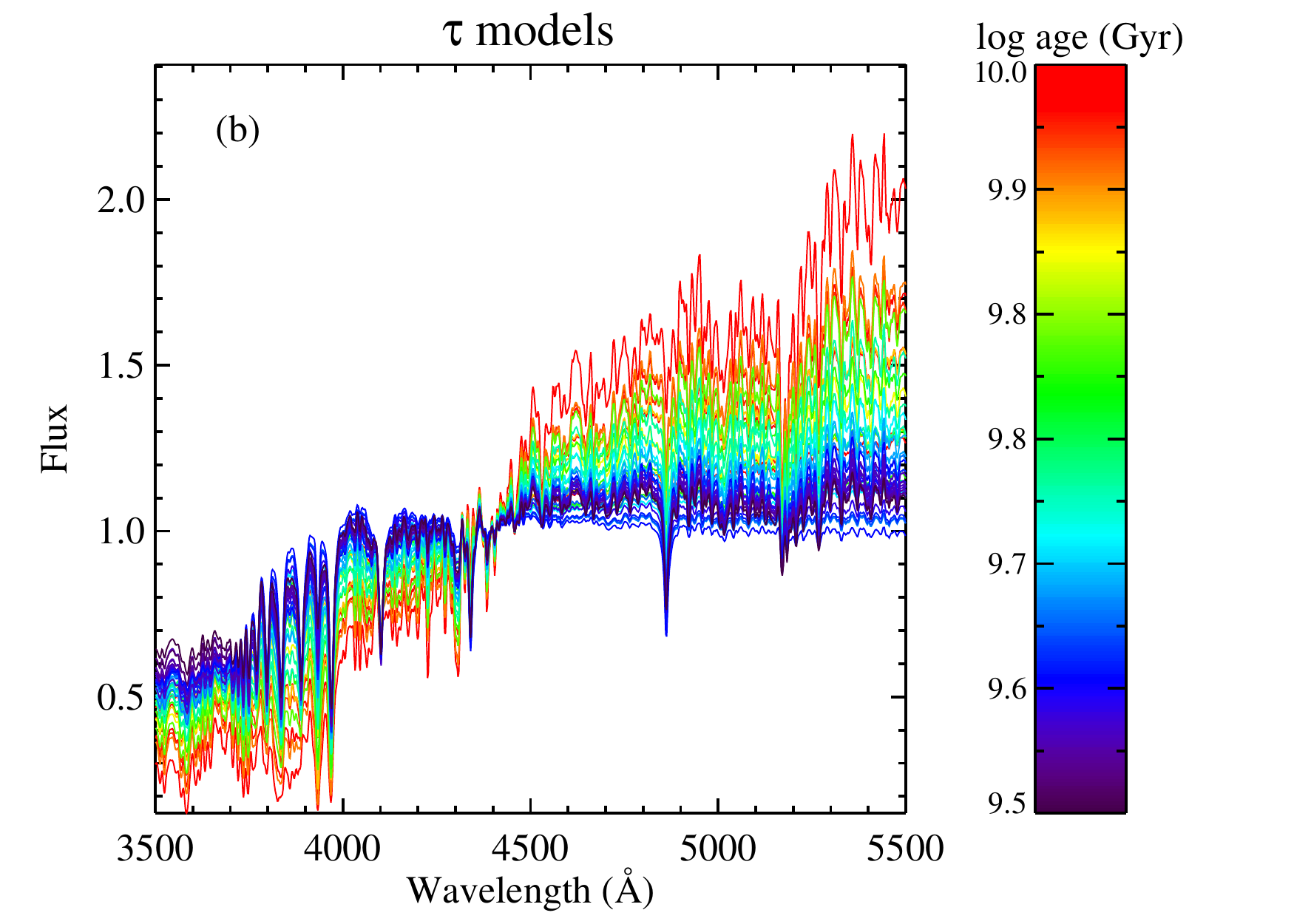}
\plotone{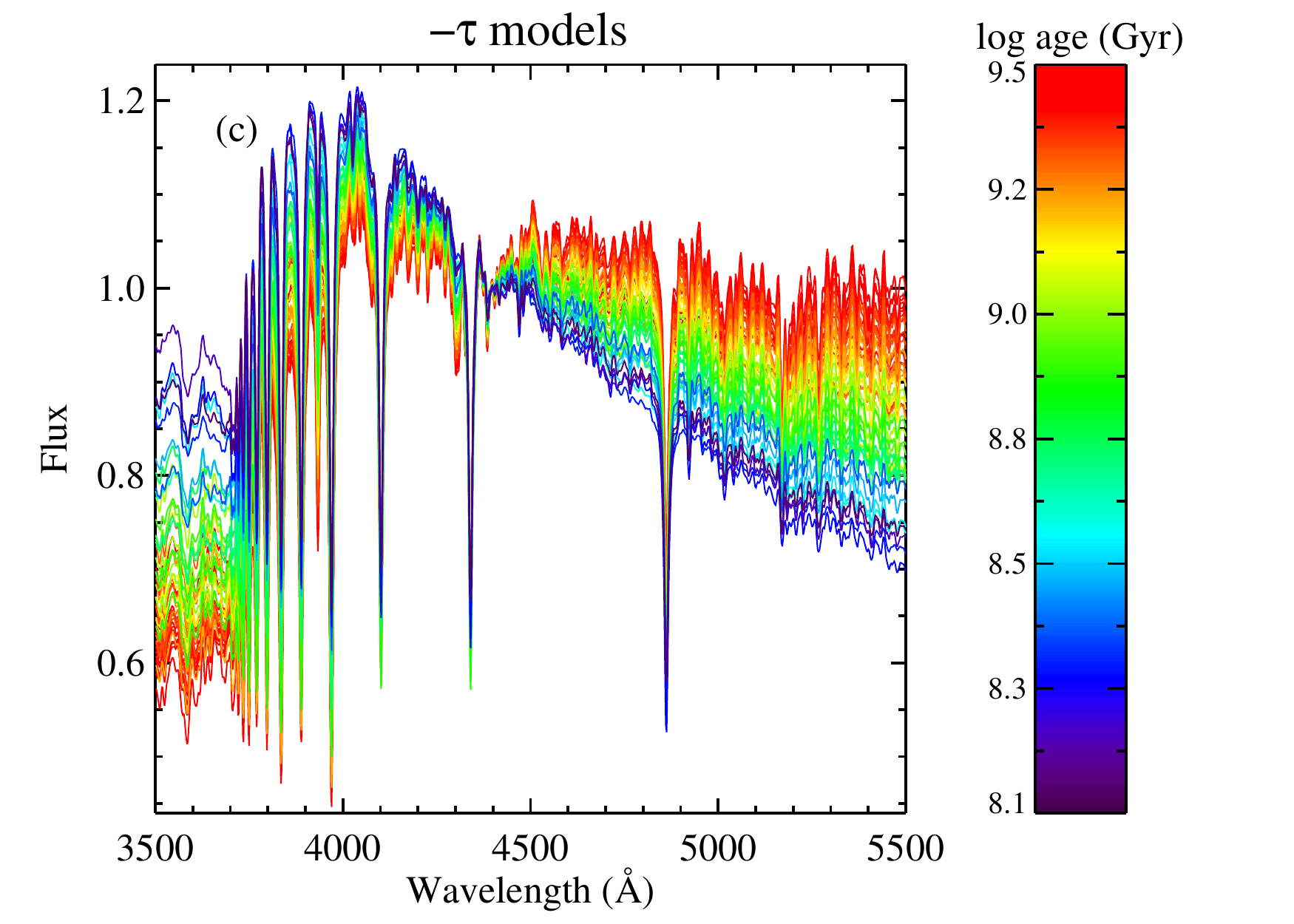}
\epsscale{1}
\caption{Our template spectra. (a) Models with constant star formation rates.  (b)  Model spectra from declining star formation rates.  (c)  Model spectra generated from increasing star formation rates.  \label{example_templates}}
\end{figure*}

\subsection{How Well Can Ages and Metallicities be Recovered?}

While there are numerous spectral synthesis codes, few have actually compared SSP decomposition to fitting more complicated SFHs.  We have therefore run a series of tests on the GANDALF \citep{Sarzi06} routine to find the optimal way to measure star formation histories.  The simplest method is to find the best fitting simple stellar population (SSP).  More complicated is to use a linear combination of multiple bursts of star formation (mSSP).  Finally, one can use continuous star formation, usually parametrized by an exponential decay ($\tau$SP).

We construct templates using the GALEXV code \citep{Bruzual03}.  In particular, we constructed  models with exponentially increasing or decreasing star formation, and models with constant star formation.  We then ask what is the best way to measure a flux-weighted average age --- find an optimal combination of SSP models or find the single best complex star formation history?  We plot our templates in Figure~\ref{example_templates}.

We construct artificial input spectra with known age and metallicity, re-binned to match our observations and with Gaussian noise added so that the spectra have a SNR per pixel of 50.  For these tests, we use the wavelength range of 3500-5800 \AA\ and wavelength binning and resolution of 146 \kms (2.2 \AA) and 5.3 \AA\ (the same as our data cubes).  

\begin{figure*}
\plottwo{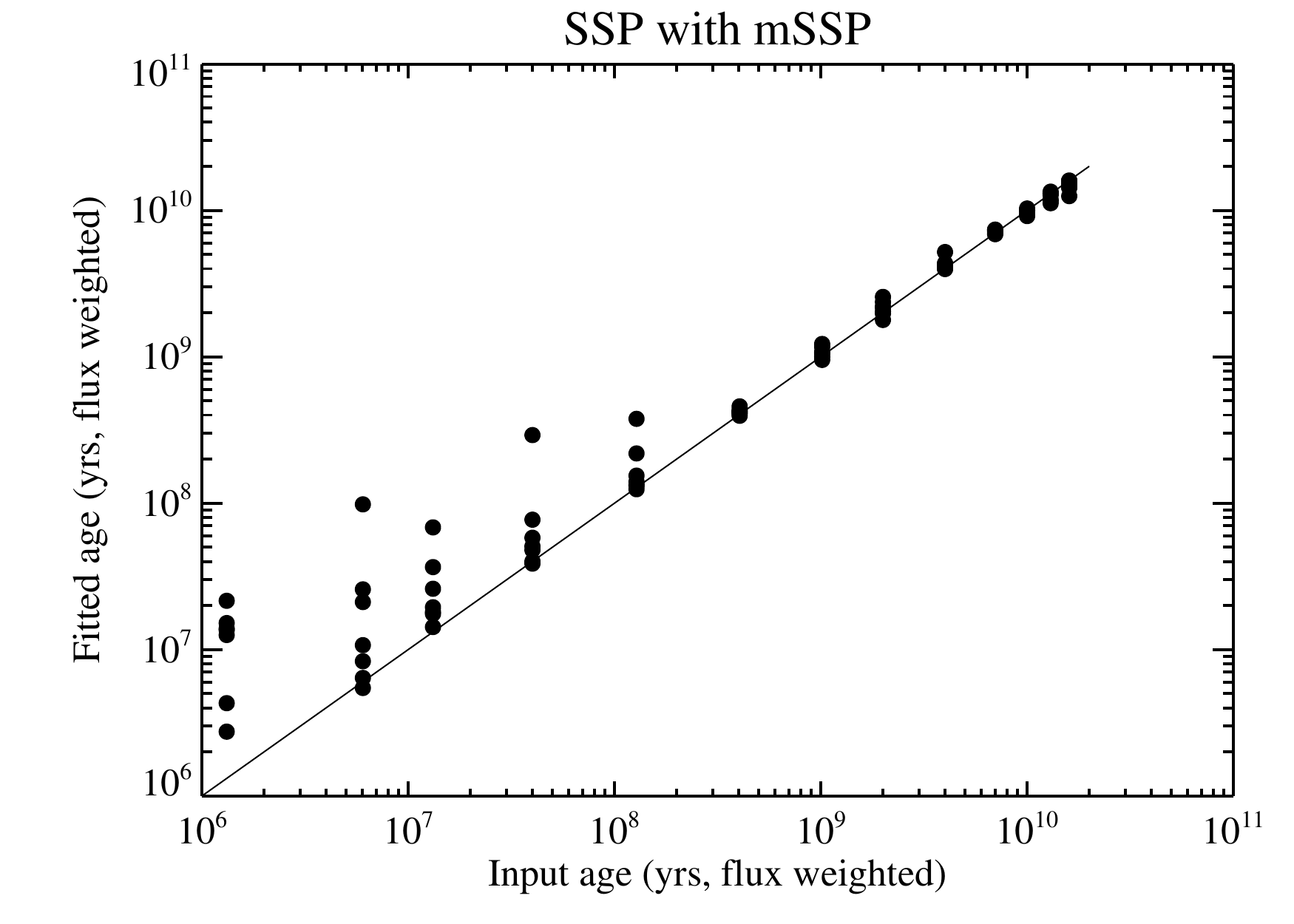}{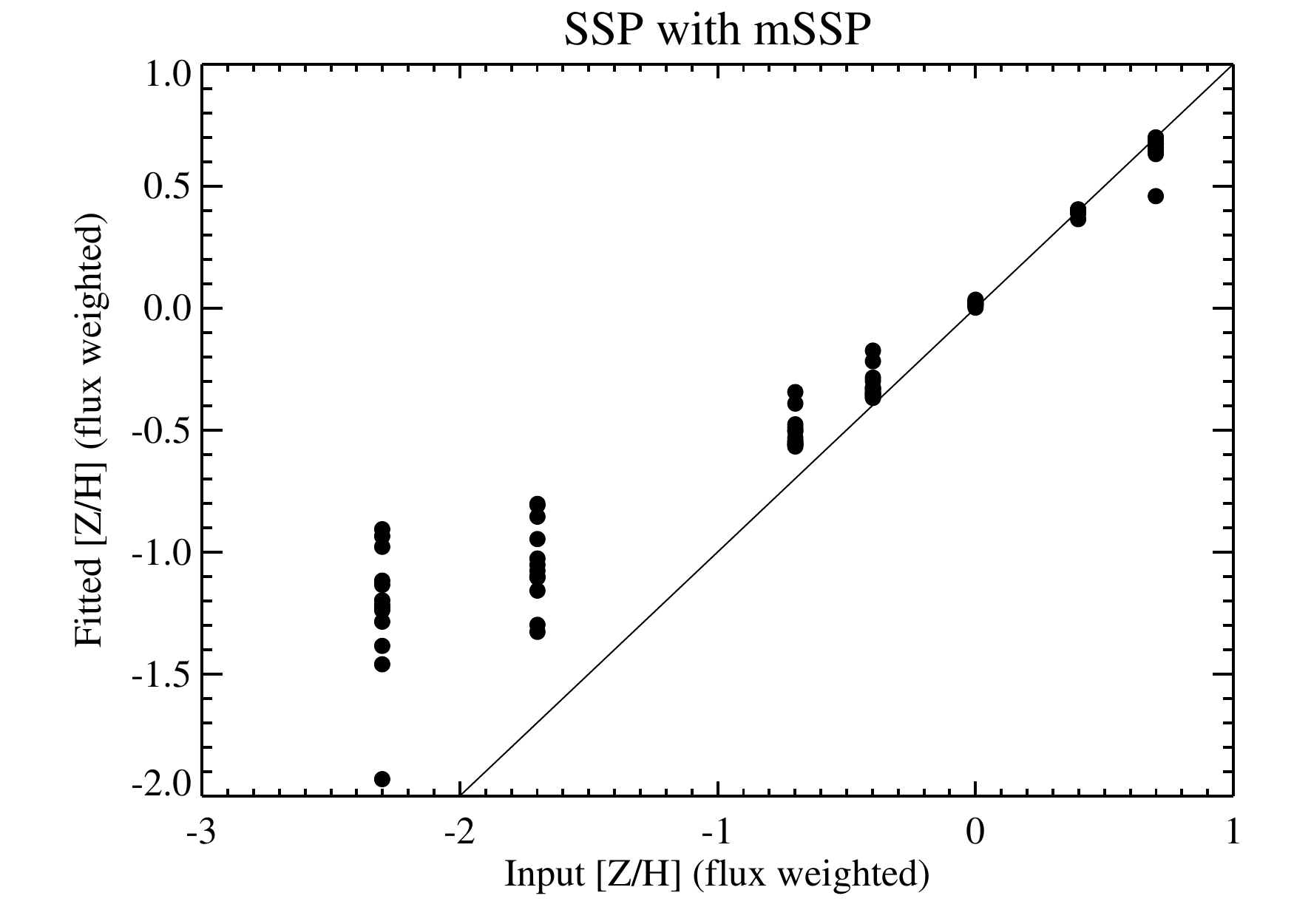}
\caption{Results of fitting single SSP spectra with mSSP models.   \label{sspwssp}}
\end{figure*}

In Figure~\ref{sspwssp}, we show our results of using GANDALF to fit an SSP spectrum with a linear combination of multiple SSP (mSSP) template spectra.  We point out that this is an idealized fitting scenario.  The input spectra and template spectra are from the same source, we have not added any emission lines and there is no template mismatch.  For old ages and high metallicities, the fits are excellent.  However, the fits diverge when the metallicity is sub-solar or the flux-weighted age is less than 100 Myr.  This is a fairly damning experiment for trying to decompose SFHs with mSSP as it implies just the presence of Gaussian noise is enough to trick the code into using more templates than are actually necessary.  Inspecting the $\chi^2$\ per degree-of-freedom for these fits, we find all the fits fell between 0.9 and 1.1.  There is no particular correlation between the $\chi^2$\ values and how well the fits matched the input age and metallicity.  

\begin{figure*}
\plottwo{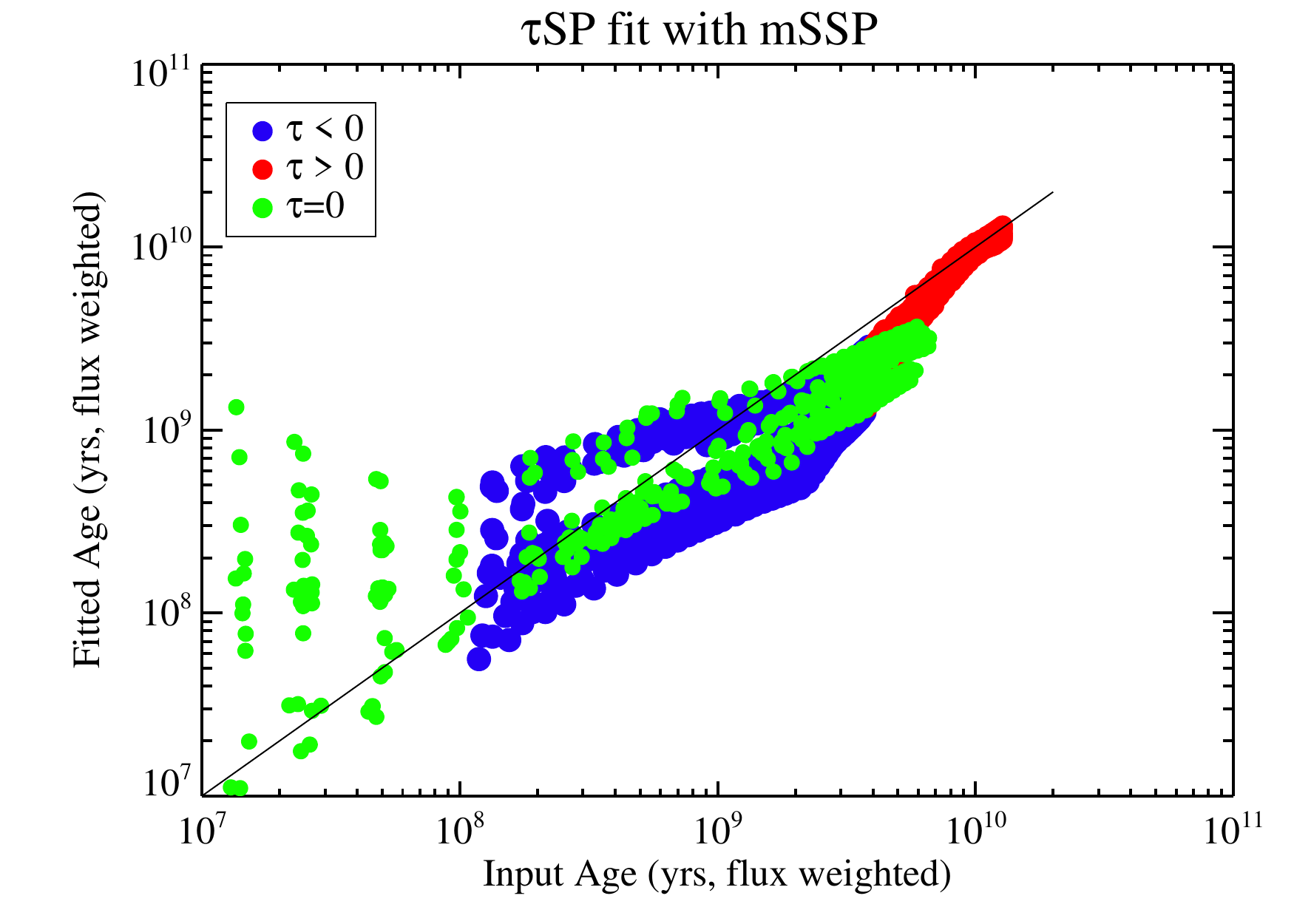}{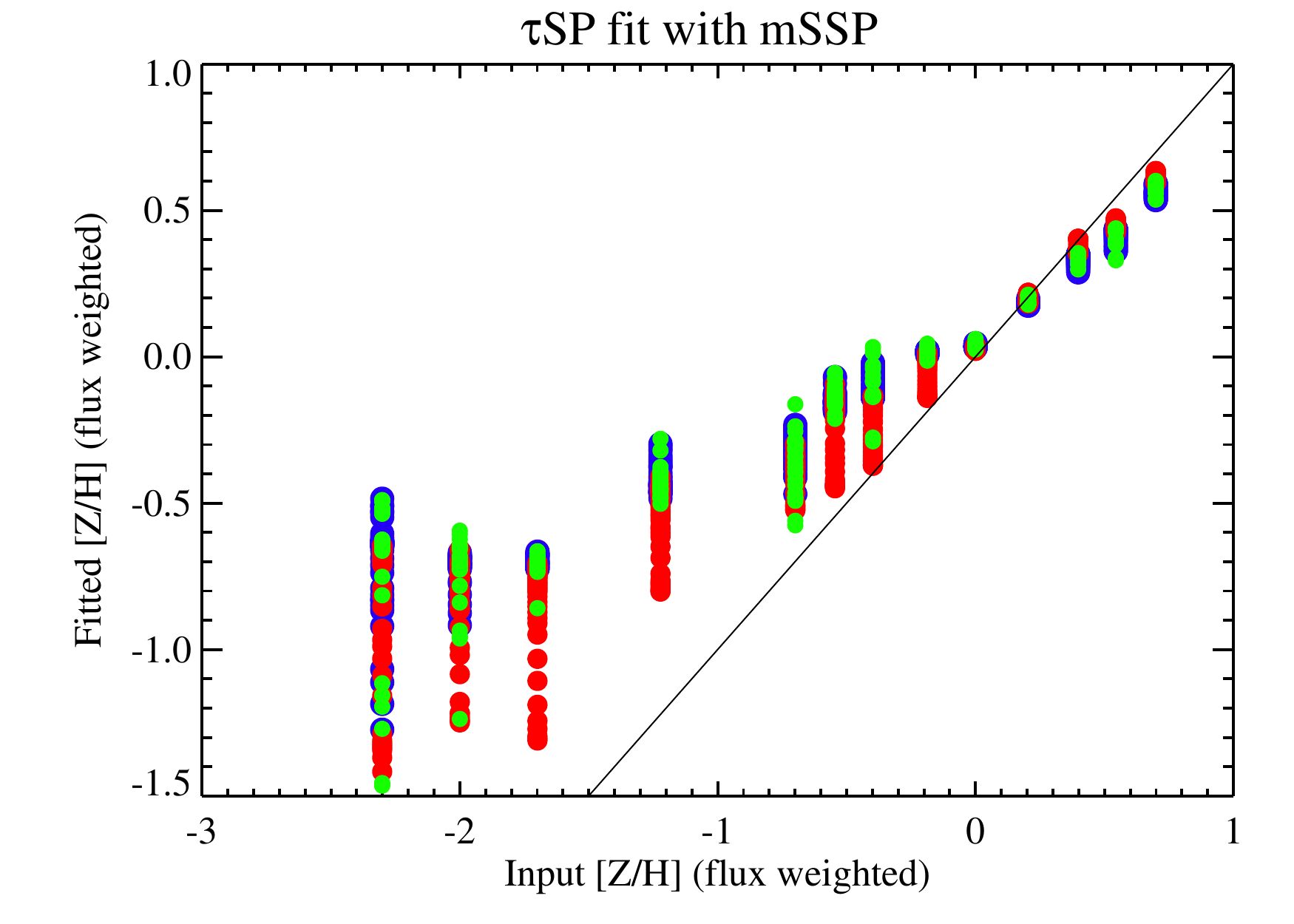}
\caption{Results from fitting continuous star formation histories with mSSP models.  The results are reasonable for declining star formation histories and metal rich populations, but diverge significantly for metal poor and younger populations.\label{tauwssp} }
\end{figure*}

Next, we test how well mSSPs can fit $\tau$SP spectra.  Our results are plotted in Figure~\ref{tauwssp}, where we fit increasing, decreasing, and constant star formation histories.  As before, the mSSP work well for old metal-rich populations.  The fits are biased to returning young ages and high metallicities for the majority of input spectra.  This is caused by GANDALF's propensity to often place a small amount of weight on the oldest stars.  Again, the $\chi^2$\ per degree-of-freedom fell between 0.88 and 1.13 for these fits.

The results of decomposing an observed spectrum into multiple SSP models is also very wavelength dependent.  If we limit the SSP decomposition to the region redward of 4040 \AA, we find the resulting average age can be two or more Gyr different from fitting the full spectrum.  This is a well known effect where the Ca II H and K lines are very sensitive to young populations \citep{Smith09}.  

\citet{Marmol-Queralto2011} use a similar analysis and also investigate varying the SNR.  They also find that decomposing with SSP models is difficult, with flux-weighted age errors of $\sim50\%$ for old stars and metallicity errors of $\sim100\%$.

\begin{figure*}
\plottwo{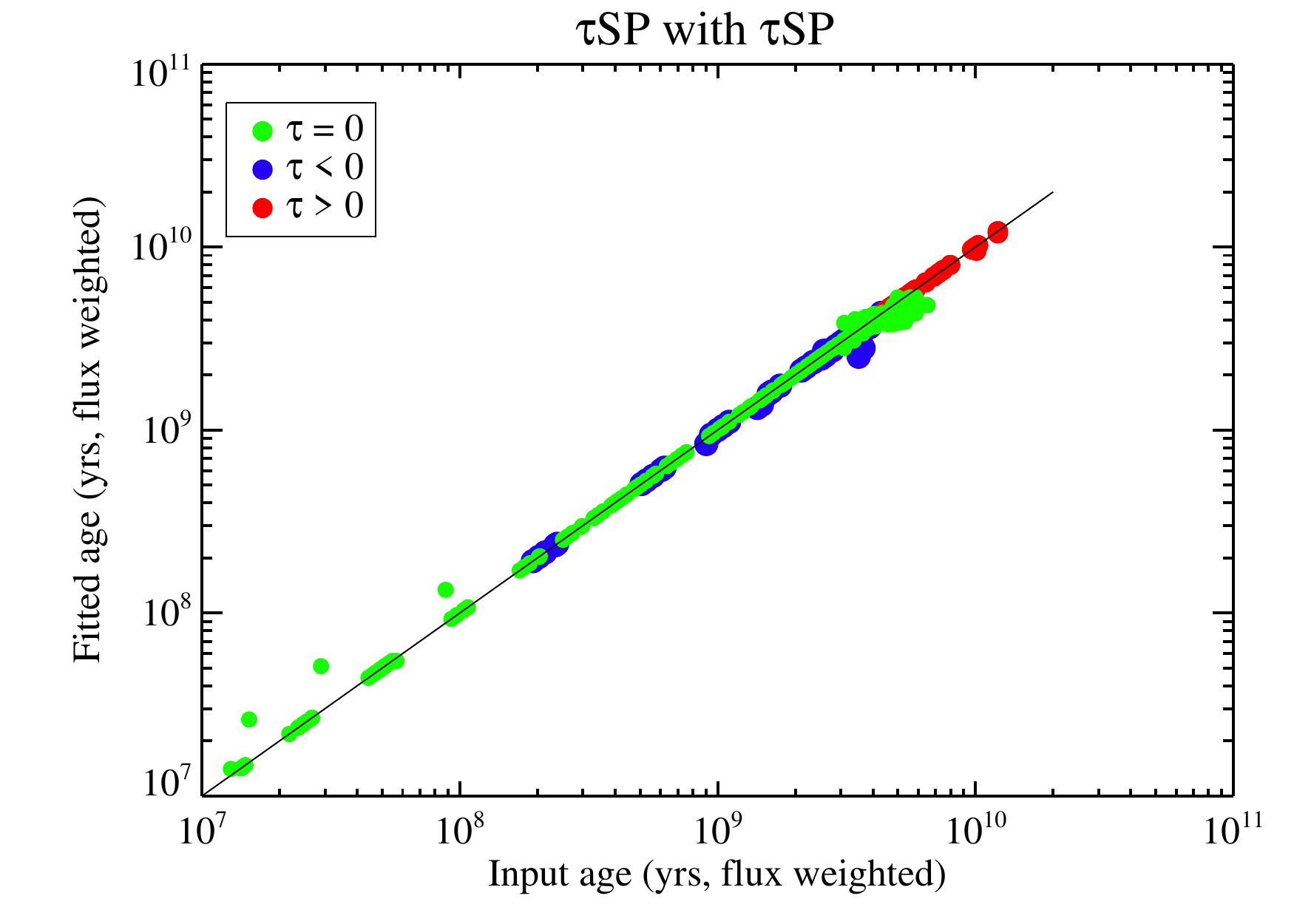}{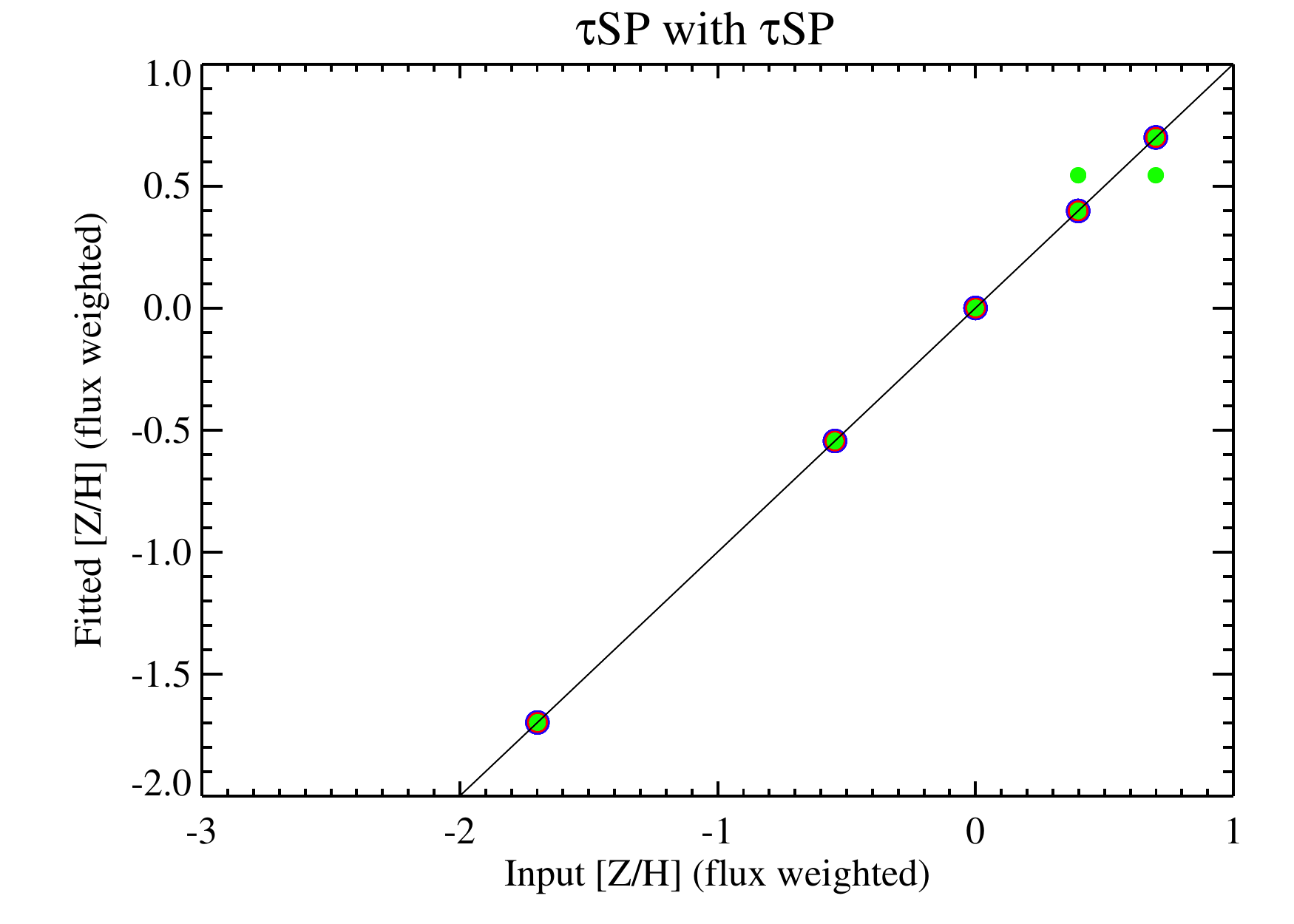}
\caption{ Fitting extended star formation histories with the best single extended SFH template.  Solid lines show the perfect-fit case.  Unlike fitting with a linear combination of SSP spectra, we do an excellent job of recovering the input parameters.  \label{tauwtau}}
\end{figure*}

We must emphasize that this is not a failure on the part of the GANDALF code--if anything we are guilty of not using GANDALF the way it was intended.  GANDALF was written primarily to make better emission line measurements while correcting for underlying stellar absorption features.

\begin{figure*}
\epsscale{.35}
\plotone{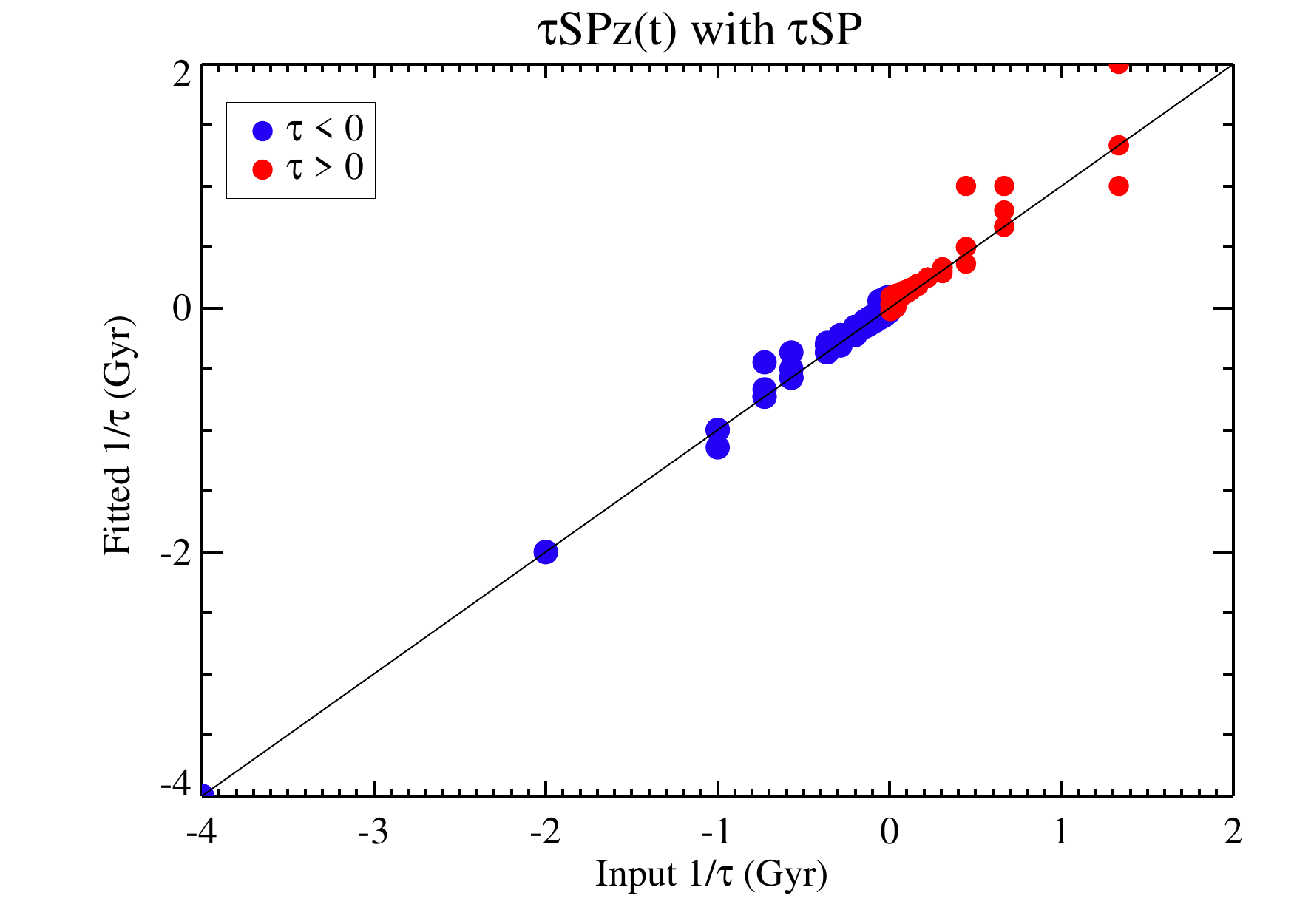}
\plotone{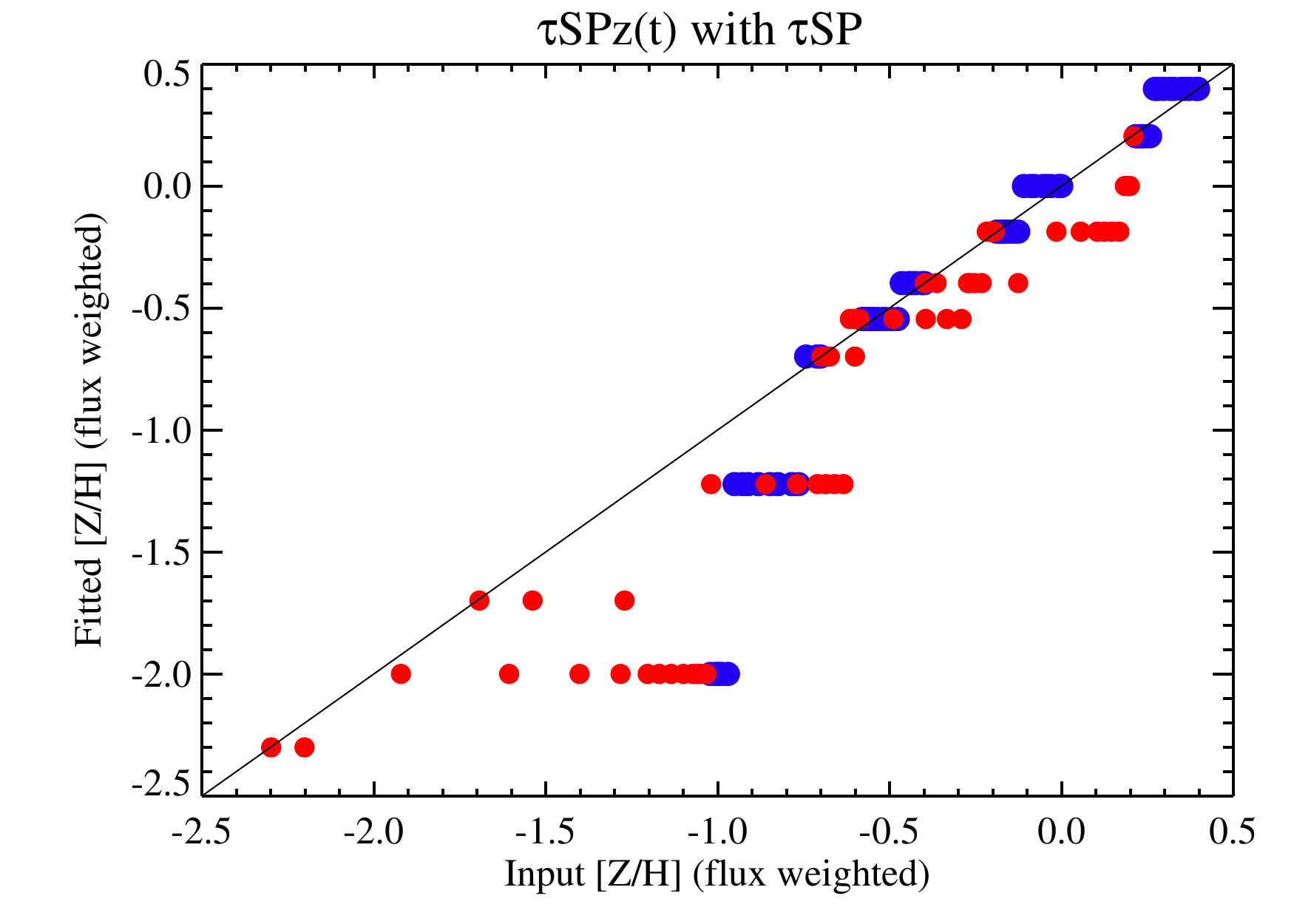}
\plotone{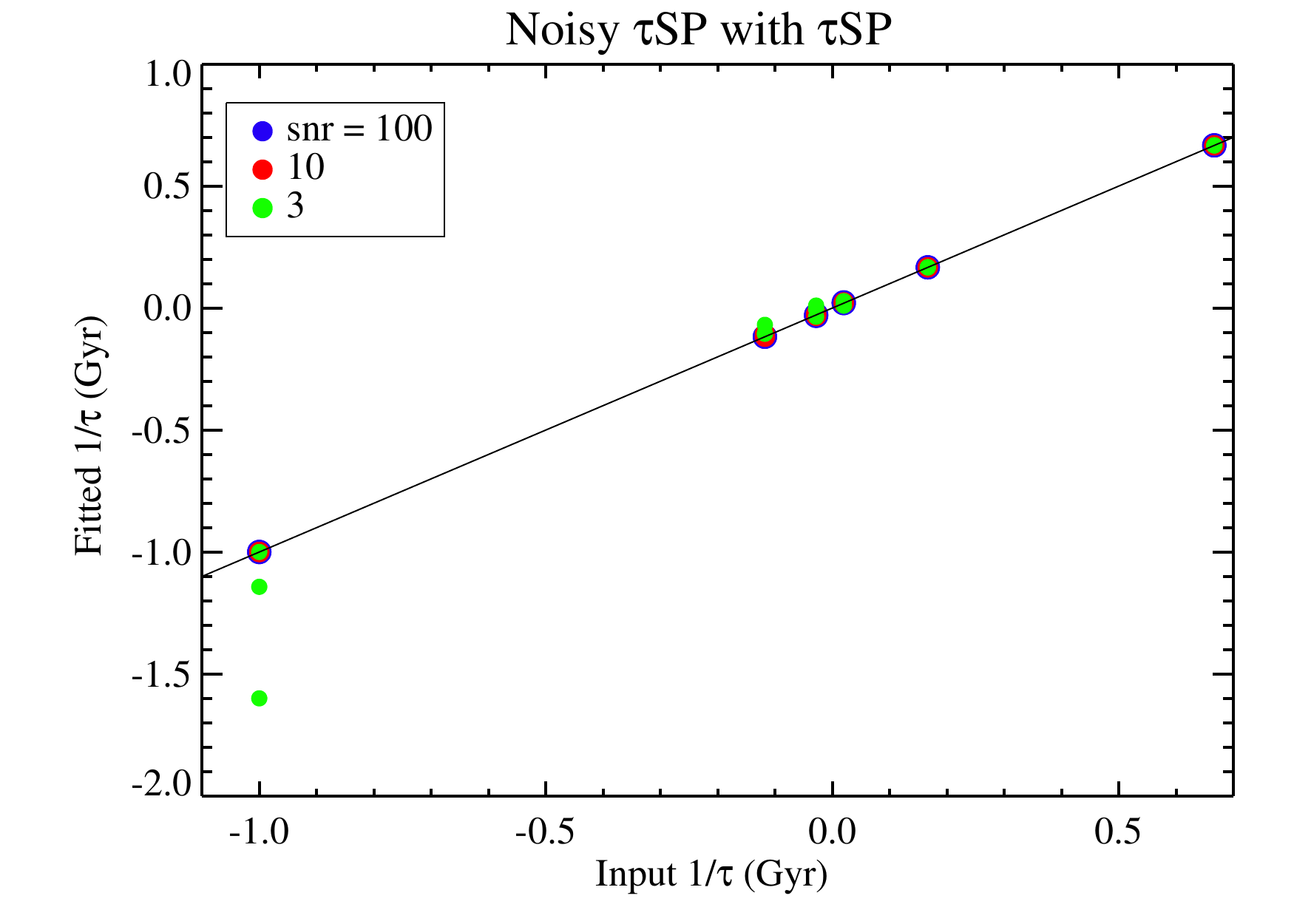}
\epsscale{1}
\caption{Left:  Results from fitting star formation histories which have experienced significant chemical evolution.  While we fail to recover the correct metallicities, our fitted ages are still quite robust.  Right:  Fitting star formation histories which are not perfectly smooth exponentials.  Our fits only fail when the recent star formation is very bursty.  \label{tauzwtau}}
\end{figure*}

Figure~\ref{tauwtau} shows the results of fitting extended star formation histories with the {\emph{best single}} template.  Unlike fitting a combination of SSPs, this technique is very robust against varying SNR and wavelength ranges.  The major disadvantage is that the fits should be expected to return poor fits for galaxies which have bursty star formation histories.  We also do not include any metallicity evolution, assuming all the stars have a similar metallicity.  While this sounds like a rather drastic assumption, \citet{Williams08} use resolved stars in the outskirts of M81 and find that over the entire star formation history of their observed region the average metallicity only varies by $\sim0.5$\ dex.  Figure~\ref{tauzwtau} shows results from a simulation where we added simple chemical evolution to the spectra to be fit.  As expected, our fitting procedure does a poor job matching the metallicity, however, we still do an excellent job recovering the ages.  Figure~\ref{tauzwtau} also shows the results of fitting star formation histories that are not perfect exponentials.  In these cases, we added varying amounts of noise to smooth exponential star formation histories with bursts of one-hundredth, one-tenth, and one-third the smooth SFR.  Our fitting method returned excellent results, failing only when there was a recent strong burst of star formation.

\begin{figure*}
\plottwo{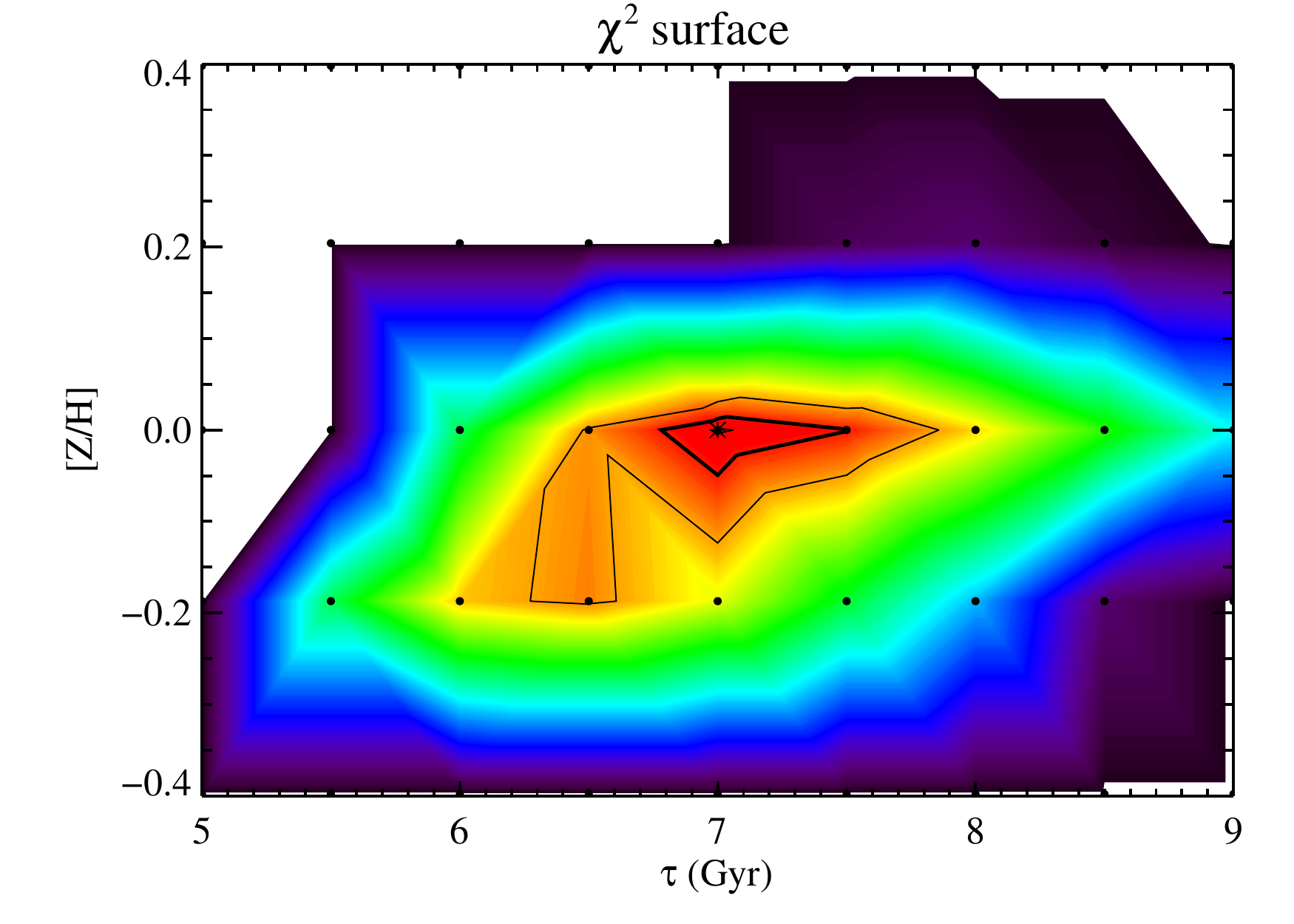}{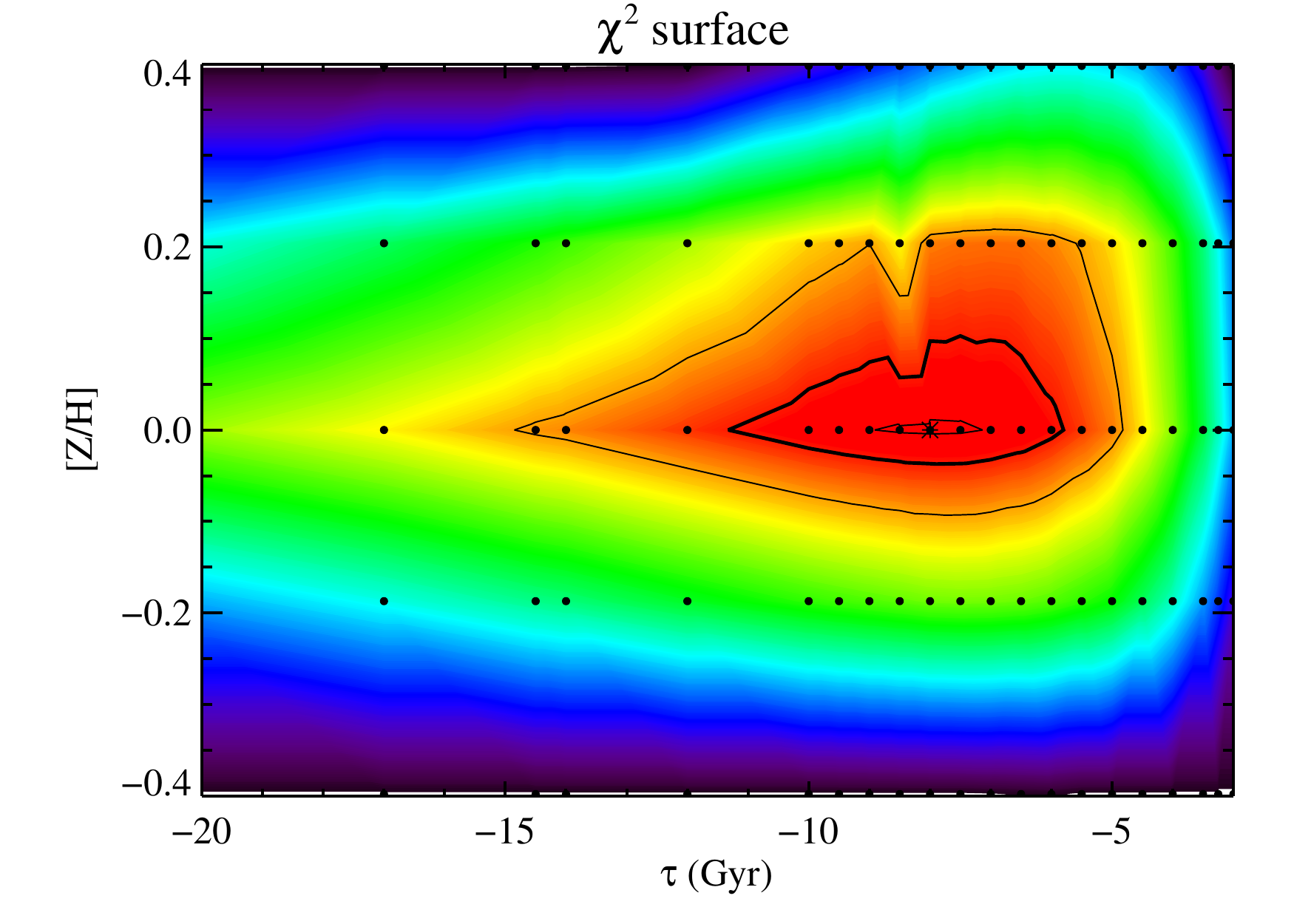}\\
\plottwo{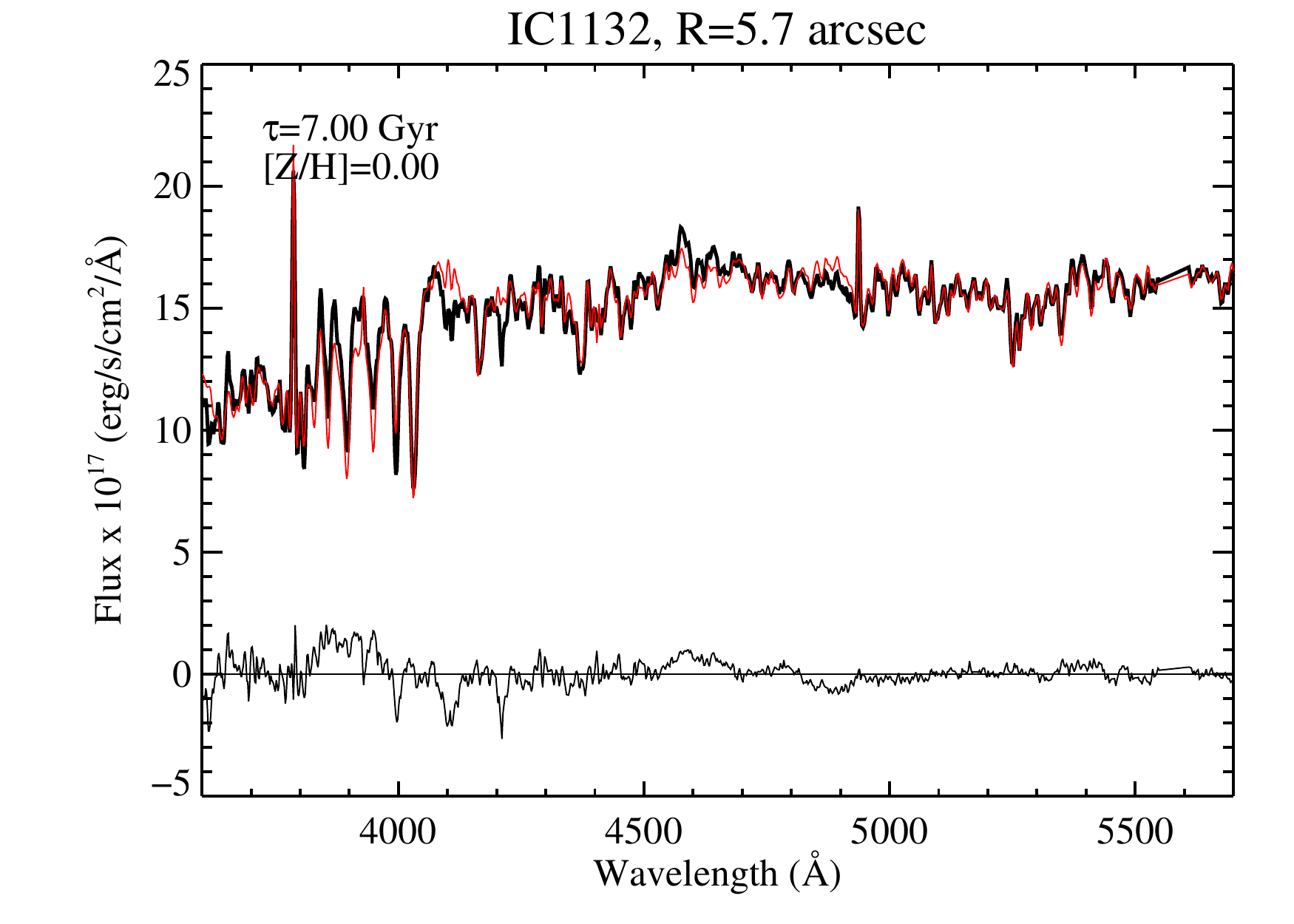}{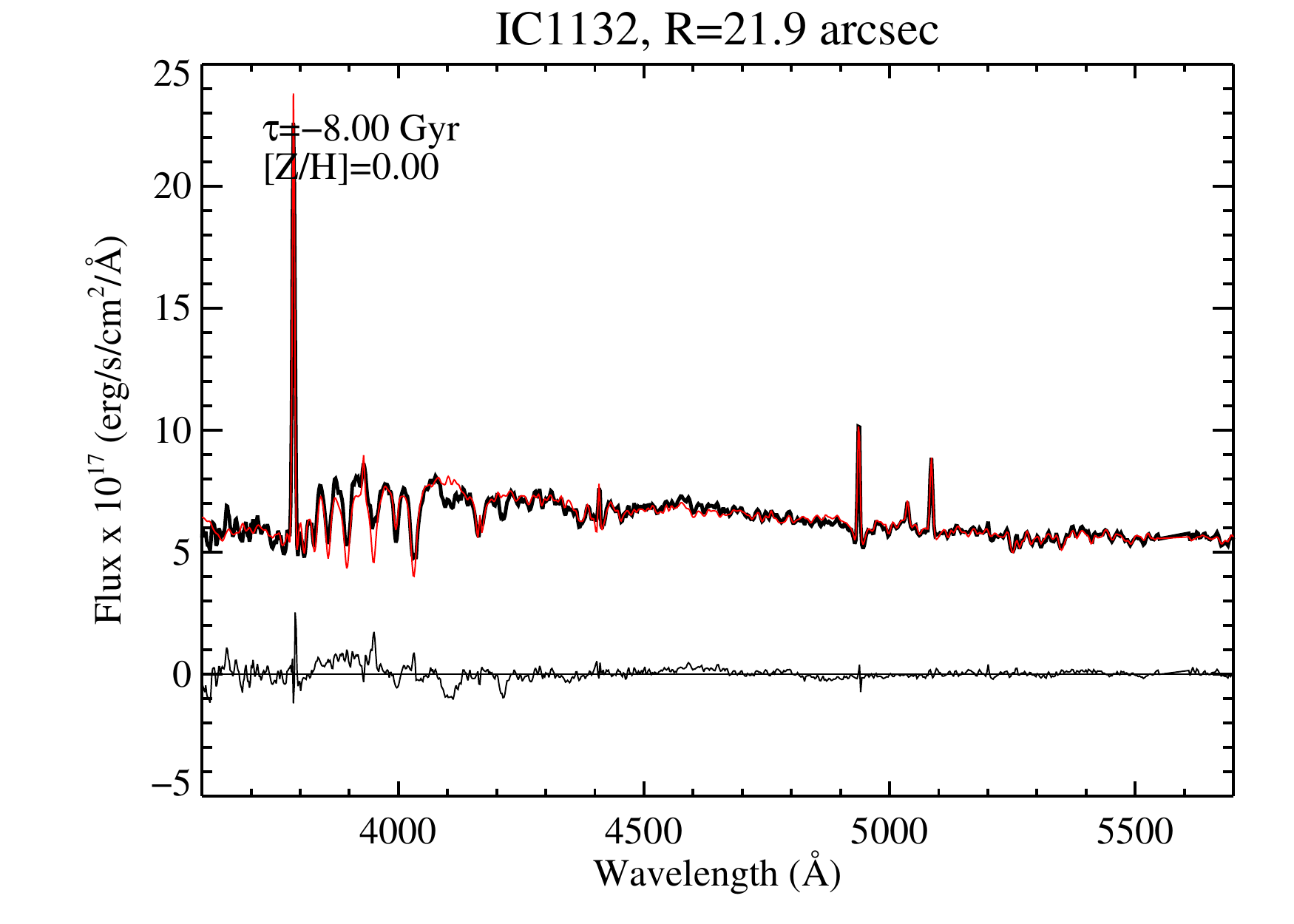}
\caption{Example of our stellar population fitting results.  Top panels show the $\chi^2$ surface generated using GANDALF and a set of exponential star formation histories.  Contours mark the $\Delta\chi^2=2.3,20$ (thick), and 50 levels.  Model points are marked with black circles and the best fit models are marked with stars.  The lower panels show the best fit model (stellar spectrum plus Gaussian emission lines) in red along with the raw spectrum and residuals in black. \label{example_fit}}
\end{figure*}

Based on our experience fitting HST derived star formation histories, we find our technique has systematic error of $\sim$0.6 Gyr, and that the $\Delta\chi^2=20$ returns a reasonable uncertainty \citep{Yoachim10c}.  Examples of our preferred fitting procedure are shown in Figure~\ref{example_fit}.  Because our residuals are dominated by systematic differences (rather than Gaussian noise), we cannot legitimately use formal $\chi^2$ uncertainties on the fitted parameters.

\begin{figure*}
\epsscale{1}
\plottwo{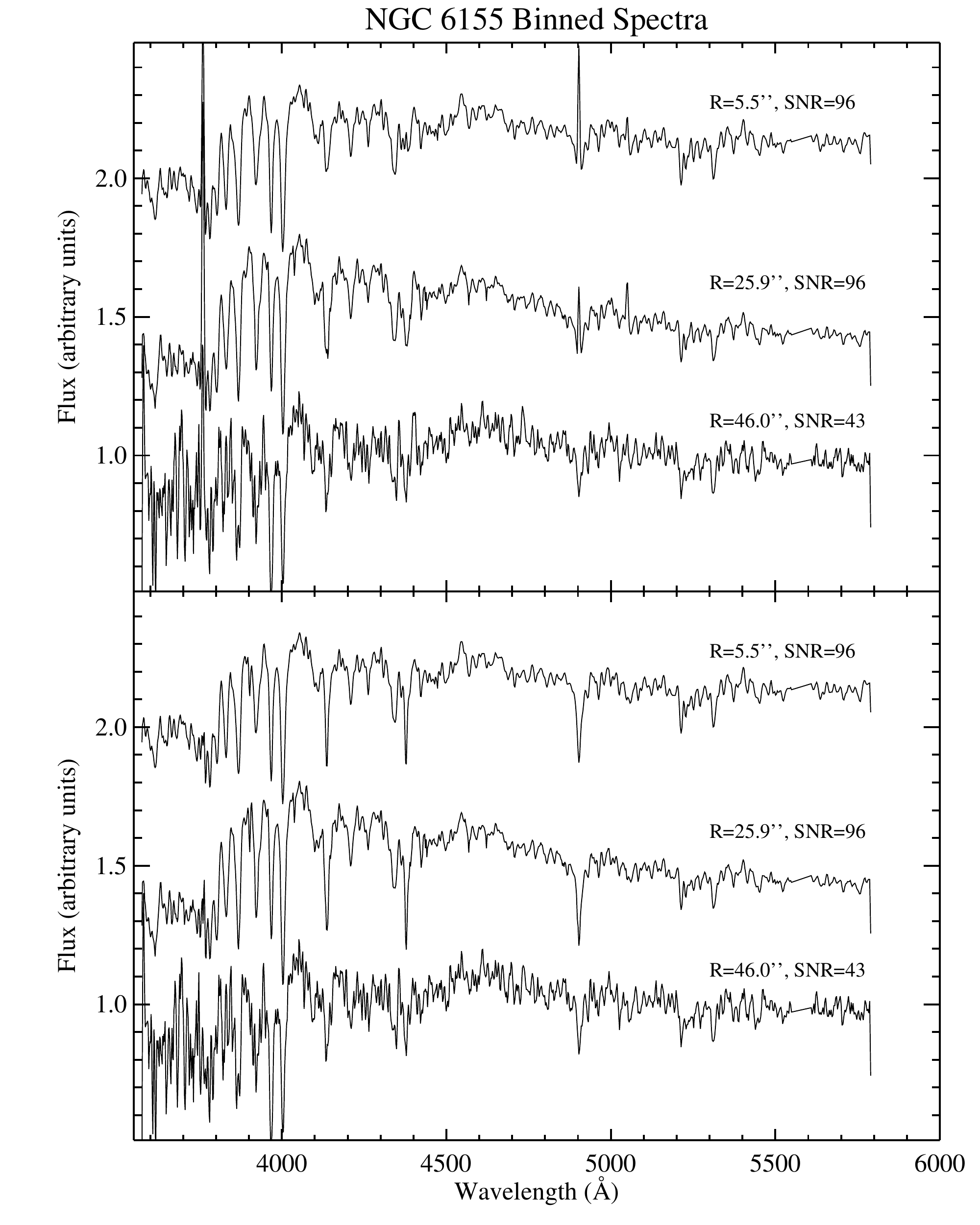}{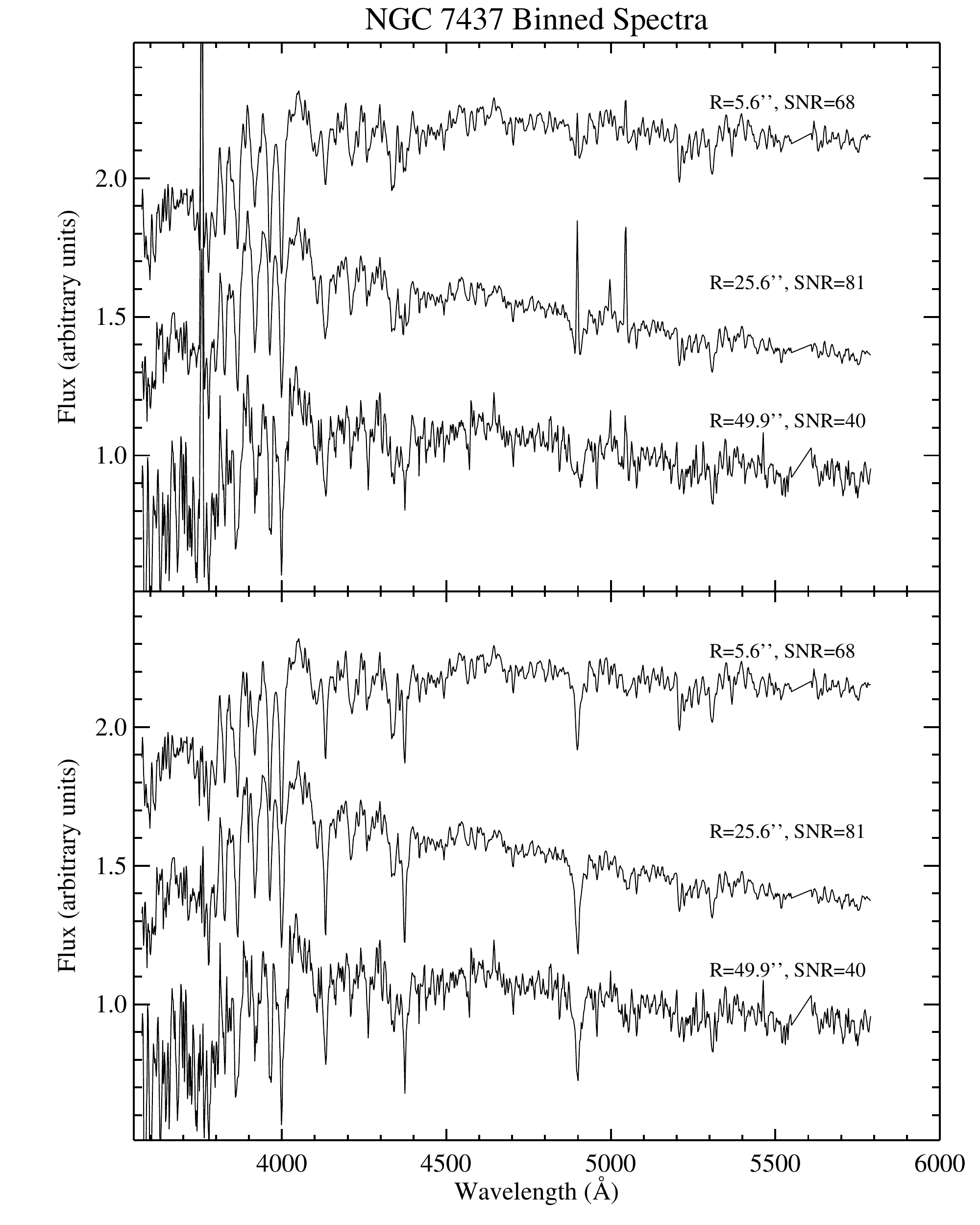}
\epsscale{1}
\caption{Examples of our spectra after binning fibers.  Top panels show the binned spectra while bottom panels show the results after subtracting off the best-fit emission lines.   \label{example_binned}}
\end{figure*}

Figure~\ref{example_binned} shows examples of our binned spectra for two galaxies.  In particular, we show how GANDALF can be used to remove emission lines to recover the underlying stellar features.  Based on our previous experience, we only attempt to fit spectra where the SNR per pixel is greater than 40.

\clearpage

\section{Analysis}
We present the galaxies ordered by their maximum rotational velocity.


\begin{figure*}
 \begin{center}$
 \begin{array}{ccc}
   \epsscale{.35}
    \includegraphics[scale=0.3]{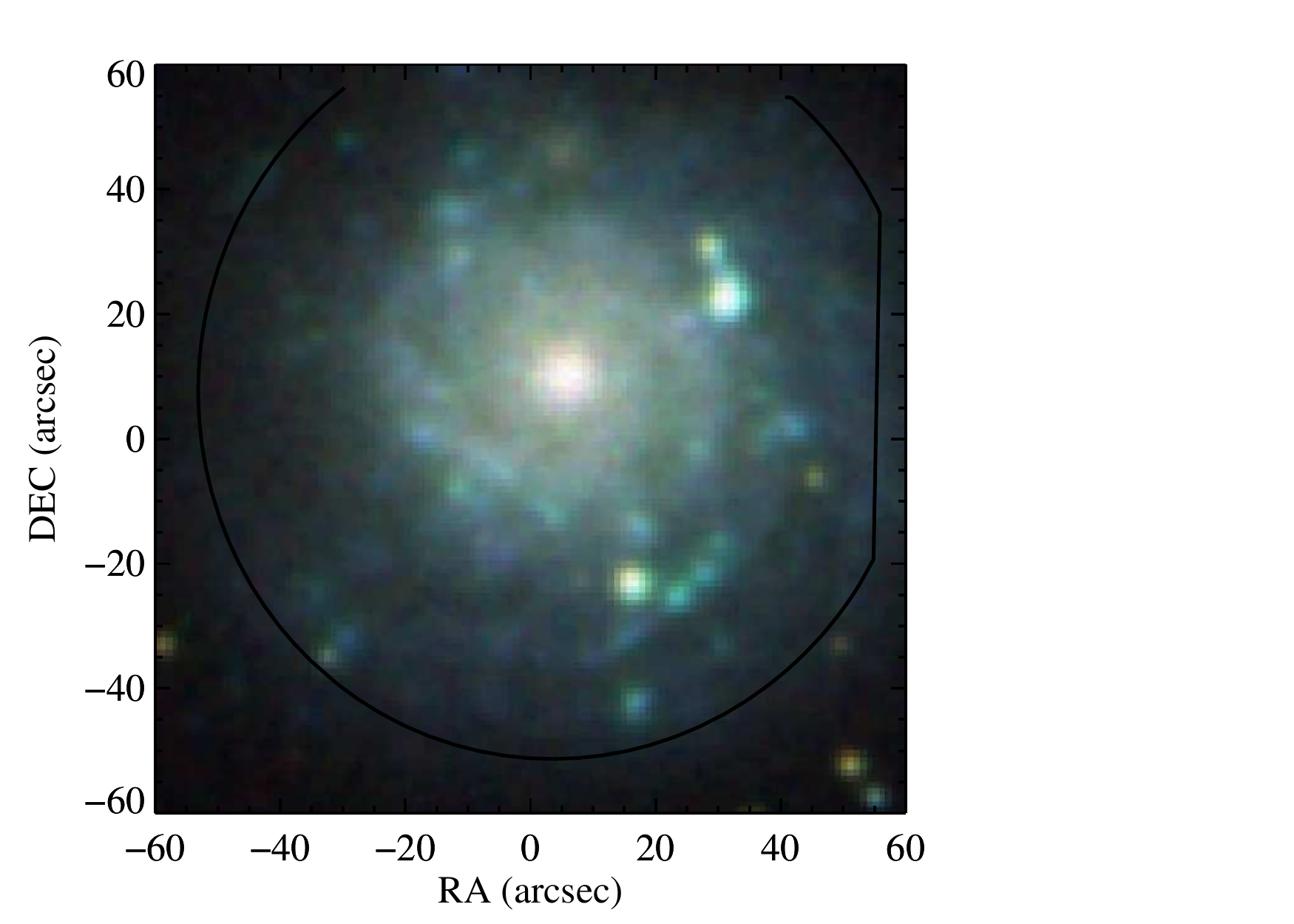}&  
    \includegraphics[scale=0.3]{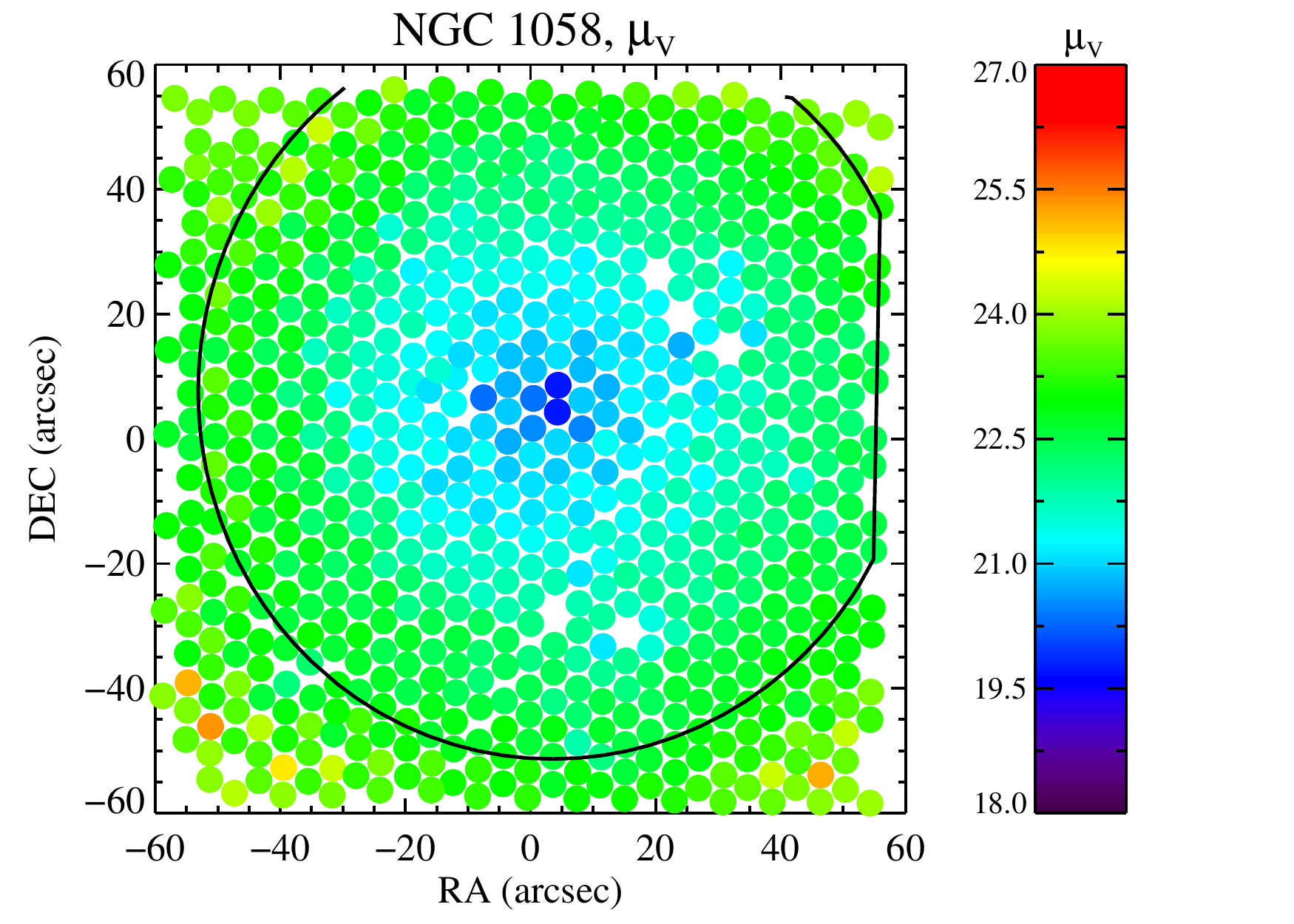}&  
    \includegraphics[scale=0.3]{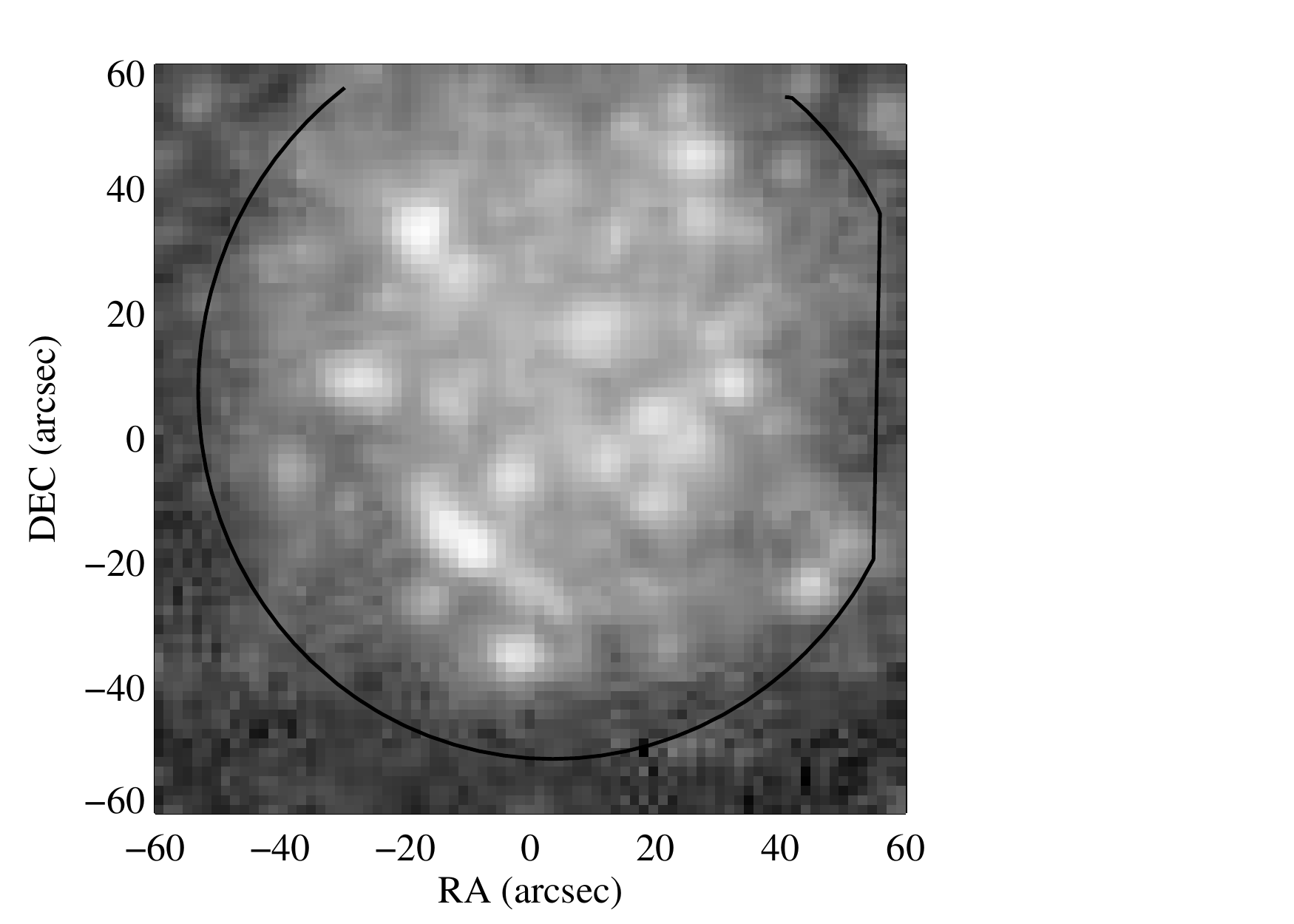}\\ 
    \includegraphics[scale=0.3]{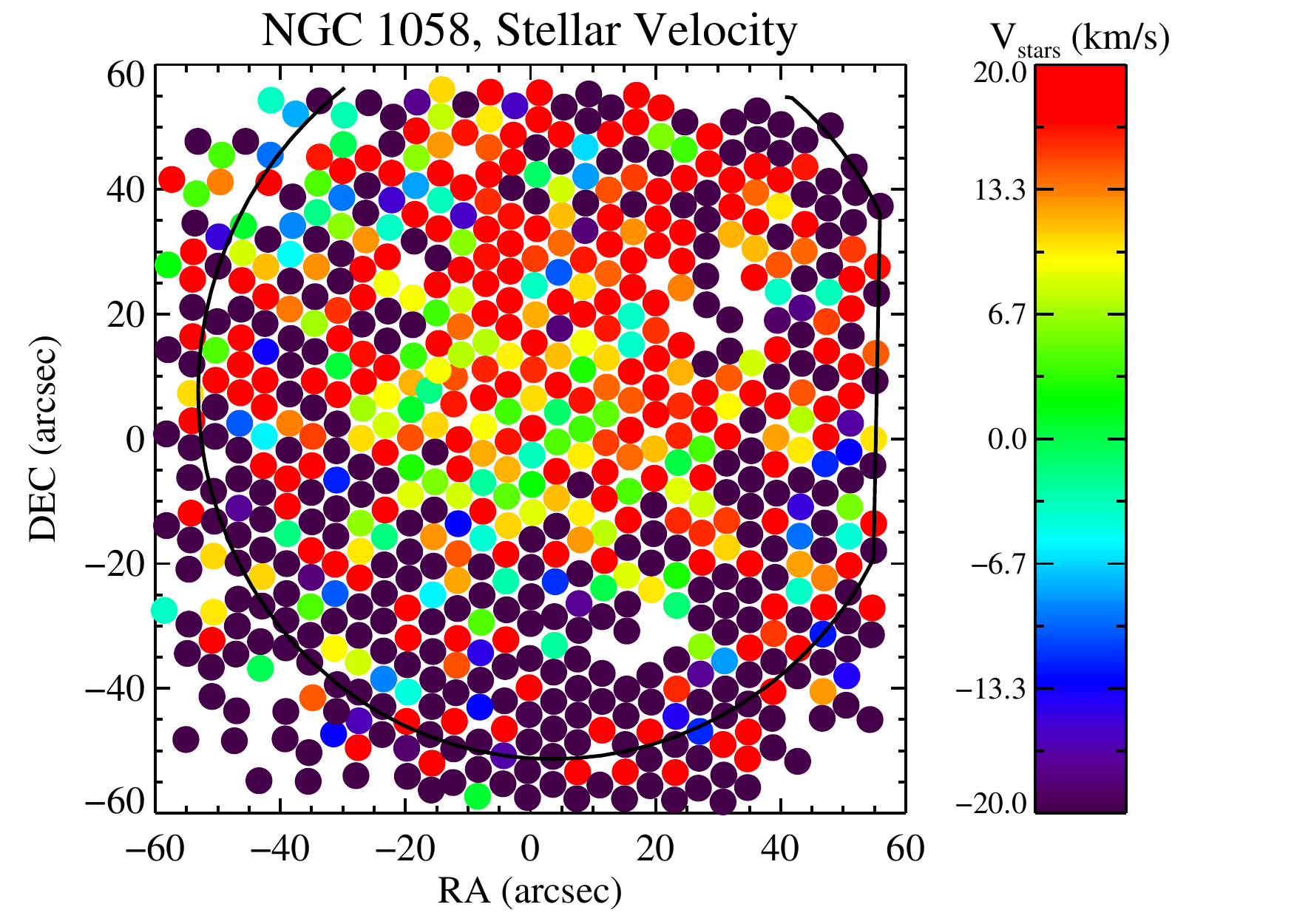}&  
    \includegraphics[scale=0.3]{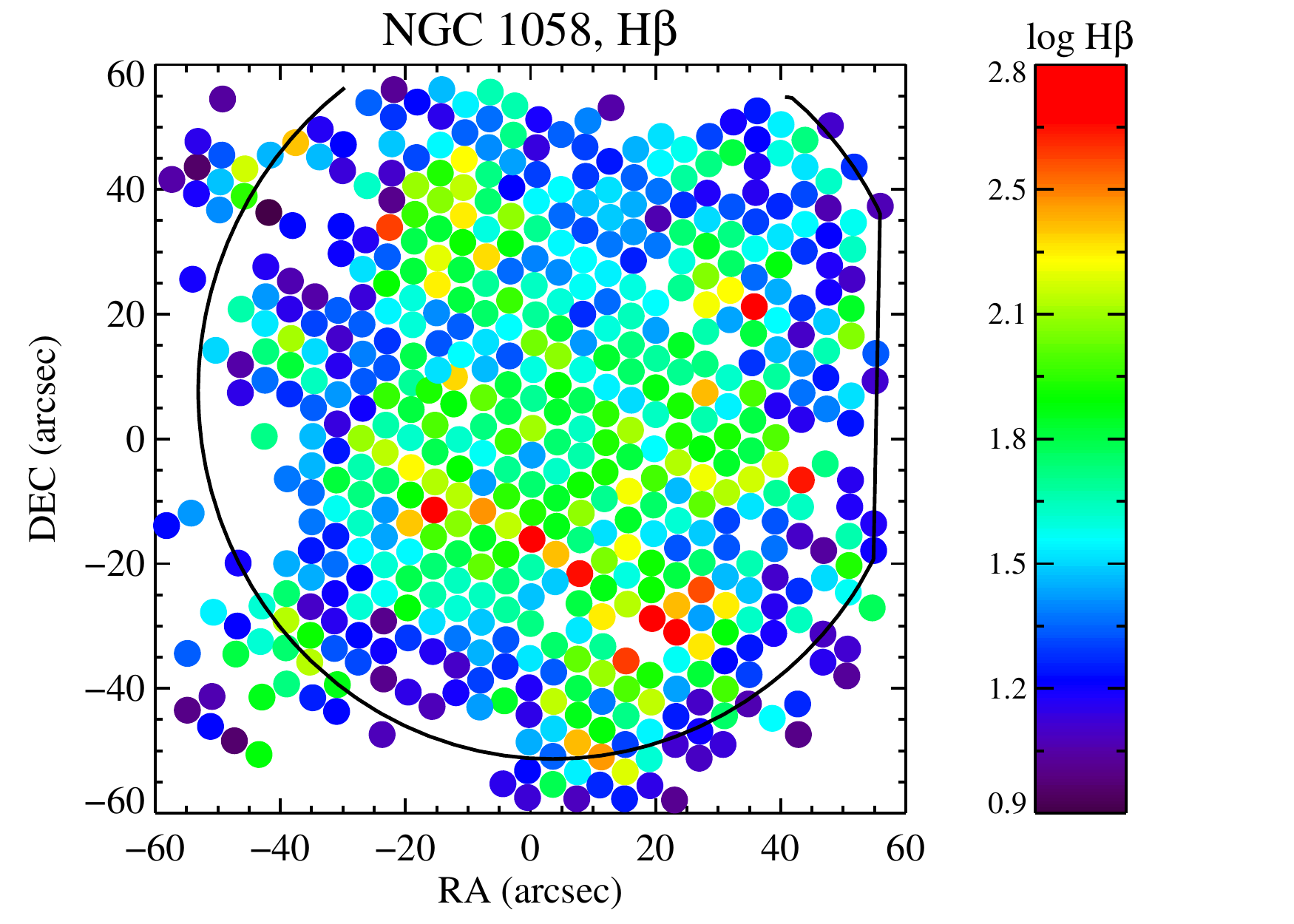}&  
    \includegraphics[scale=0.3]{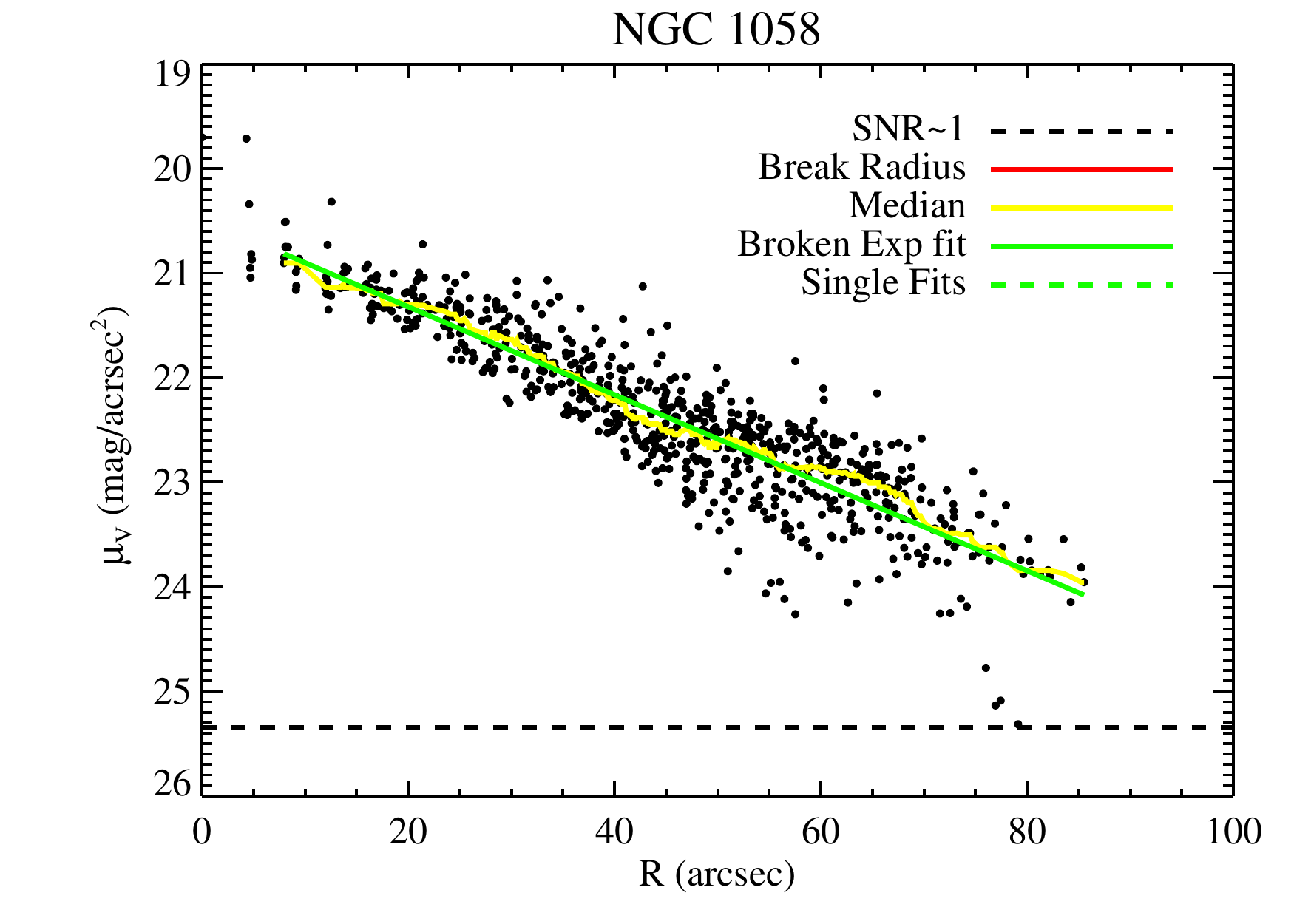}\\ 
    \includegraphics[scale=0.3]{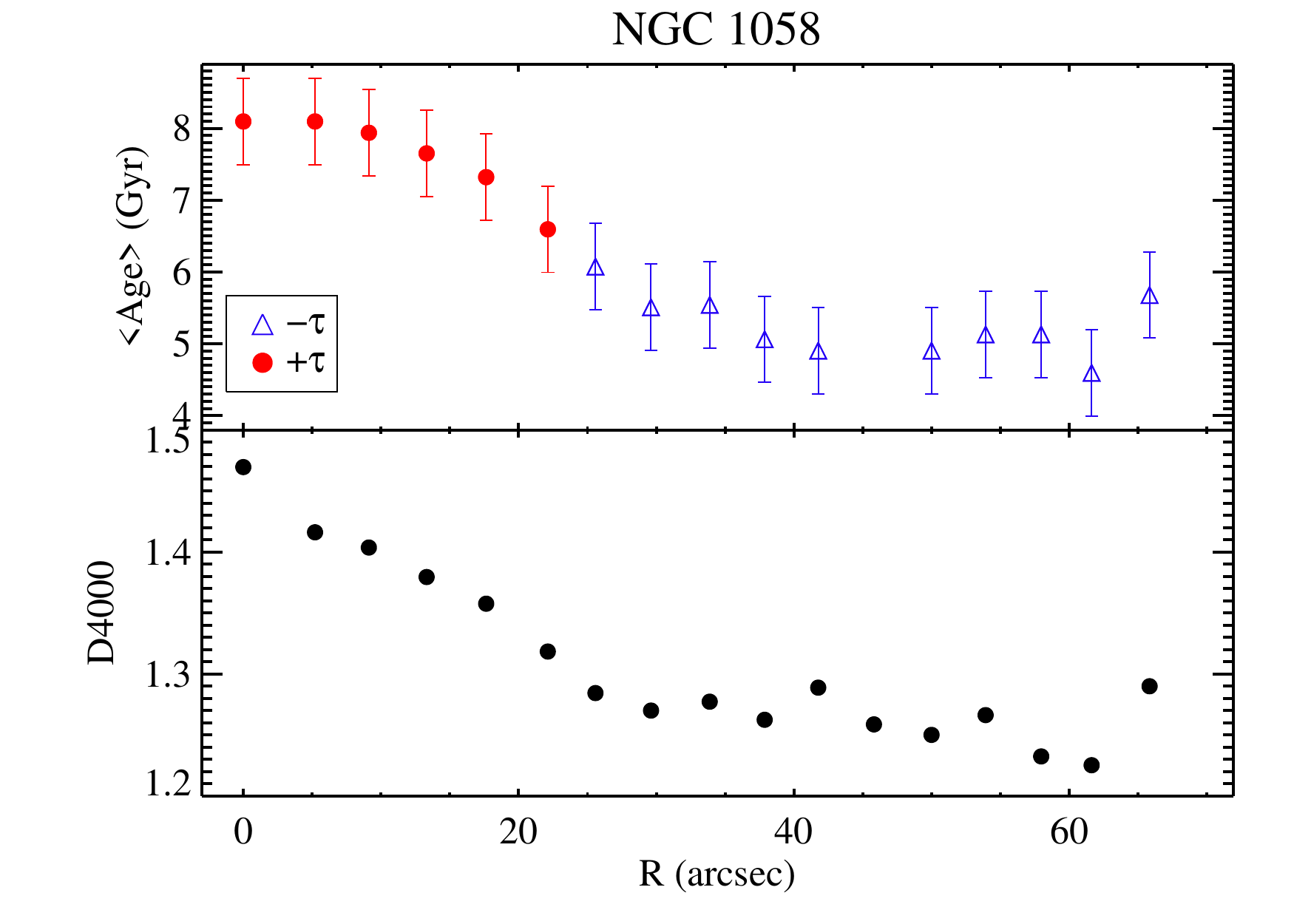}&  
    \includegraphics[scale=0.3]{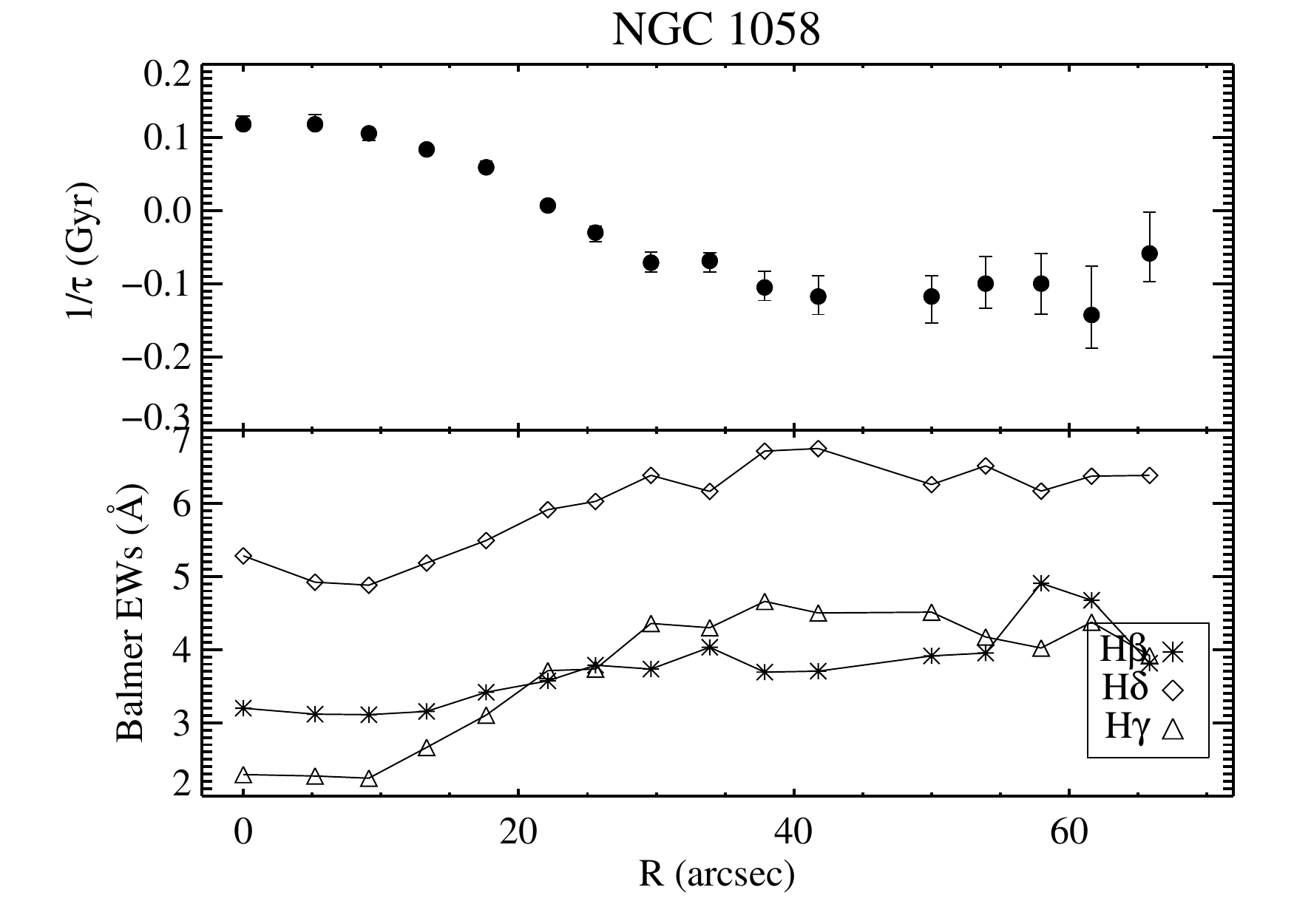}&  
    \includegraphics[scale=0.3]{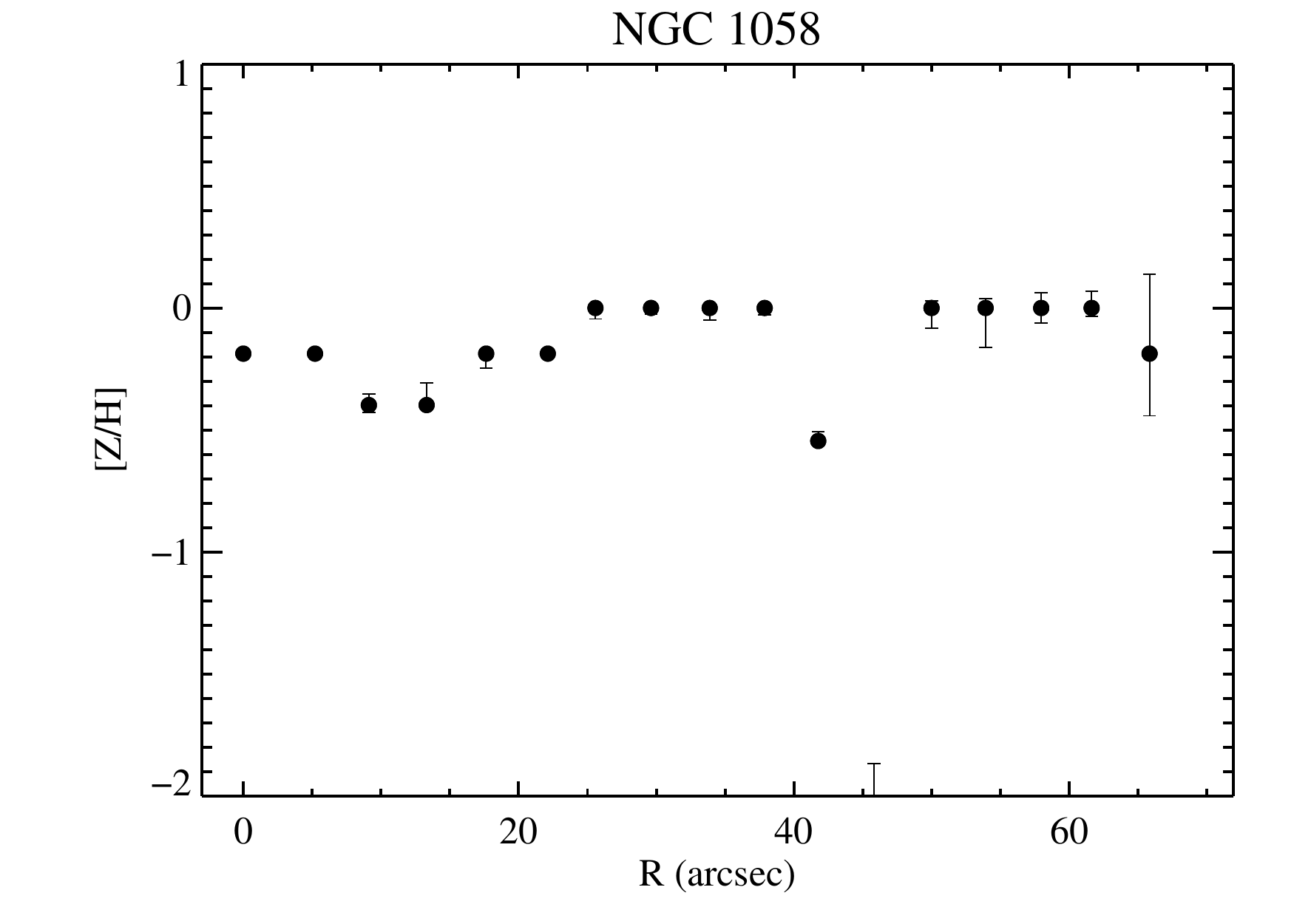}       
\end{array}$
    \epsscale{1}
 \end{center}
\caption{NGC 1058.  Upper left:  SDSS or DSS image of the target.  Upper Middle:  $V$-band surface brightness measured from Mitchell Spectrograph spectra.  Fibers below a signal-to-noise threshold have been masked along with those containing bright foreground stars.  Upper Right:  Near UV image from GALEX (if available).  Middle Left:  Stellar velocity field measured with pPXF.  Middle Middle:  \hb\ emission line flux after correcting for stellar absorption.  Middle Right:  Radial surface brightness profile measured from fiber magnitudes.  A yellow line marks a running median, the solid green line shows the best fit broken exponential.  The broken green lines shows single exponentials fit inside and outside the break radius and extrapolated to the full radial range.  Lower Left:  best-fit mass-weighted stellar ages from binned spectra.  Red points are used for models that converged on exponentially decreasing SFHs, while blue points are used for increasing SFHs.  We also show the age sensitive D4000 index for comparison (The D4000 index has not been corrected for dust effects).  Lower Middle:  The best fit values of the star formation timescale 1/$\tau$, along with the emission-corrected Balmer line absorption equivalent widths.  Lower Right:  Best-fit flux-weighted average metallicities.  In most figures, a red line marks the surface brightness break radius.  For galaxies without a break radius, we mark the $\mu_V = 23$ mag/$\square$\arcsec\ contour with a black line.  Error bars are derived from the $\Delta\chi^2=20$\ contour (with a systematic error floor of 0.6 Gyr for the average age plots).  \label{NGC1058}}
\end{figure*}


\begin{figure*}
 \begin{center}$
 \begin{array}{ccc}
   \epsscale{.35}
\includegraphics[scale=0.27]{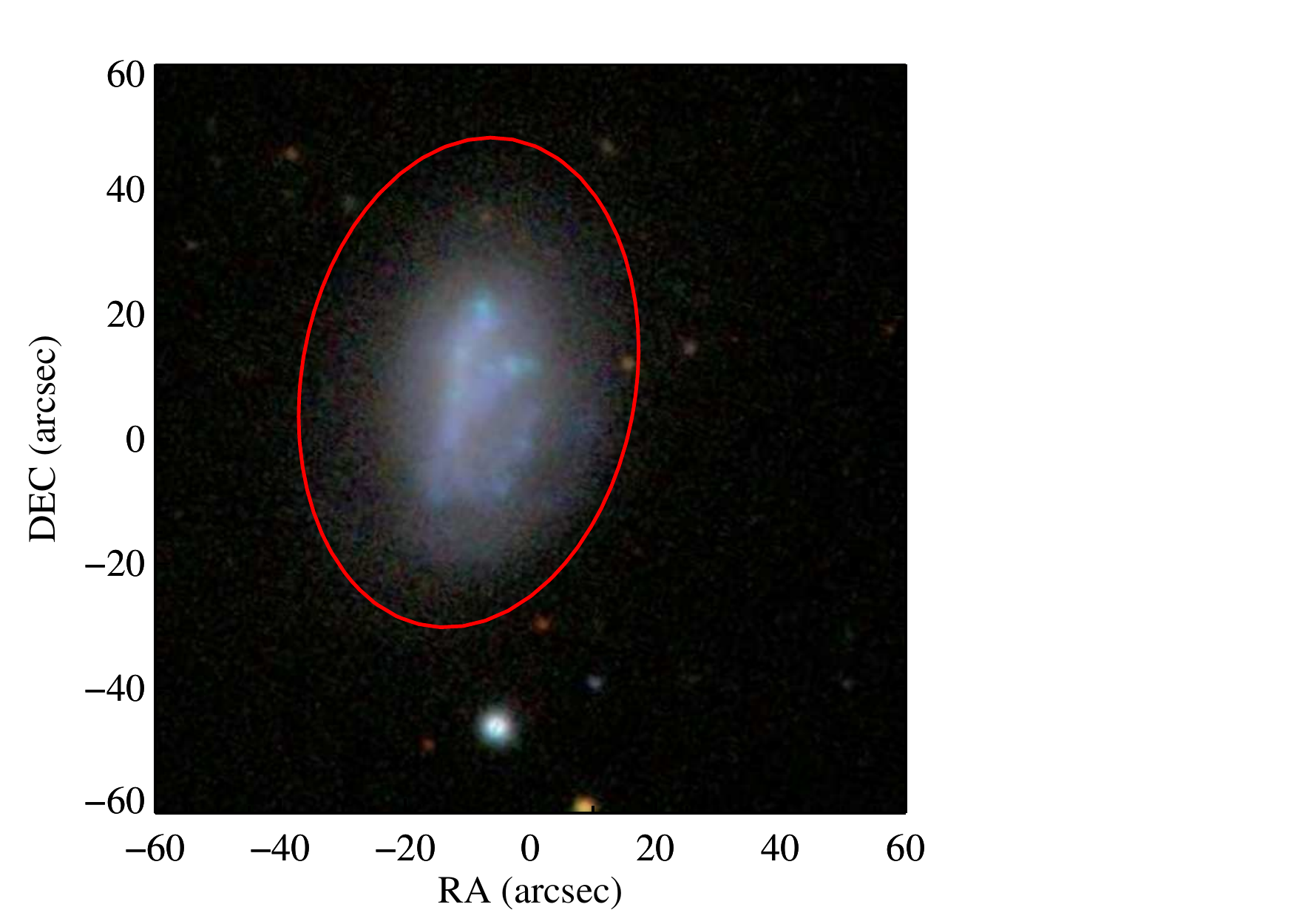}&
\includegraphics[scale=0.27]{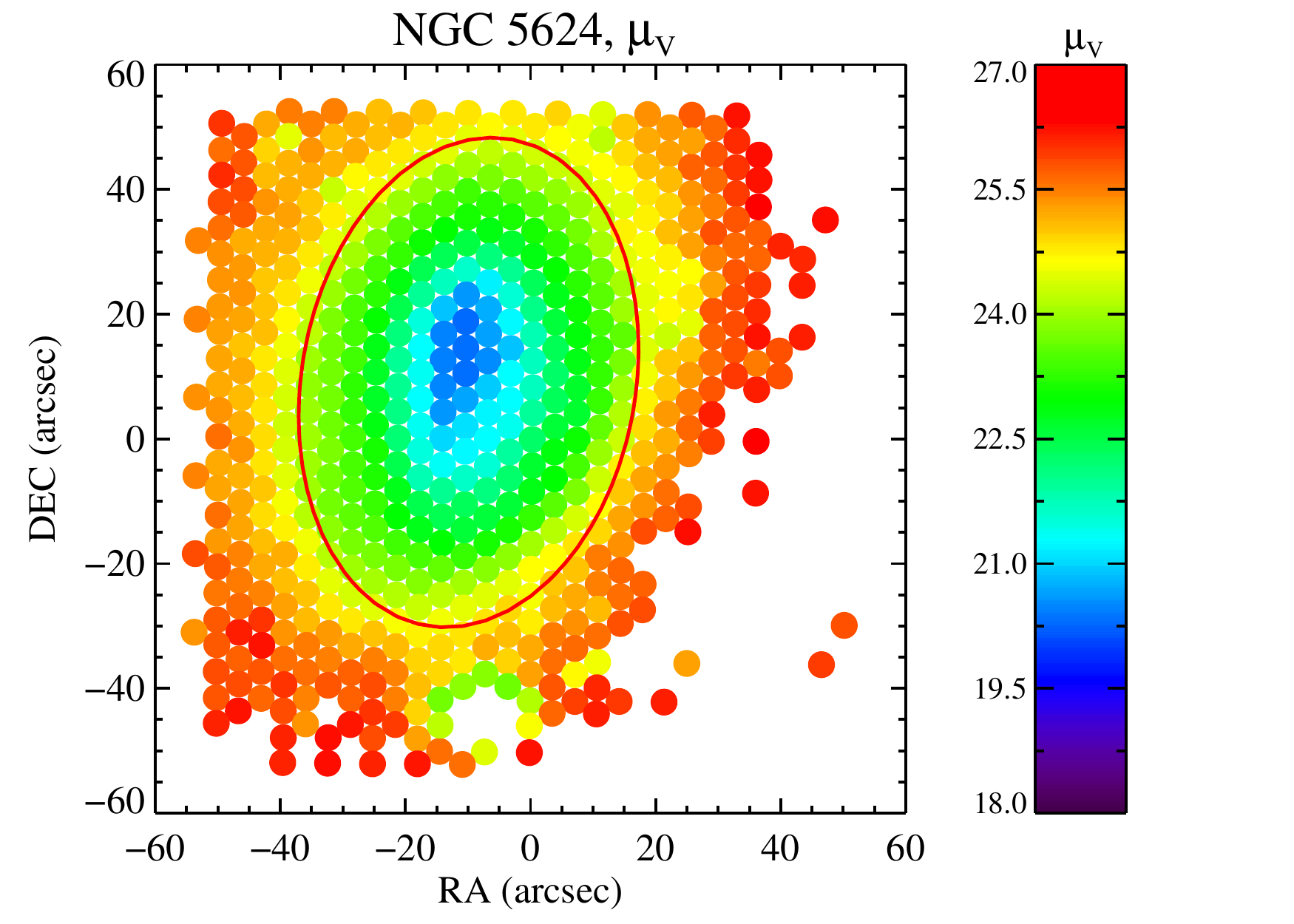}\\
\includegraphics[scale=0.27]{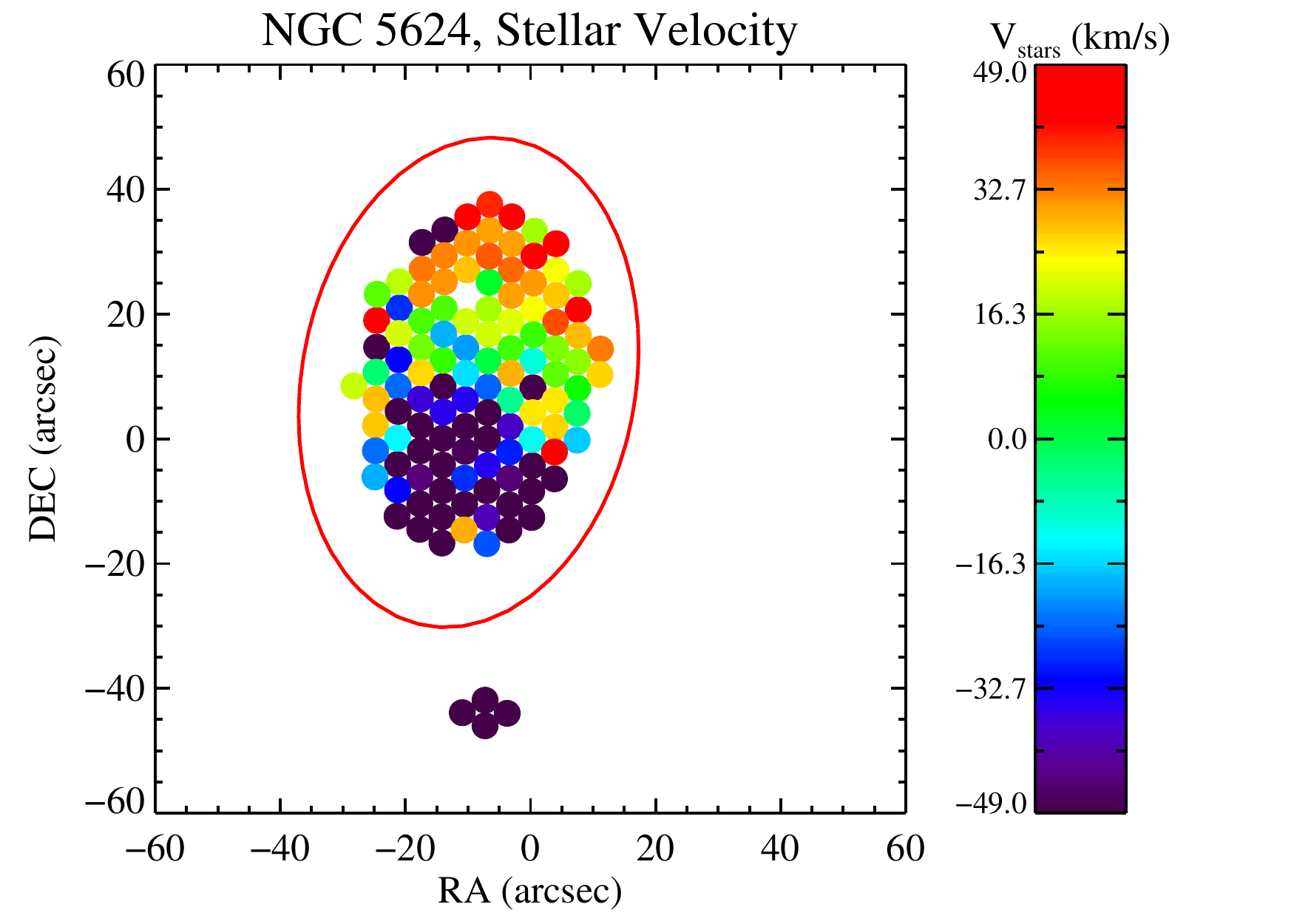}&
\includegraphics[scale=0.27]{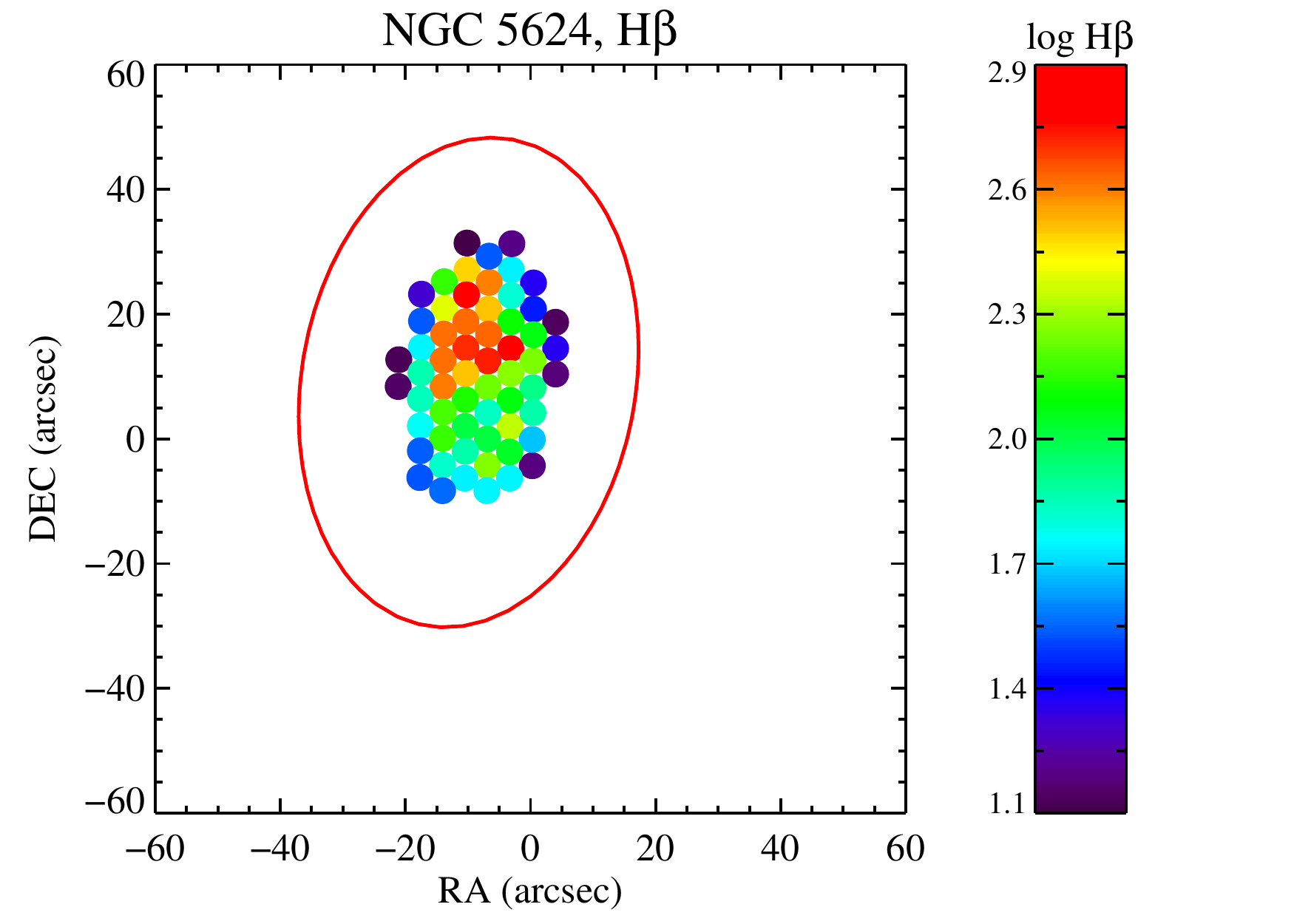}&
\includegraphics[scale=0.27]{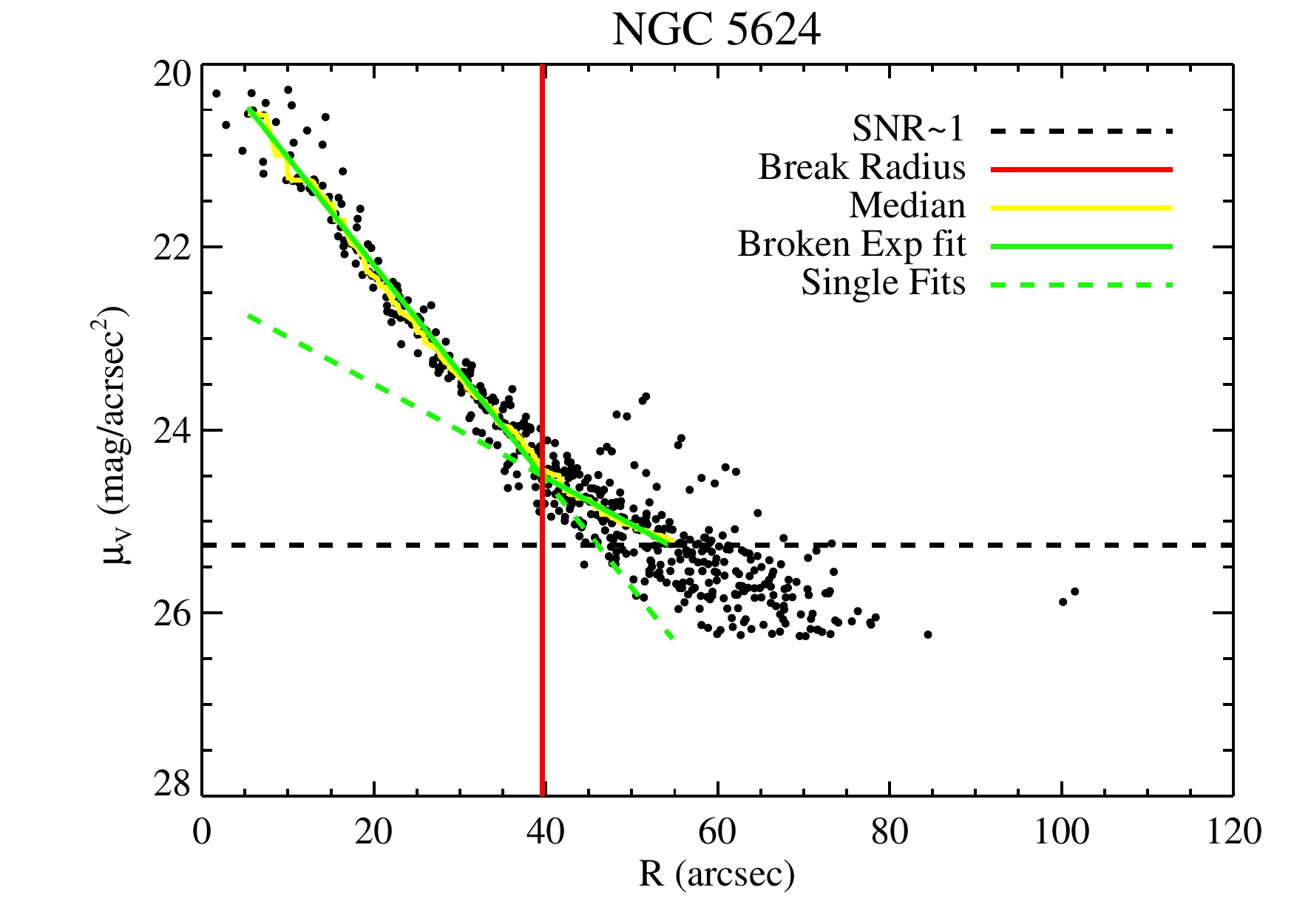}\\
\includegraphics[scale=0.27]{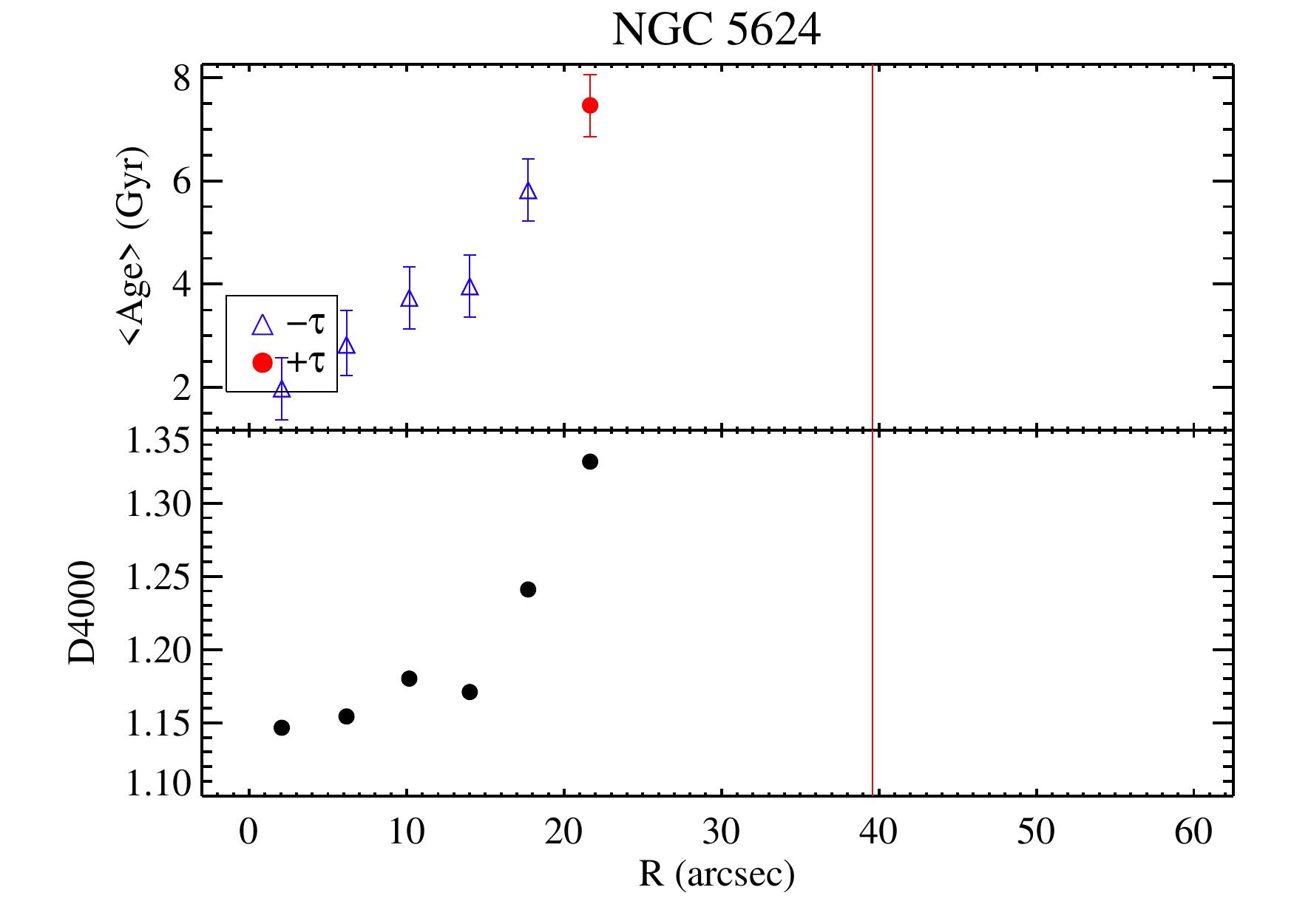}&
\includegraphics[scale=0.27]{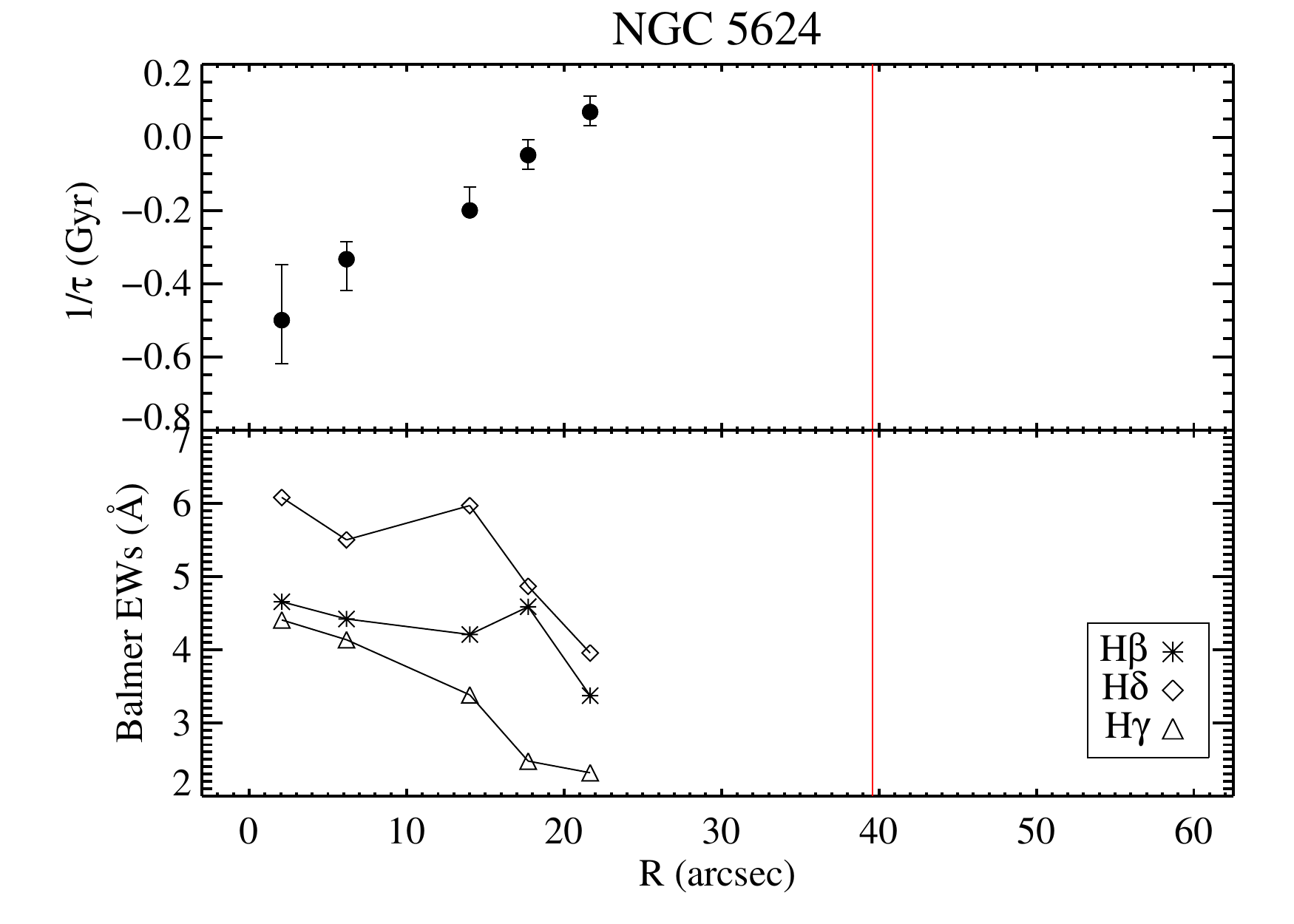}&
\includegraphics[scale=0.27]{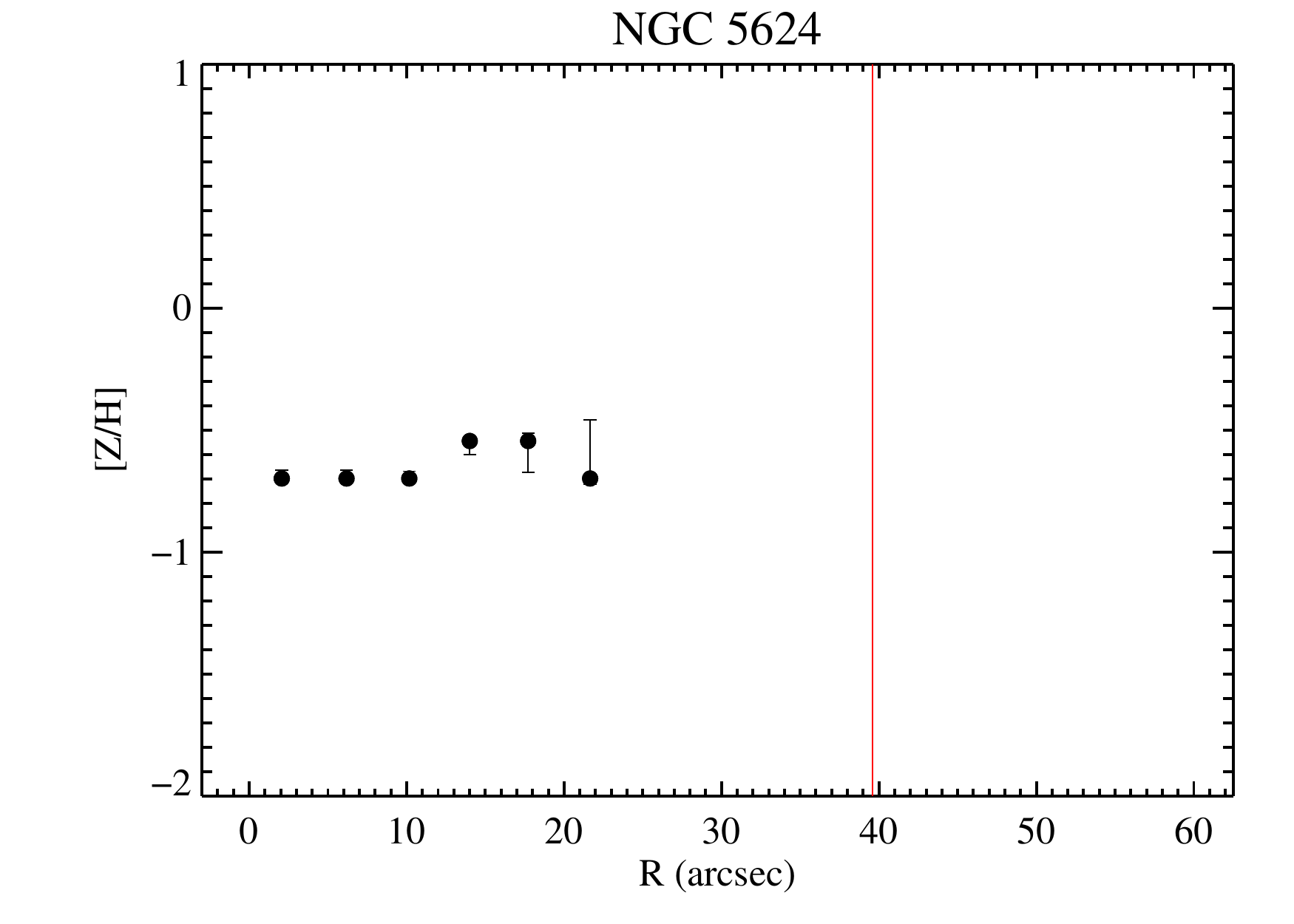}   
\end{array}$
    \epsscale{1}
 \end{center}
\caption{NGC 5624.  Same as Figure~\ref{NGC1058}. \label{NGC5624}}
\end{figure*}


\begin{figure*}
 \begin{center}$
 \begin{array}{ccc}
   \epsscale{.35}
\includegraphics[scale=0.27]{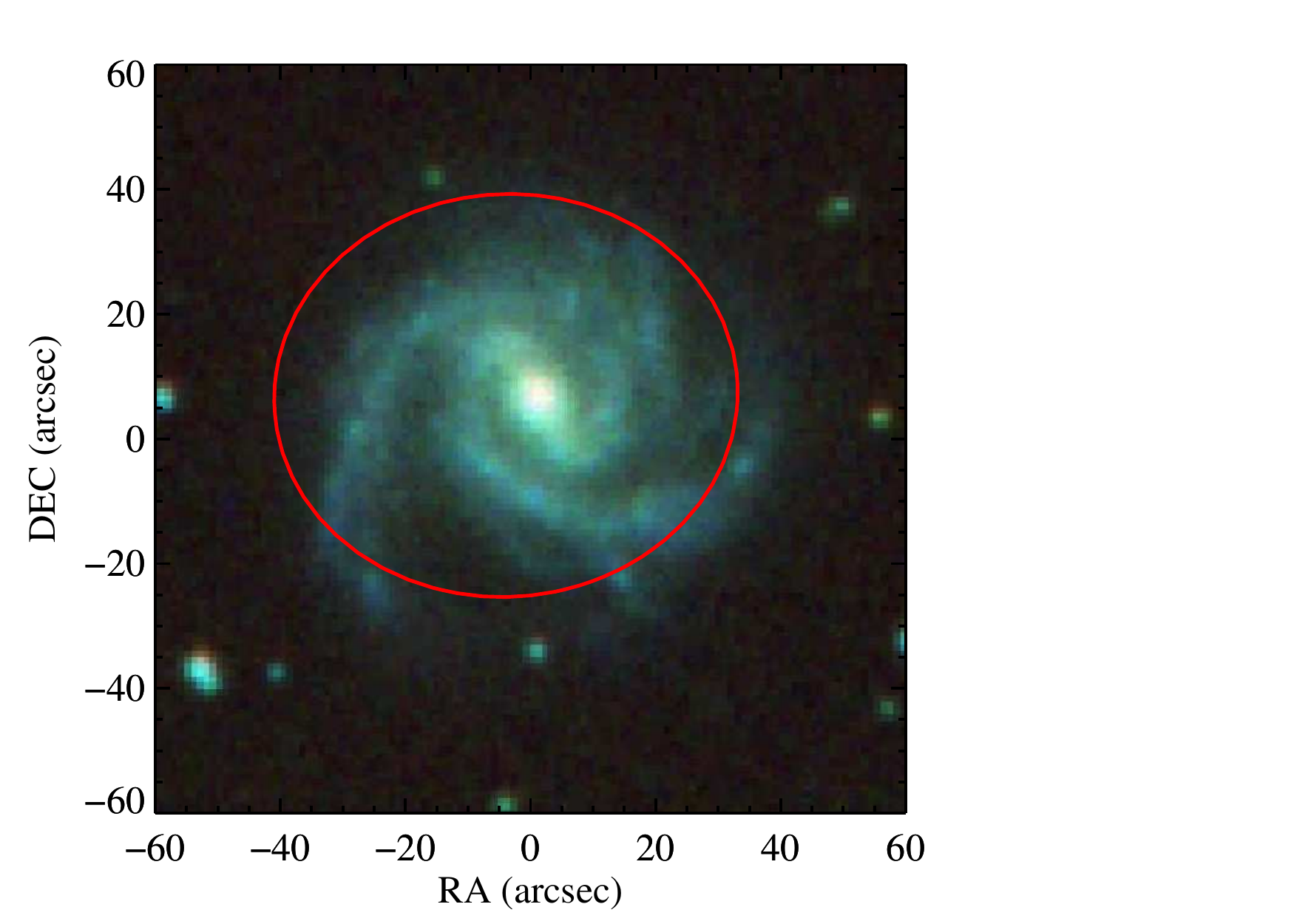}&
\includegraphics[scale=0.27]{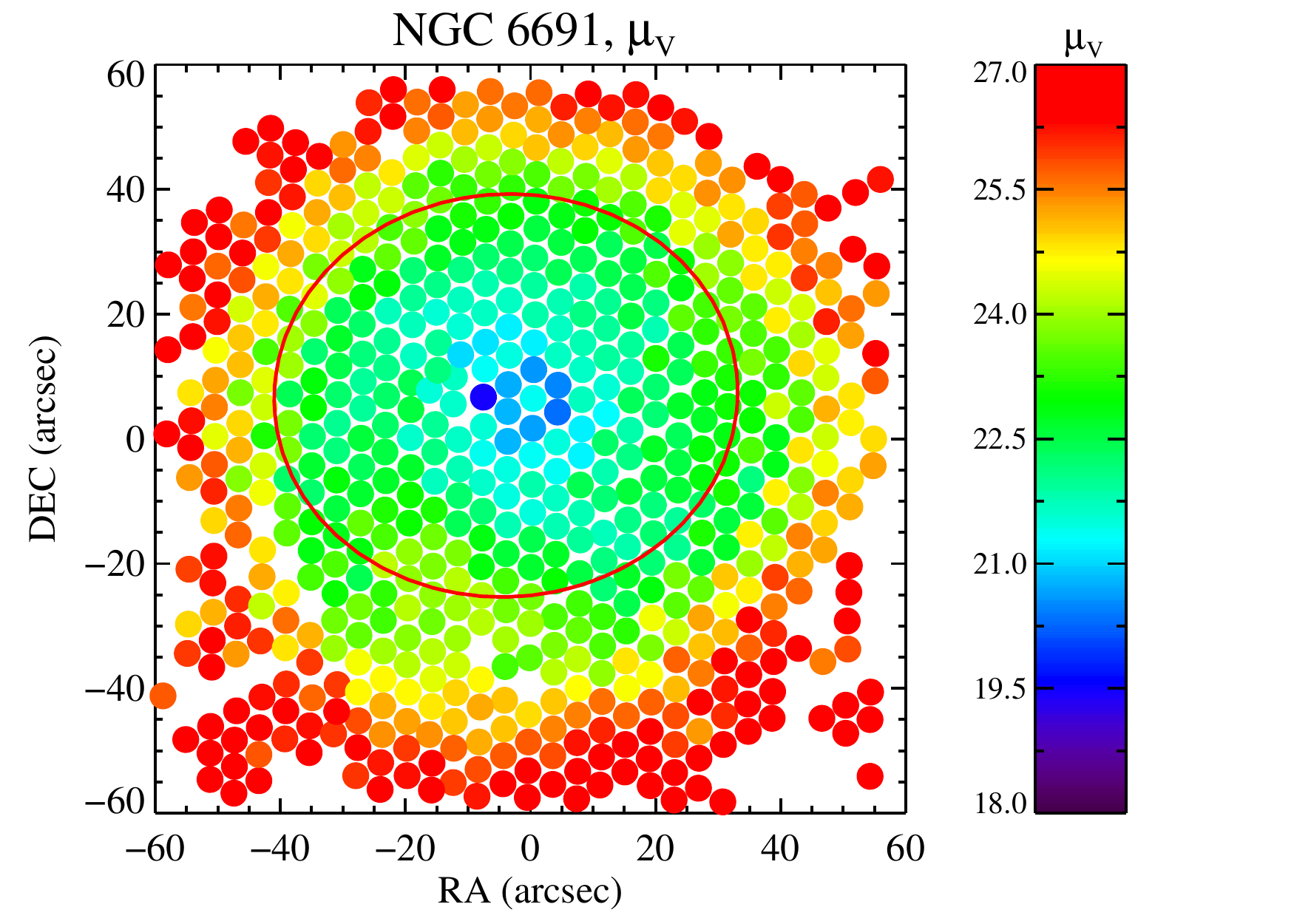}\\
\includegraphics[scale=0.27]{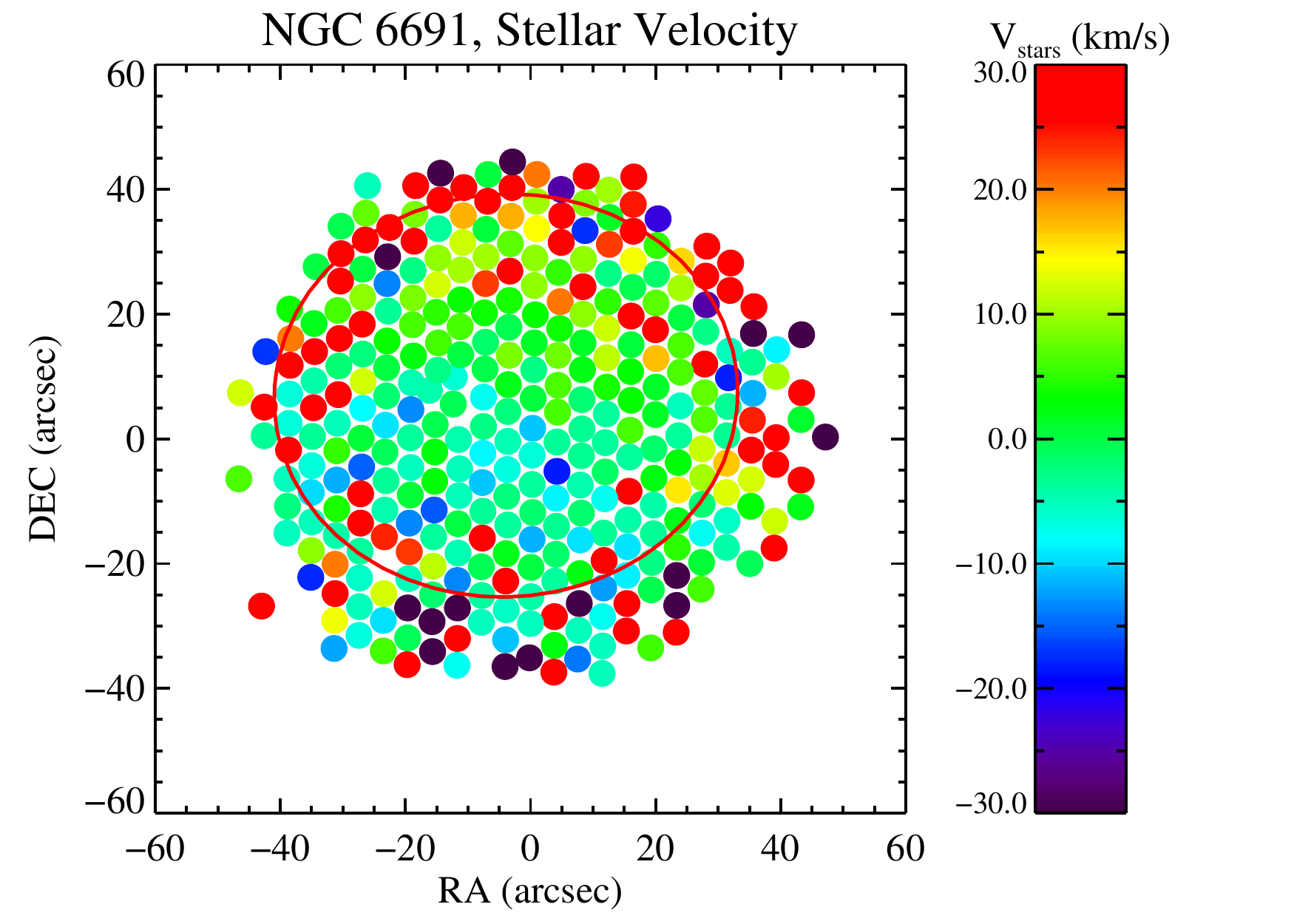}&
\includegraphics[scale=0.27]{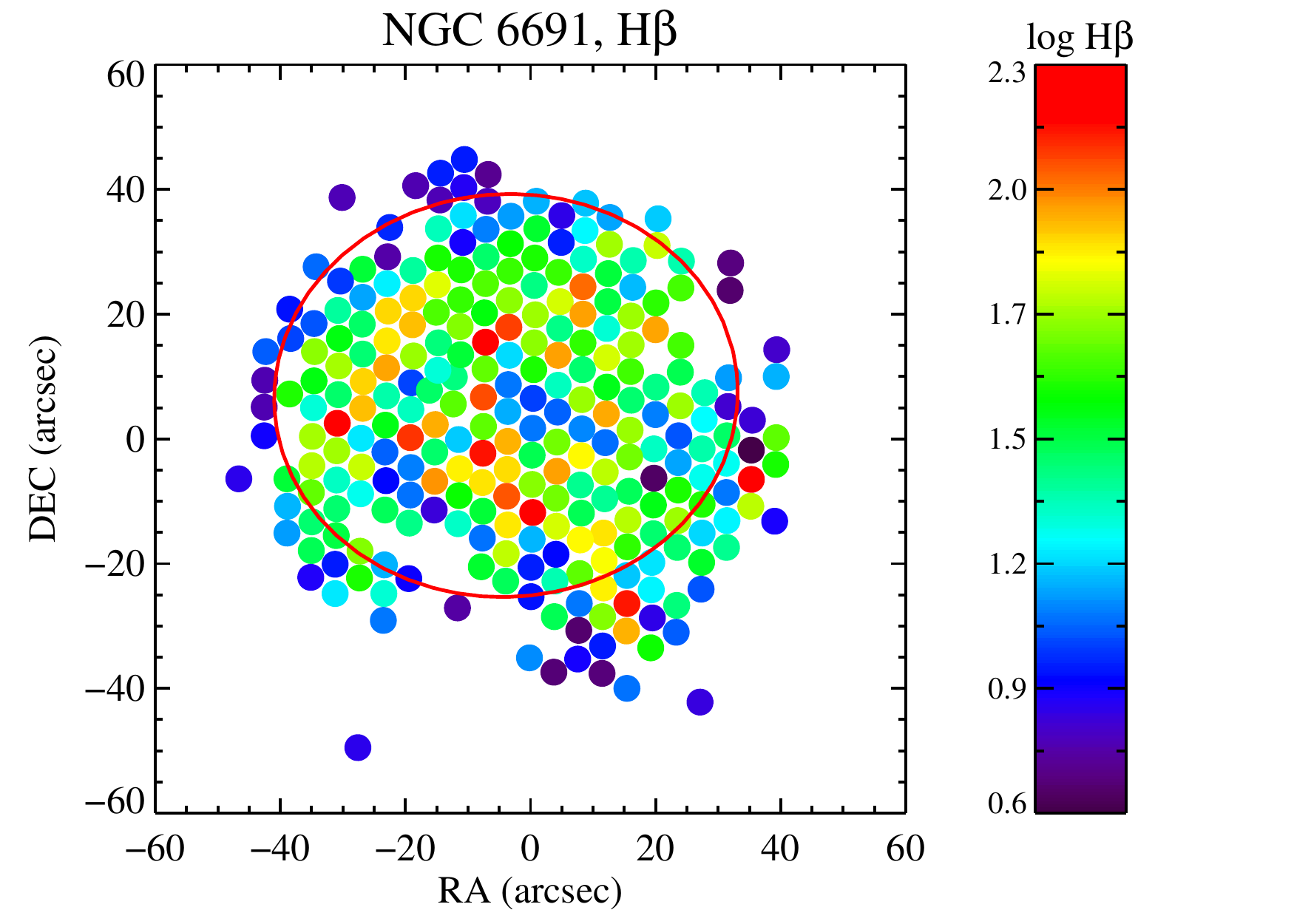}&
\includegraphics[scale=0.27]{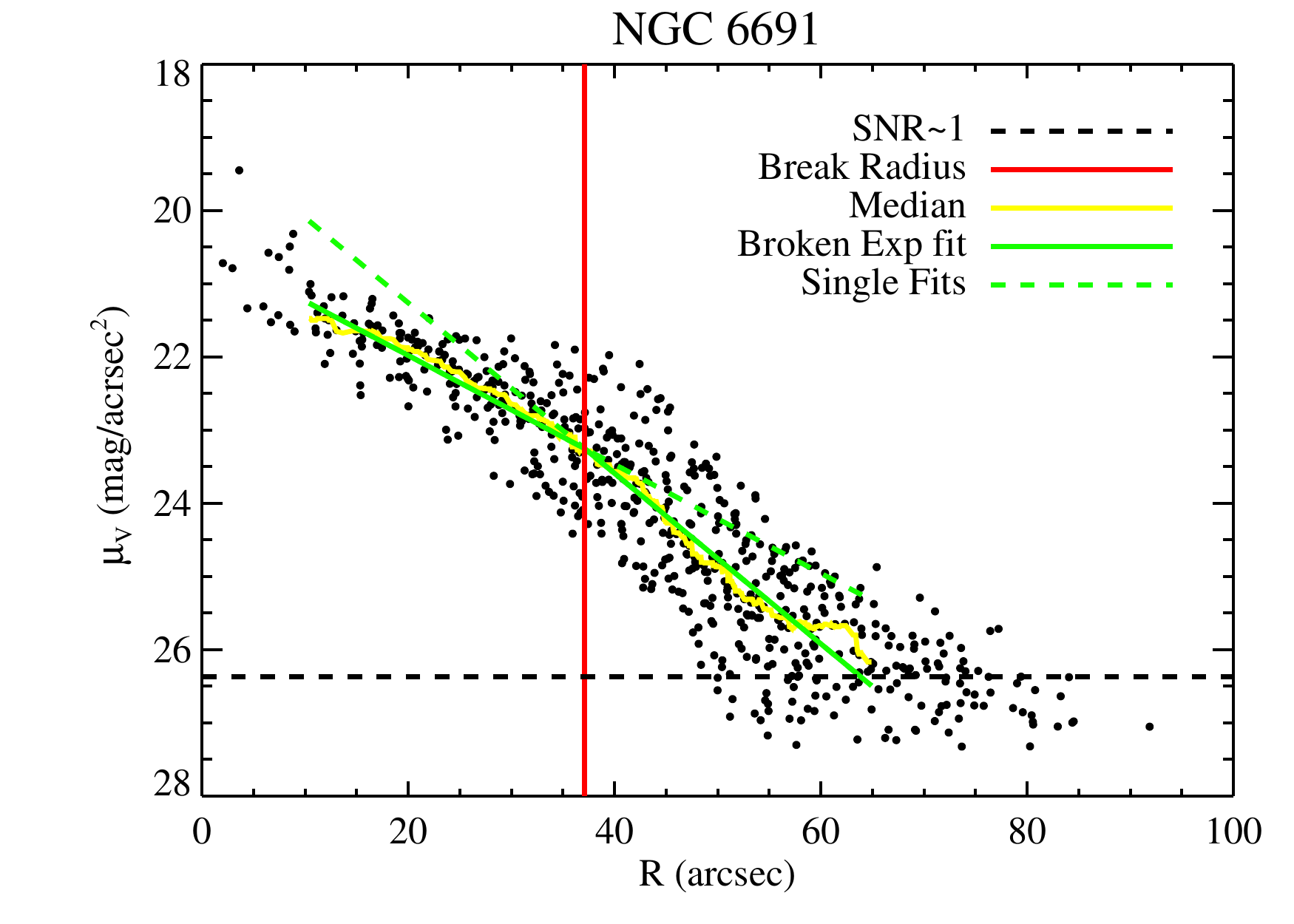}\\
\includegraphics[scale=0.27]{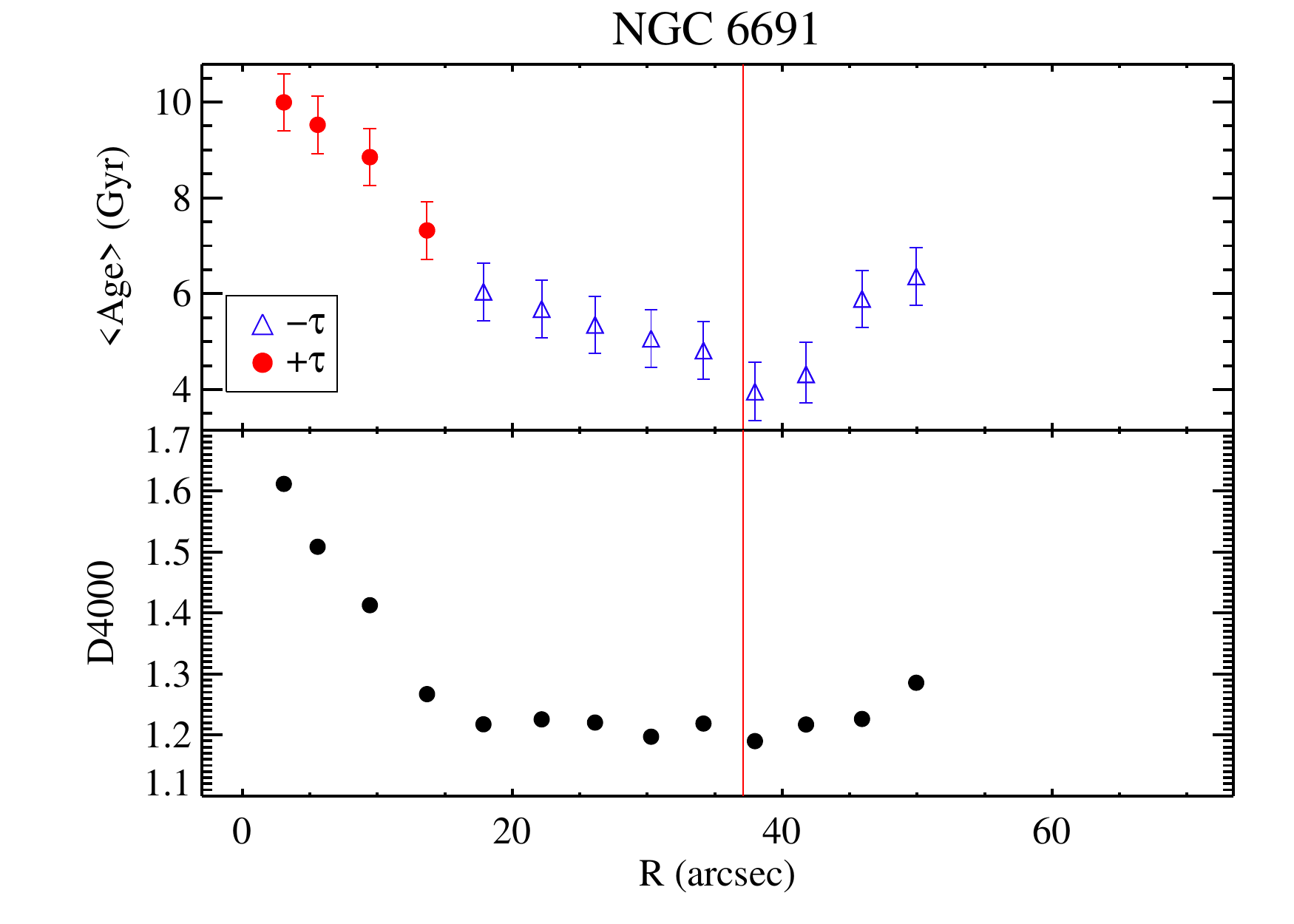}&
\includegraphics[scale=0.27]{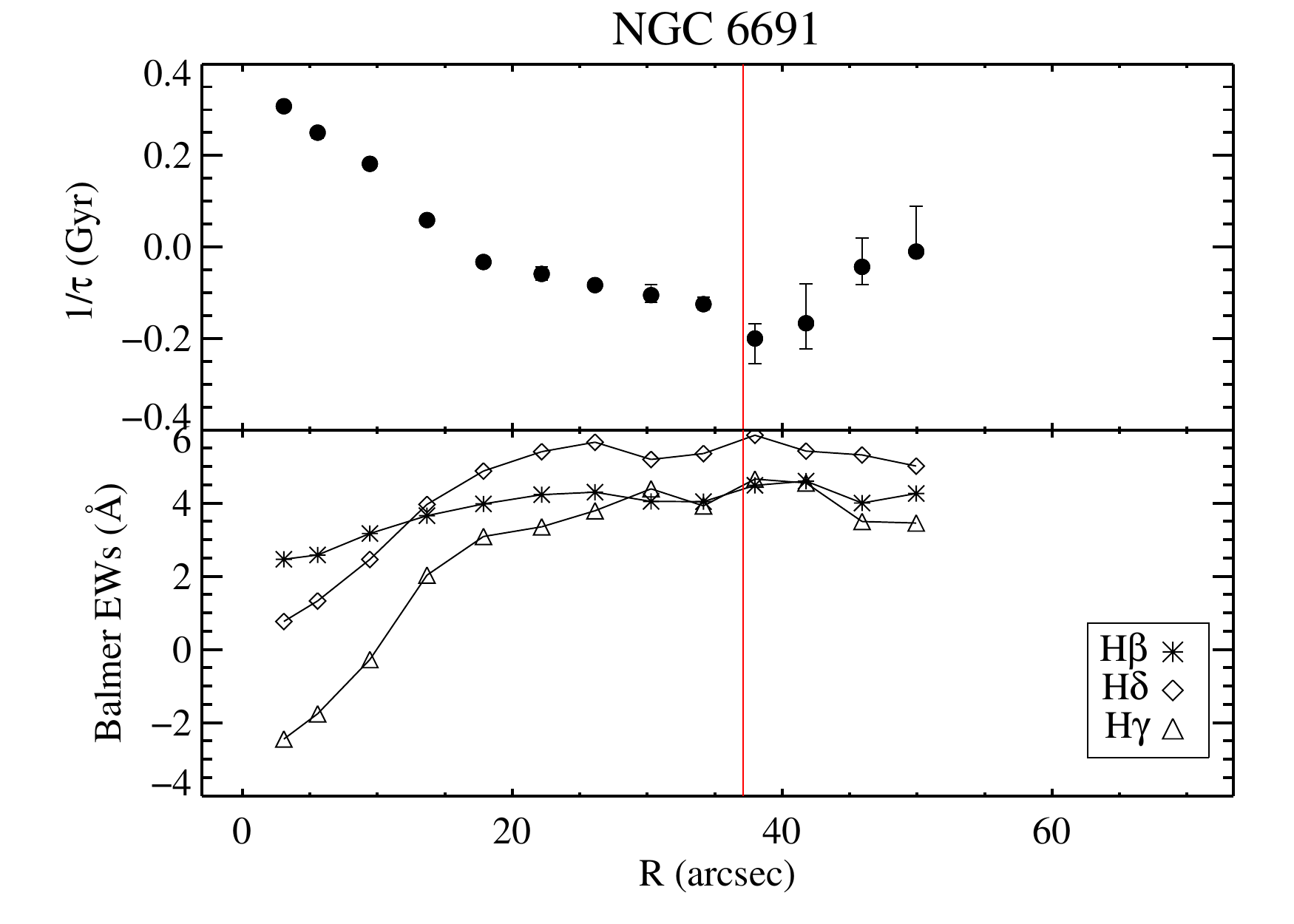}&
\includegraphics[scale=0.27]{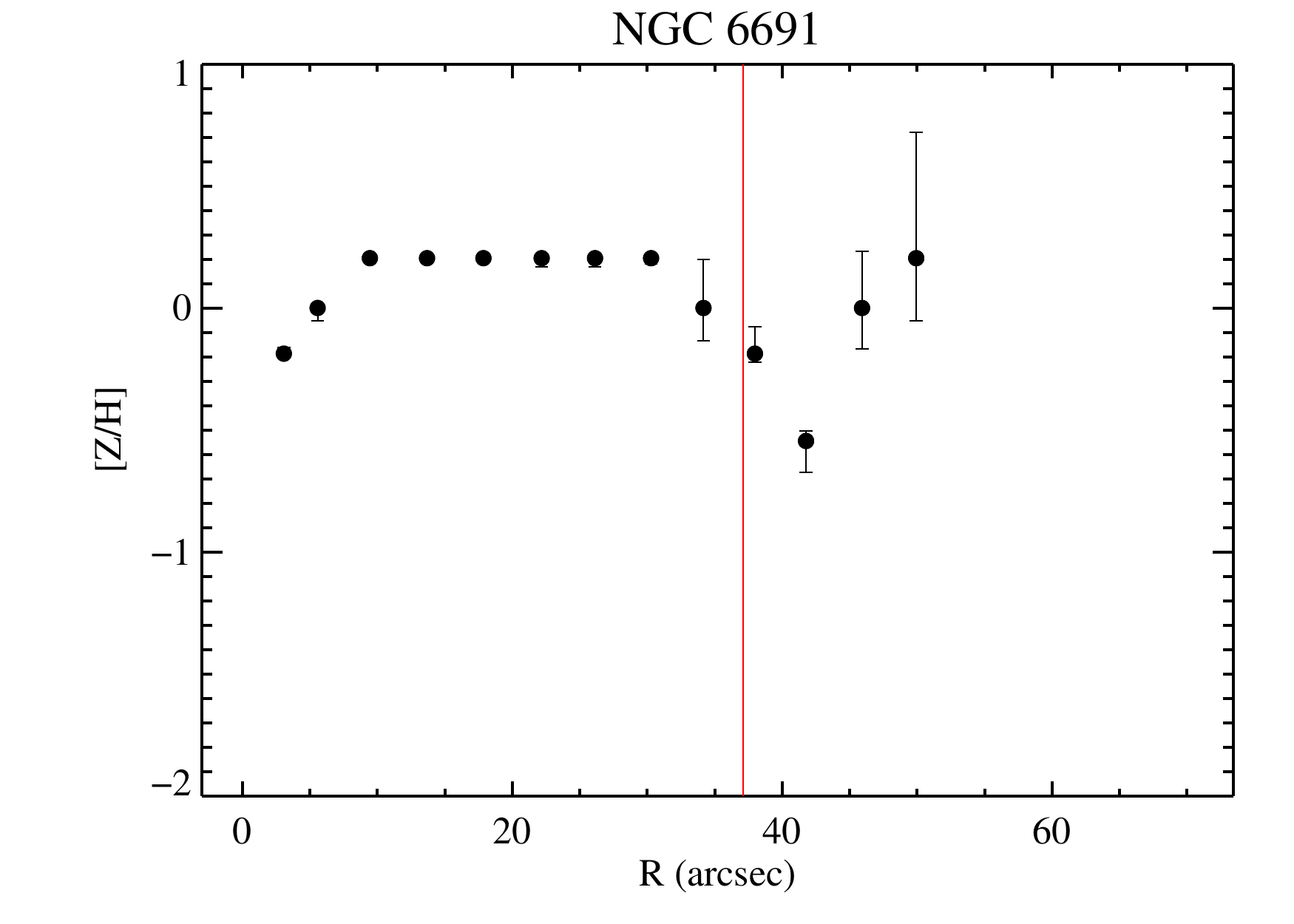}   
\end{array}$
    \epsscale{1}
 \end{center}
\caption{NGC 6691.  Same as Figure~\ref{NGC1058}. \label{NGC6691}}
\end{figure*}


\begin{figure*}
 \begin{center}$
 \begin{array}{ccc}
   \epsscale{.35}
\includegraphics[scale=0.27]{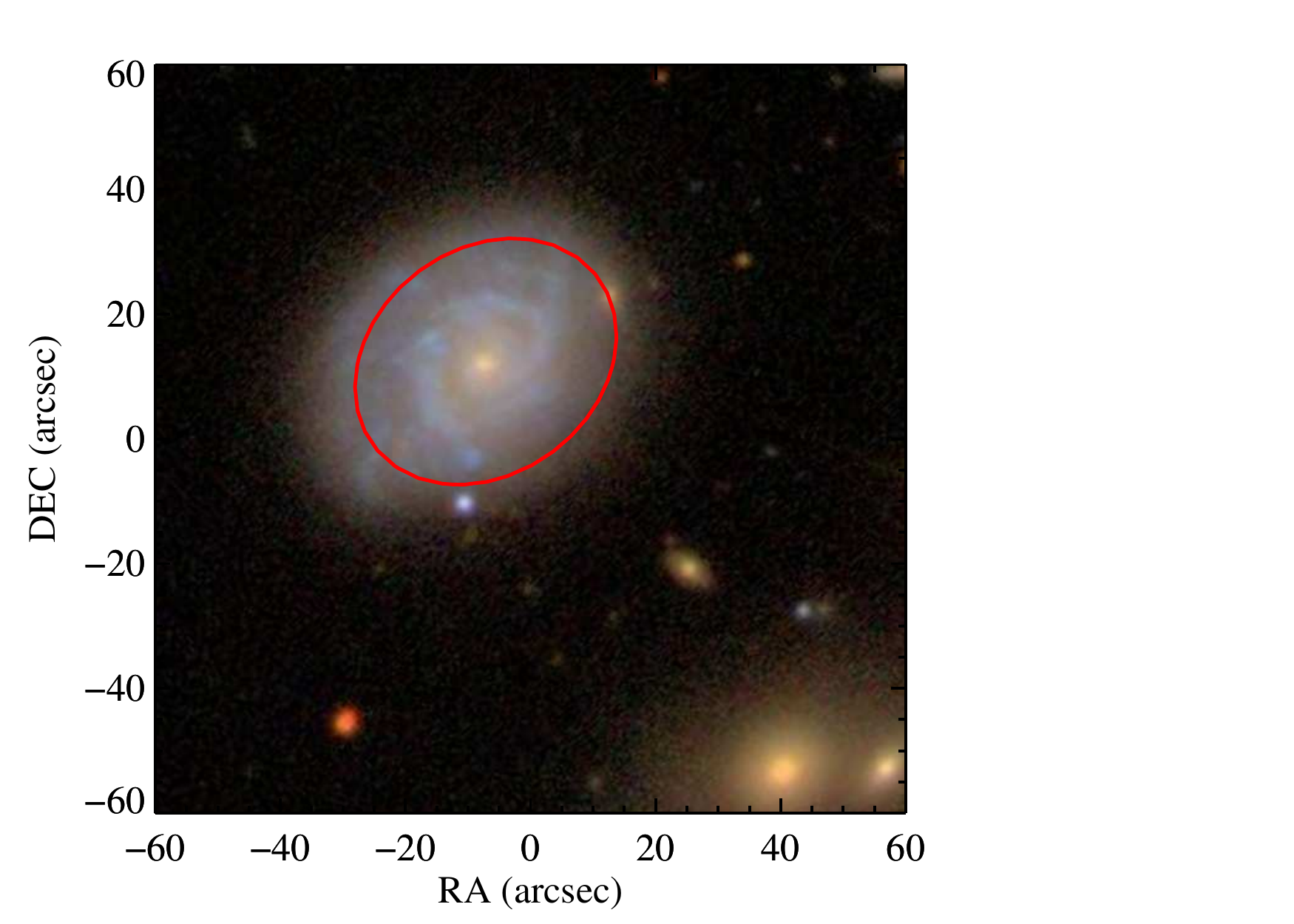}&
\includegraphics[scale=0.27]{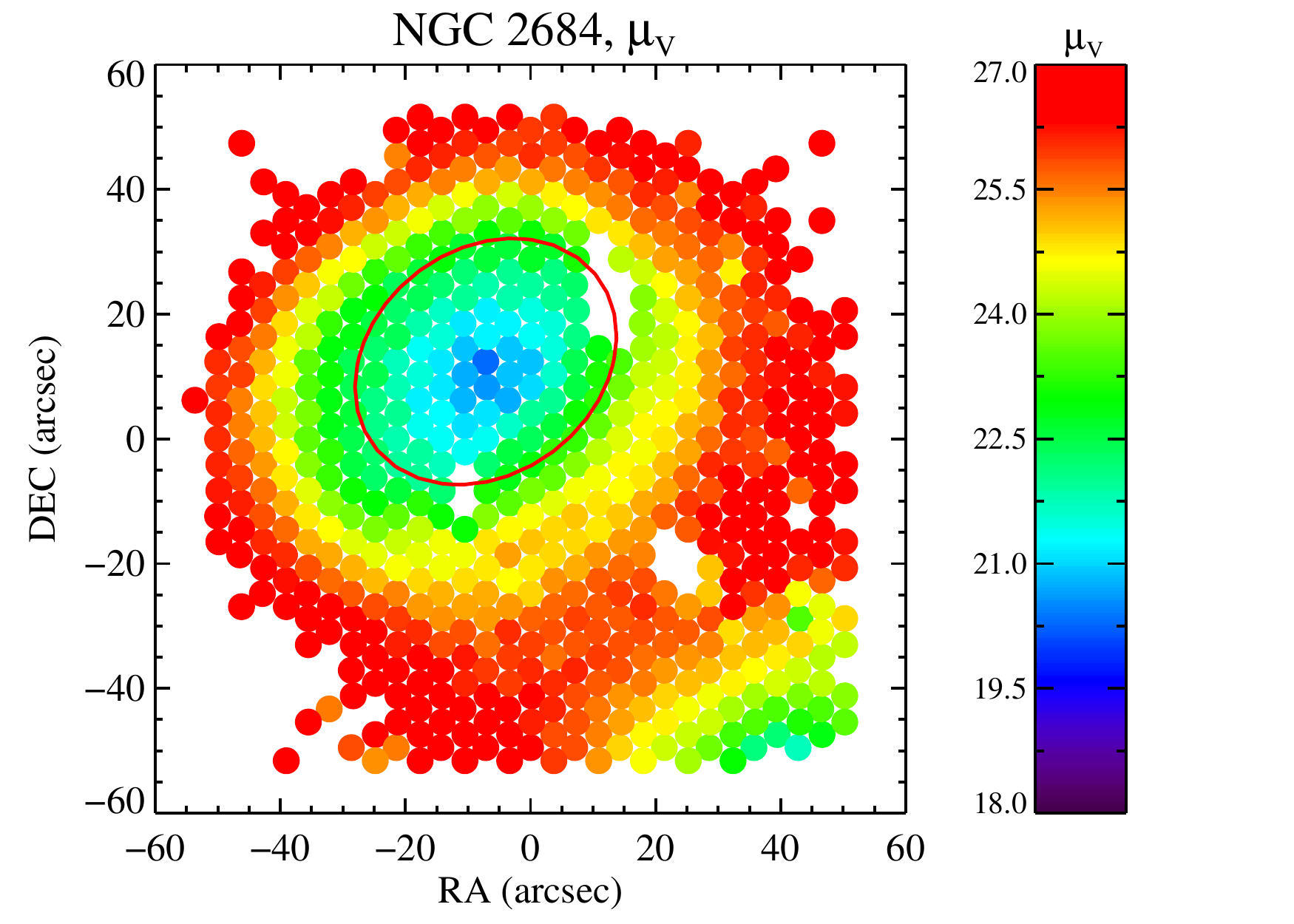}\\
\includegraphics[scale=0.27]{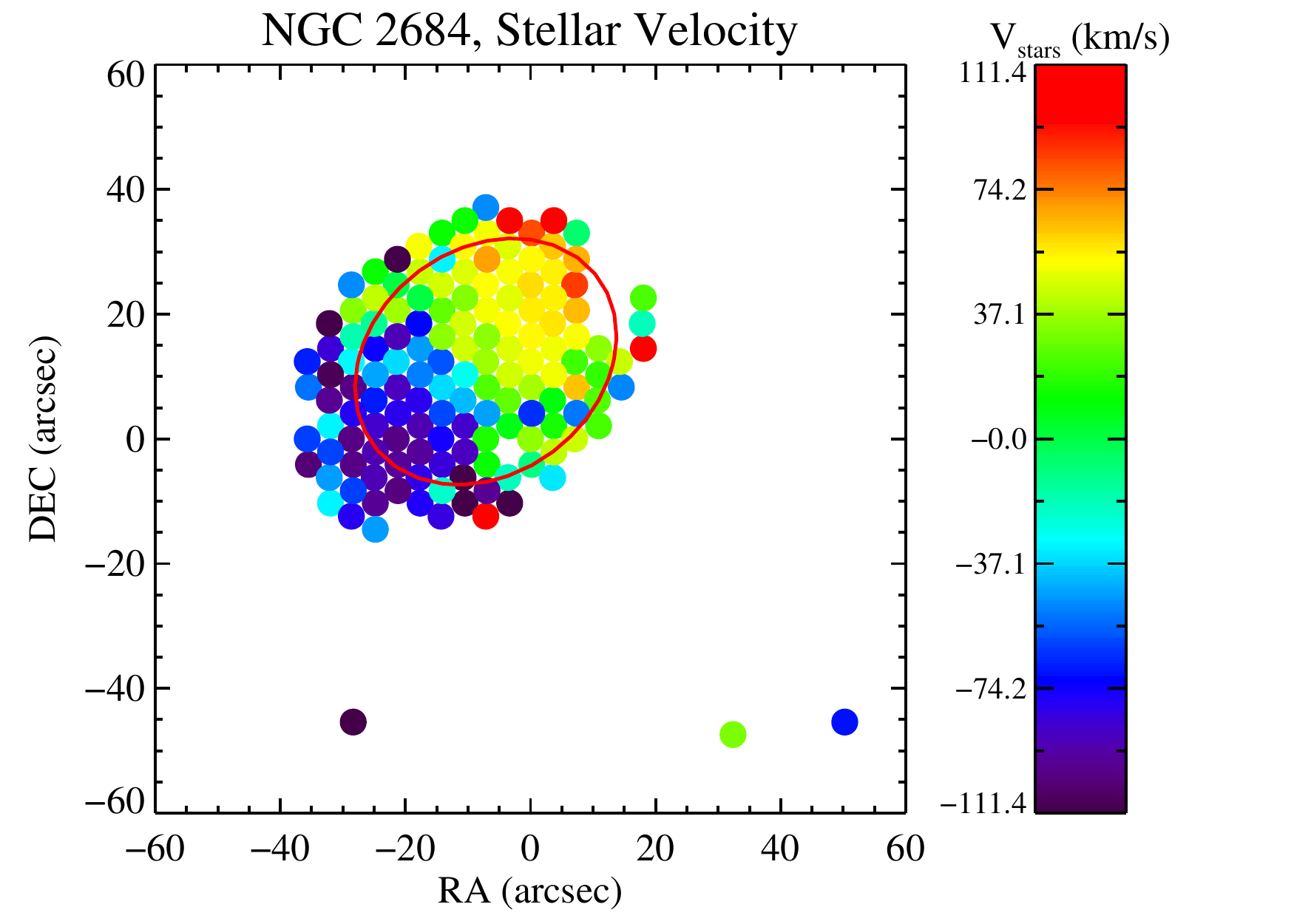}&
\includegraphics[scale=0.27]{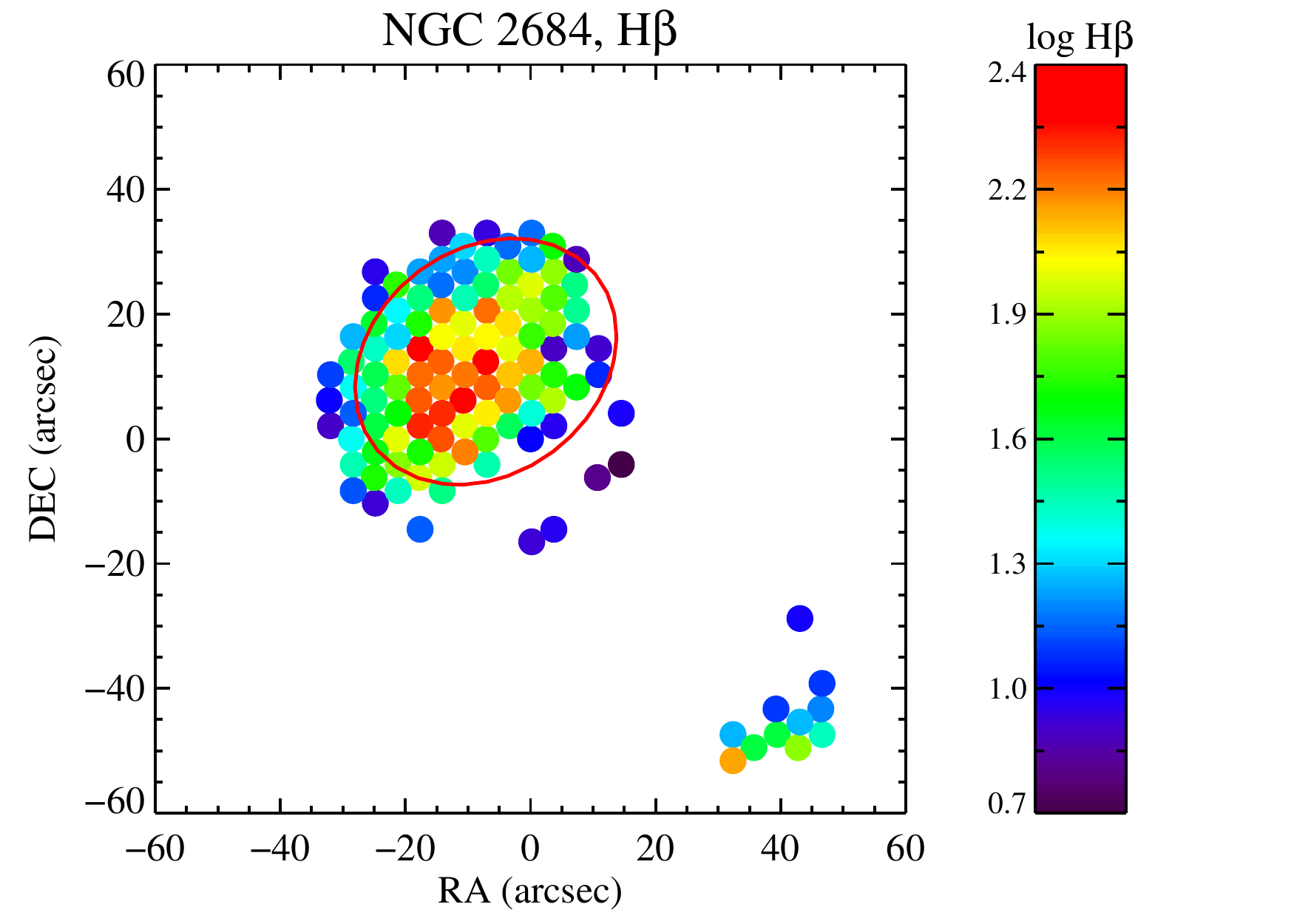}&
\includegraphics[scale=0.27]{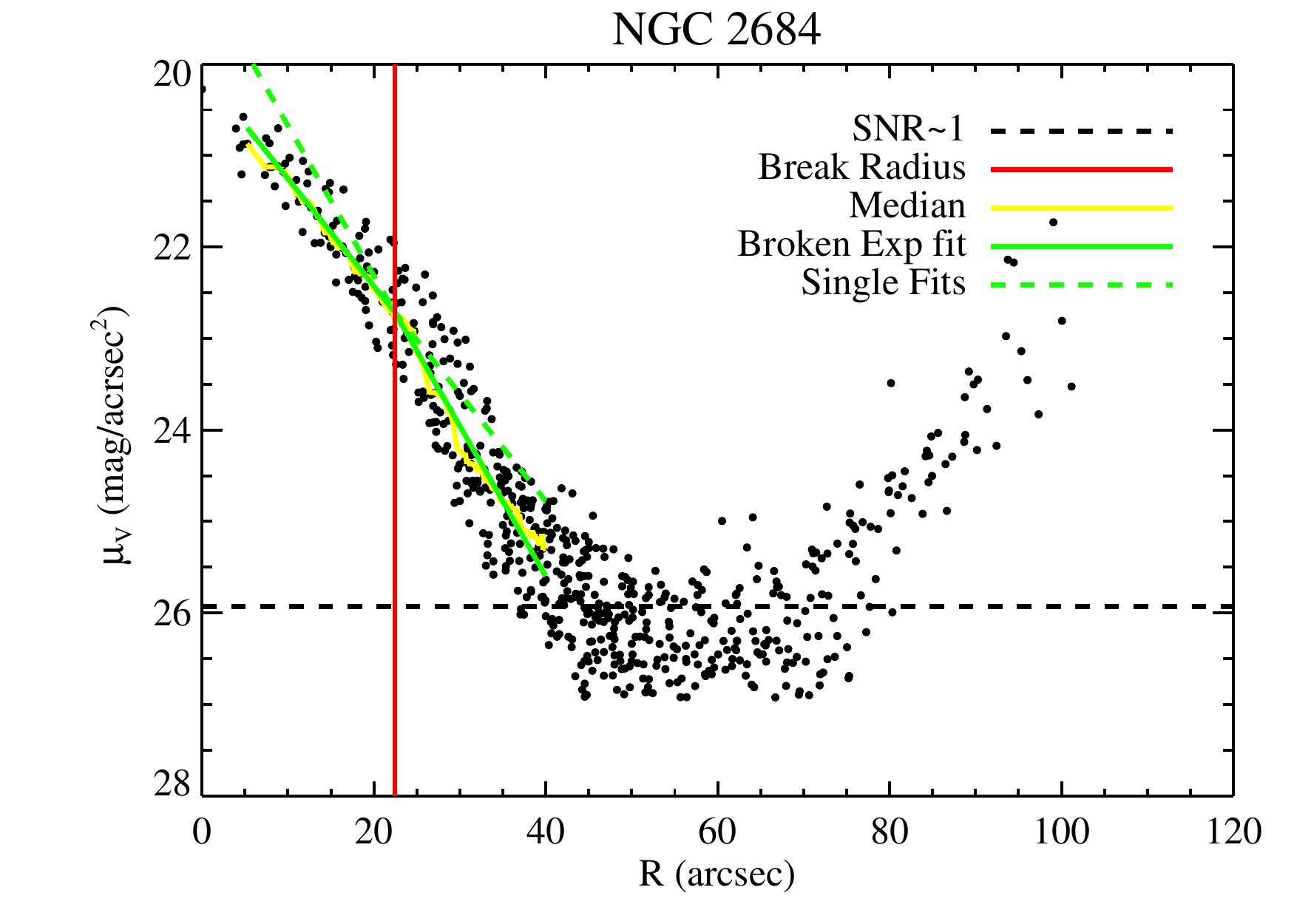}\\
\includegraphics[scale=0.27]{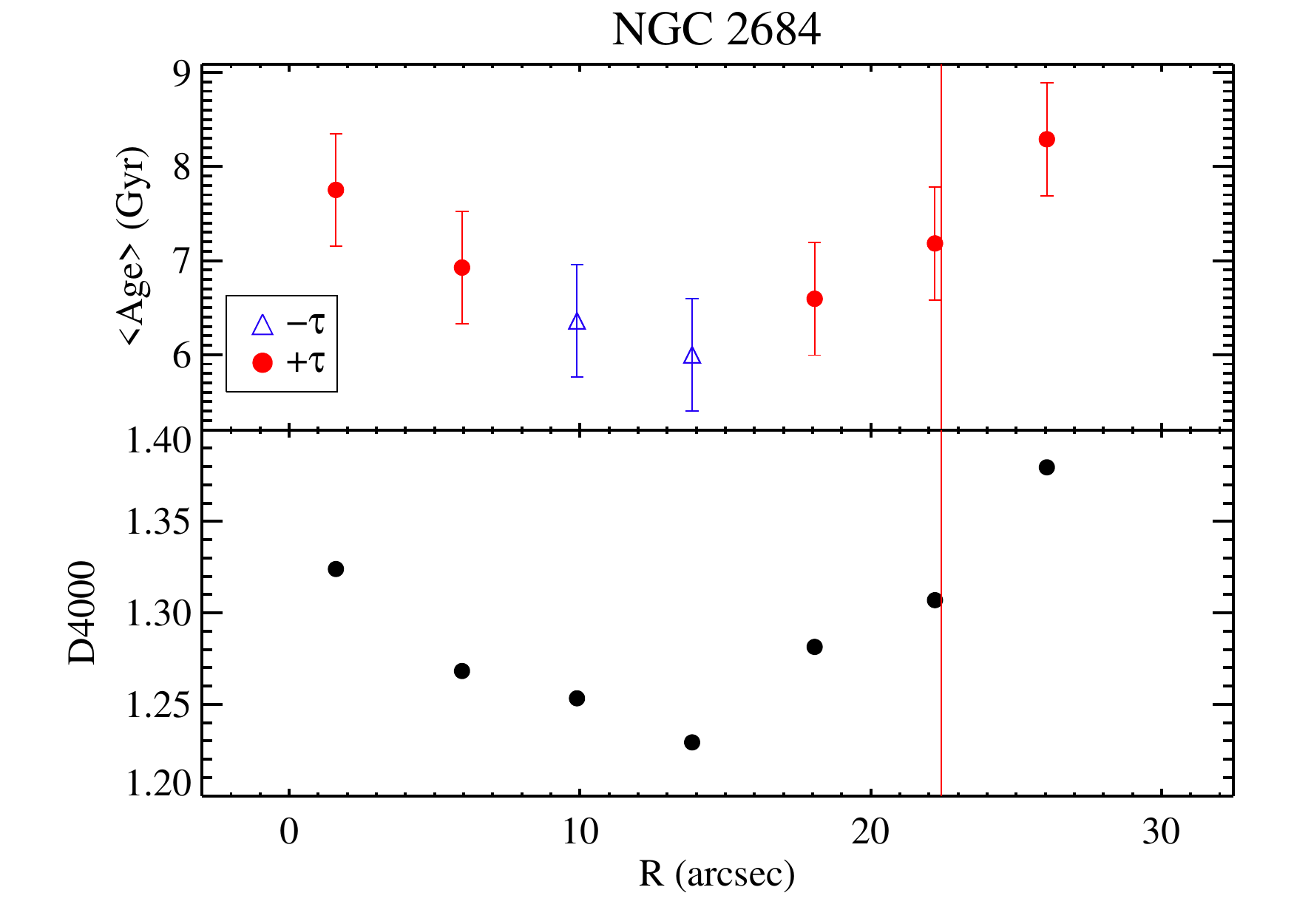}&
\includegraphics[scale=0.27]{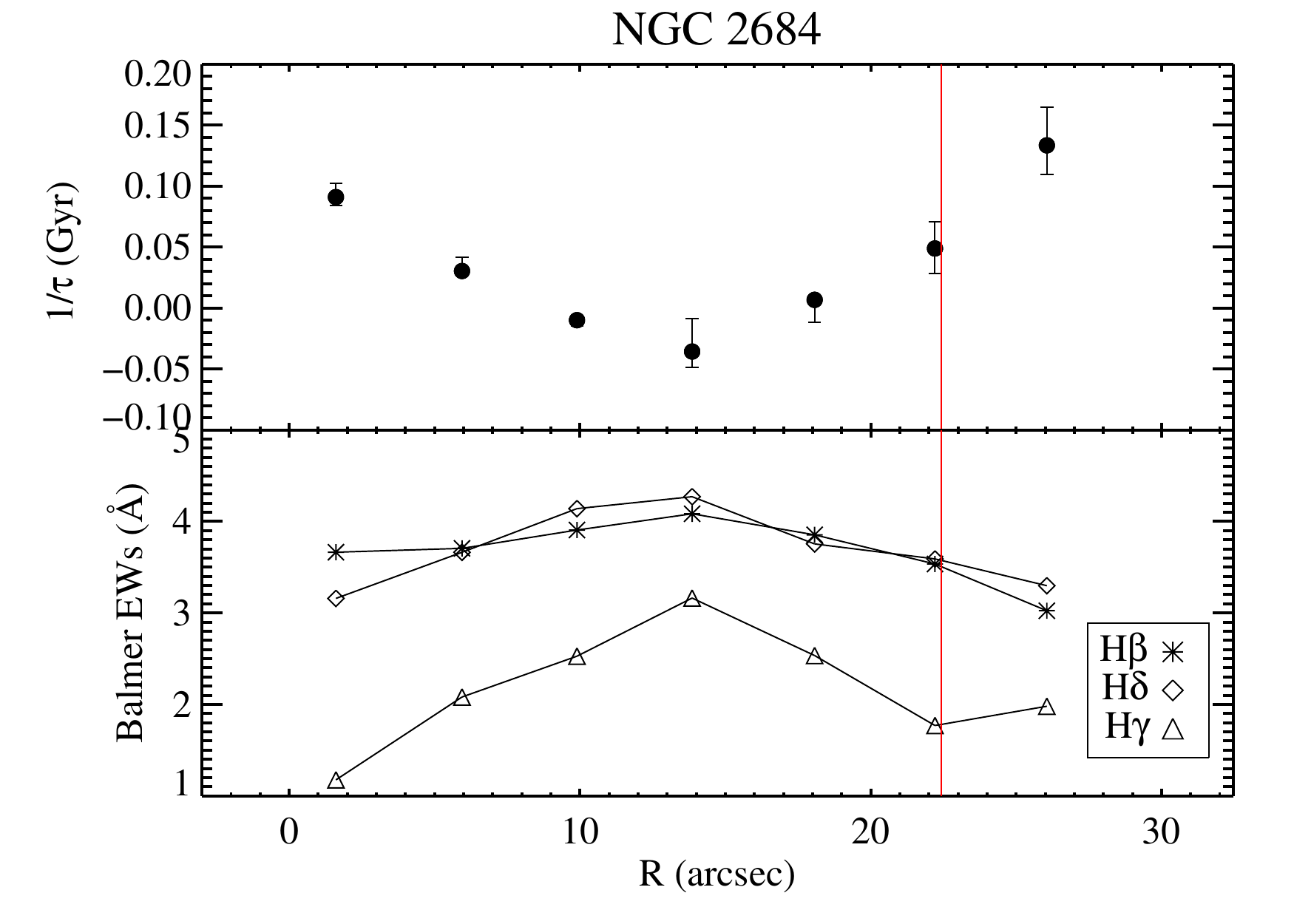}&
\includegraphics[scale=0.27]{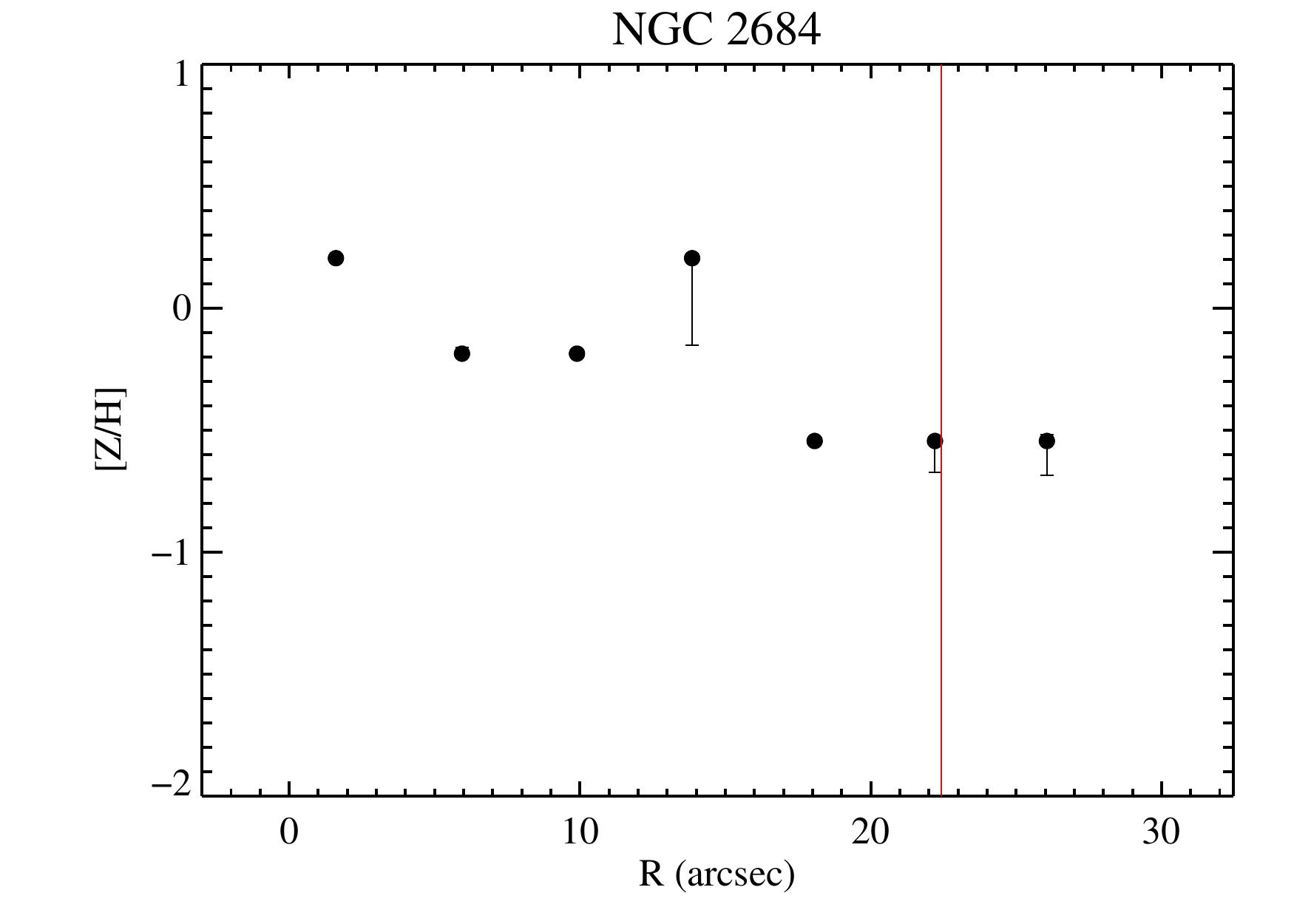}   
\end{array}$
    \epsscale{1}
 \end{center}
\caption{NGC 2684.  Same as Figure~\ref{NGC1058}. \label{NGC2684}}
\end{figure*}


\begin{figure*}
 \begin{center}$
 \begin{array}{ccc}
   \epsscale{.35}
\includegraphics[scale=0.27]{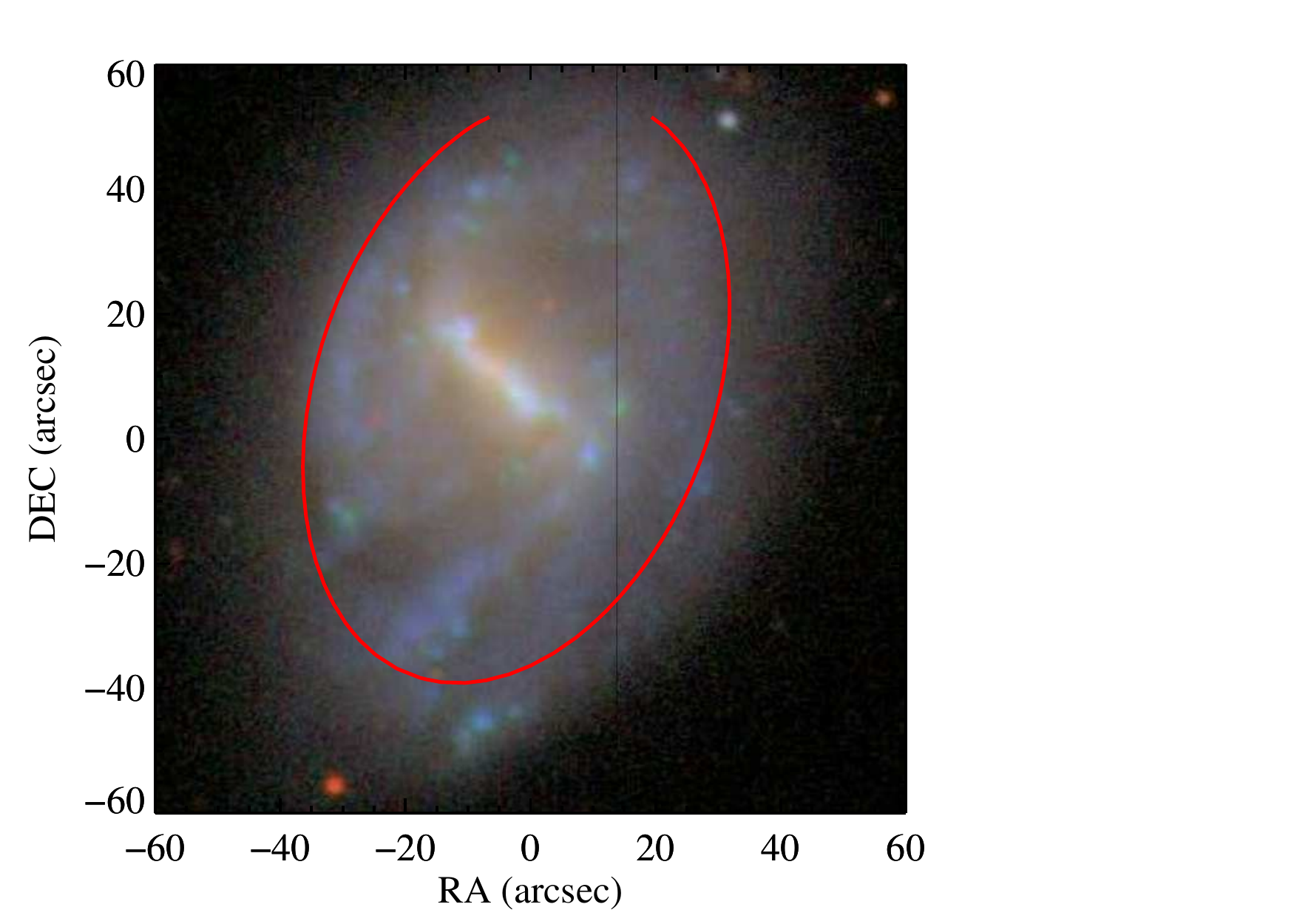}&
\includegraphics[scale=0.27]{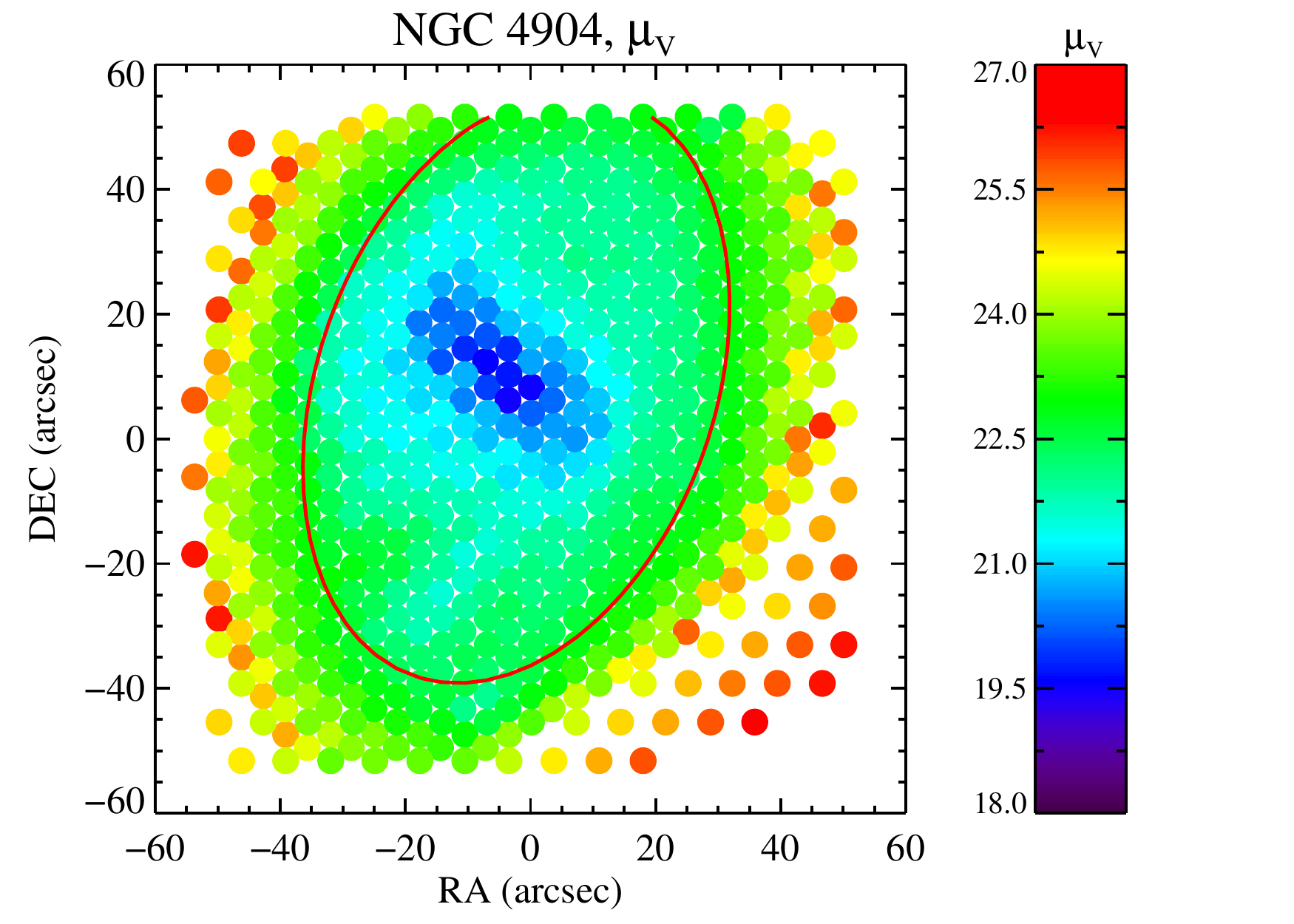}&
\includegraphics[scale=0.27]{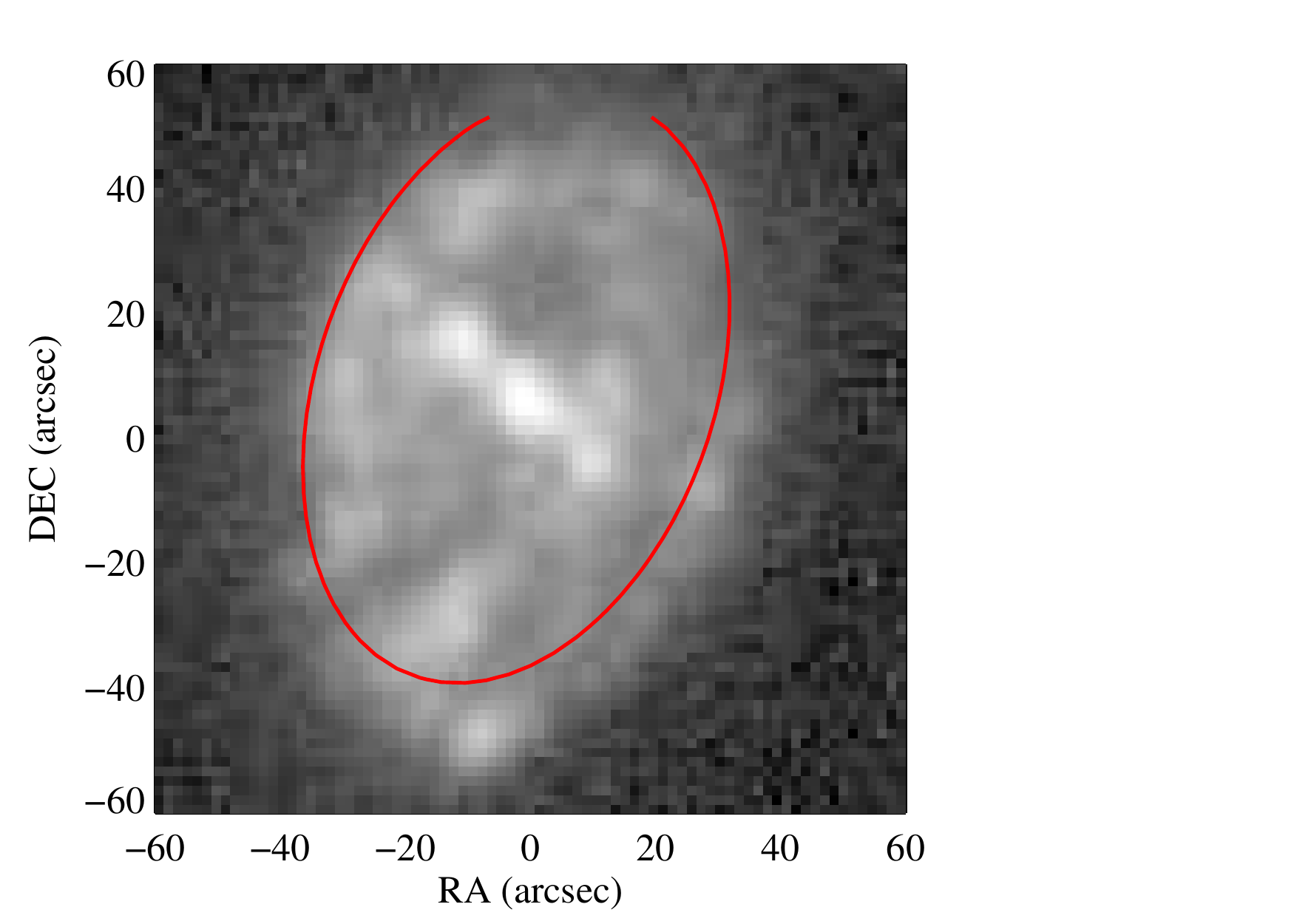}\\
\includegraphics[scale=0.27]{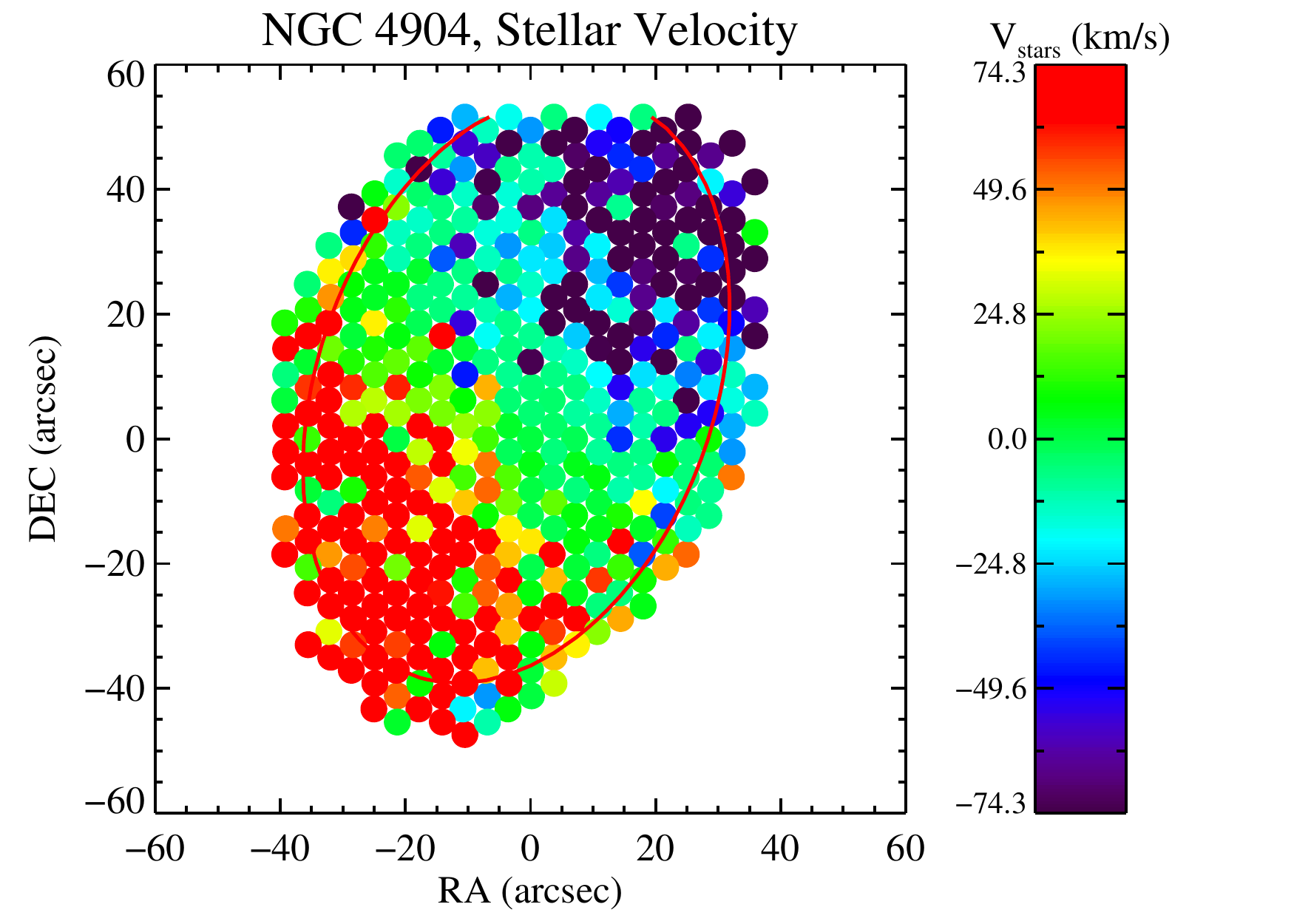}&
\includegraphics[scale=0.27]{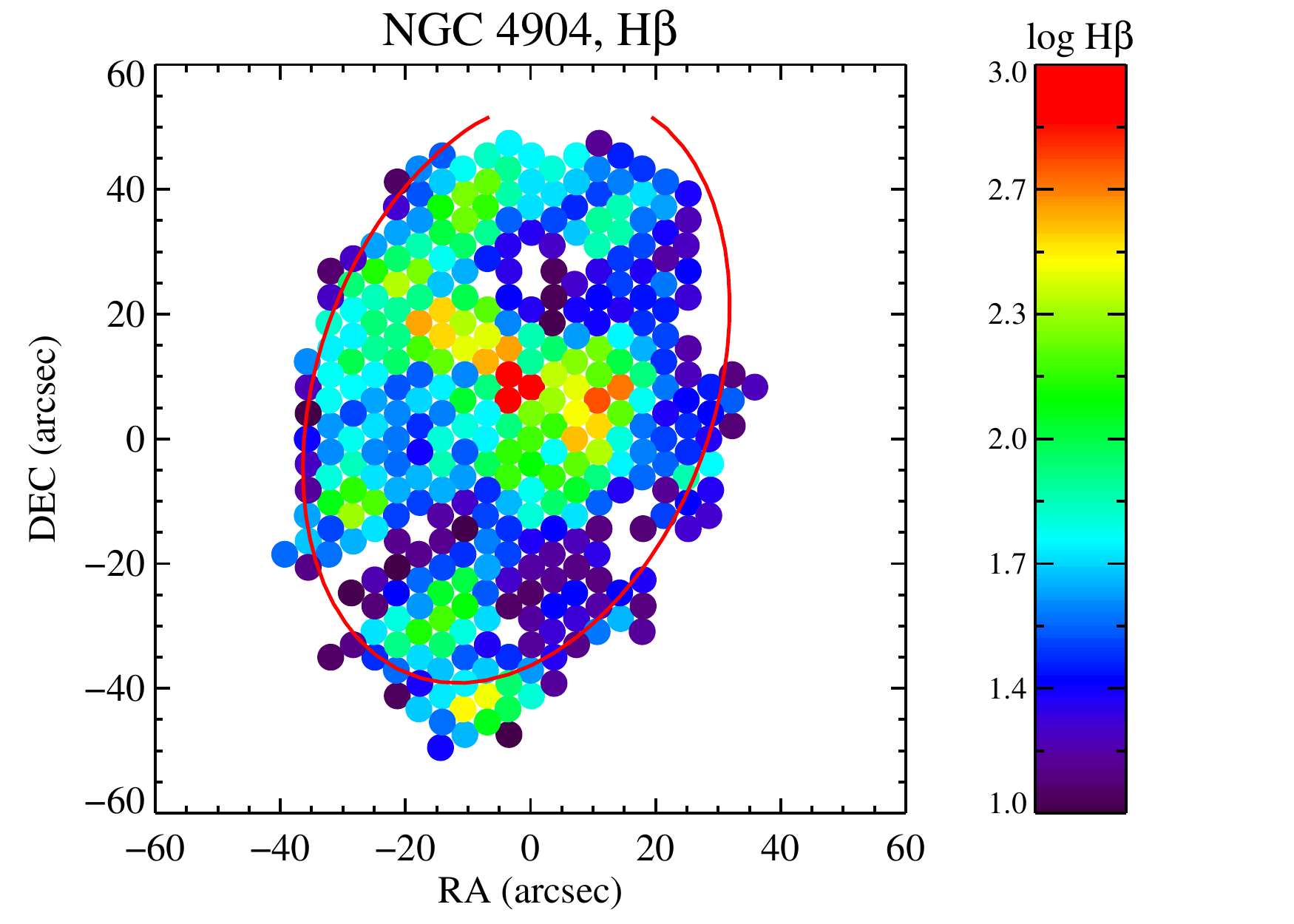}&
\includegraphics[scale=0.27]{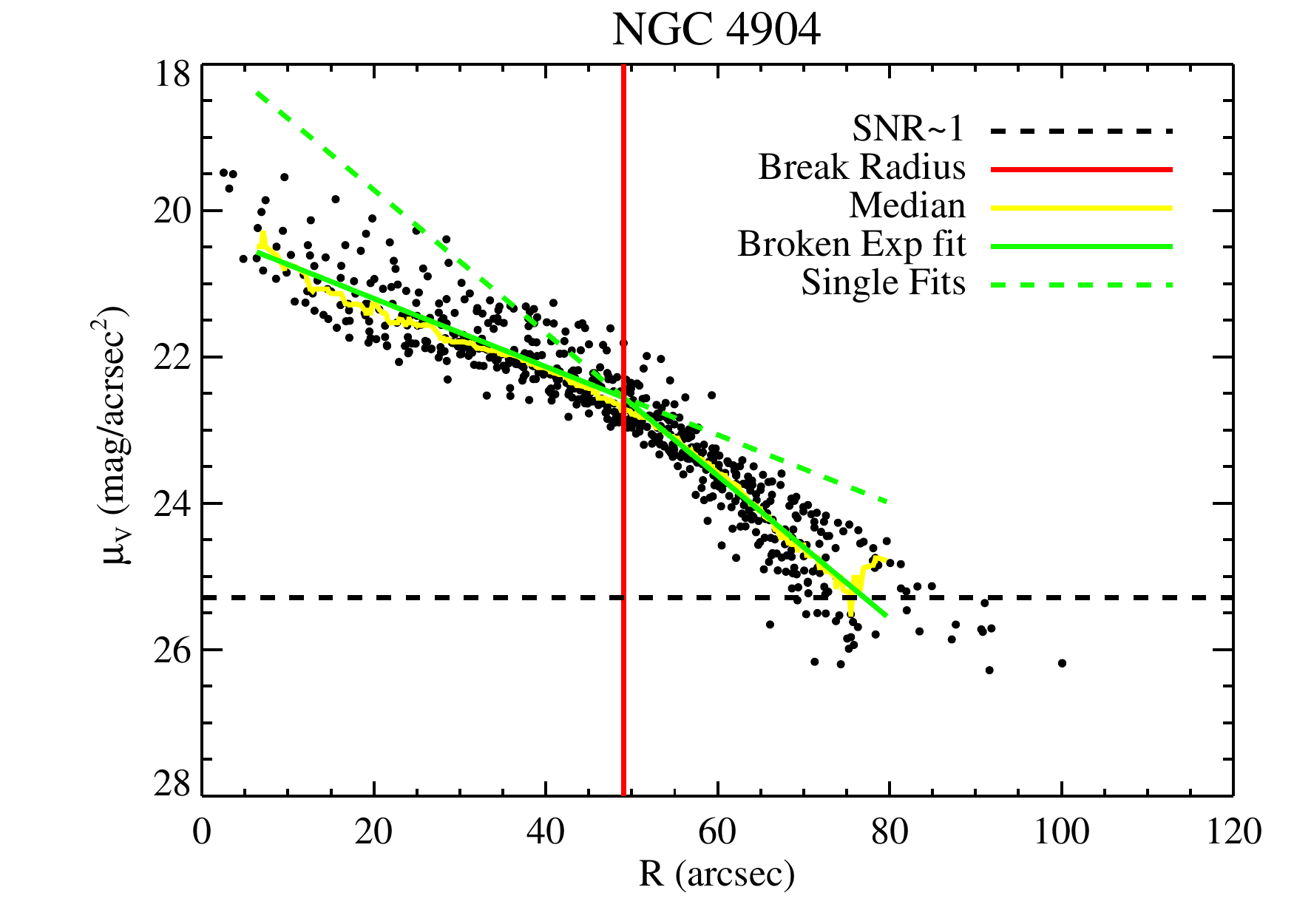}\\
\includegraphics[scale=0.27]{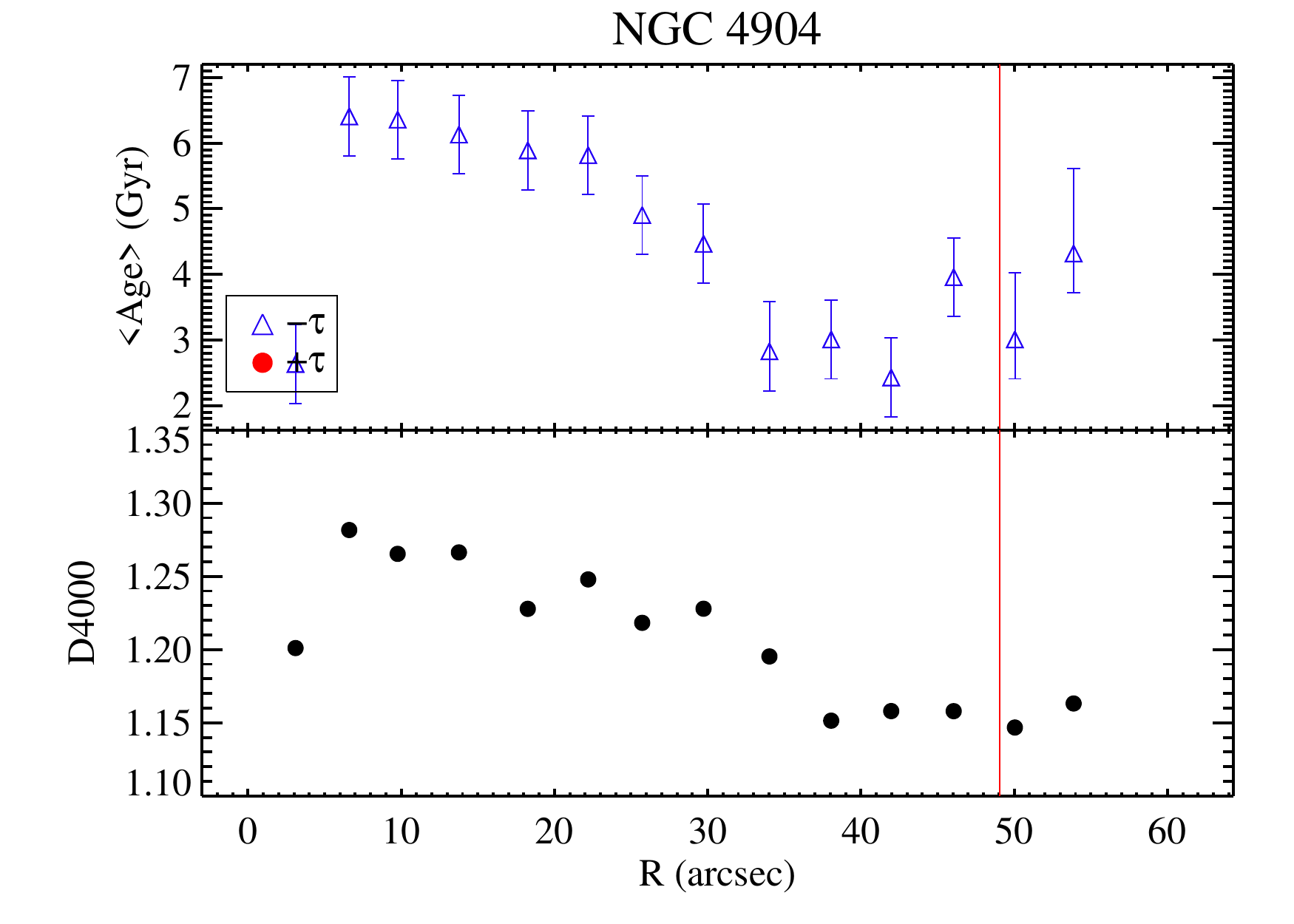}&
\includegraphics[scale=0.27]{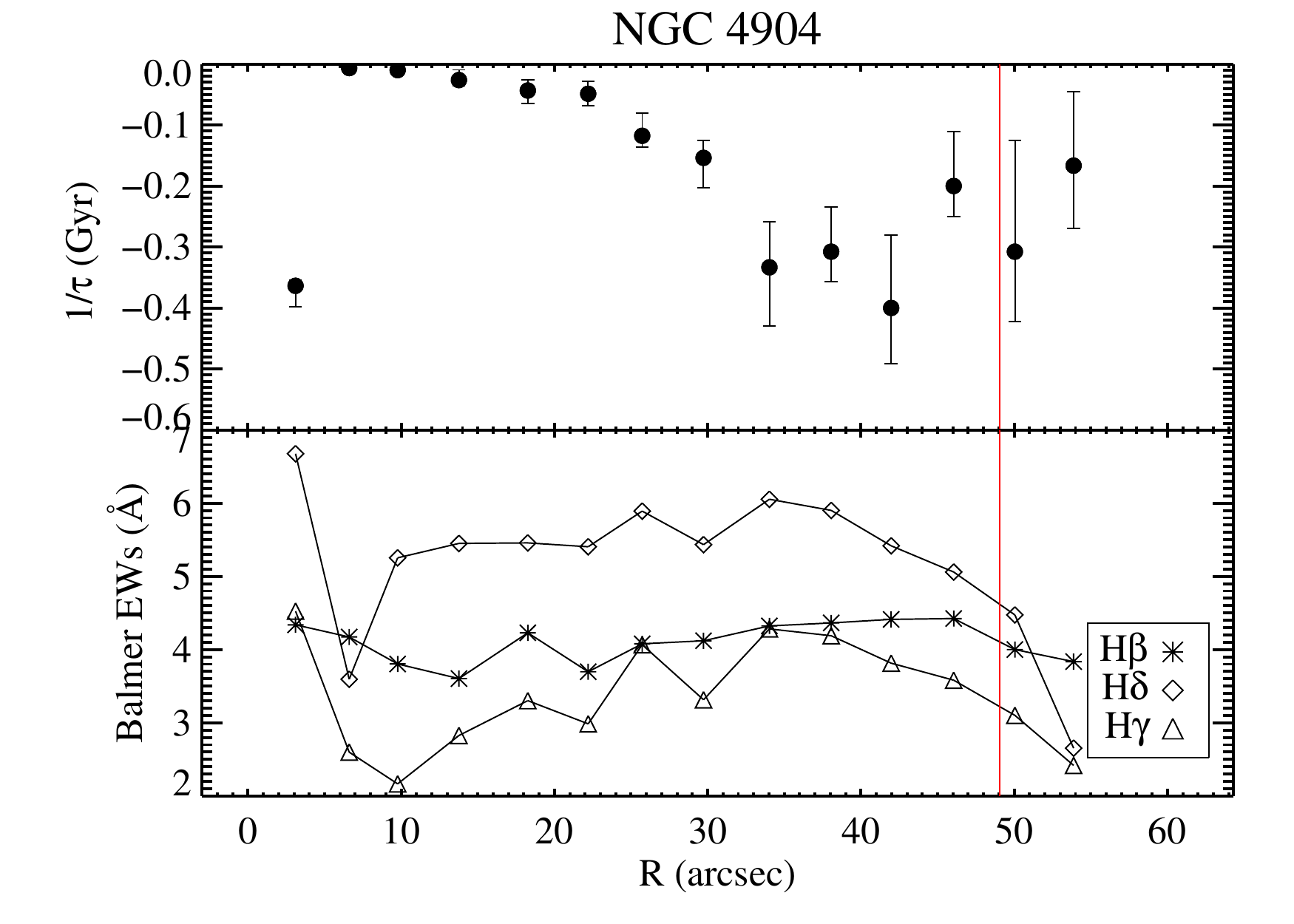}&
\includegraphics[scale=0.27]{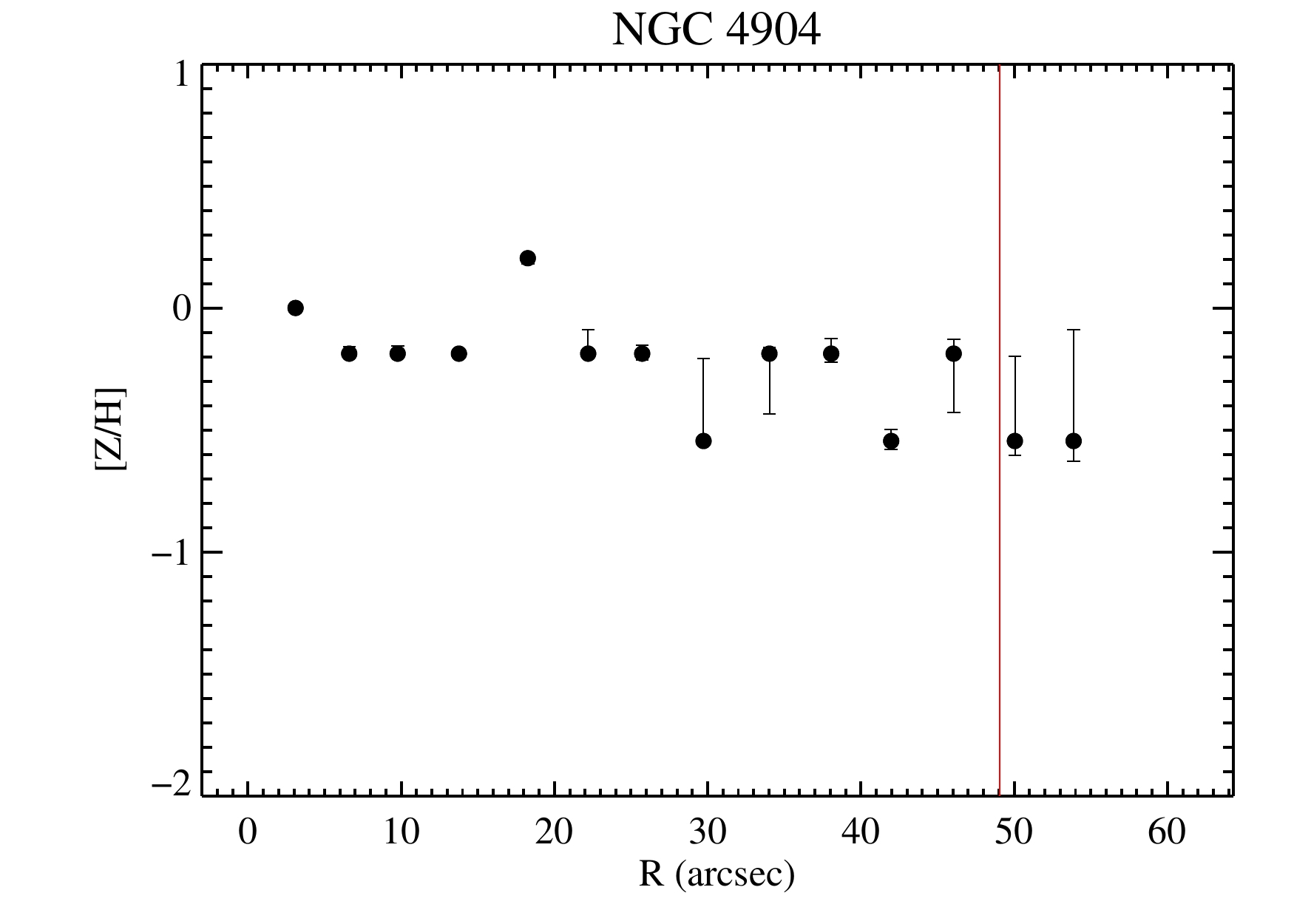}   
\end{array}$
    \epsscale{1}
 \end{center}
\caption{NGC 4904.  Same as Figure~\ref{NGC1058}. \label{NGC4904}}
\end{figure*}

\begin{figure*}
 \begin{center}$
 \begin{array}{ccc}
   \epsscale{.35}
\includegraphics[scale=0.27]{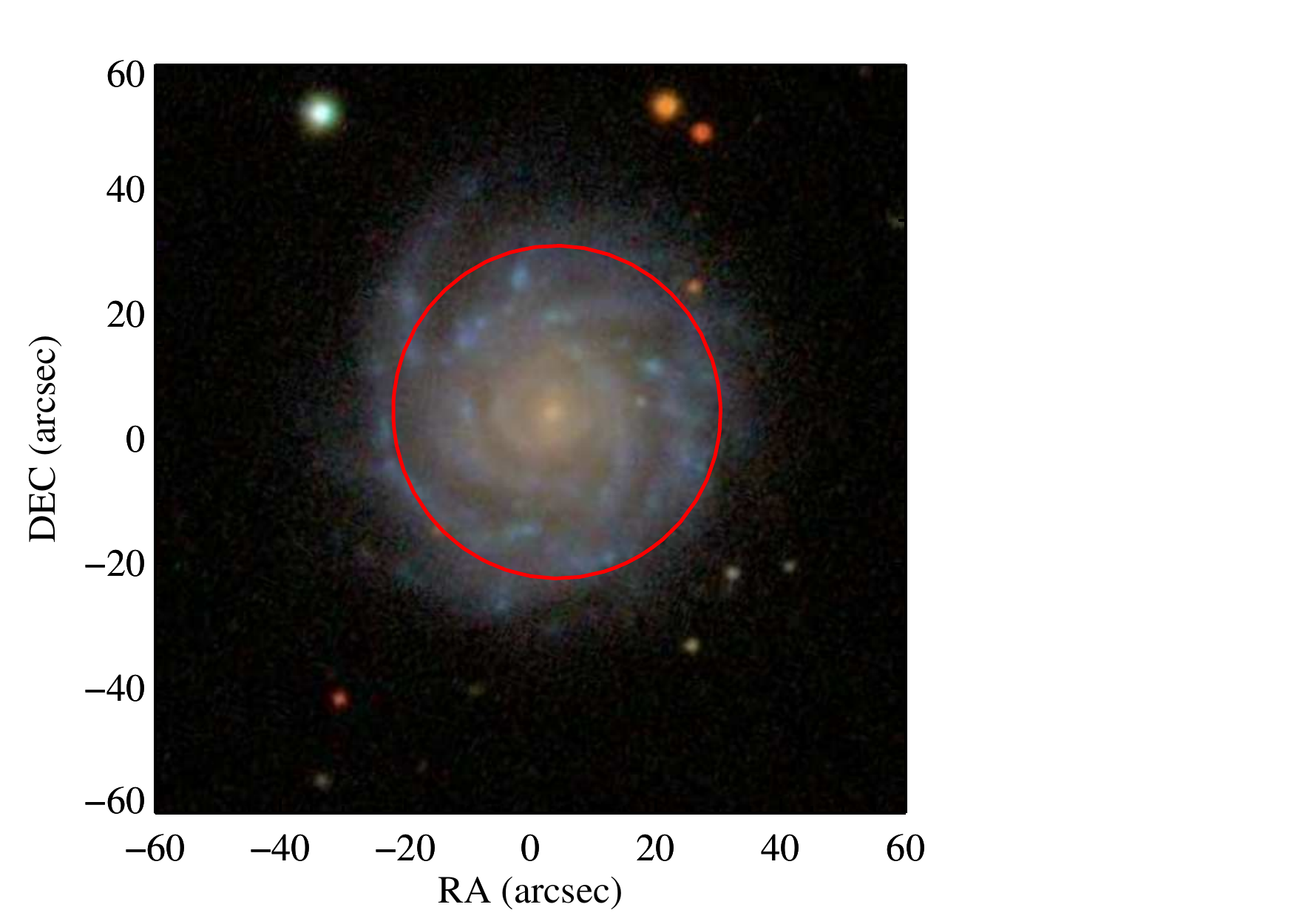}&
\includegraphics[scale=0.27]{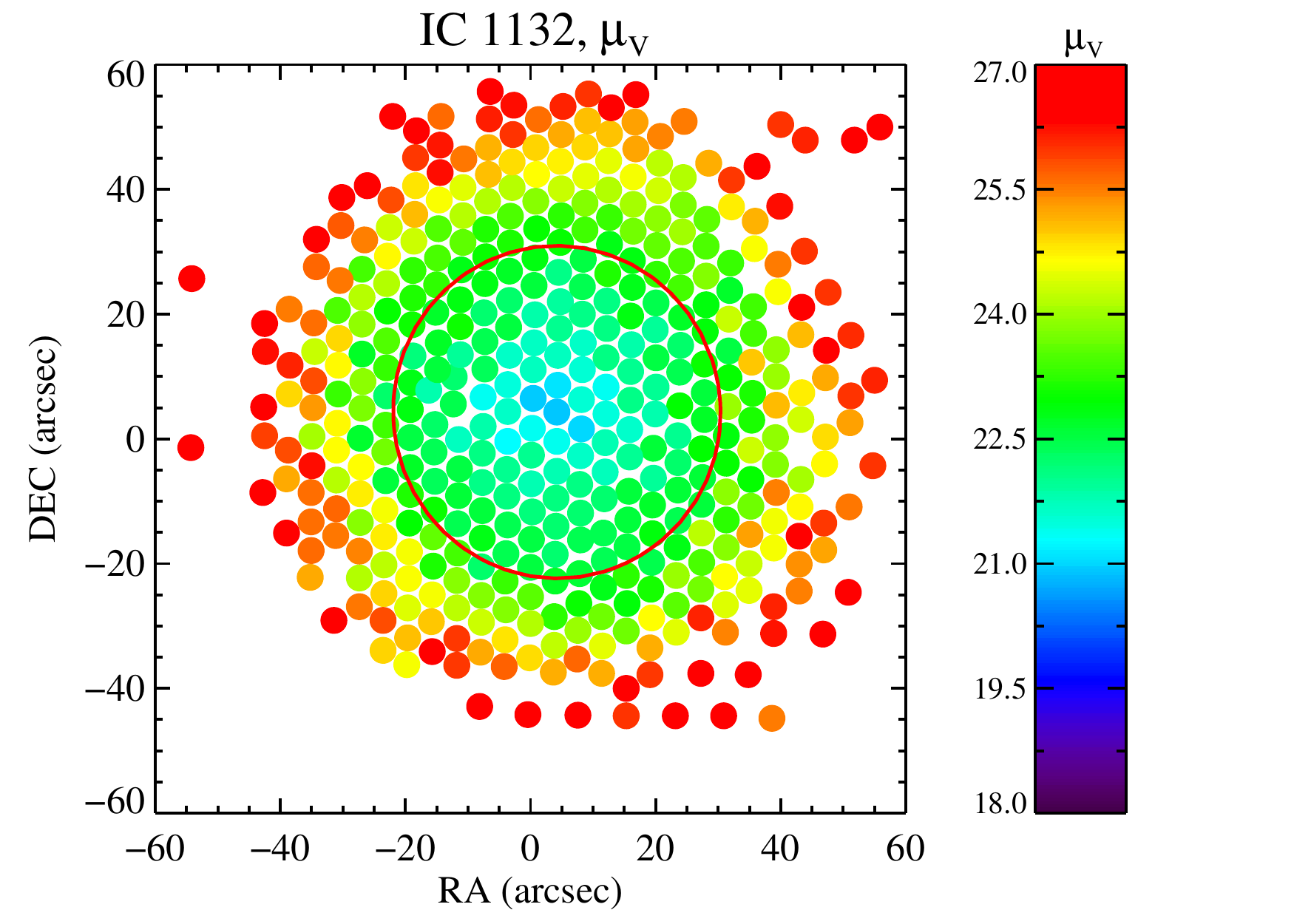}\\
\includegraphics[scale=0.27]{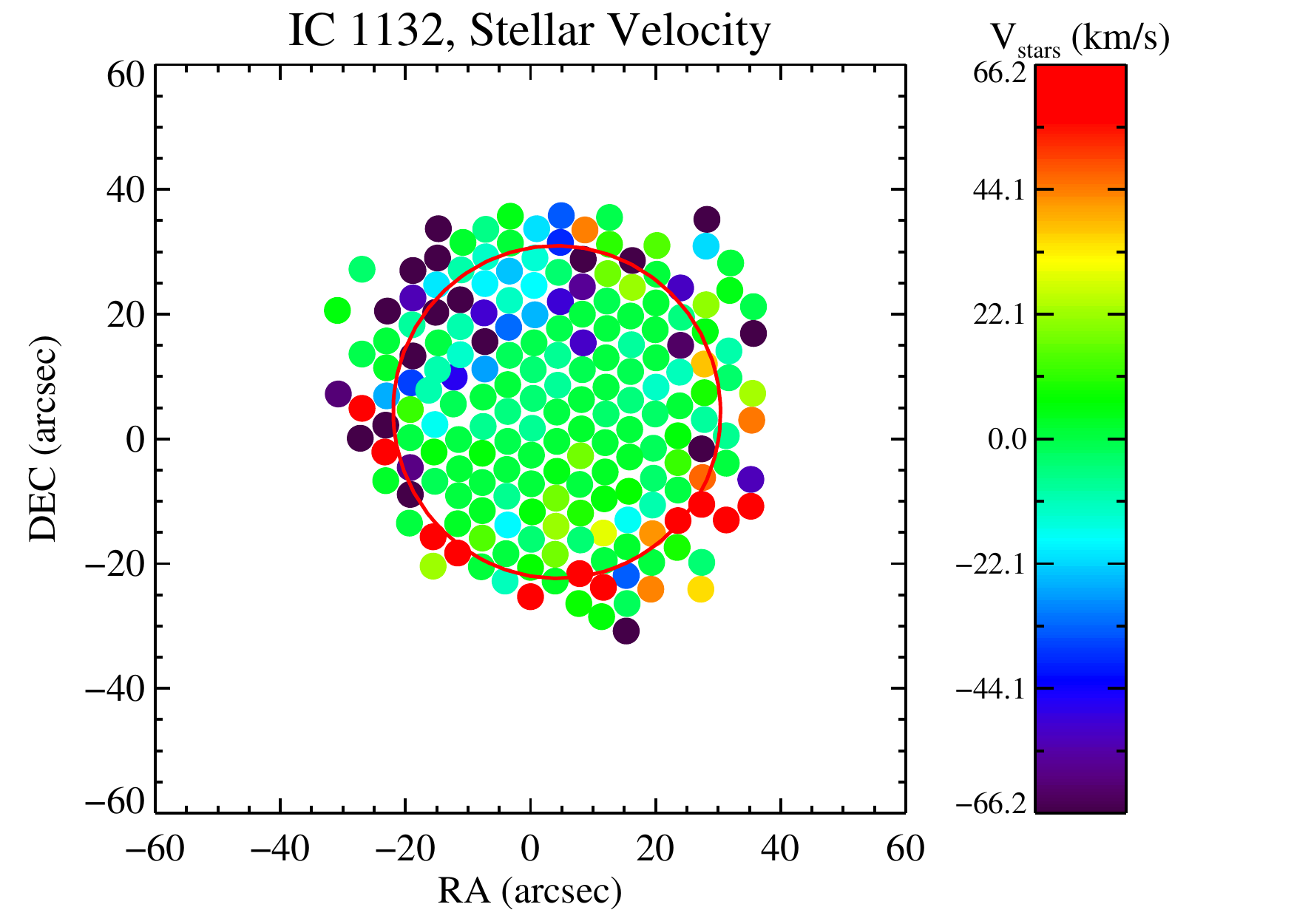}&
\includegraphics[scale=0.27]{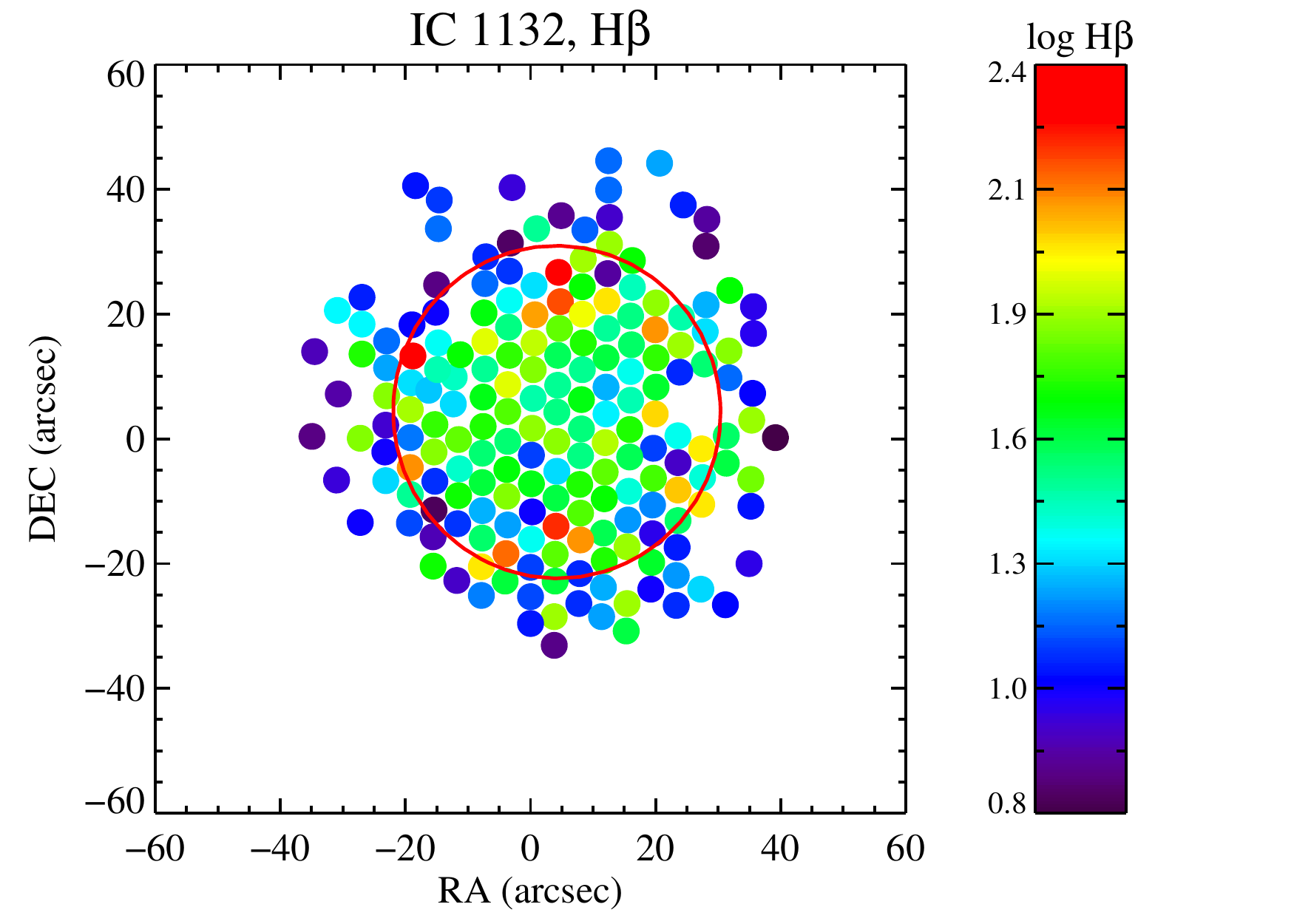}&
\includegraphics[scale=0.27]{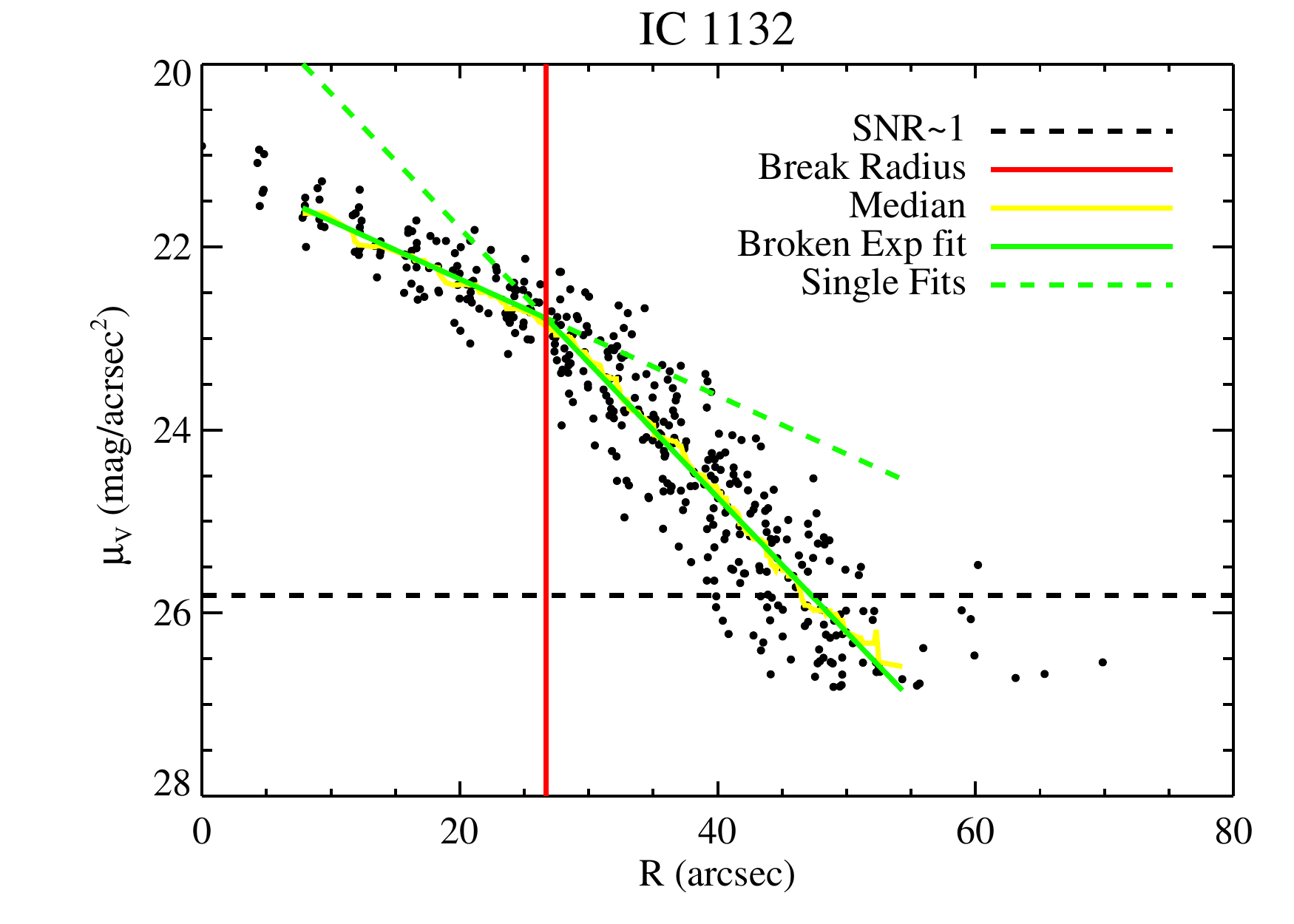}\\
\includegraphics[scale=0.27]{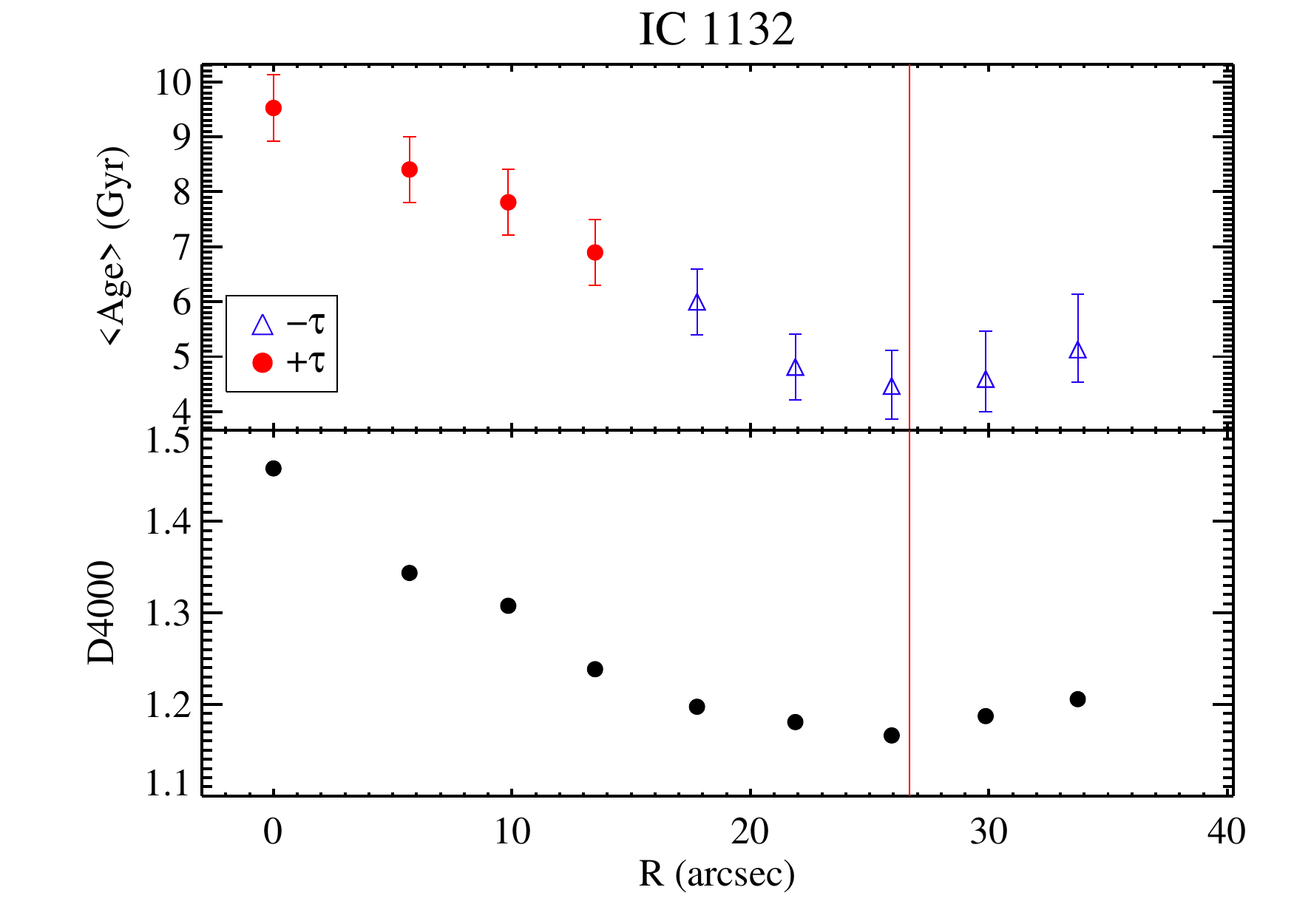}&
\includegraphics[scale=0.27]{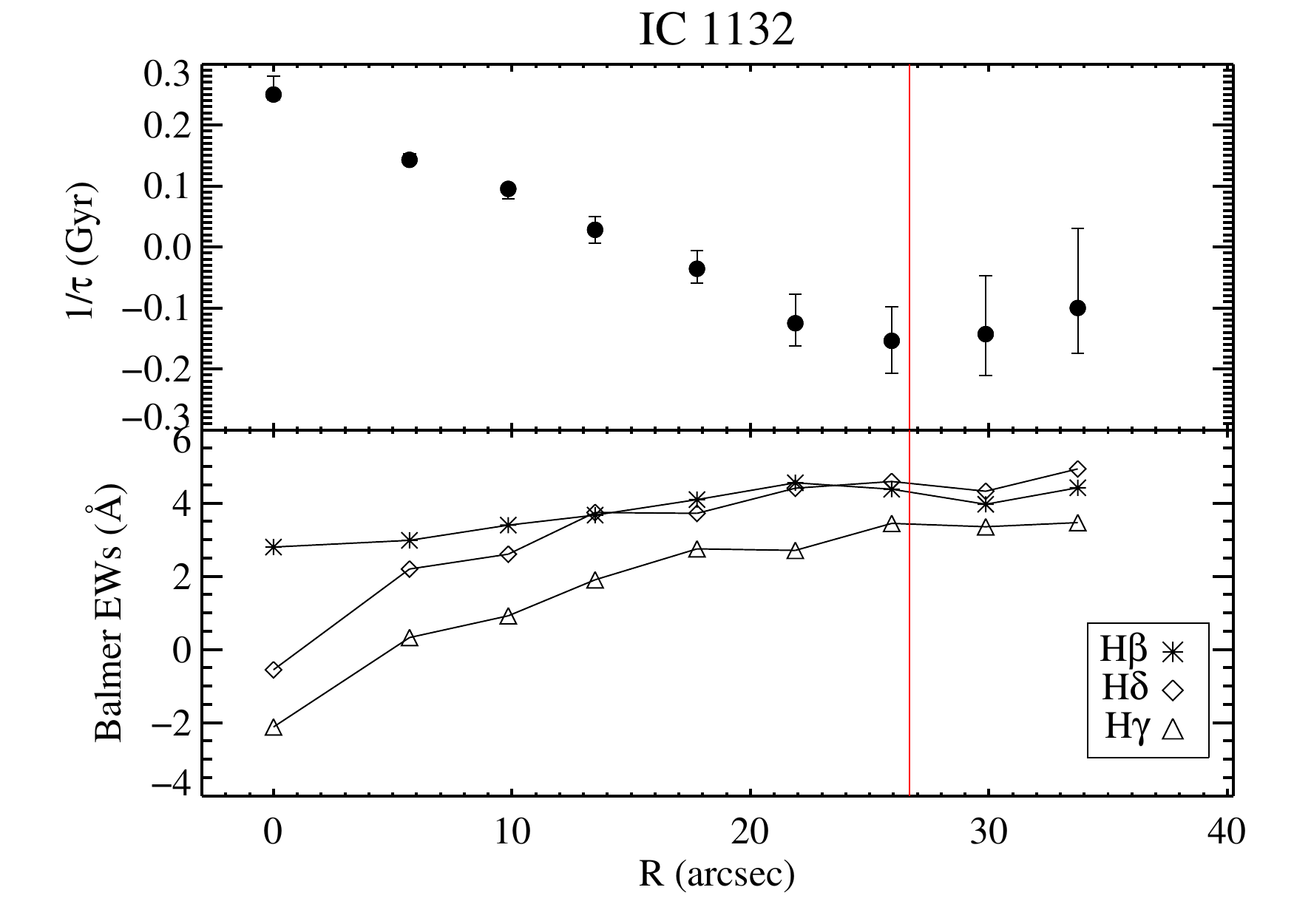}&
\includegraphics[scale=0.27]{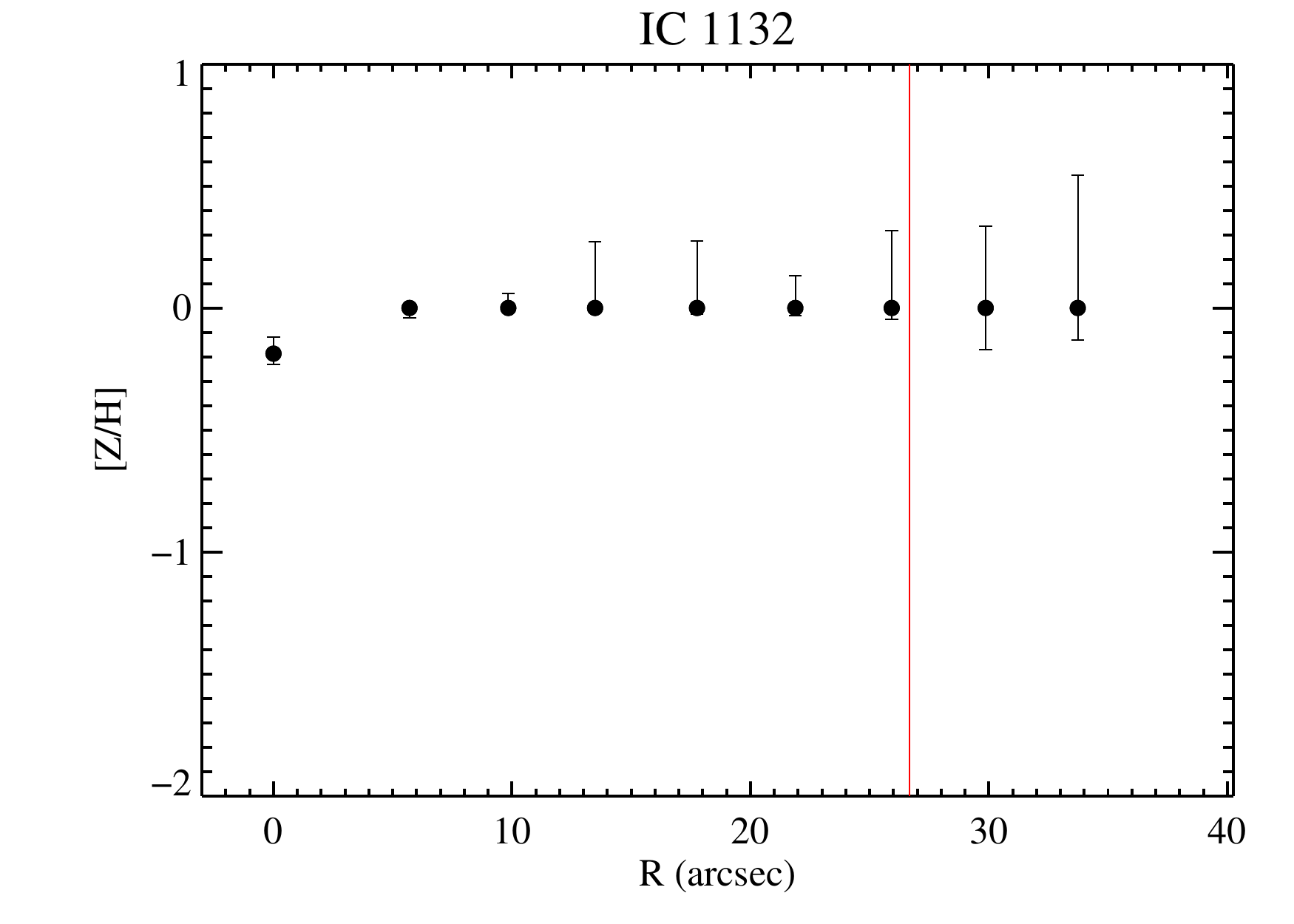}   
\end{array}$
    \epsscale{1}
 \end{center}
\caption{IC 1132.  Same as Figure~\ref{NGC1058}. \label{IC1132}}
\end{figure*}


\begin{figure*}
 \begin{center}$
 \begin{array}{ccc}
   \epsscale{.35}
\includegraphics[scale=0.27]{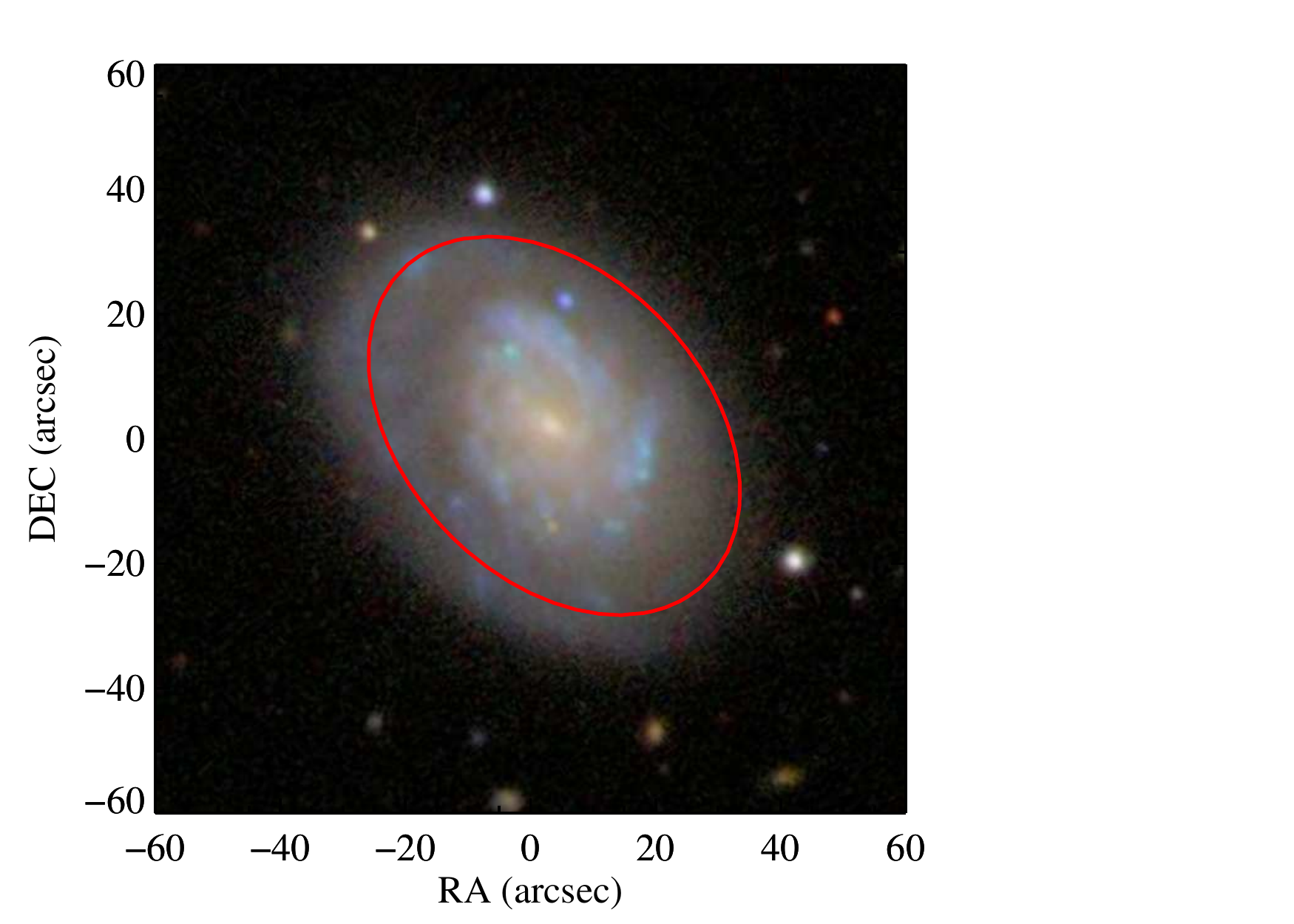}&
\includegraphics[scale=0.27]{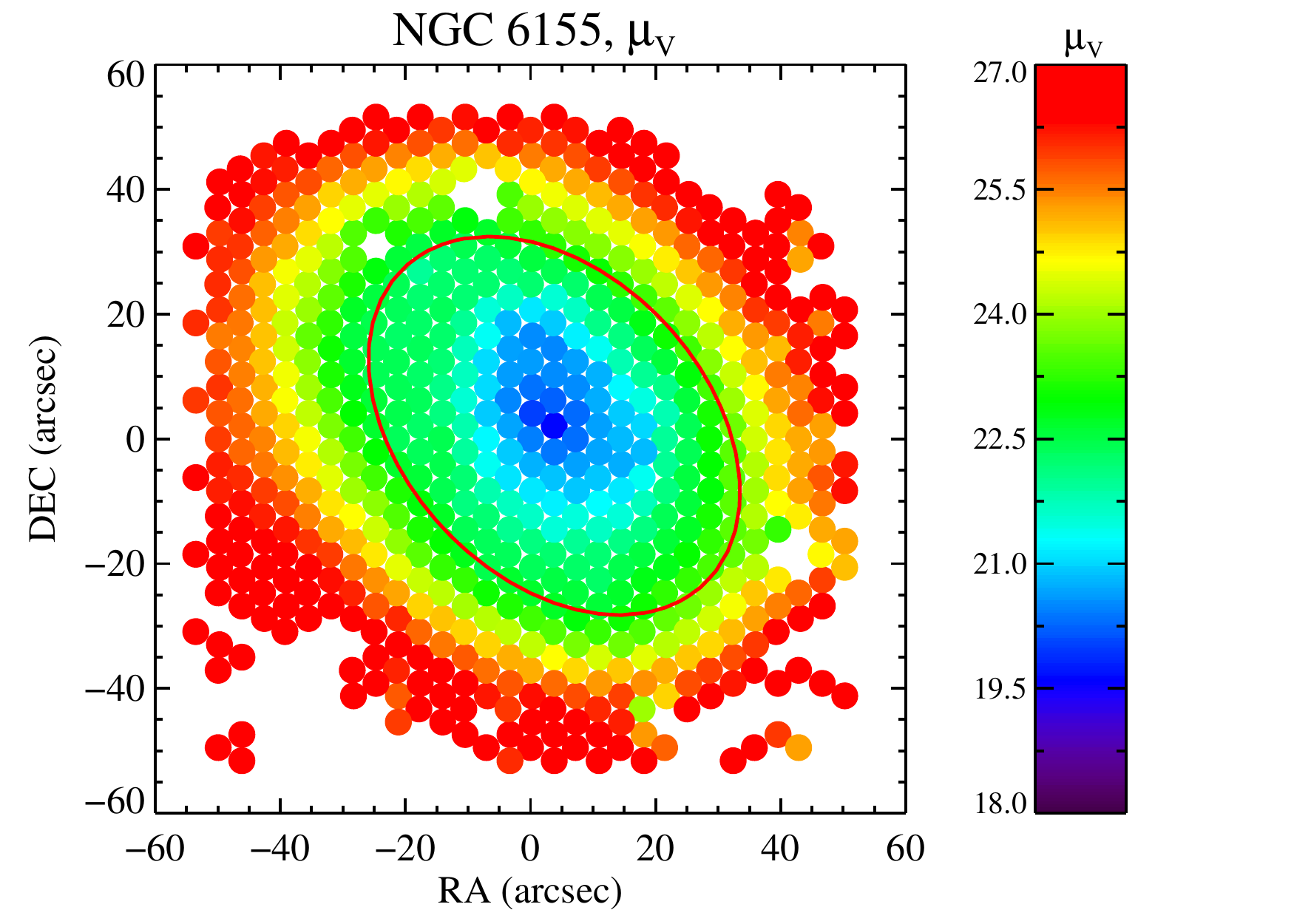}&
\includegraphics[scale=0.27]{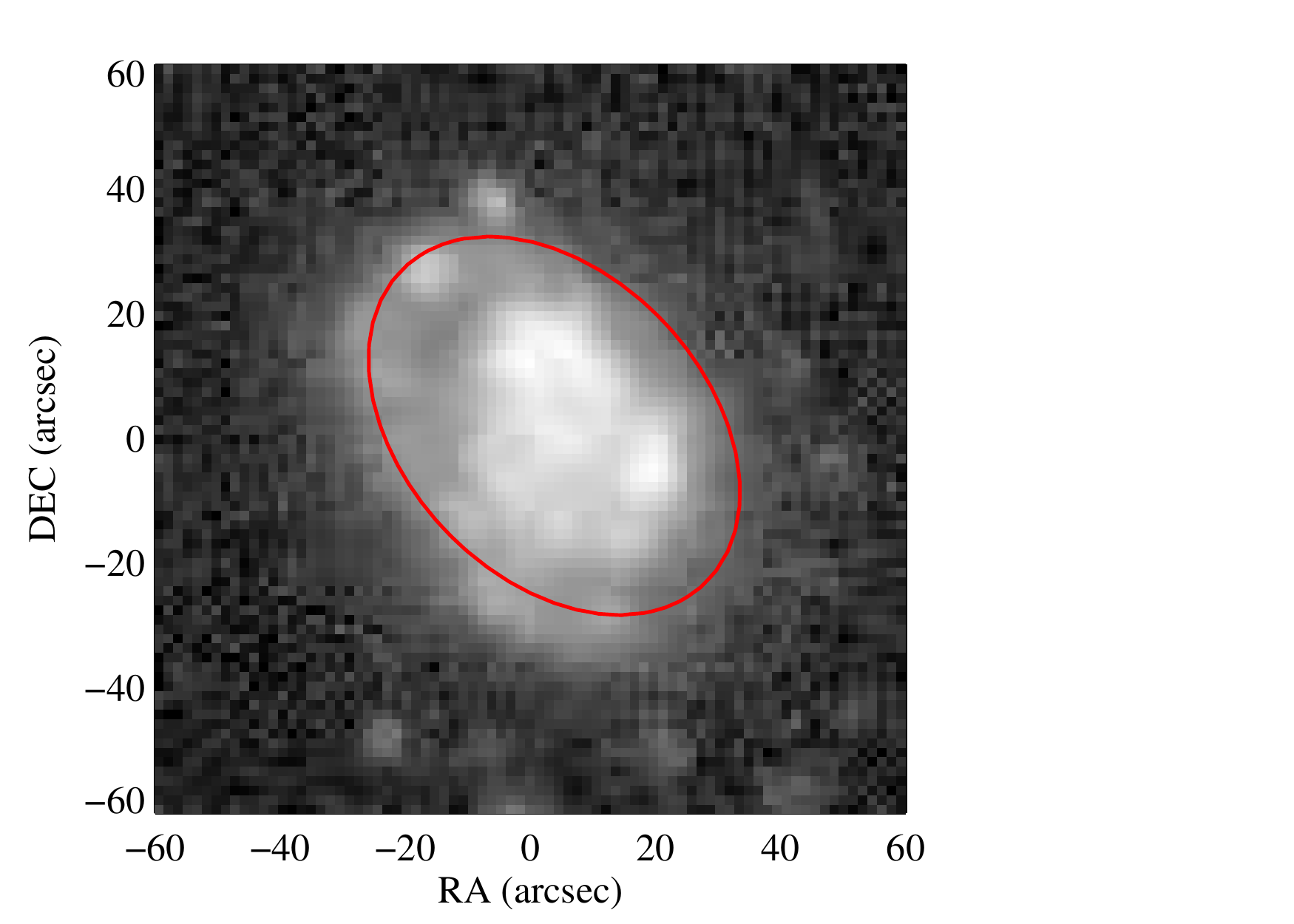}\\
\includegraphics[scale=0.27]{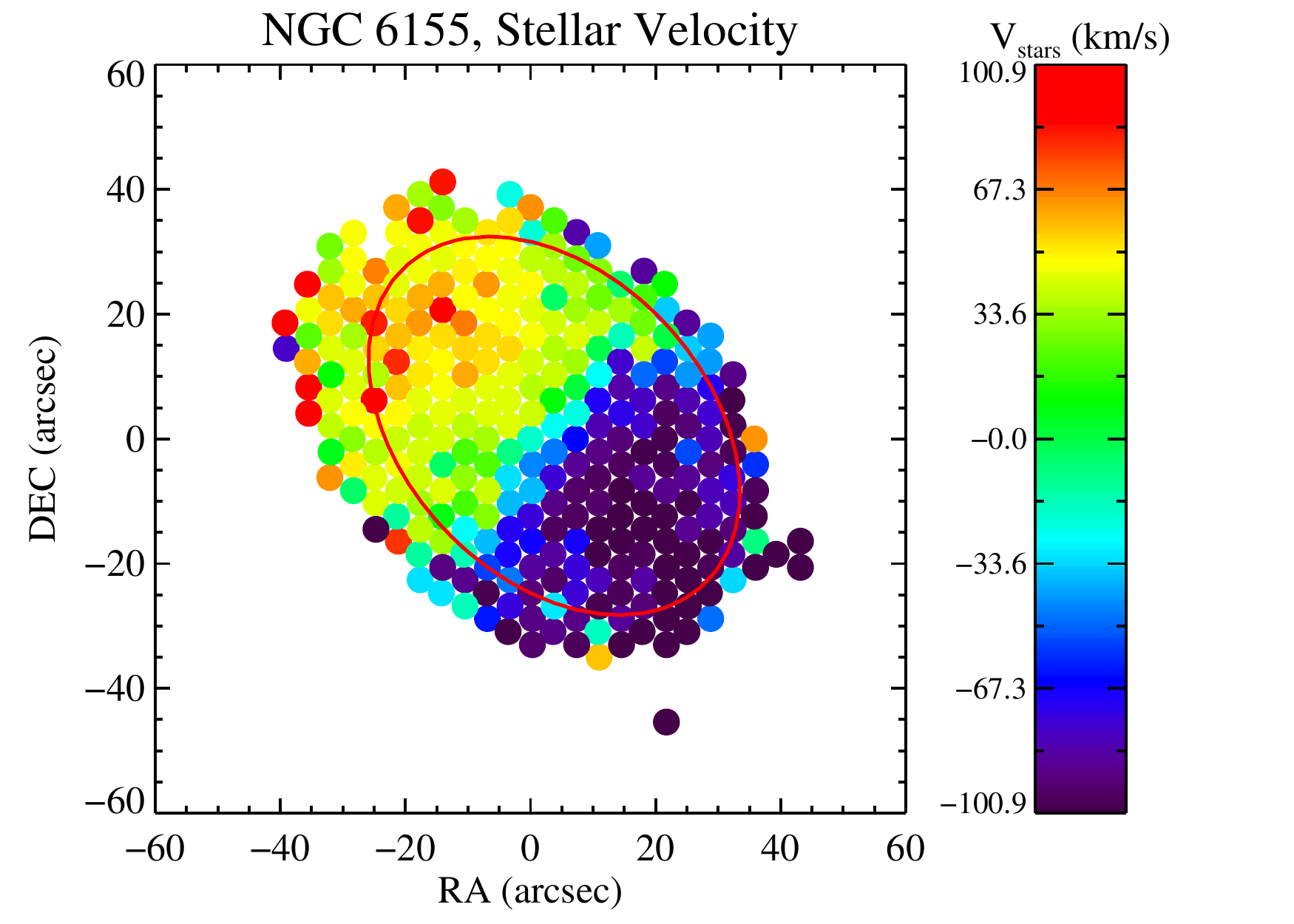}&
\includegraphics[scale=0.27]{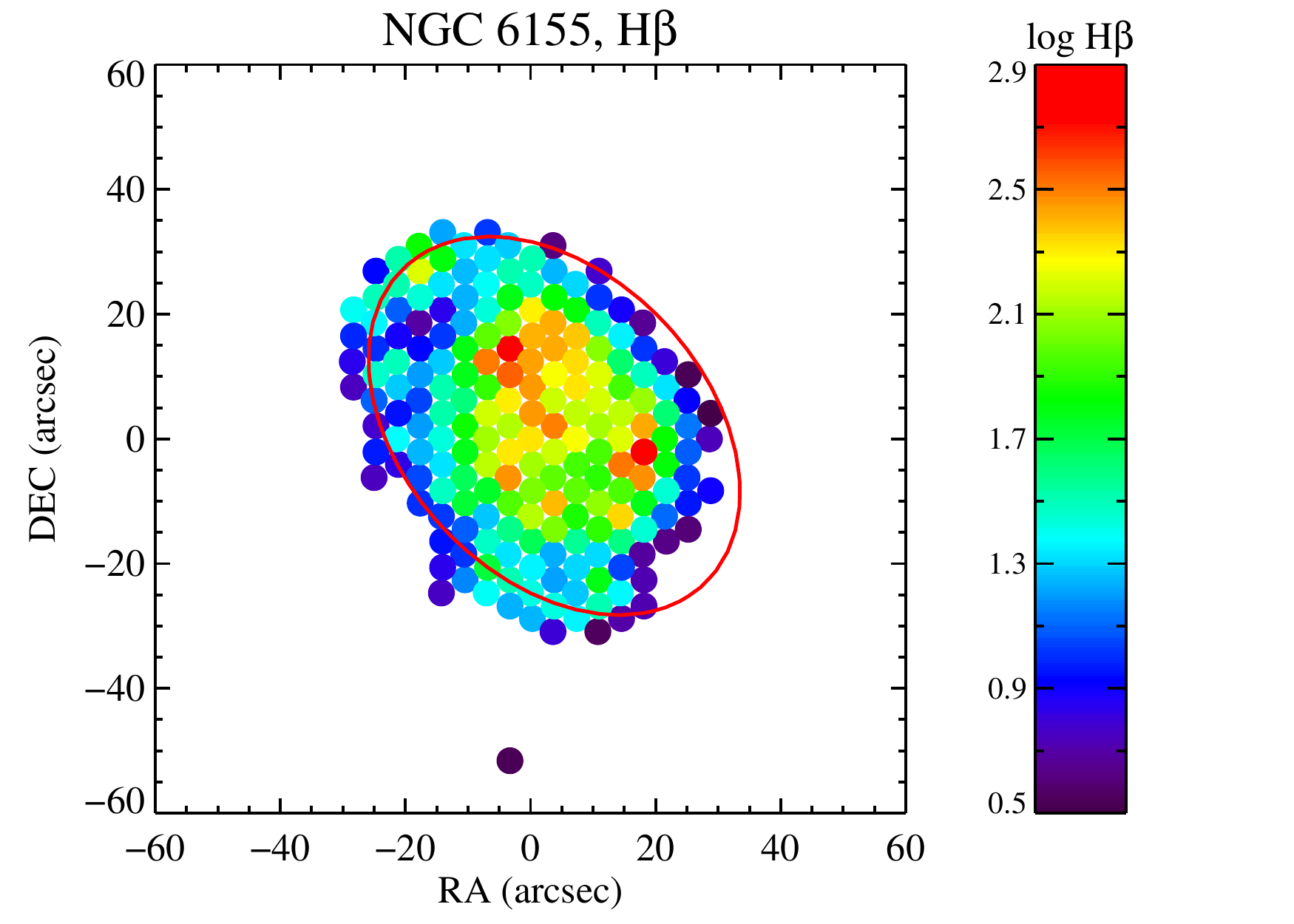}&
\includegraphics[scale=0.27]{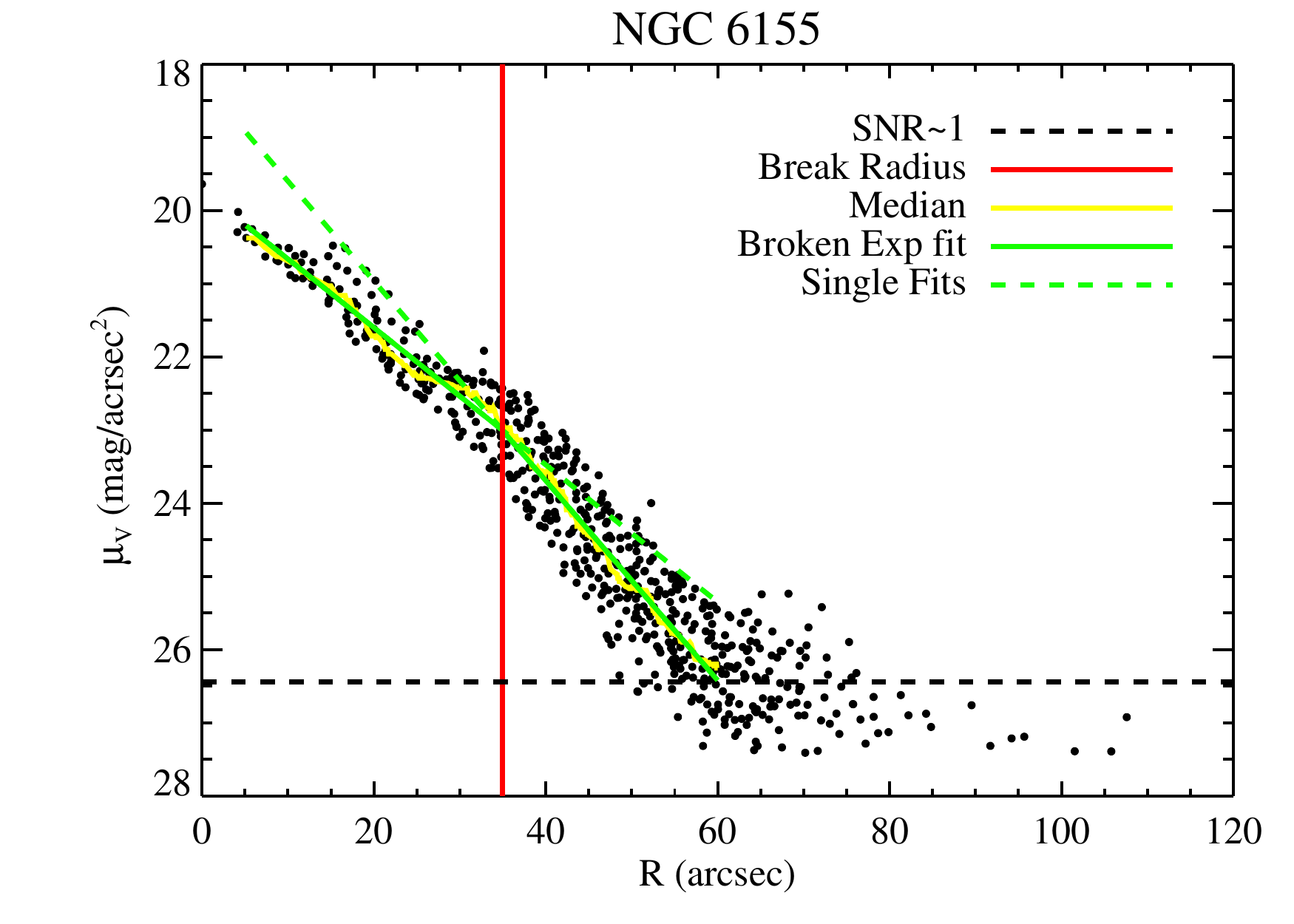}\\
\includegraphics[scale=0.27]{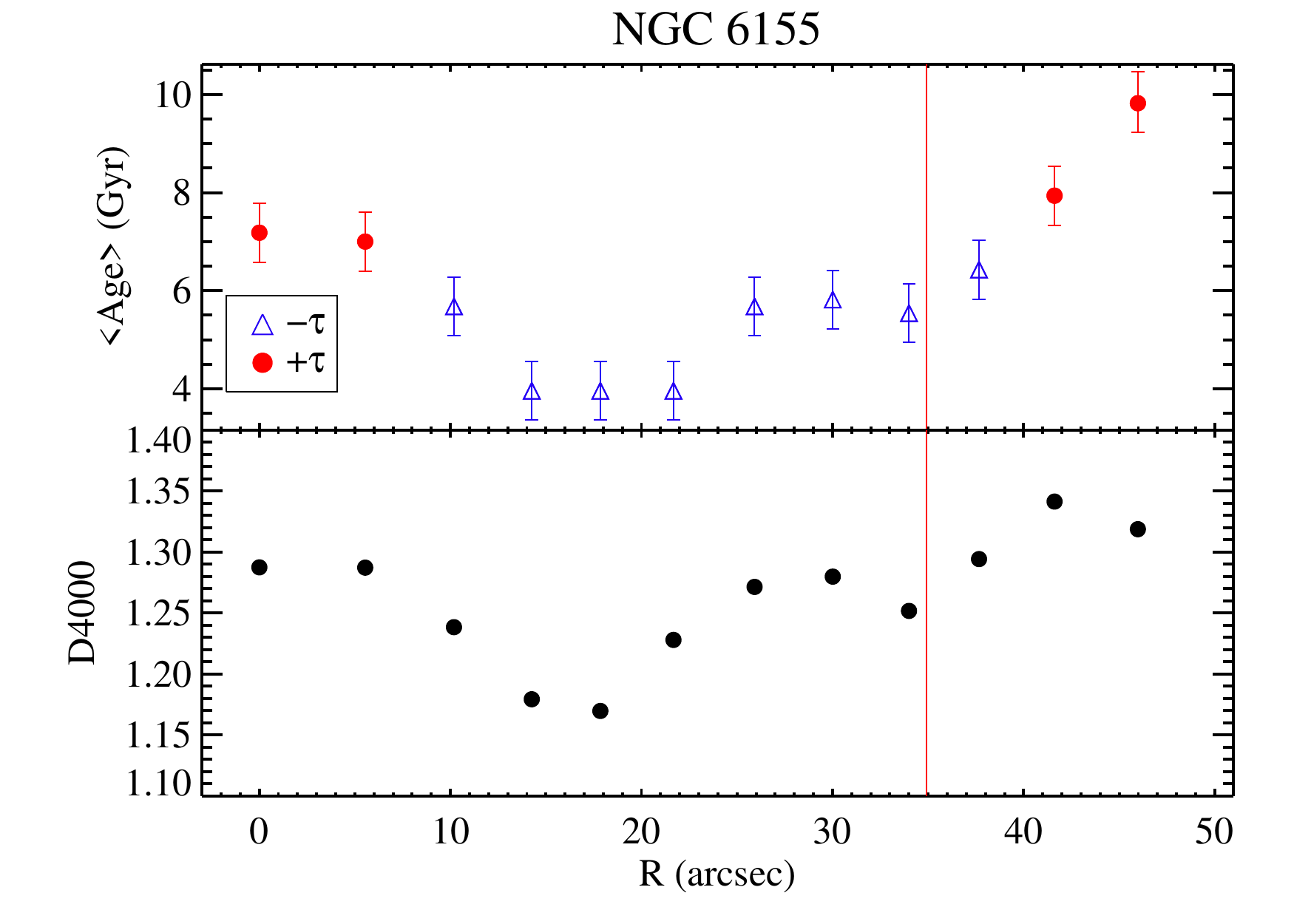}&
\includegraphics[scale=0.27]{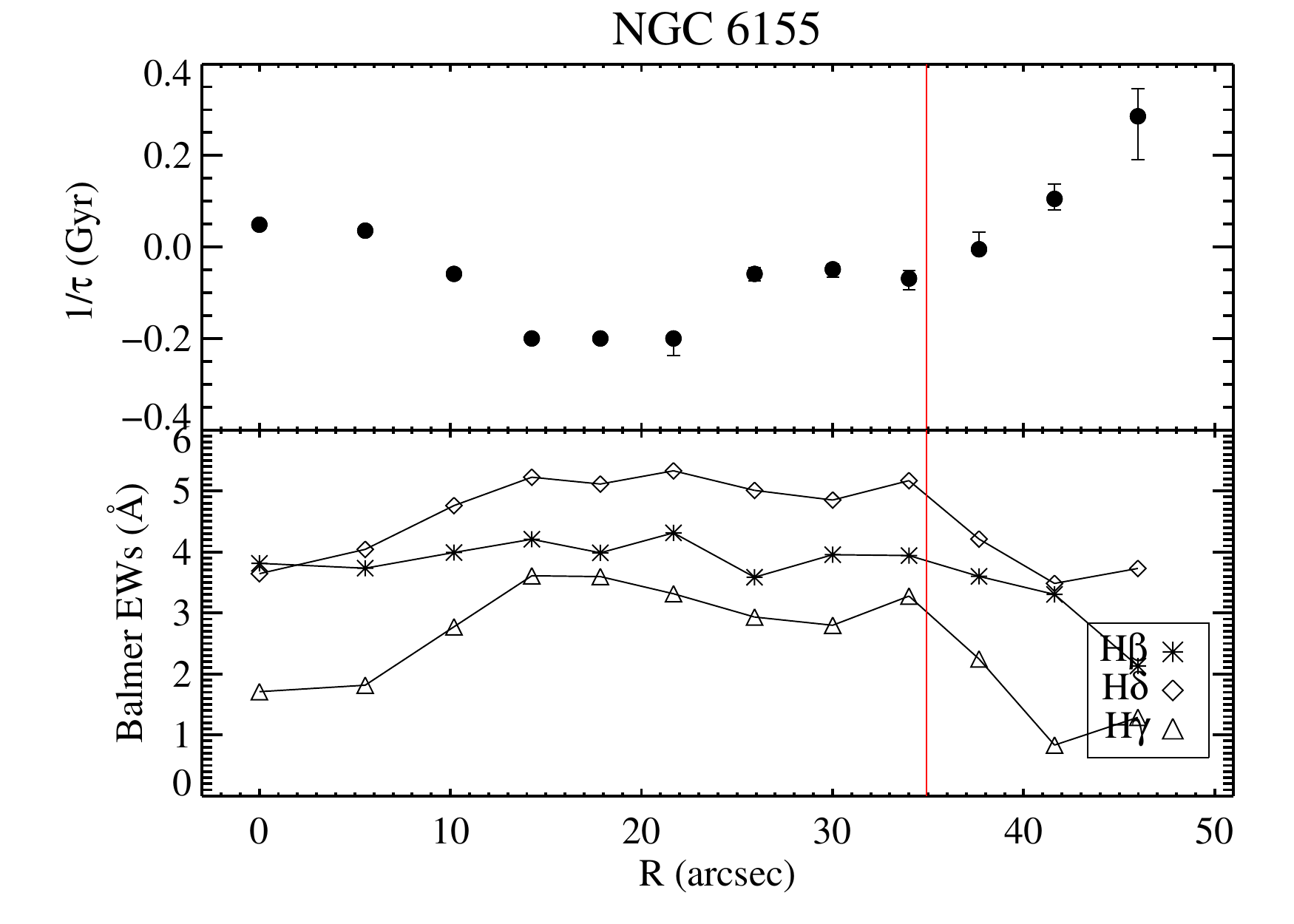}&
\includegraphics[scale=0.27]{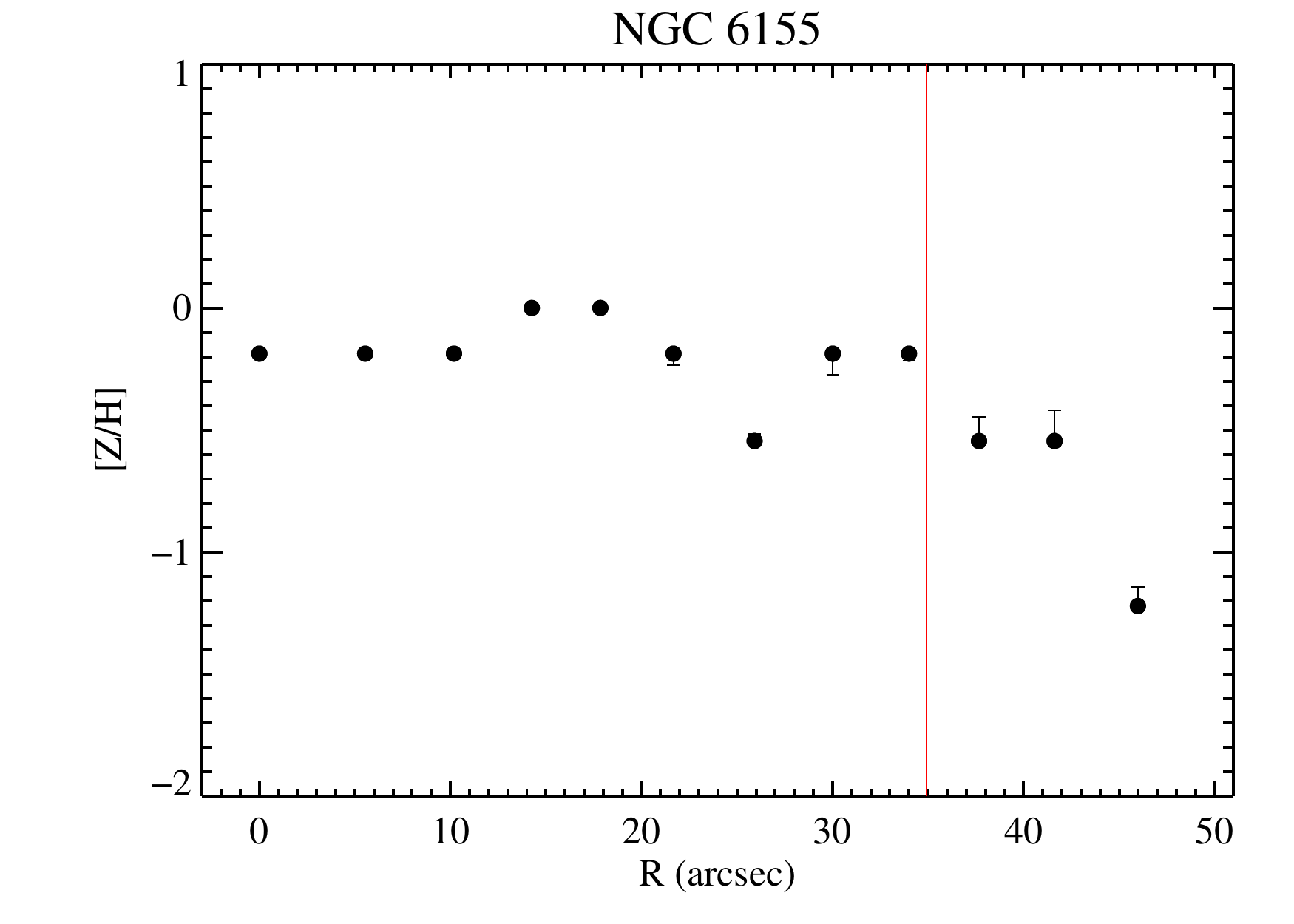} 
\end{array}$
    \epsscale{1}
 \end{center}
\caption{NGC 6155.  Same as Figure~\ref{NGC1058}. \label{NGC6155}}
\end{figure*}
\begin{figure*}
 \begin{center}$
 \begin{array}{ccc}
   \epsscale{.35}
\includegraphics[scale=0.27]{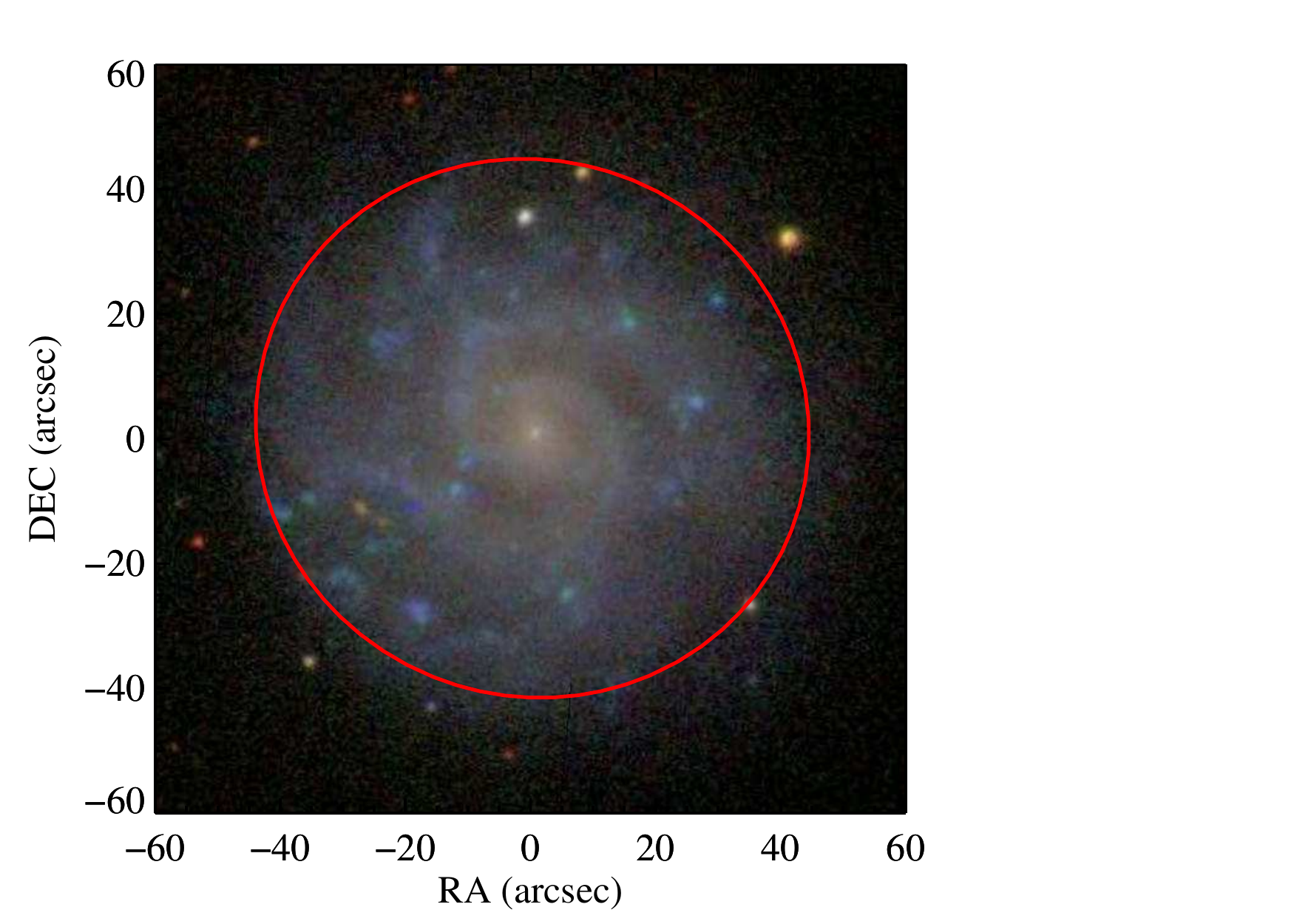}&
\includegraphics[scale=0.27]{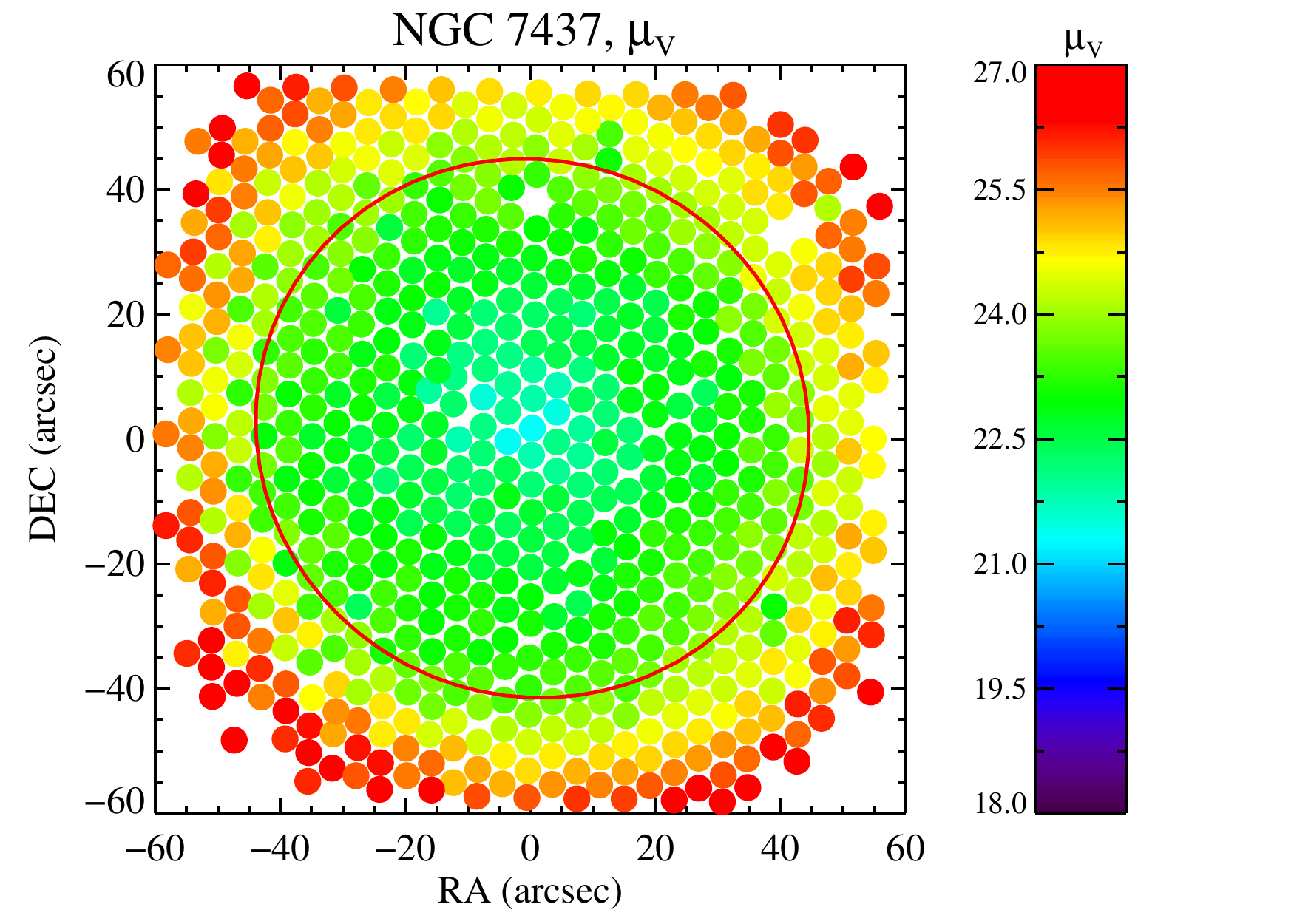}\\
\includegraphics[scale=0.27]{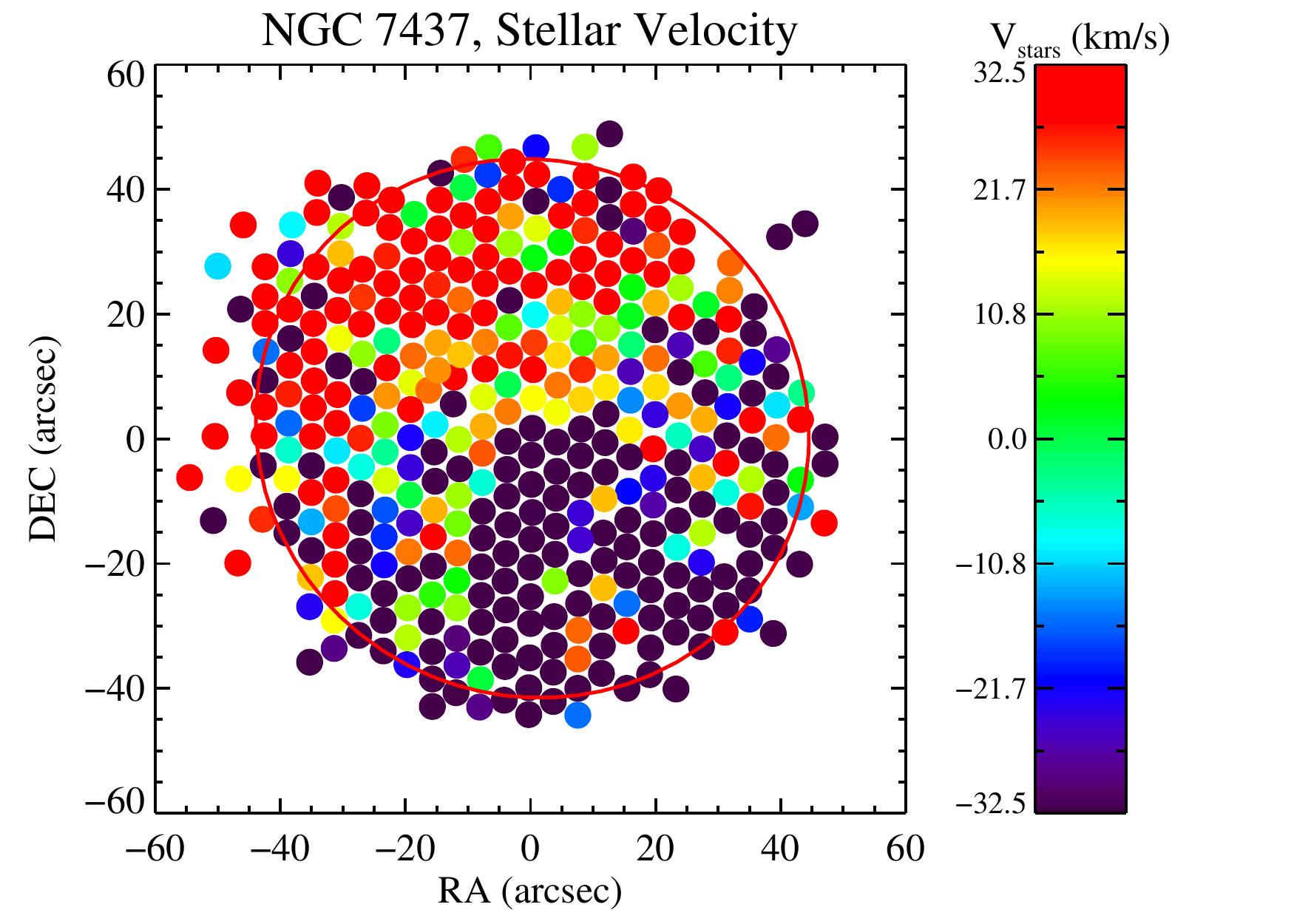}&
\includegraphics[scale=0.27]{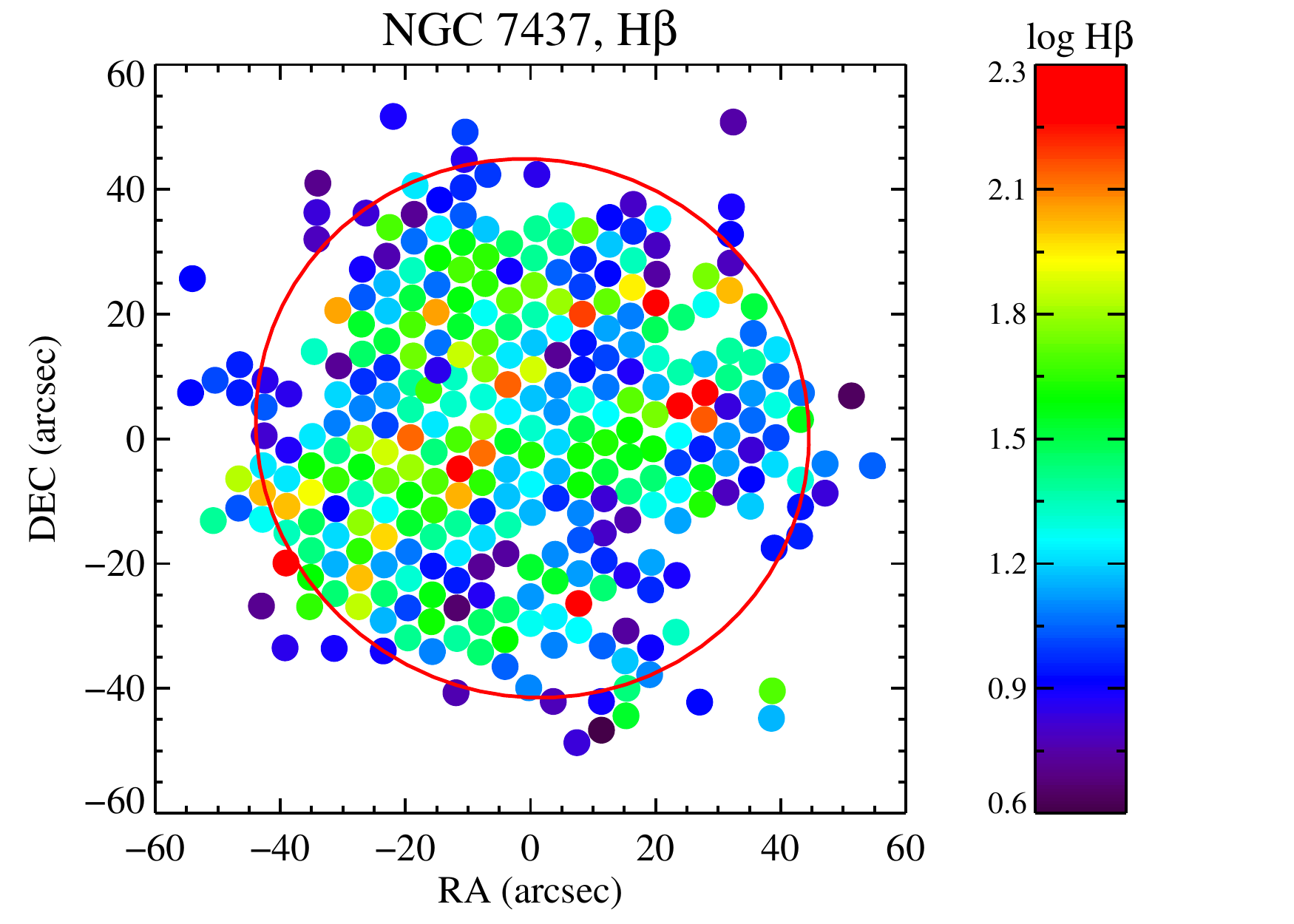}&
\includegraphics[scale=0.27]{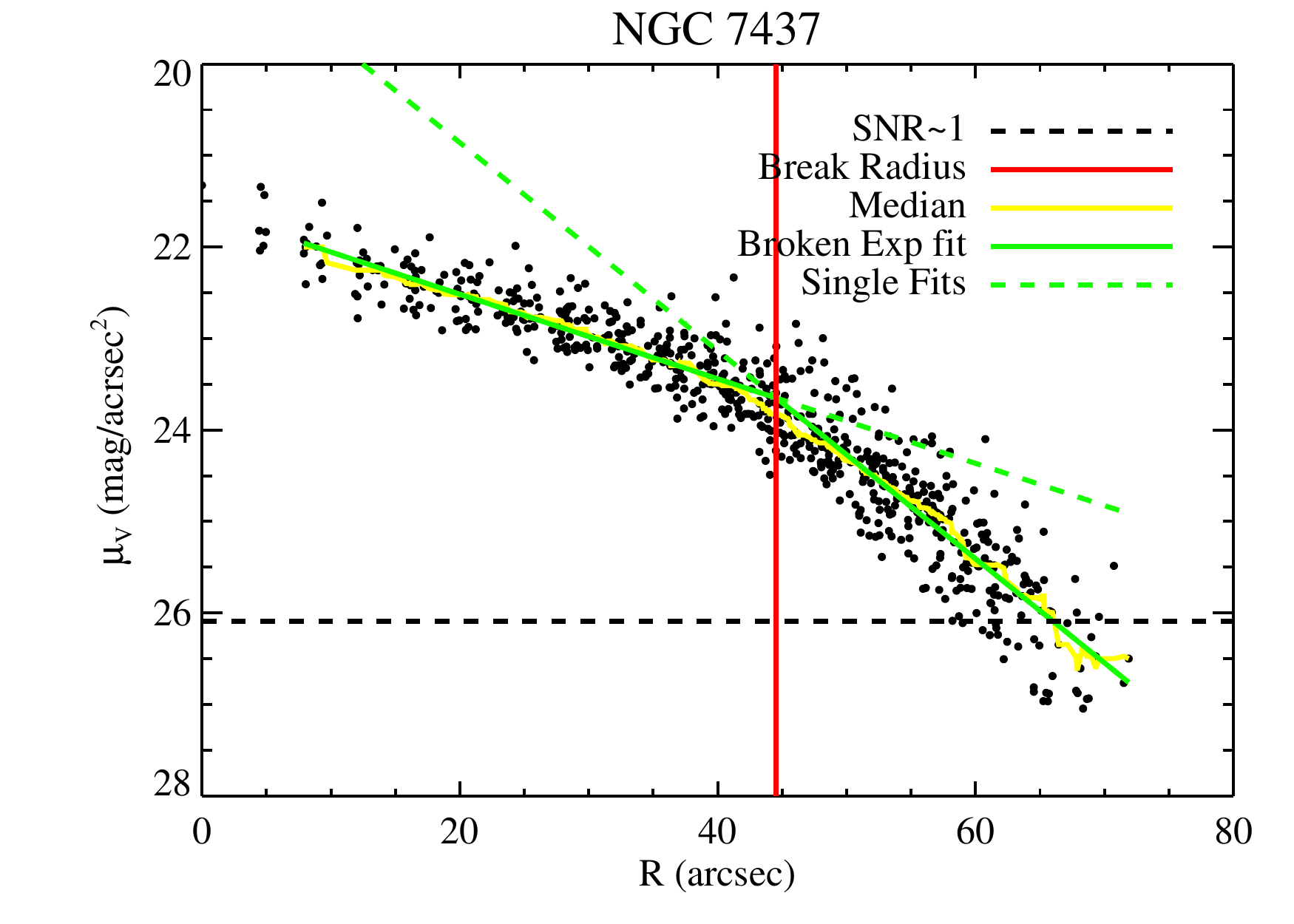}\\
\includegraphics[scale=0.27]{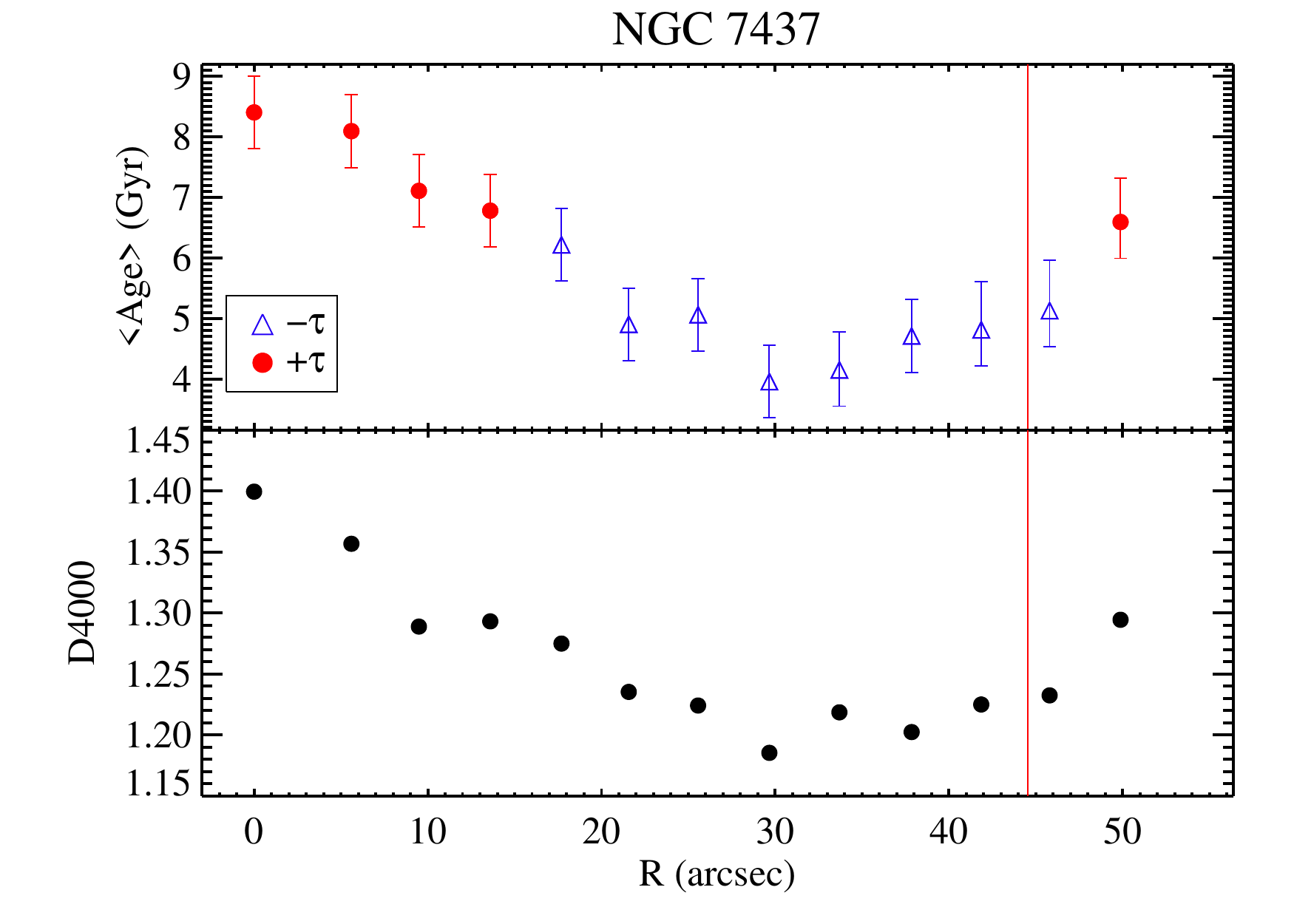}&
\includegraphics[scale=0.27]{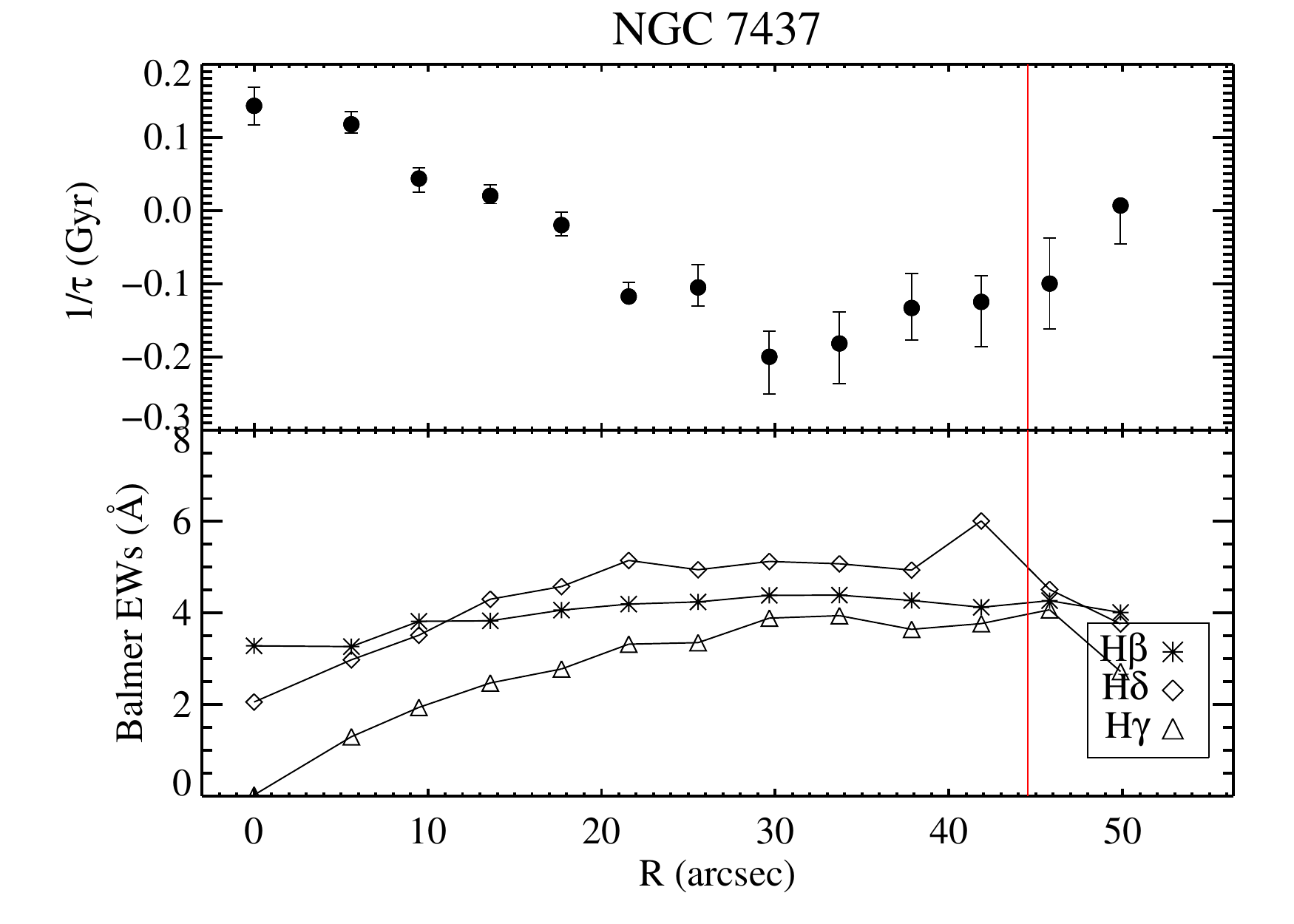}&
\includegraphics[scale=0.27]{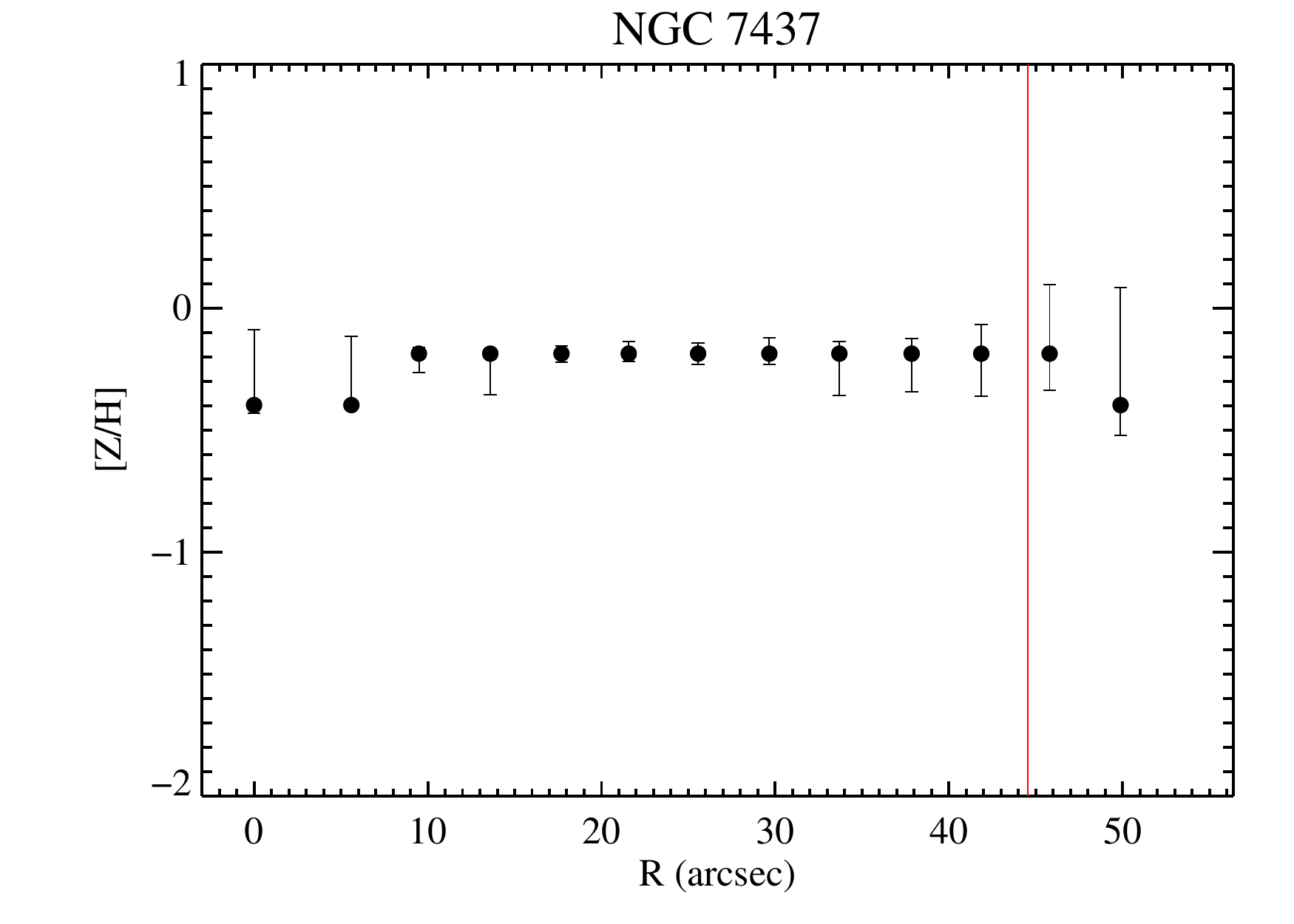}   
\end{array}$
    \epsscale{1}
 \end{center}
\caption{NGC 7437.  Same as Figure~\ref{NGC1058}. \label{NGC7437}}
\end{figure*}


\begin{figure*}
 \begin{center}$
 \begin{array}{ccc}
   \epsscale{.35}
\includegraphics[scale=0.27]{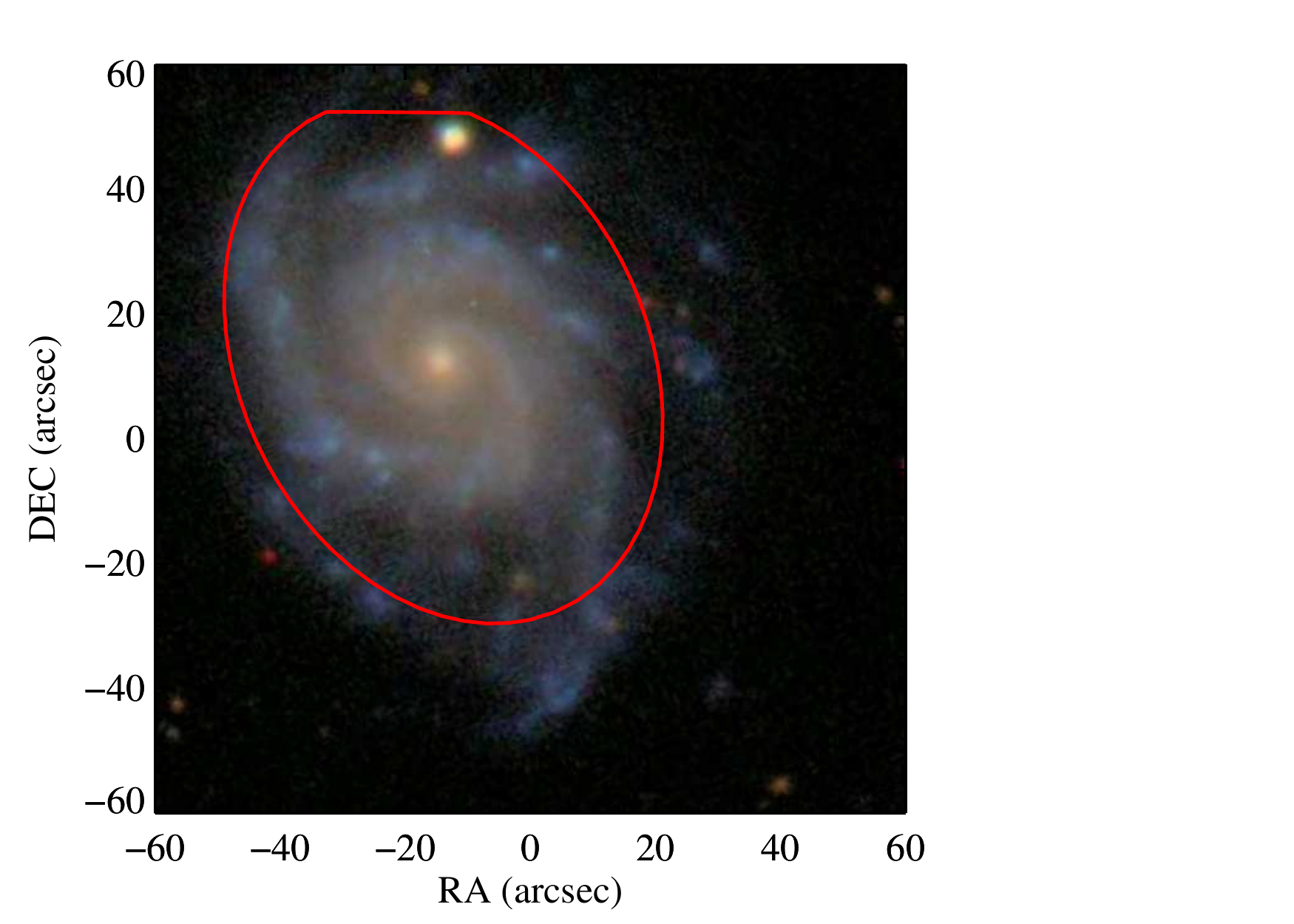}&
\includegraphics[scale=0.27]{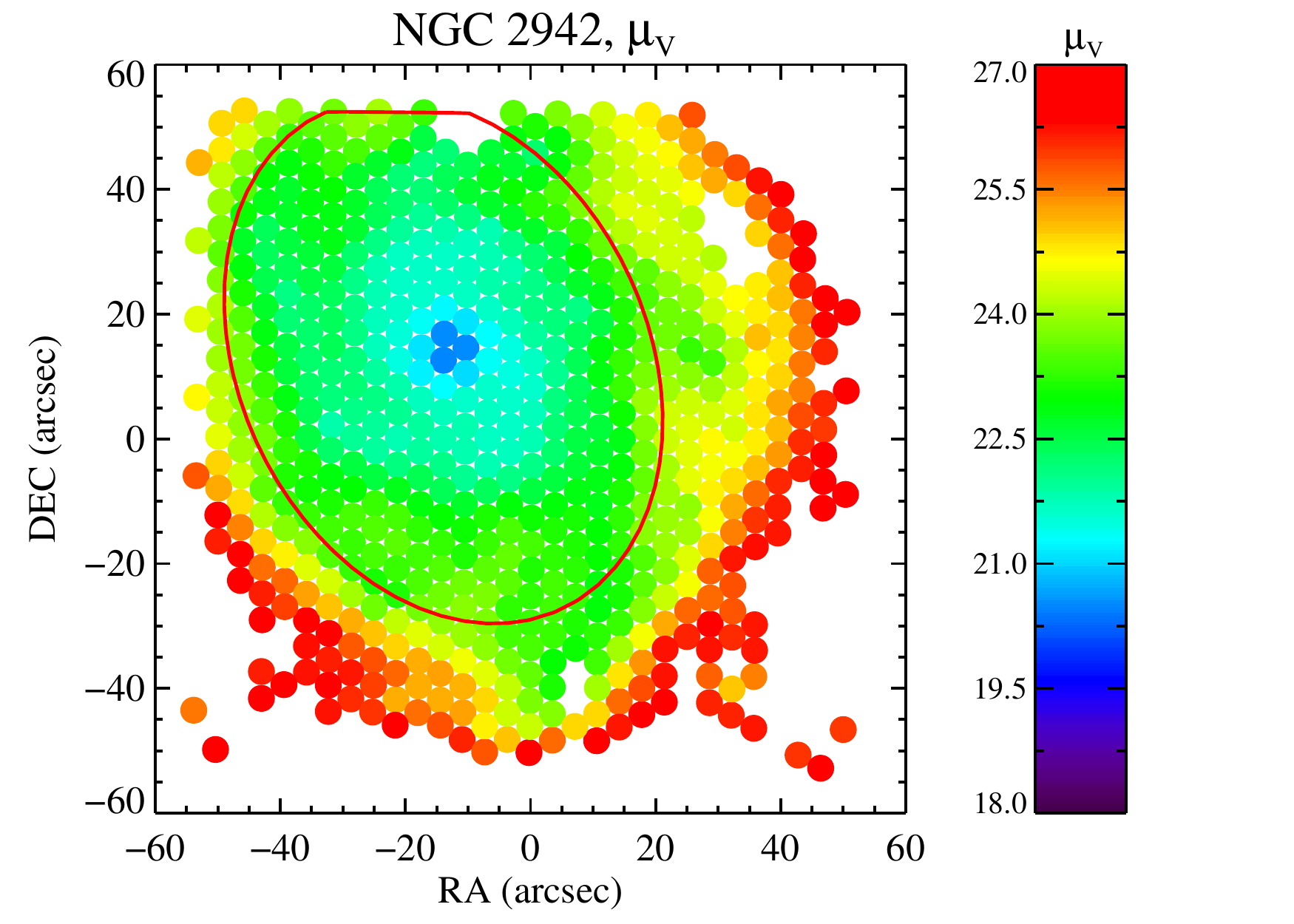}&
\includegraphics[scale=0.27]{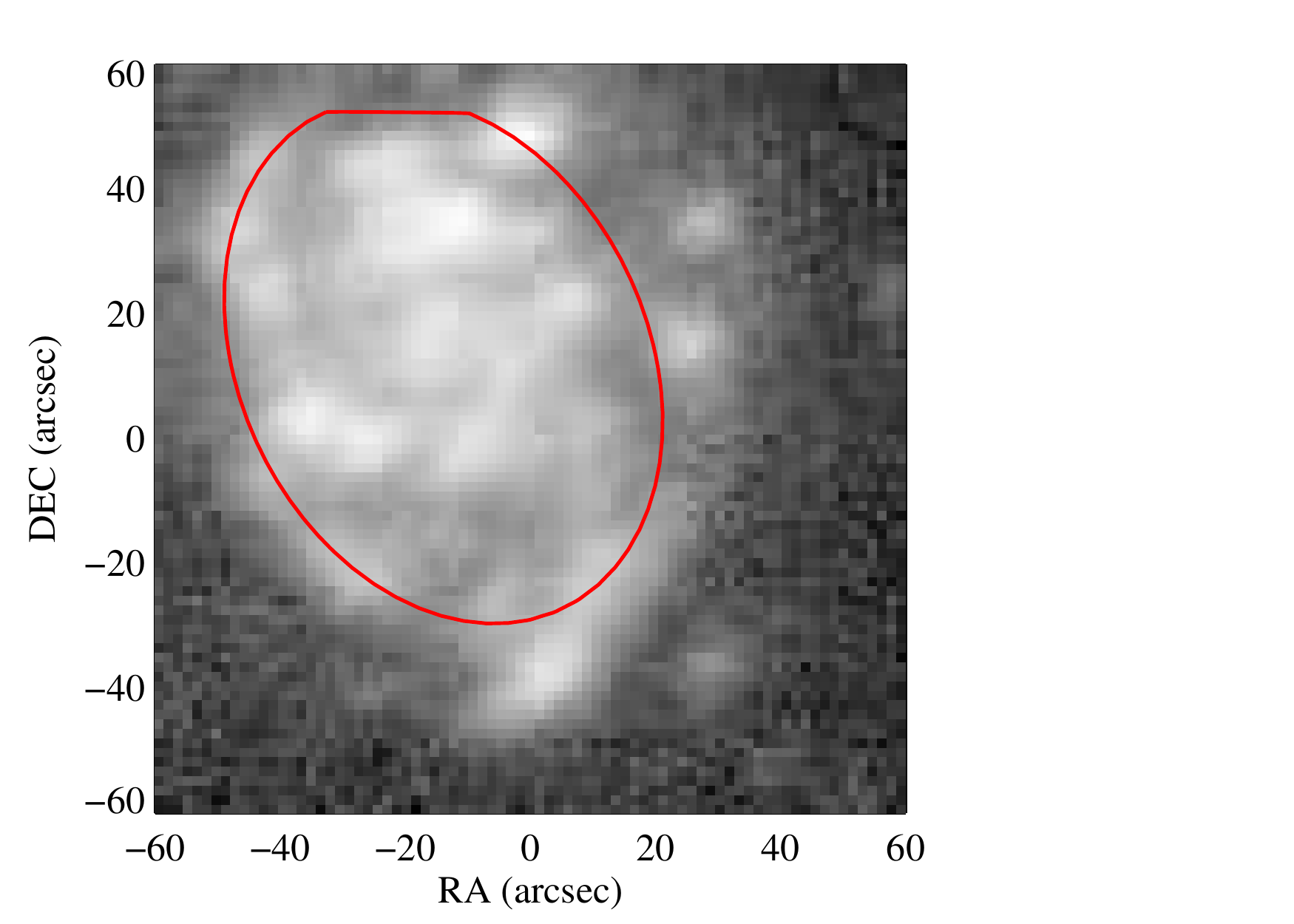}\\
\includegraphics[scale=0.27]{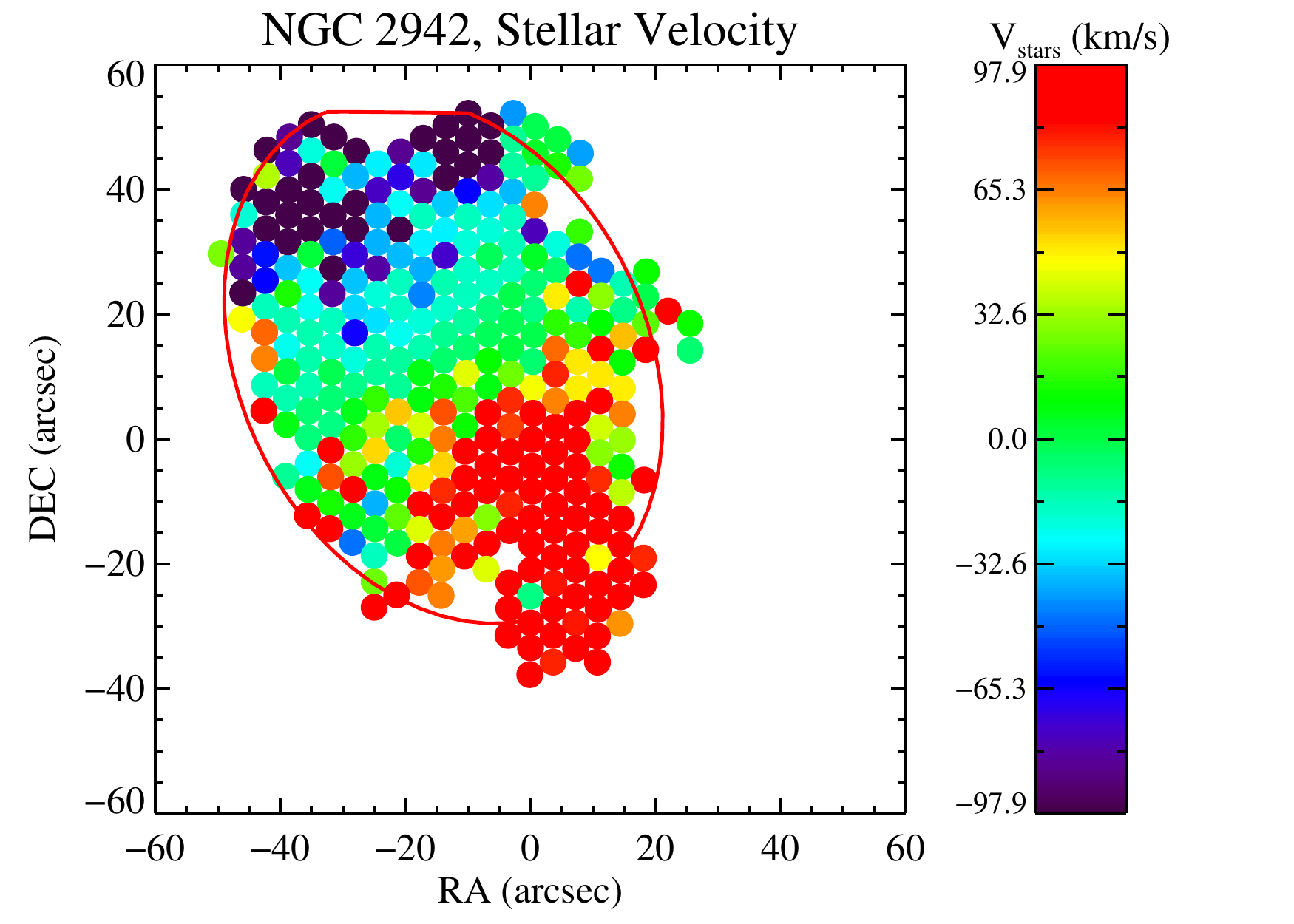}&
\includegraphics[scale=0.27]{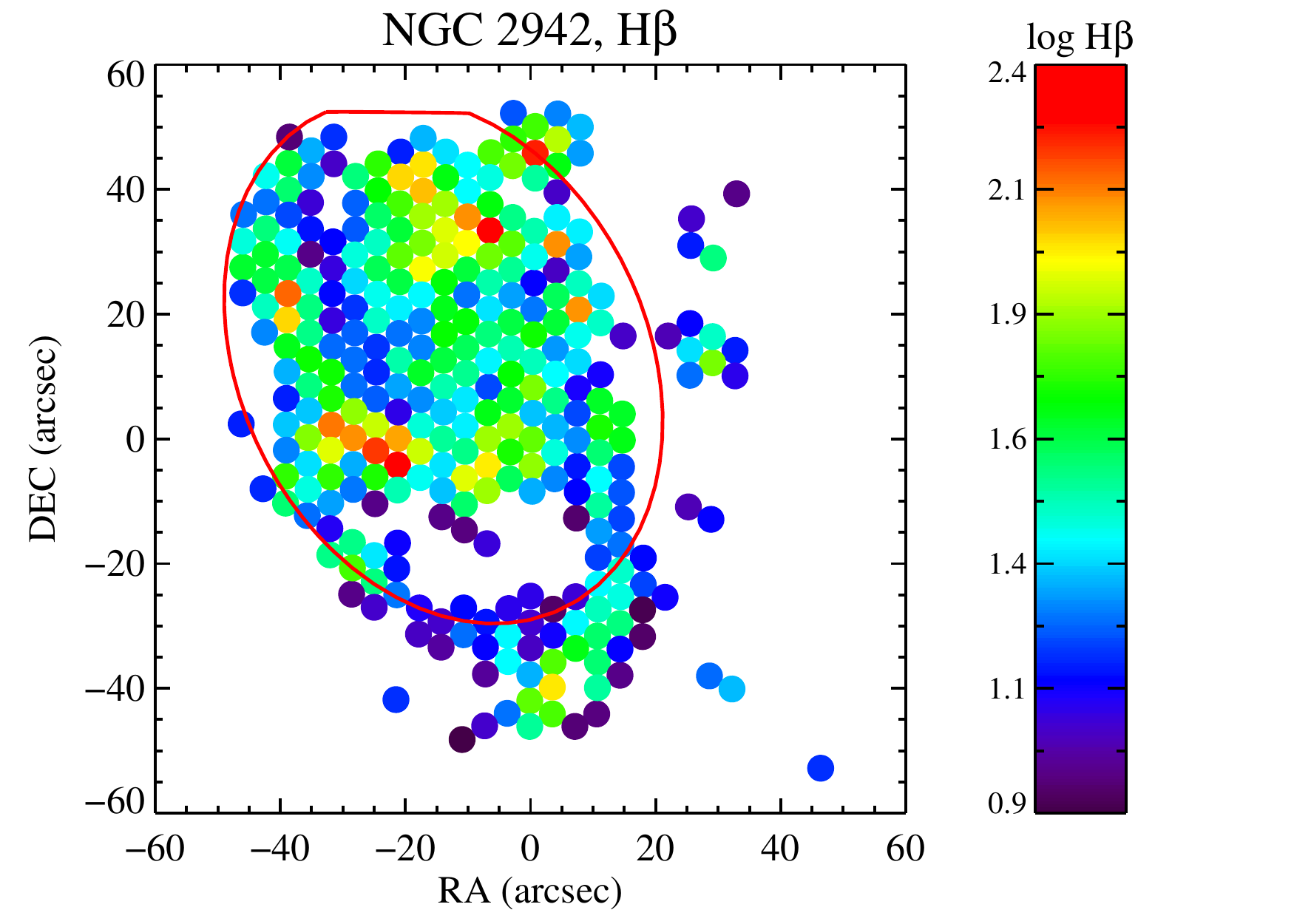}&
\includegraphics[scale=0.27]{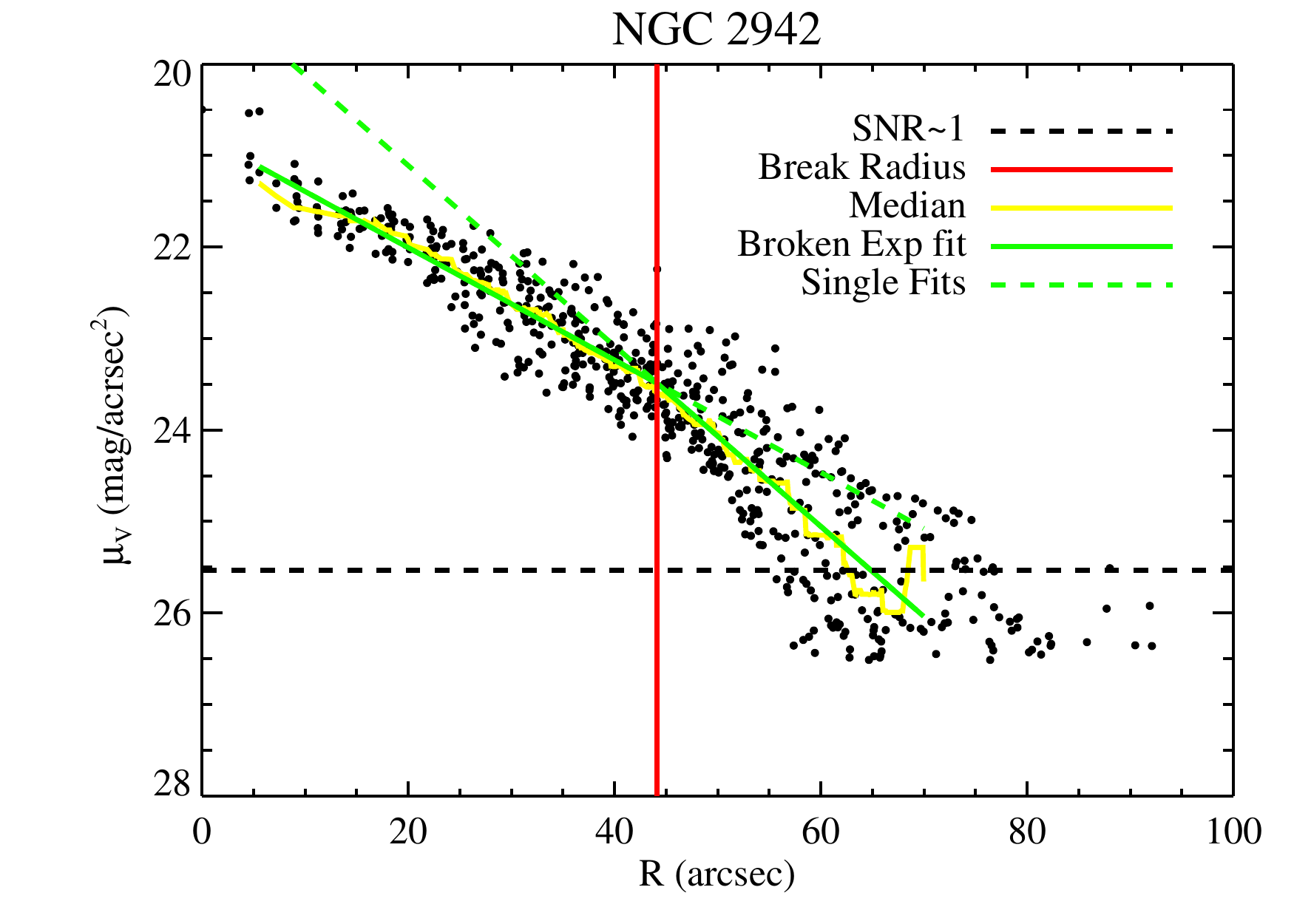}\\
\includegraphics[scale=0.27]{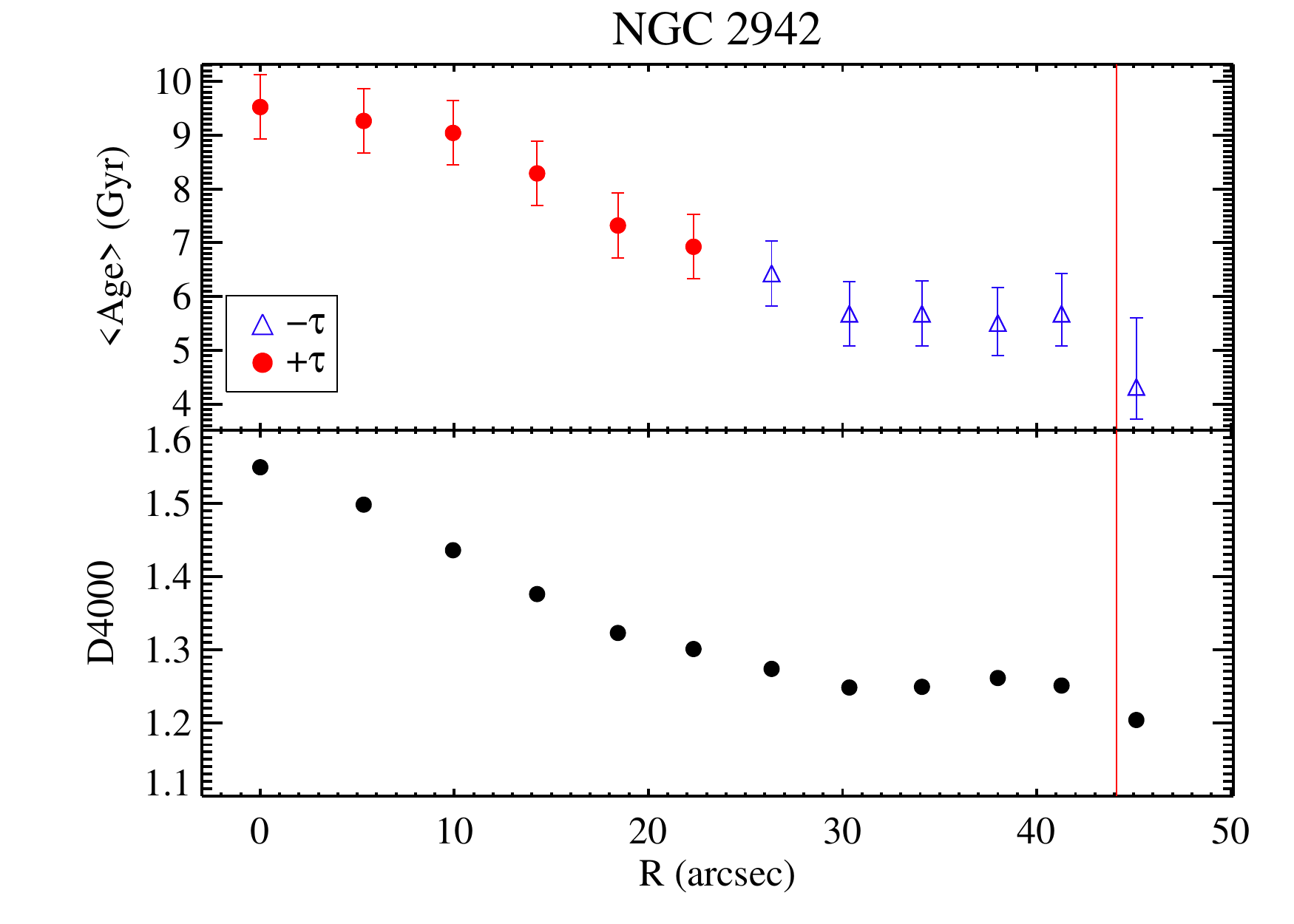}&
\includegraphics[scale=0.27]{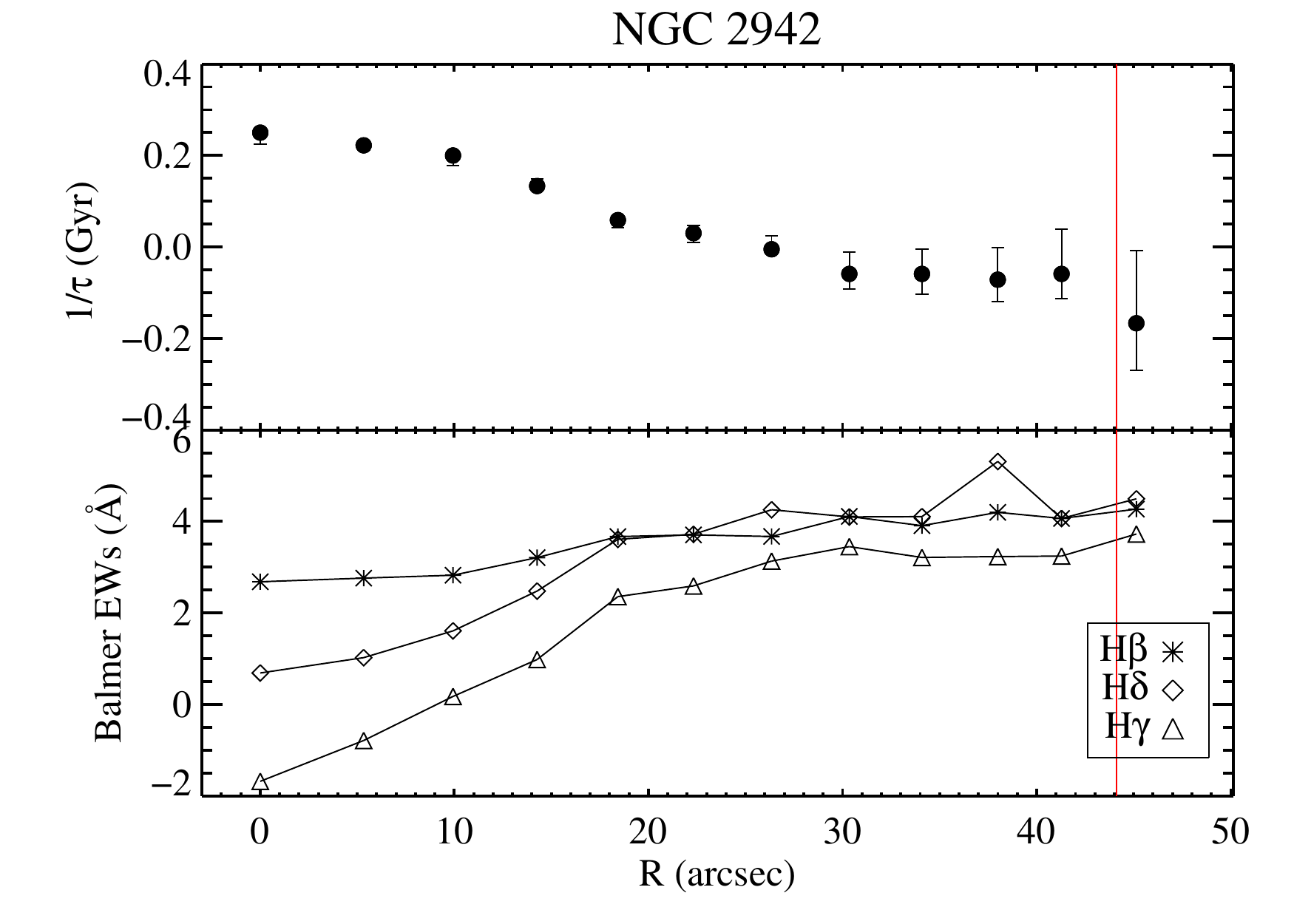}&
\includegraphics[scale=0.27]{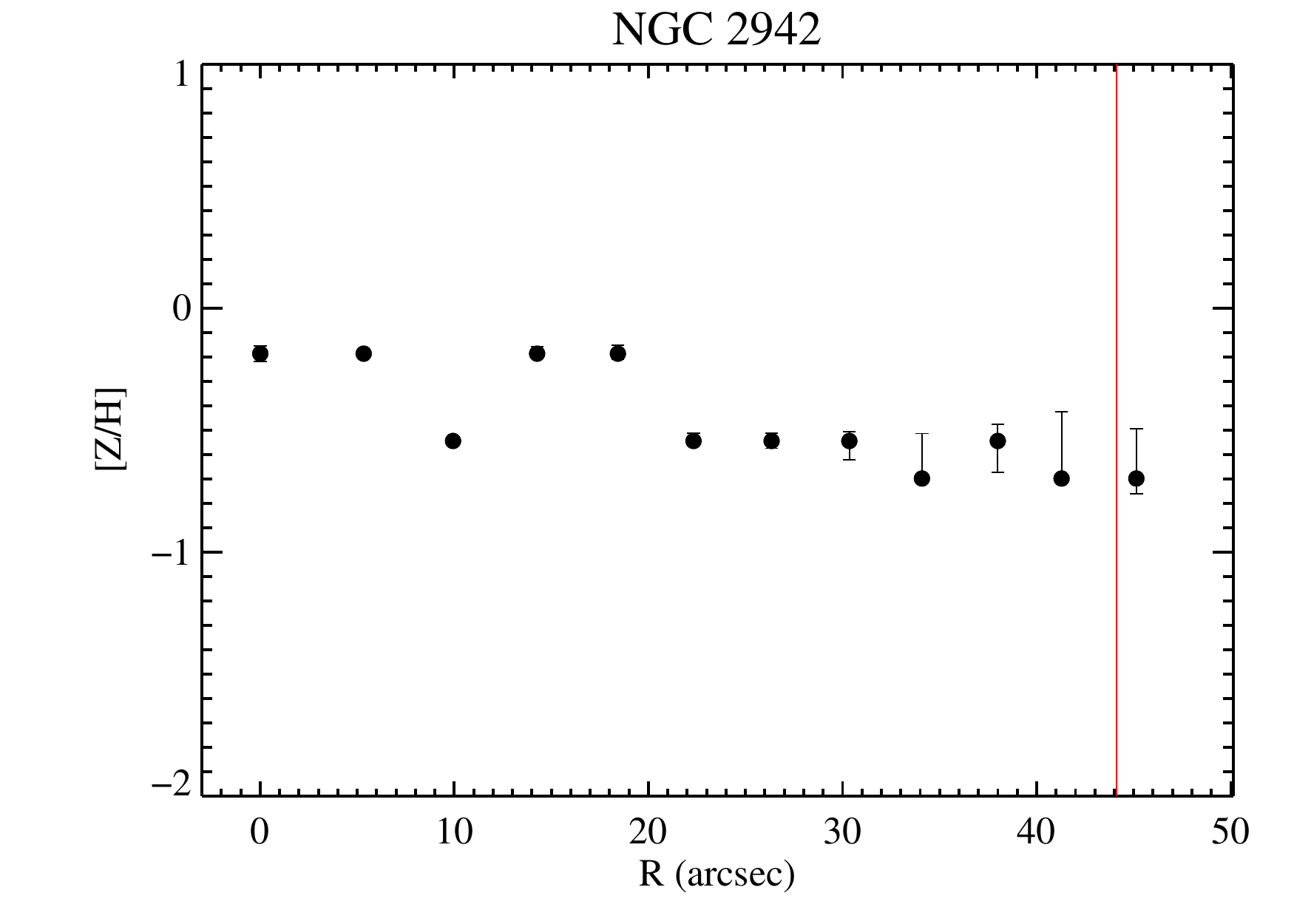}  
\end{array}$
    \epsscale{1}
 \end{center}
\caption{NGC 2942.  Same as Figure~\ref{NGC1058}. \label{NGC2942}}
\end{figure*}


\begin{figure*}
 \begin{center}$
 \begin{array}{ccc}
   \epsscale{.35}
\includegraphics[scale=0.27]{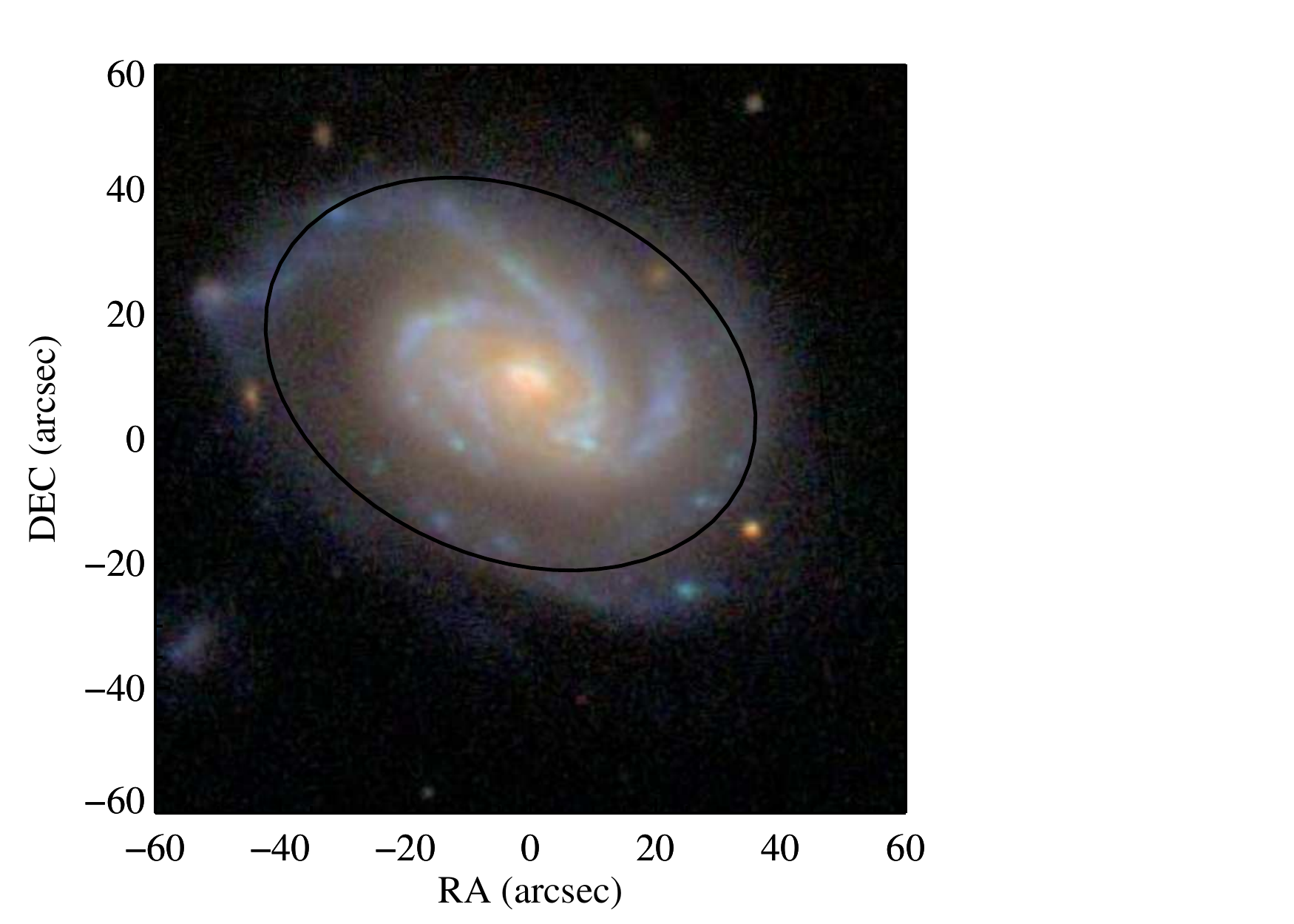}&
\includegraphics[scale=0.27]{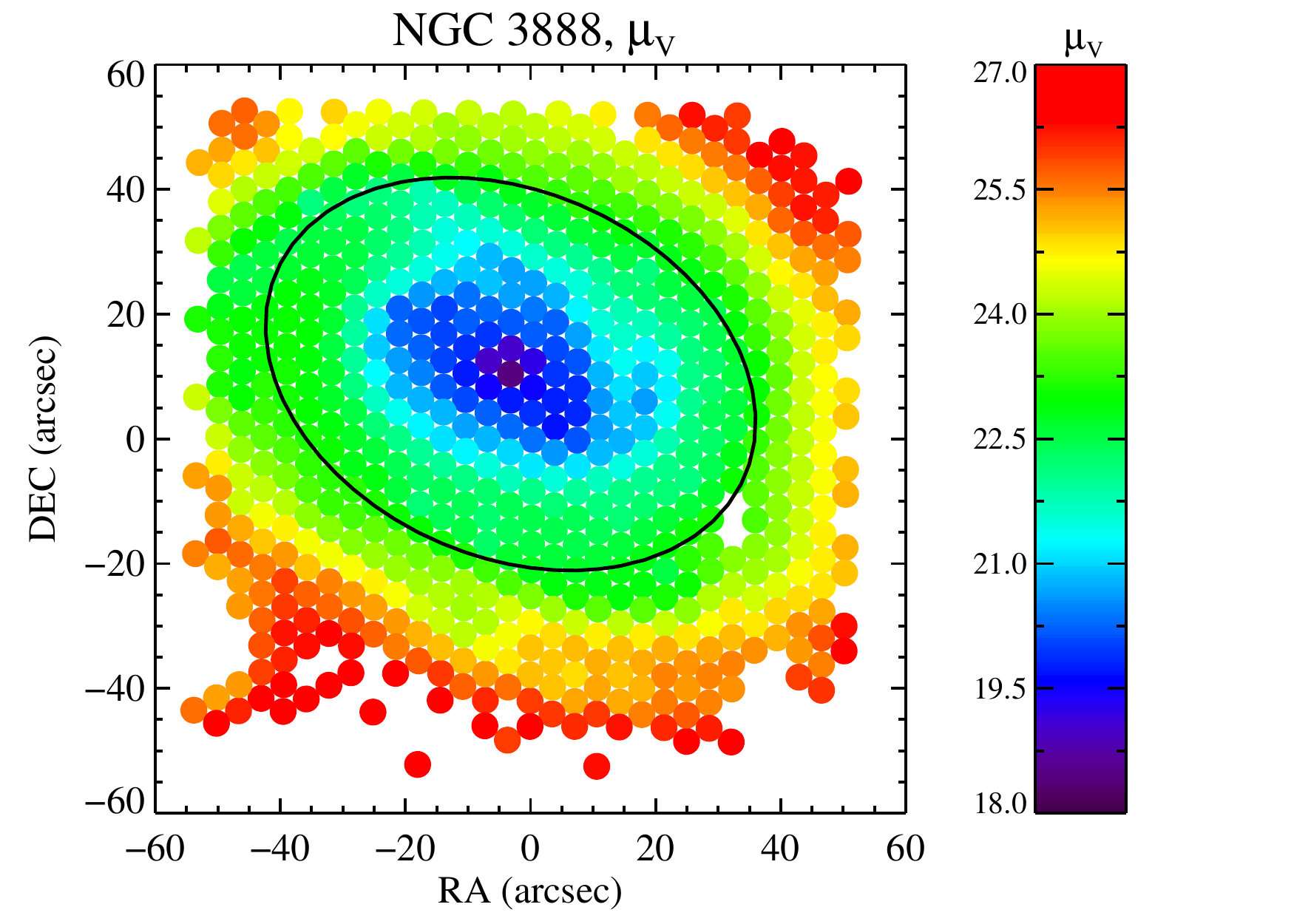}&
\includegraphics[scale=0.27]{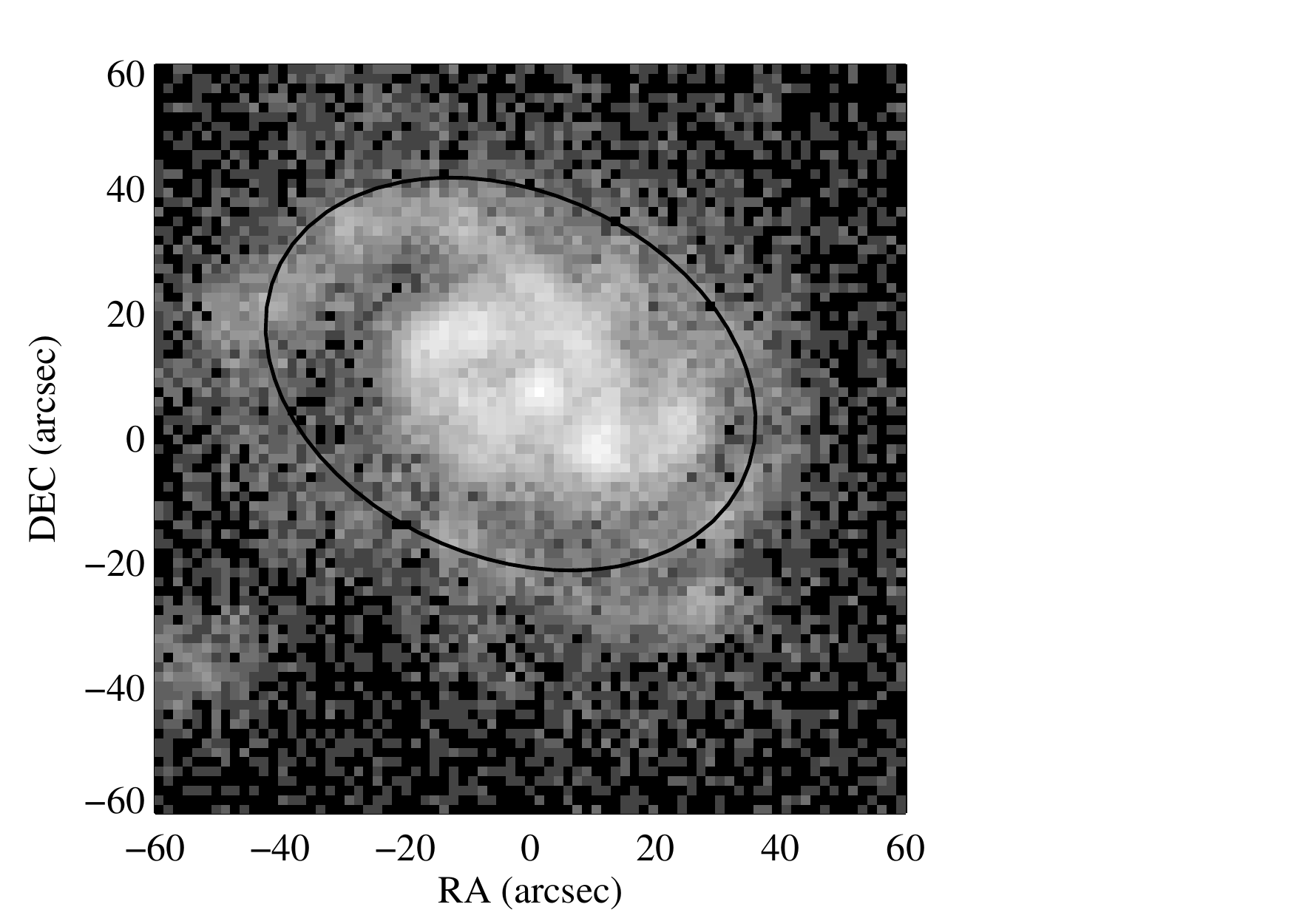}\\
\includegraphics[scale=0.27]{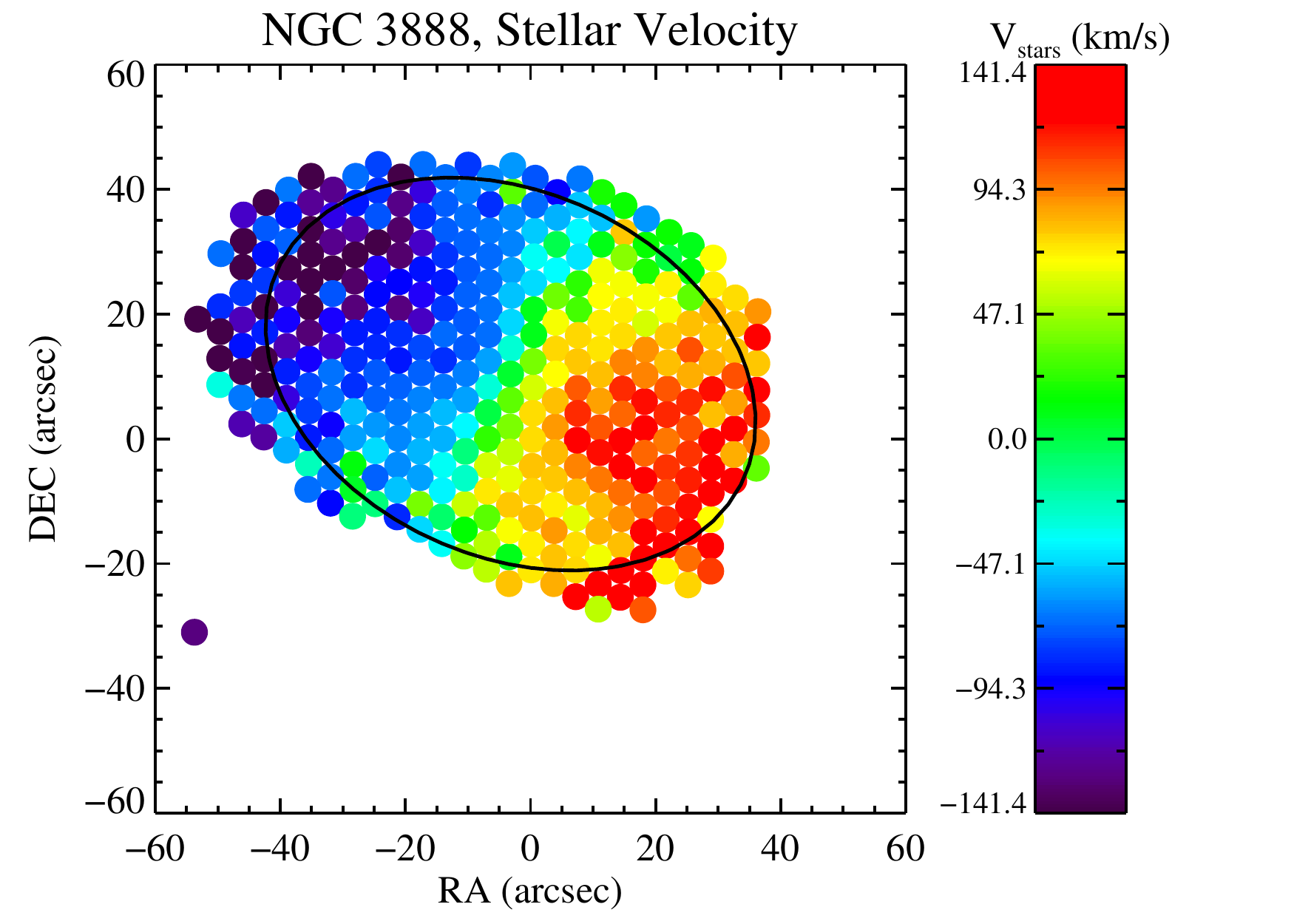}&
\includegraphics[scale=0.27]{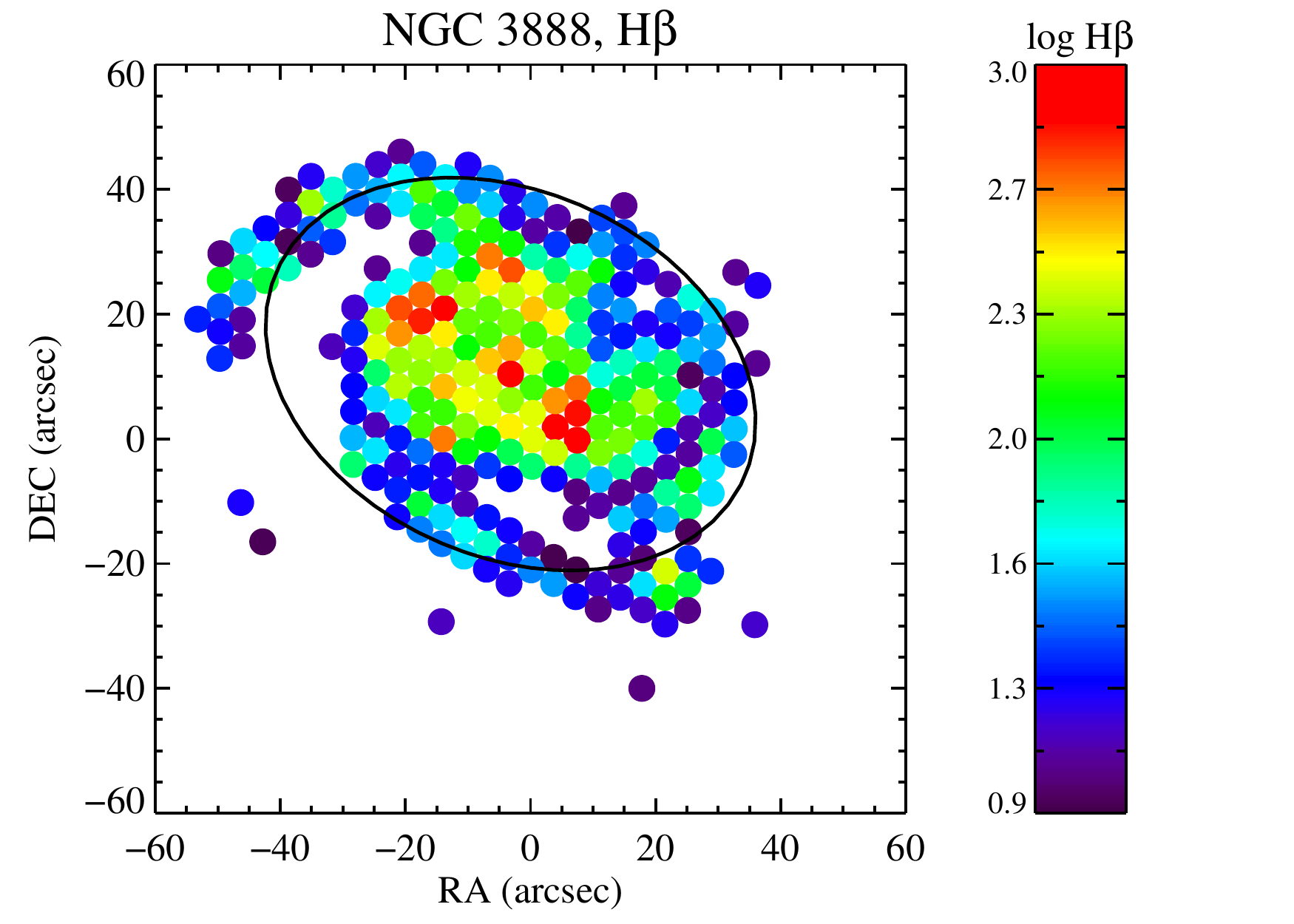}&
\includegraphics[scale=0.27]{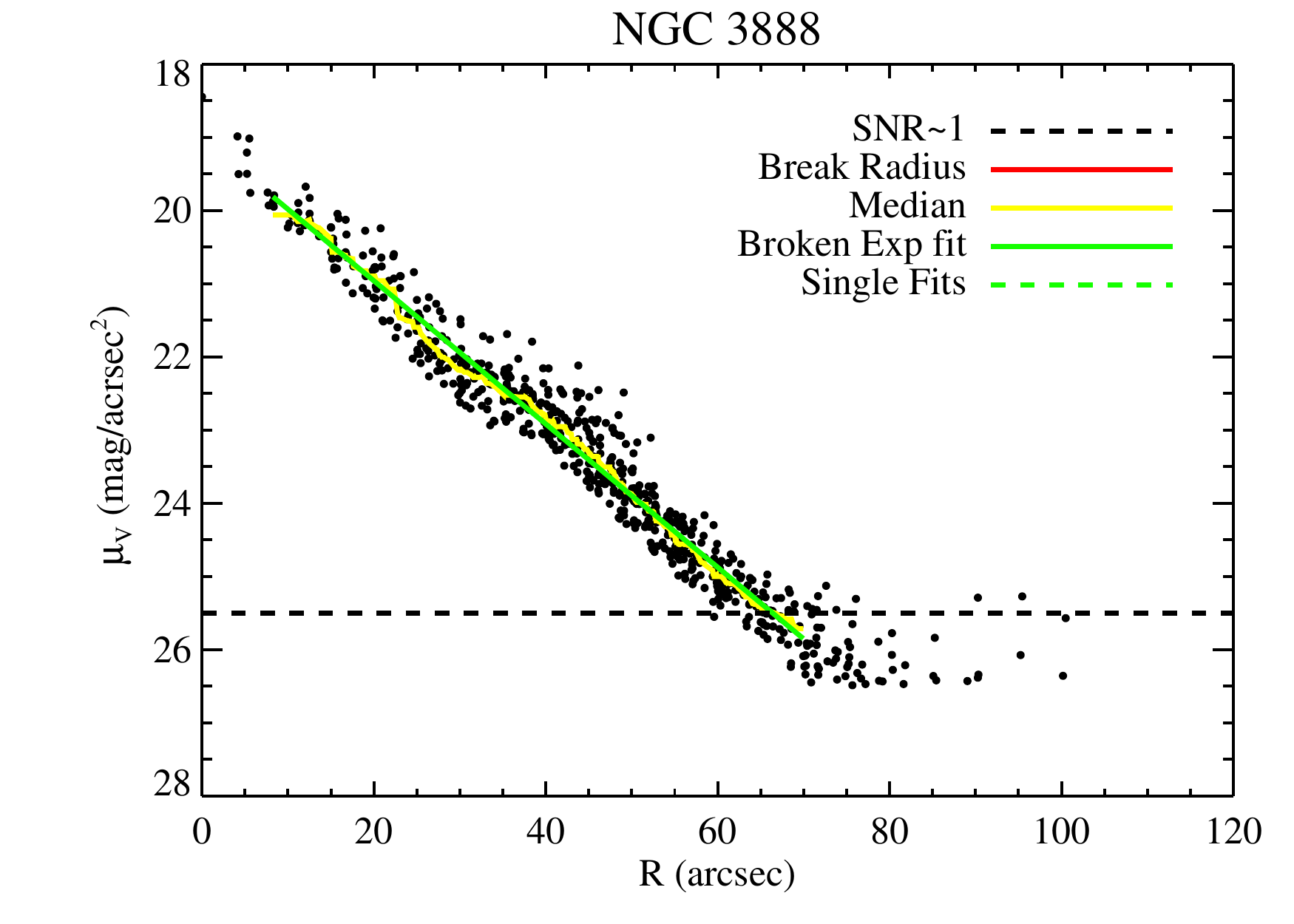}\\
\includegraphics[scale=0.27]{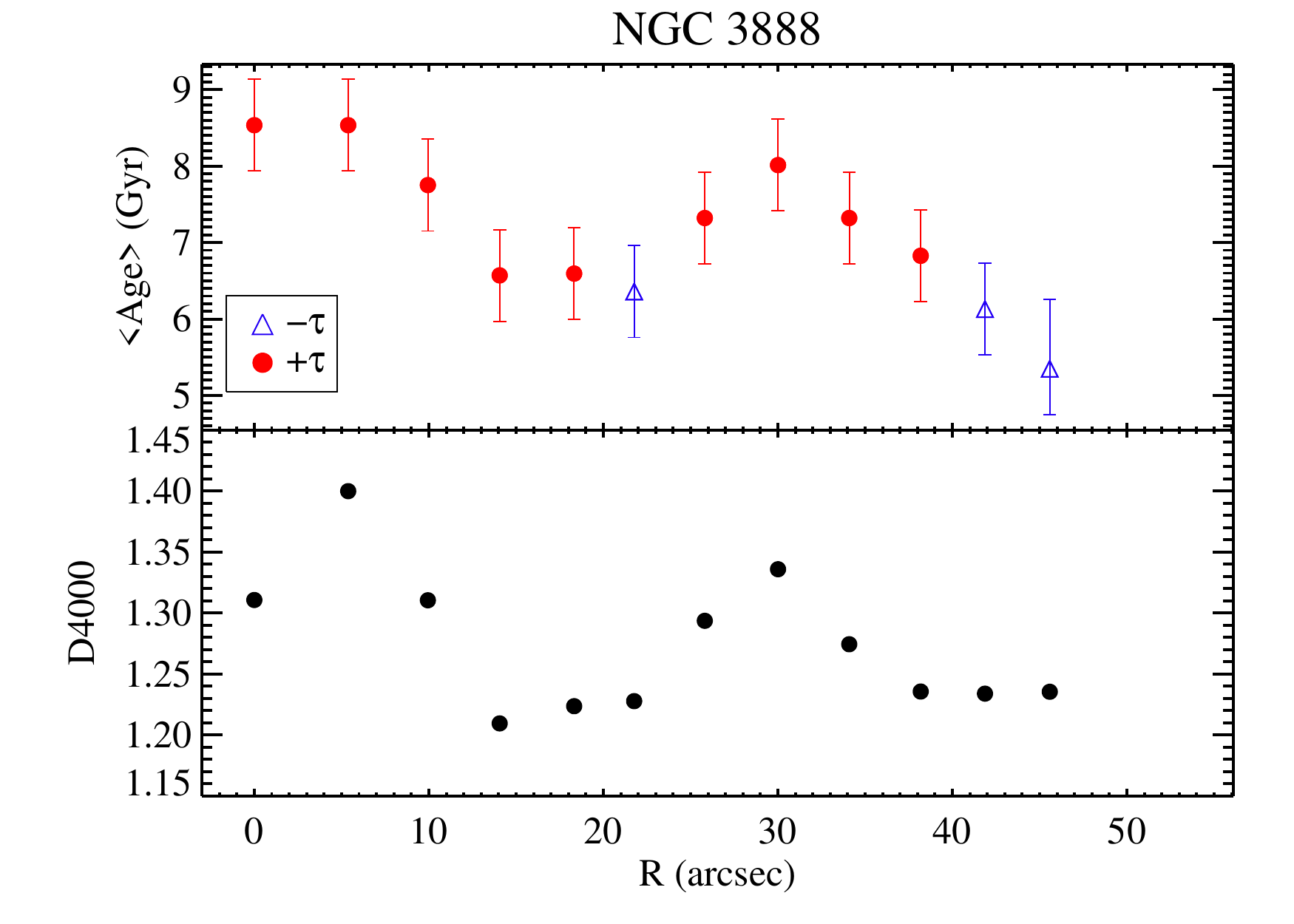}&
\includegraphics[scale=0.27]{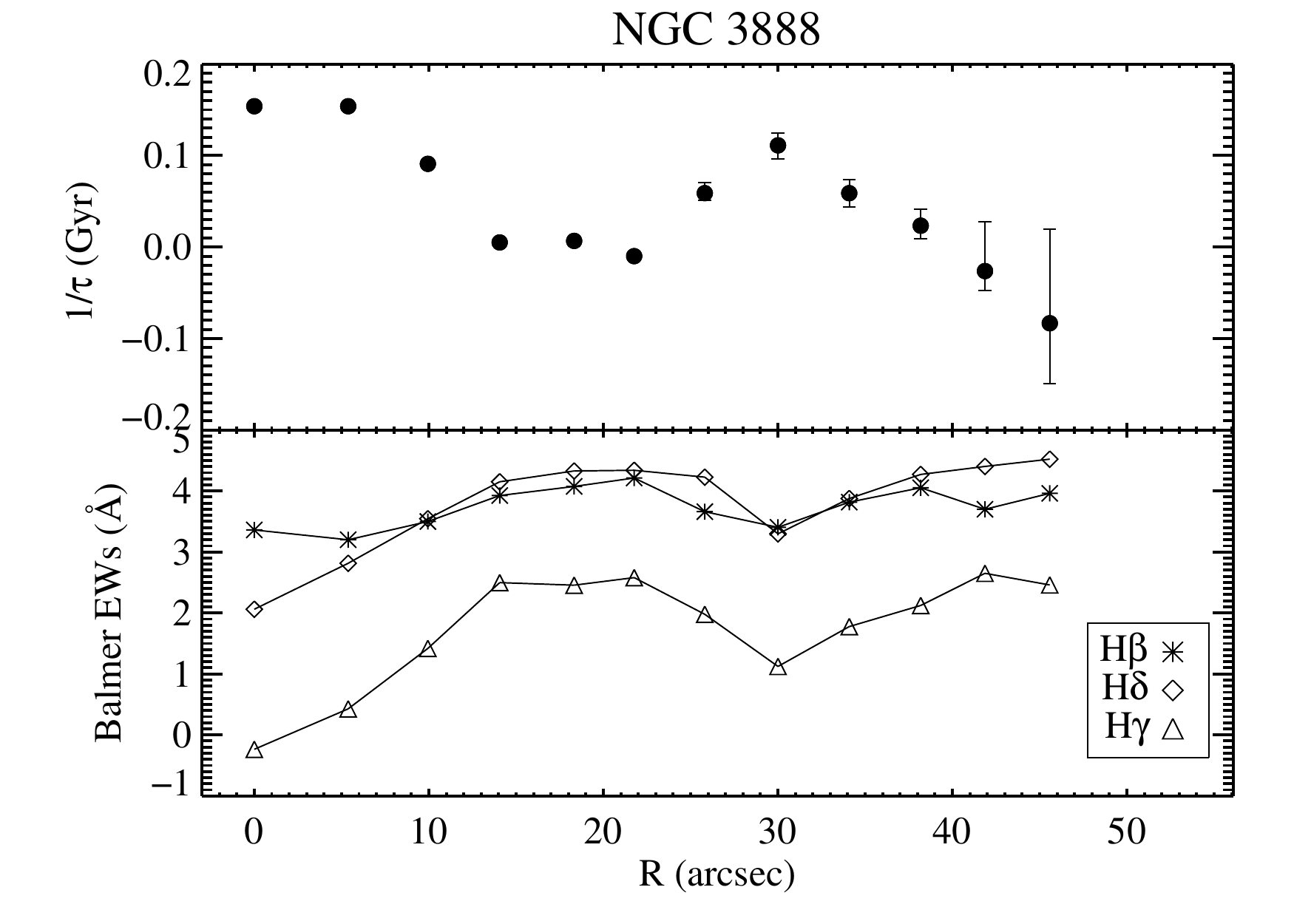}&
\includegraphics[scale=0.27]{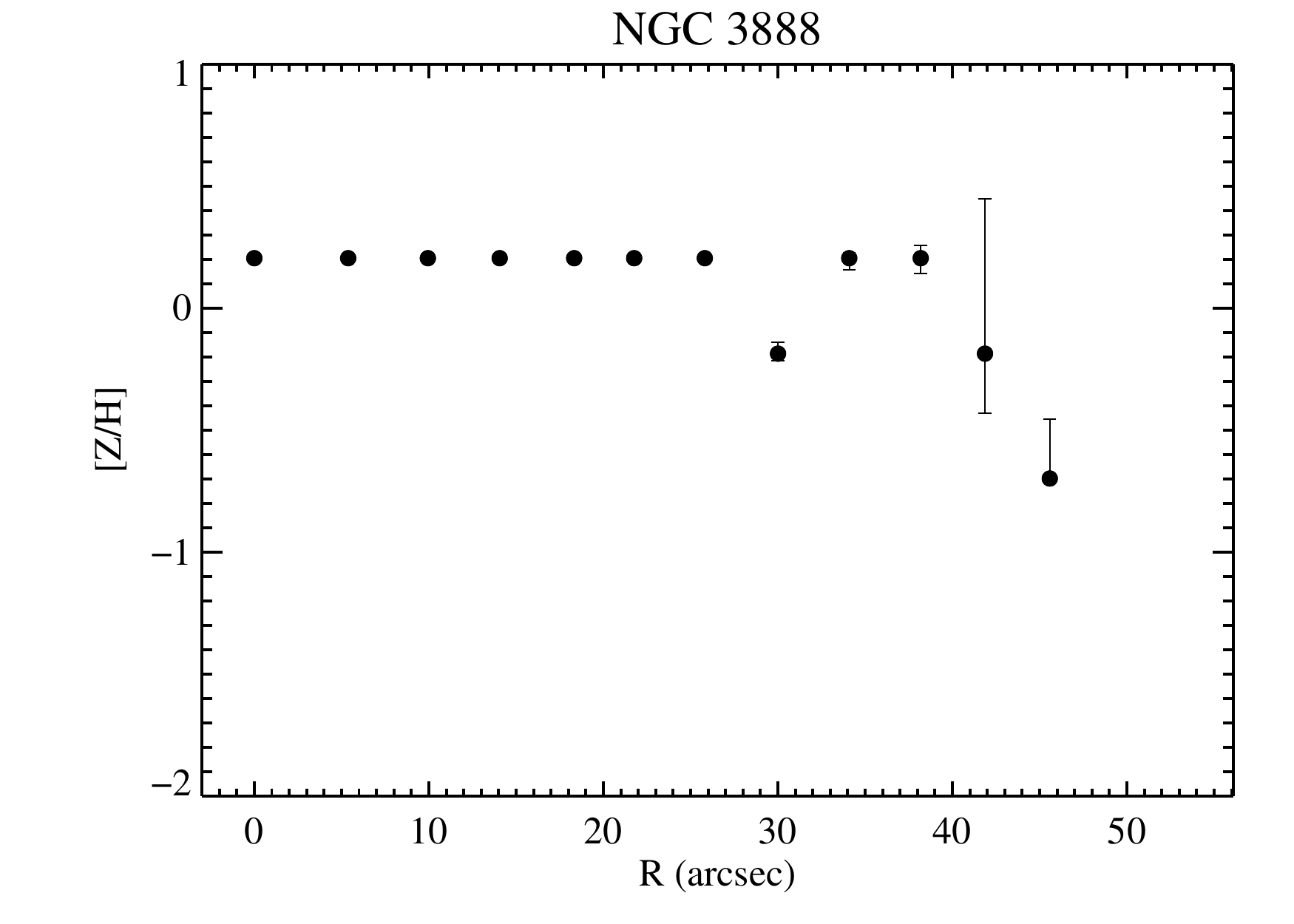}   
\end{array}$
    \epsscale{1}
 \end{center}
\caption{NGC 3888.  Same as Figure~\ref{NGC1058}.  The black ellipse marks the $mu_V=23$\ contour.  \label{NGC3888}}
\end{figure*}


\begin{figure*}
 \begin{center}$
 \begin{array}{ccc}
   \epsscale{.35}
\includegraphics[scale=0.27]{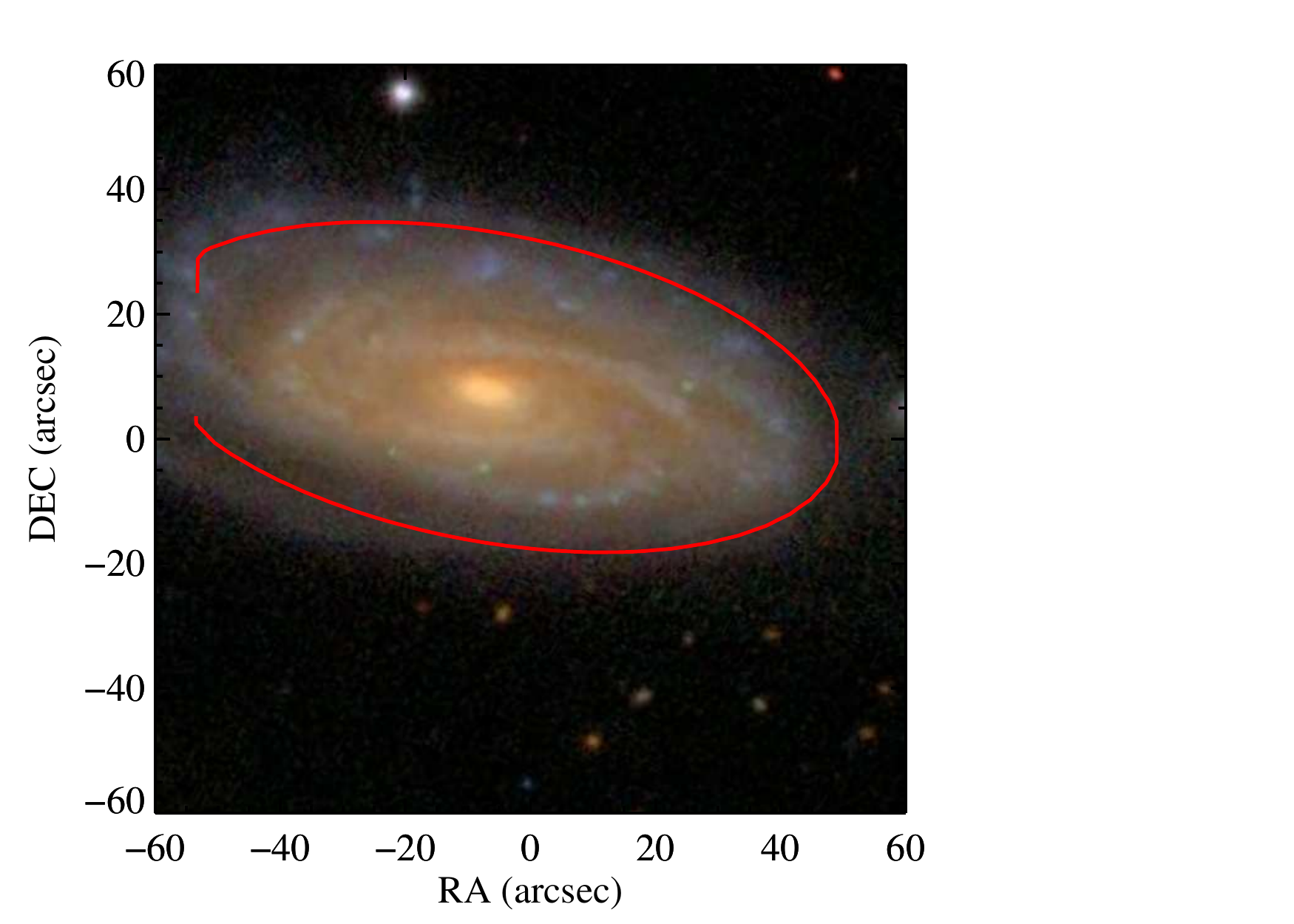}&
\includegraphics[scale=0.27]{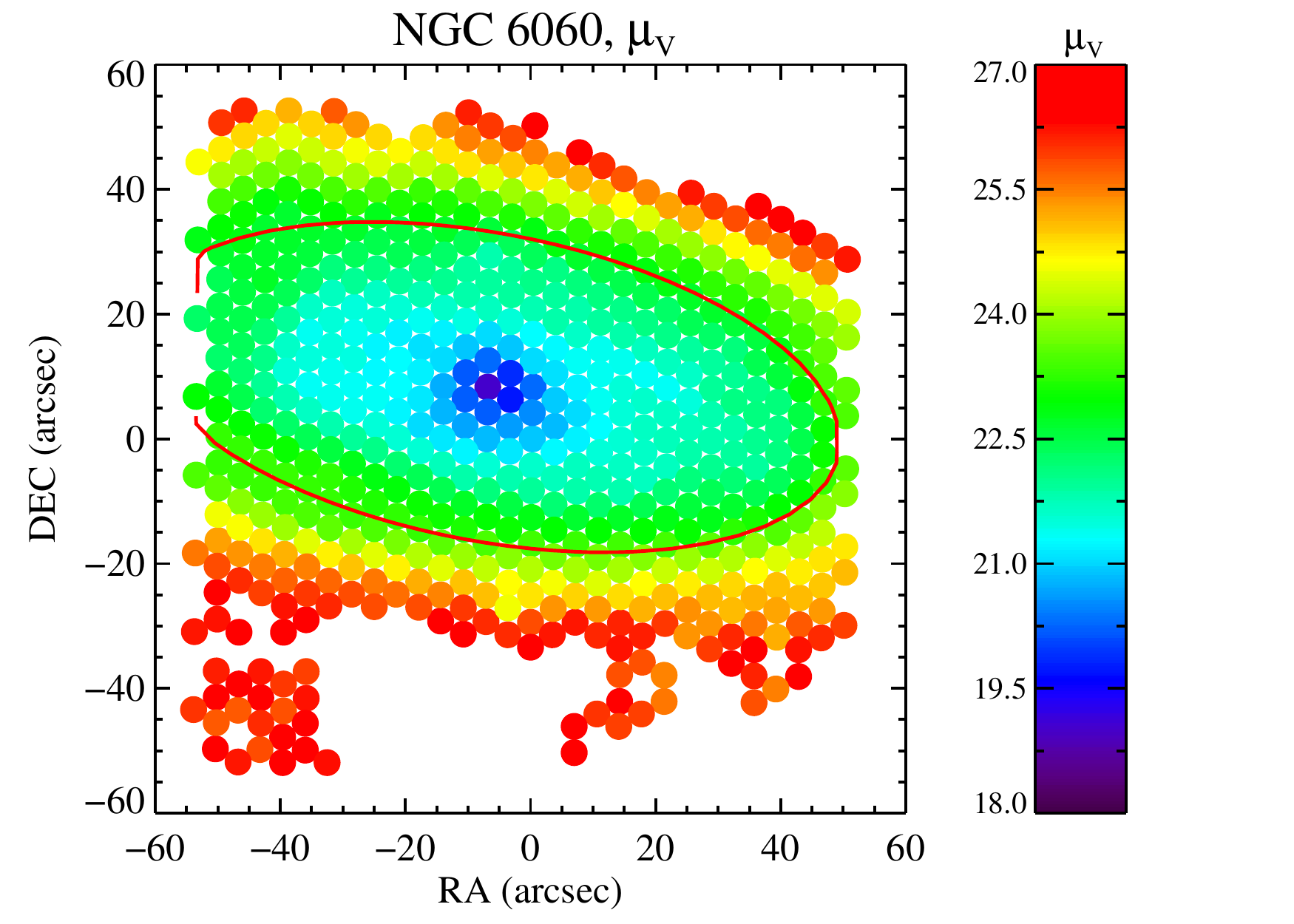}&
\includegraphics[scale=0.27]{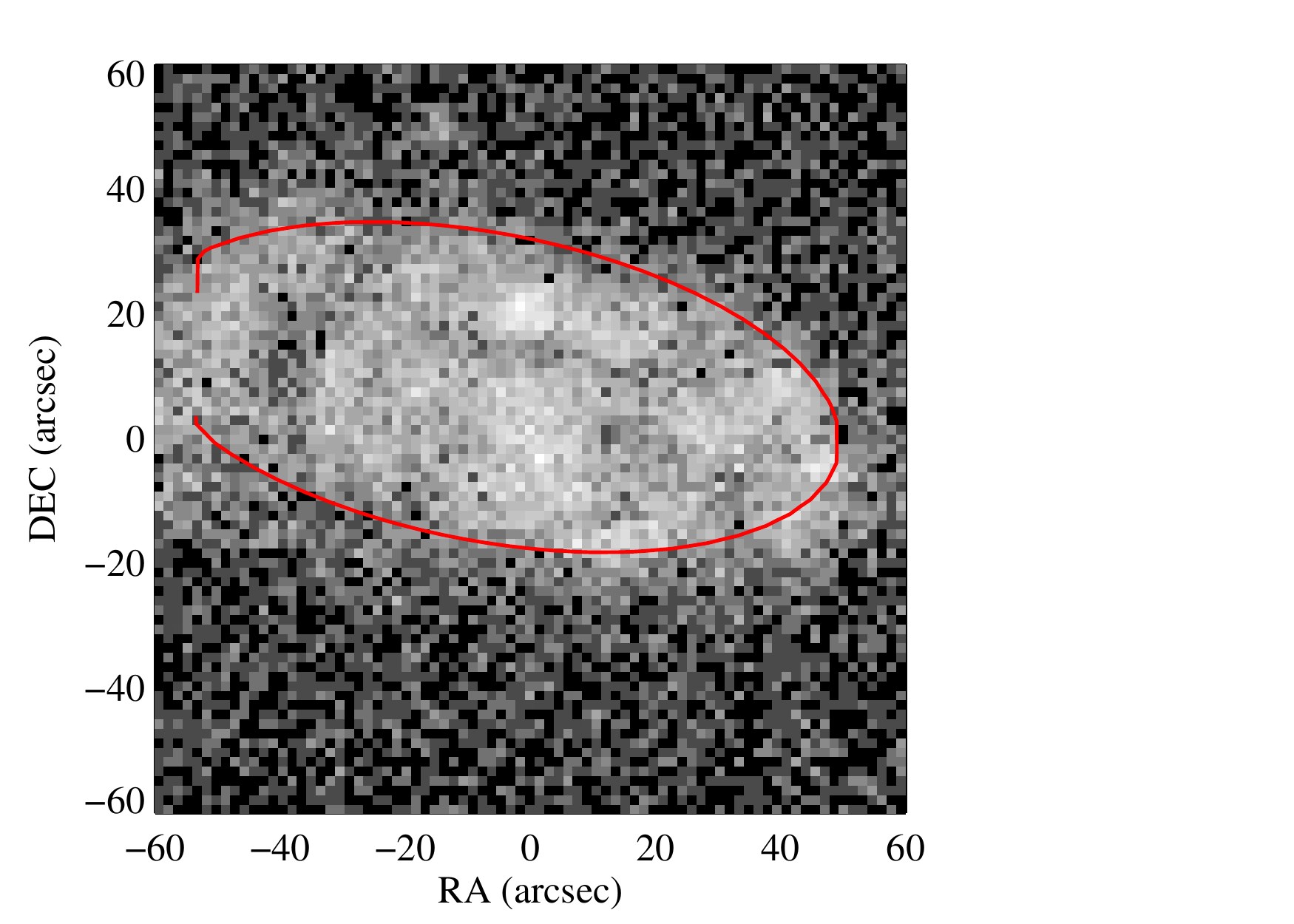}\\
\includegraphics[scale=0.27]{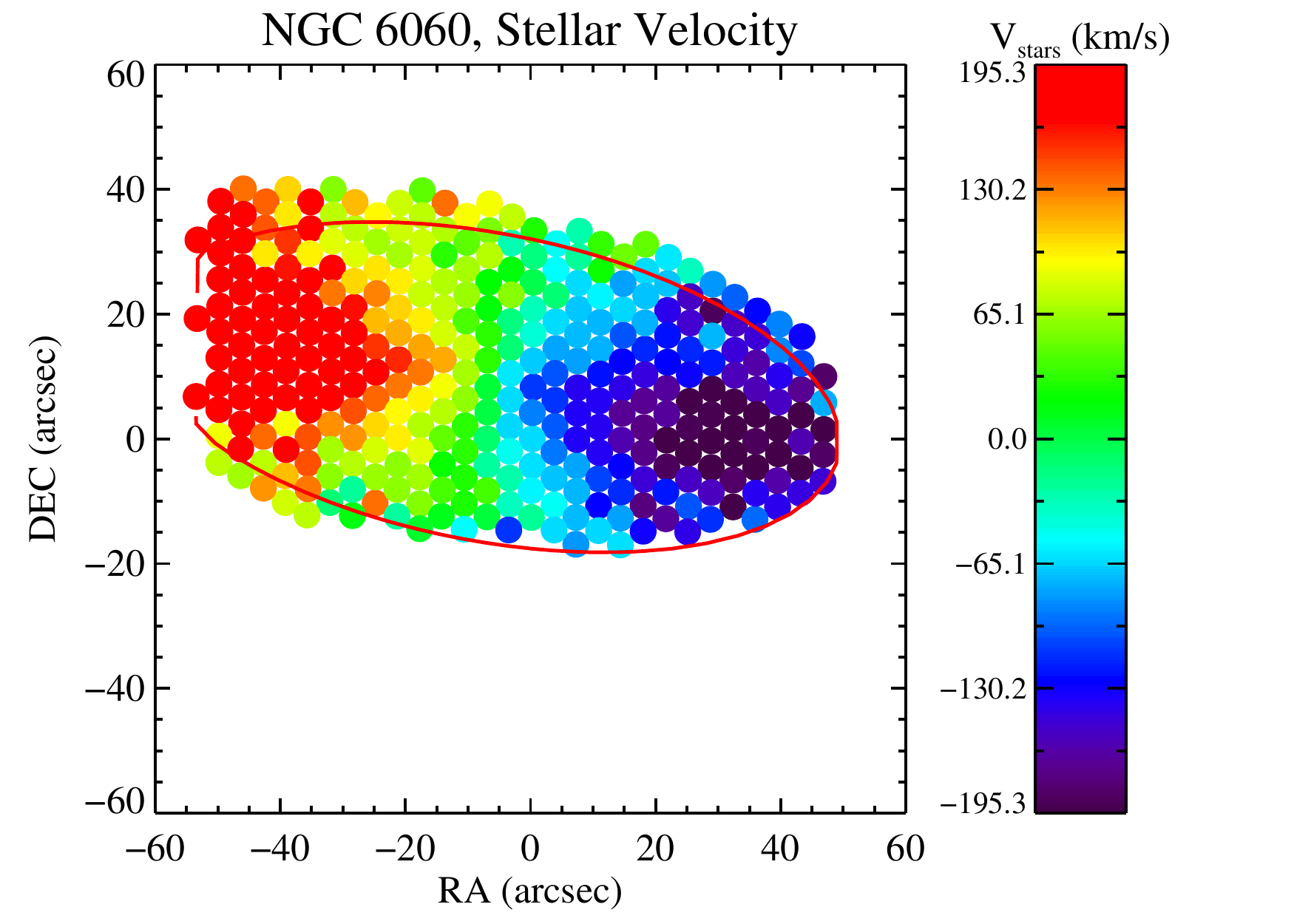}&
\includegraphics[scale=0.27]{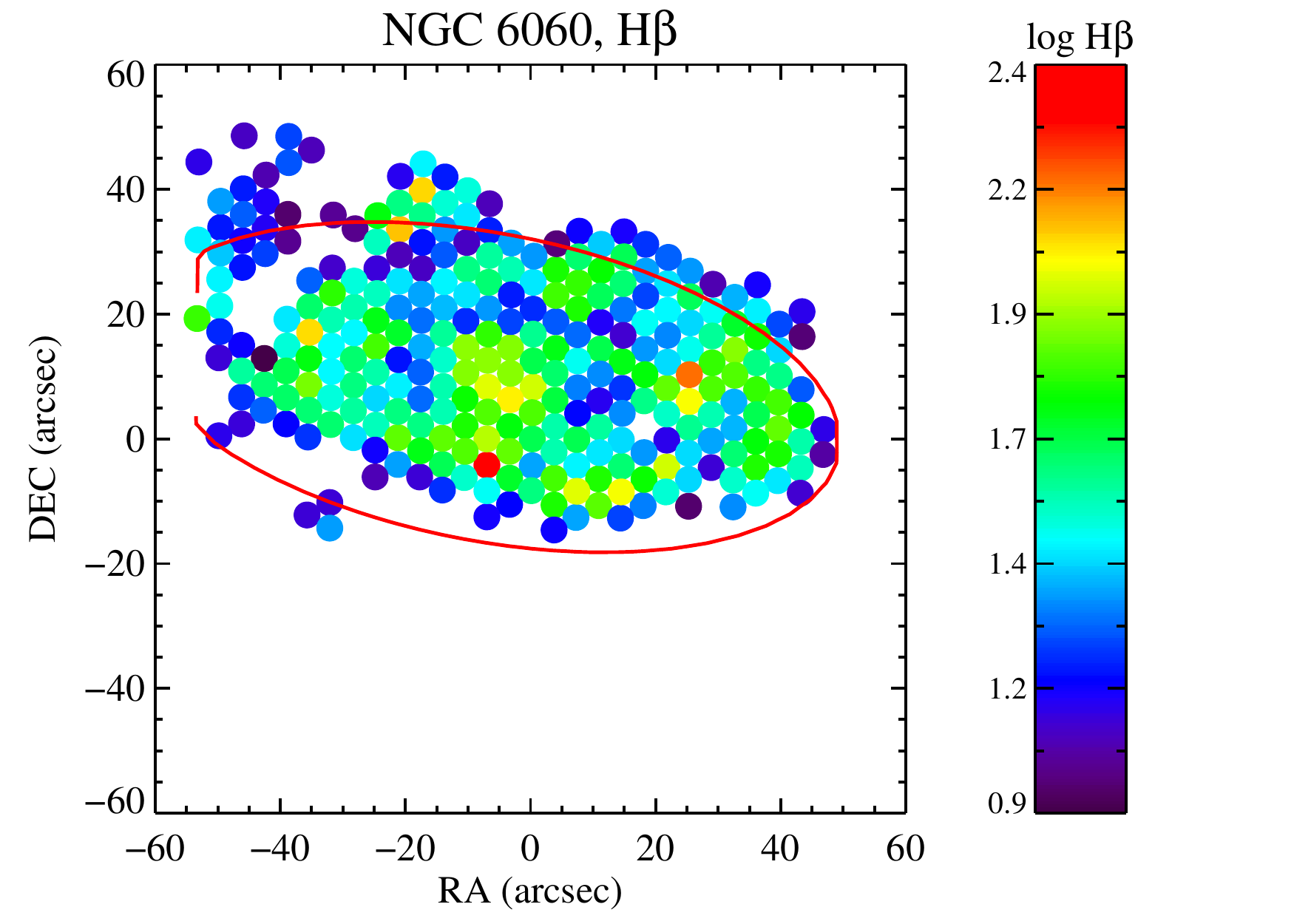}&
\includegraphics[scale=0.27]{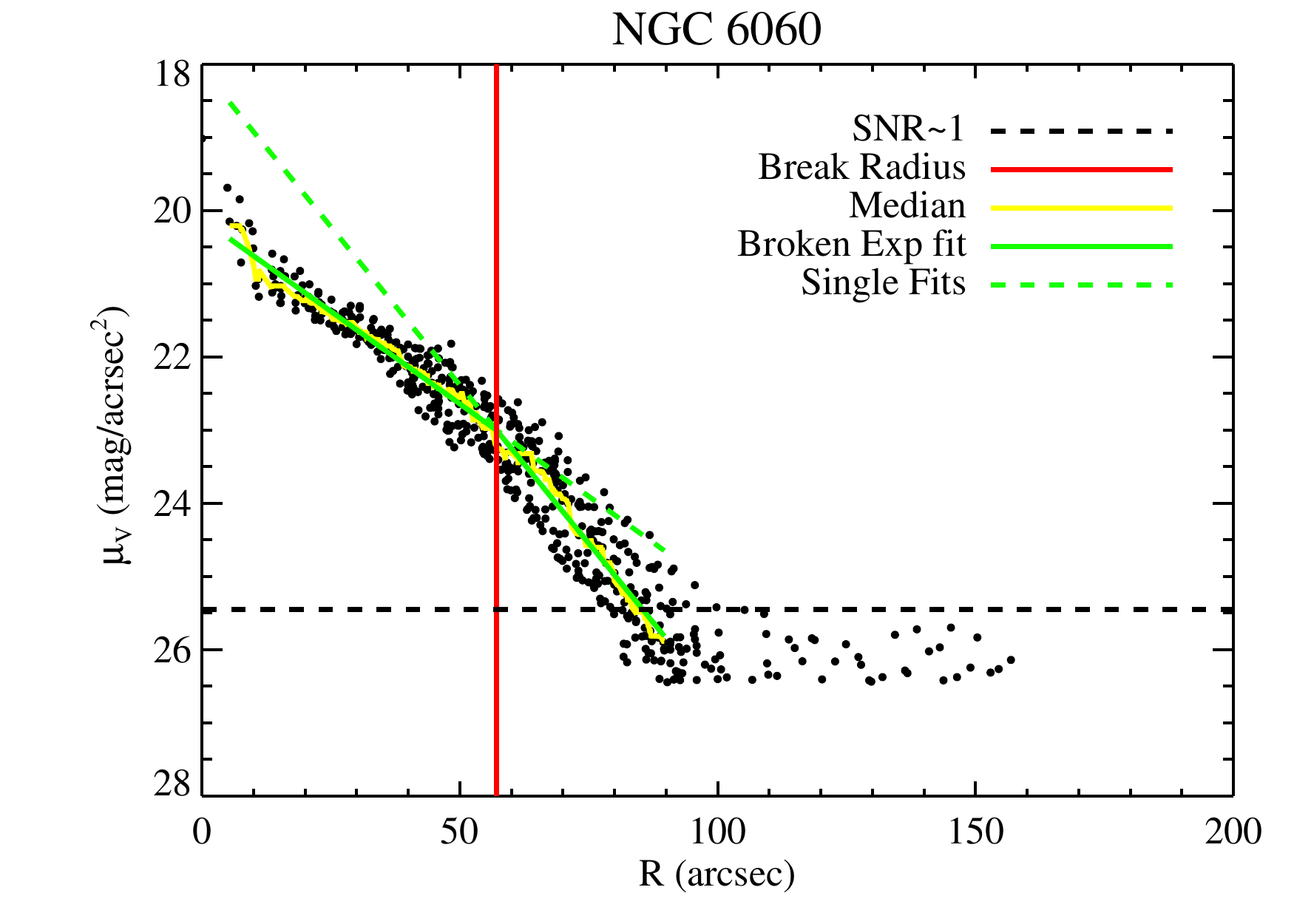}\\
\includegraphics[scale=0.27]{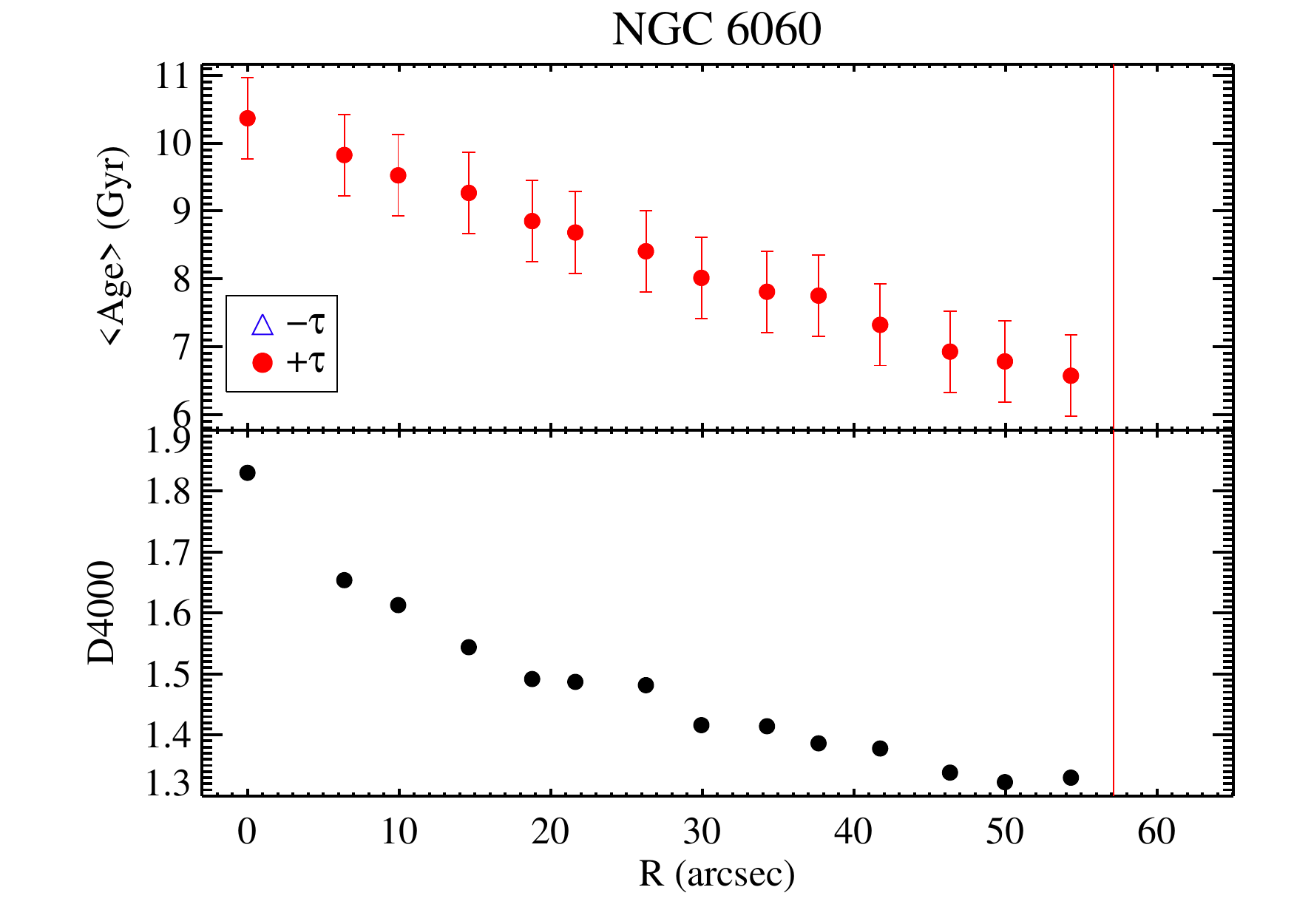}&
\includegraphics[scale=0.27]{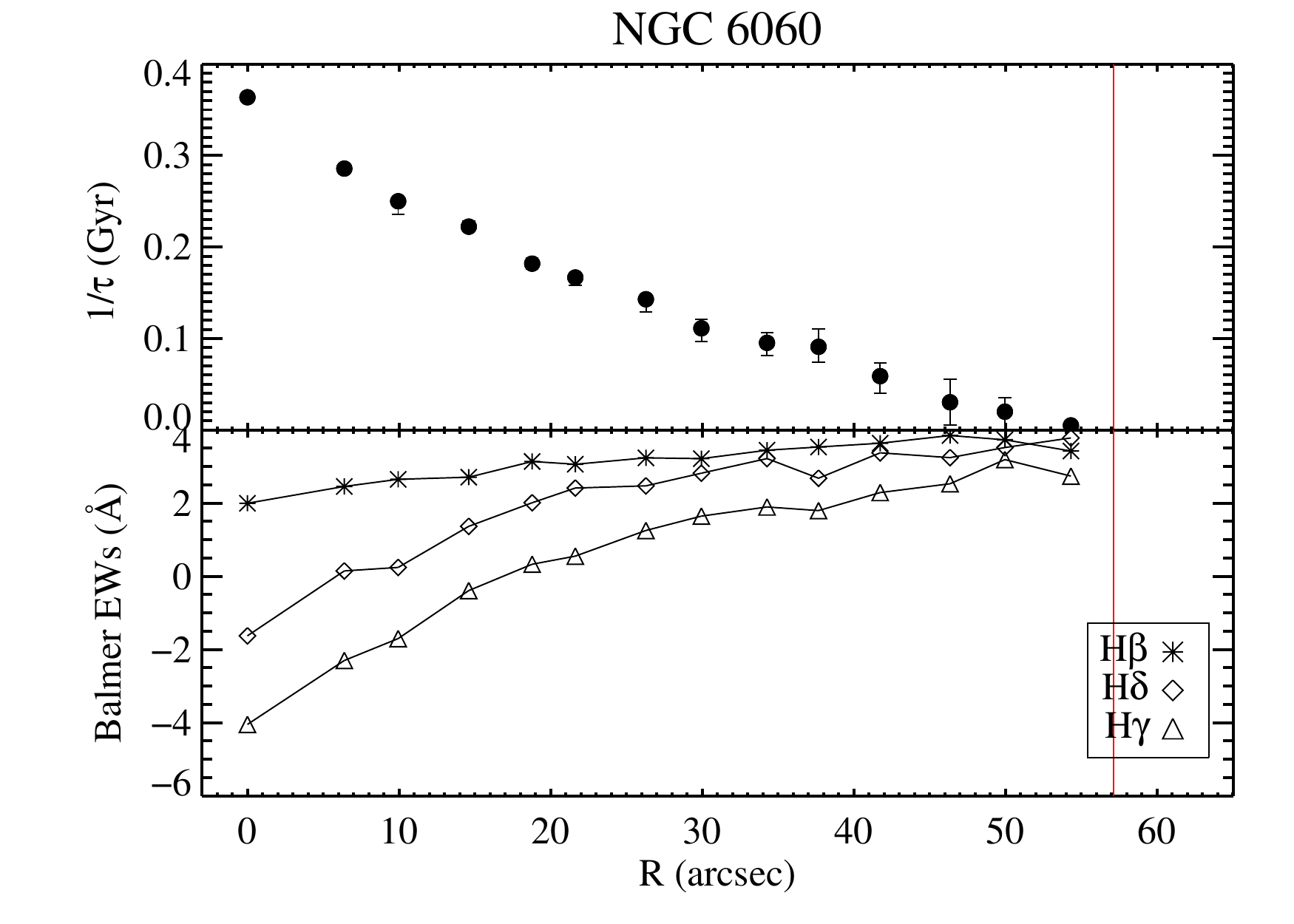}&
\includegraphics[scale=0.27]{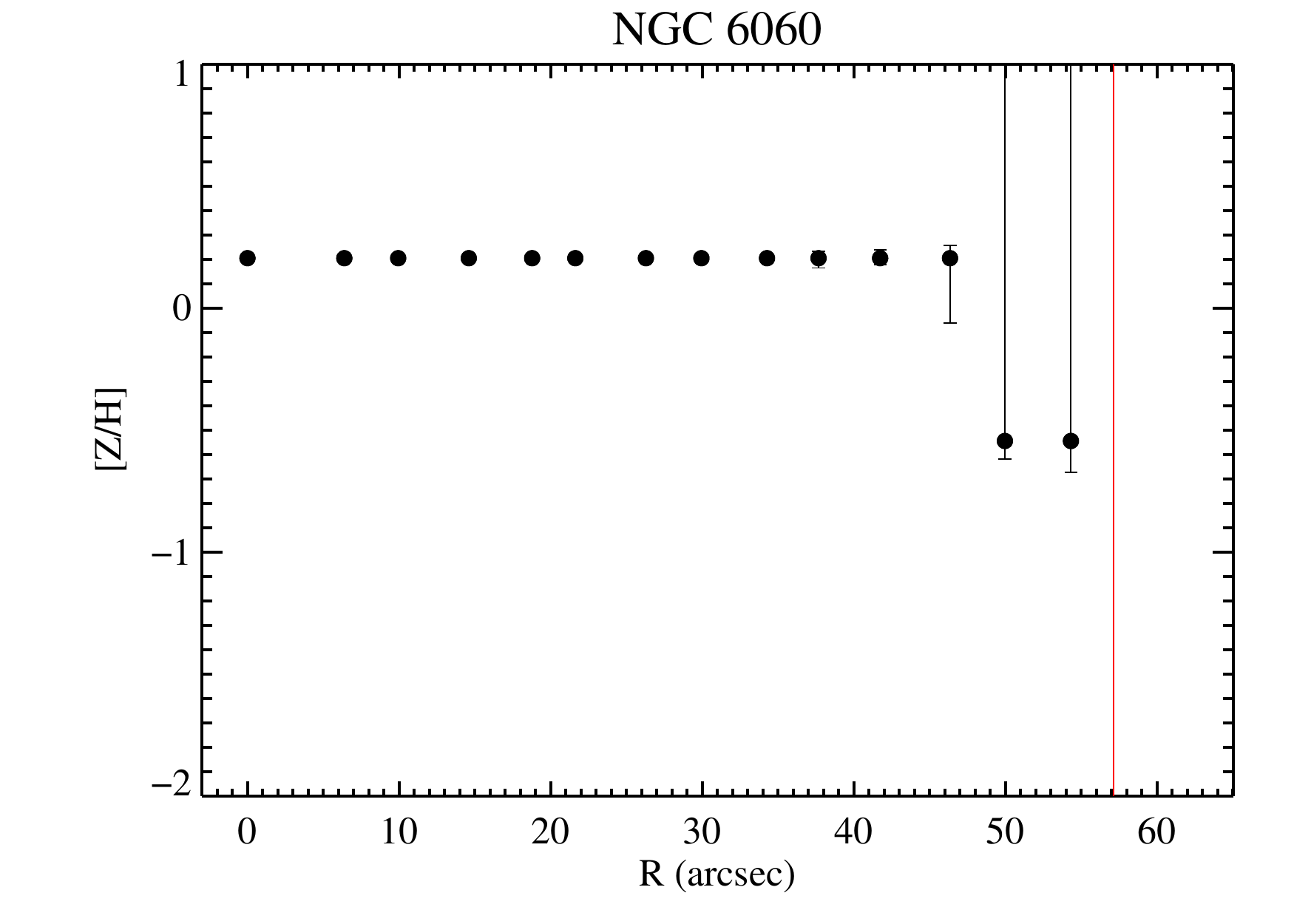}    
\end{array}$
    \epsscale{1}
 \end{center}
\caption{NGC 6060.  Same as Figure~\ref{NGC1058}. \label{NGC6060}}
\end{figure*}


\begin{figure*}
 \begin{center}$
 \begin{array}{ccc}
   \epsscale{.35}
\includegraphics[scale=0.27]{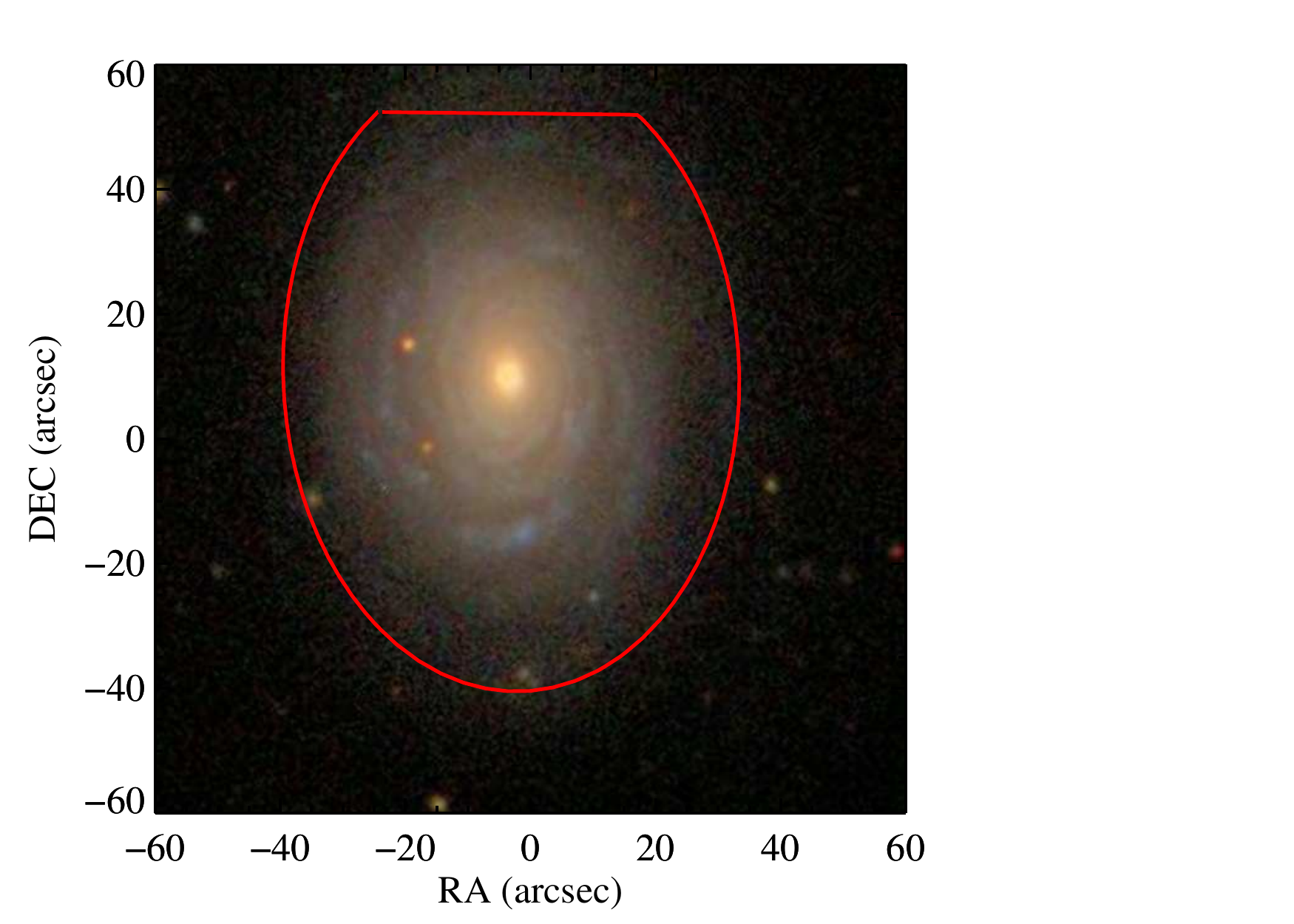}&
\includegraphics[scale=0.27]{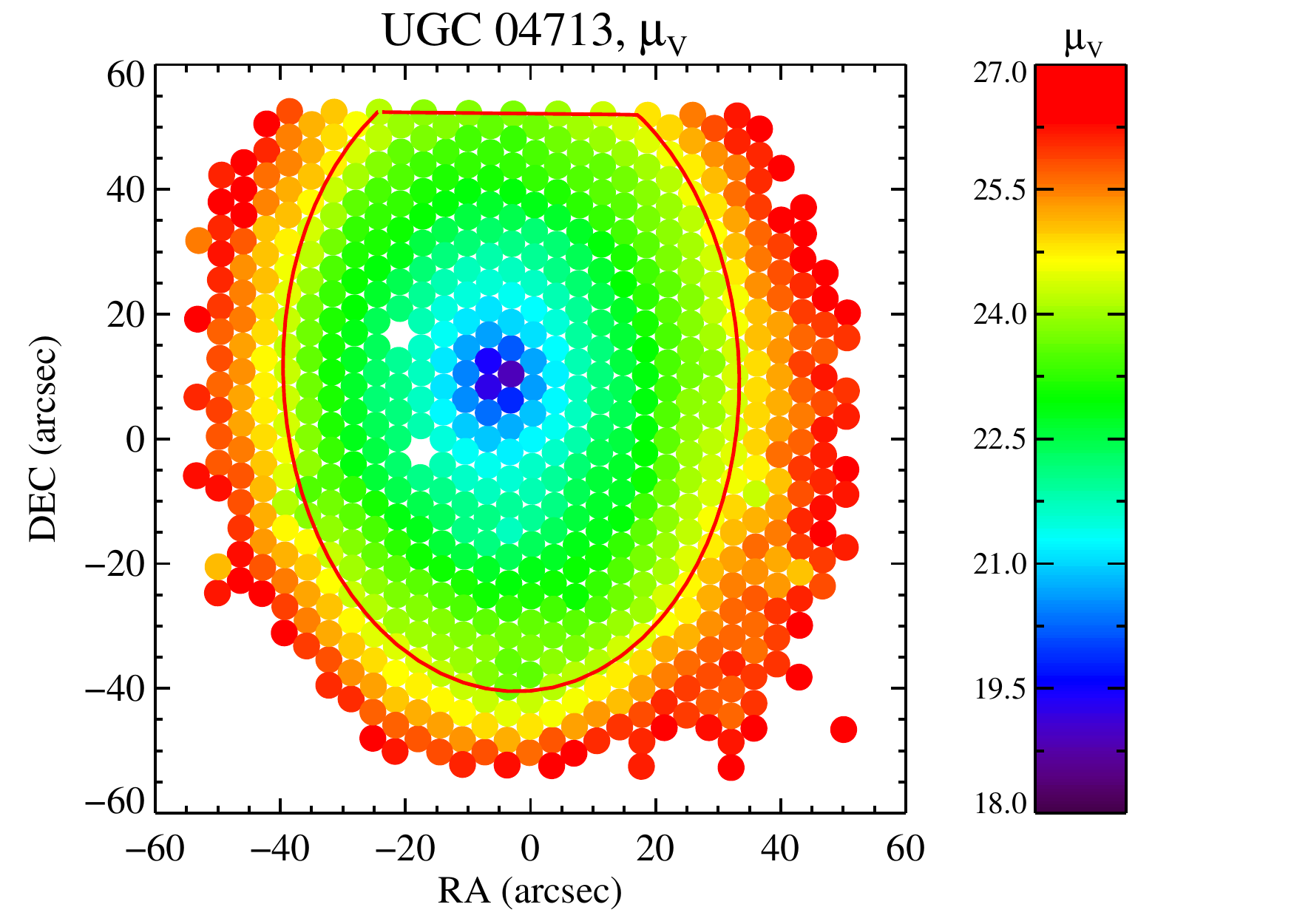}&
\includegraphics[scale=0.27]{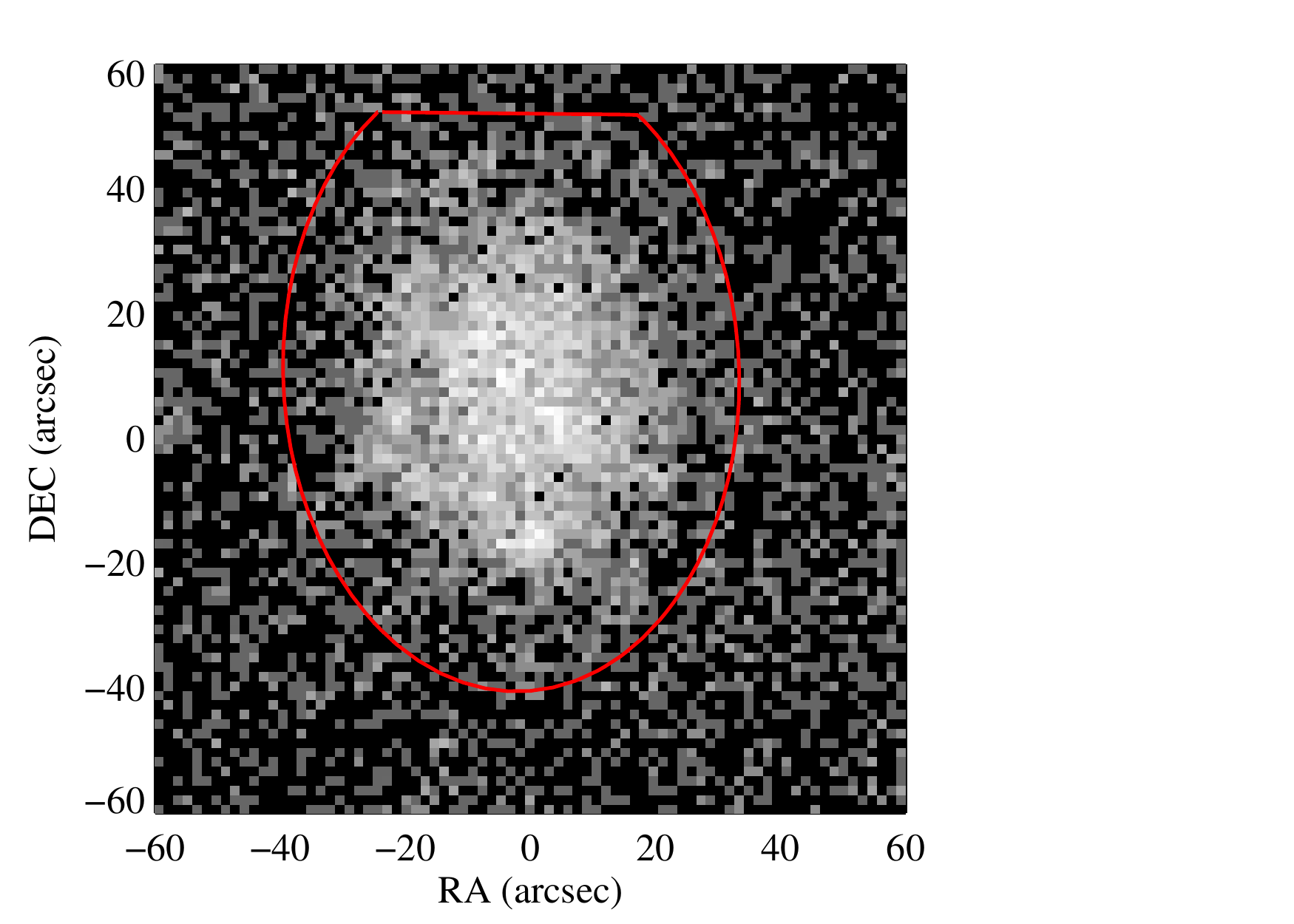}\\
\includegraphics[scale=0.27]{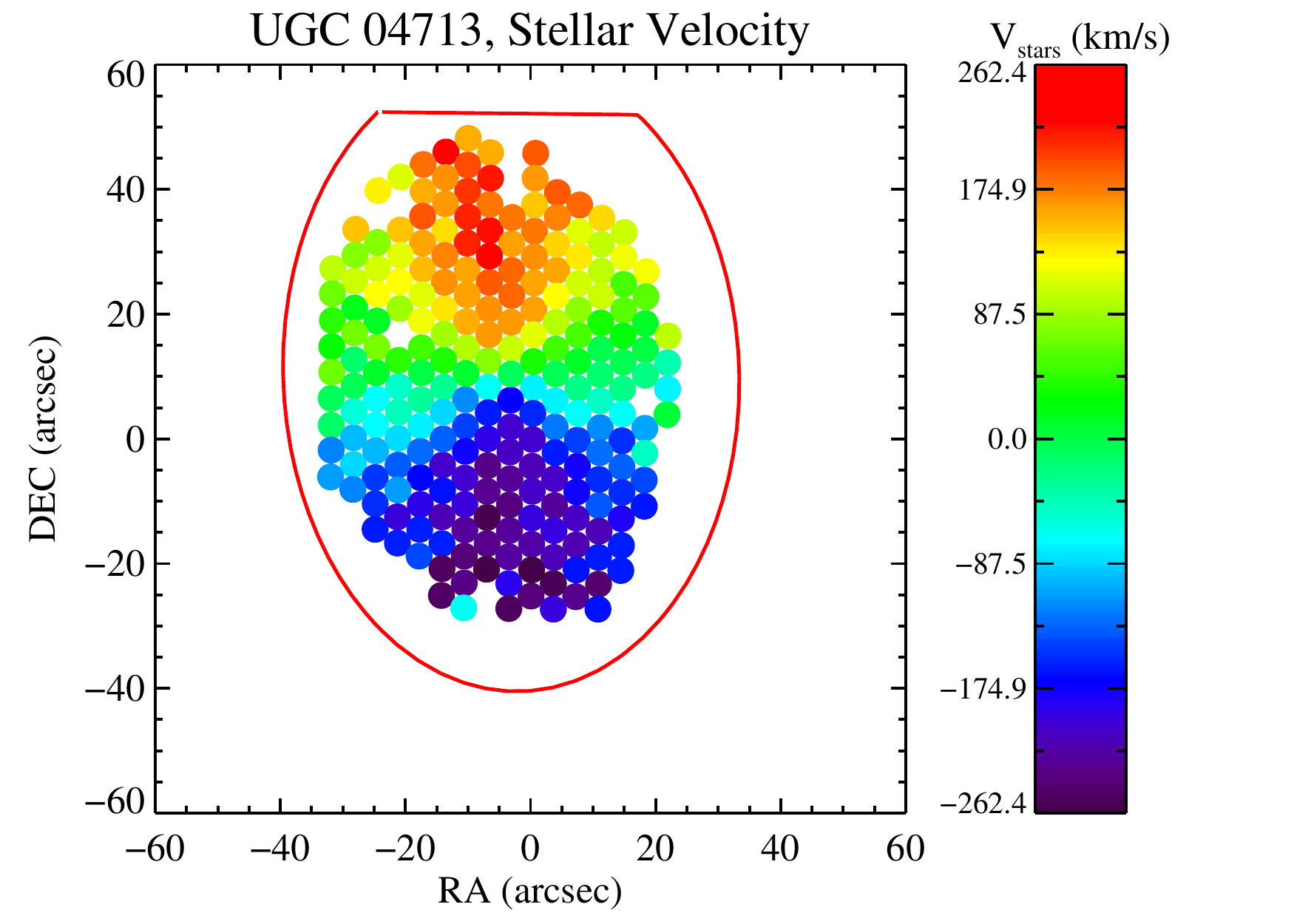}&
\includegraphics[scale=0.27]{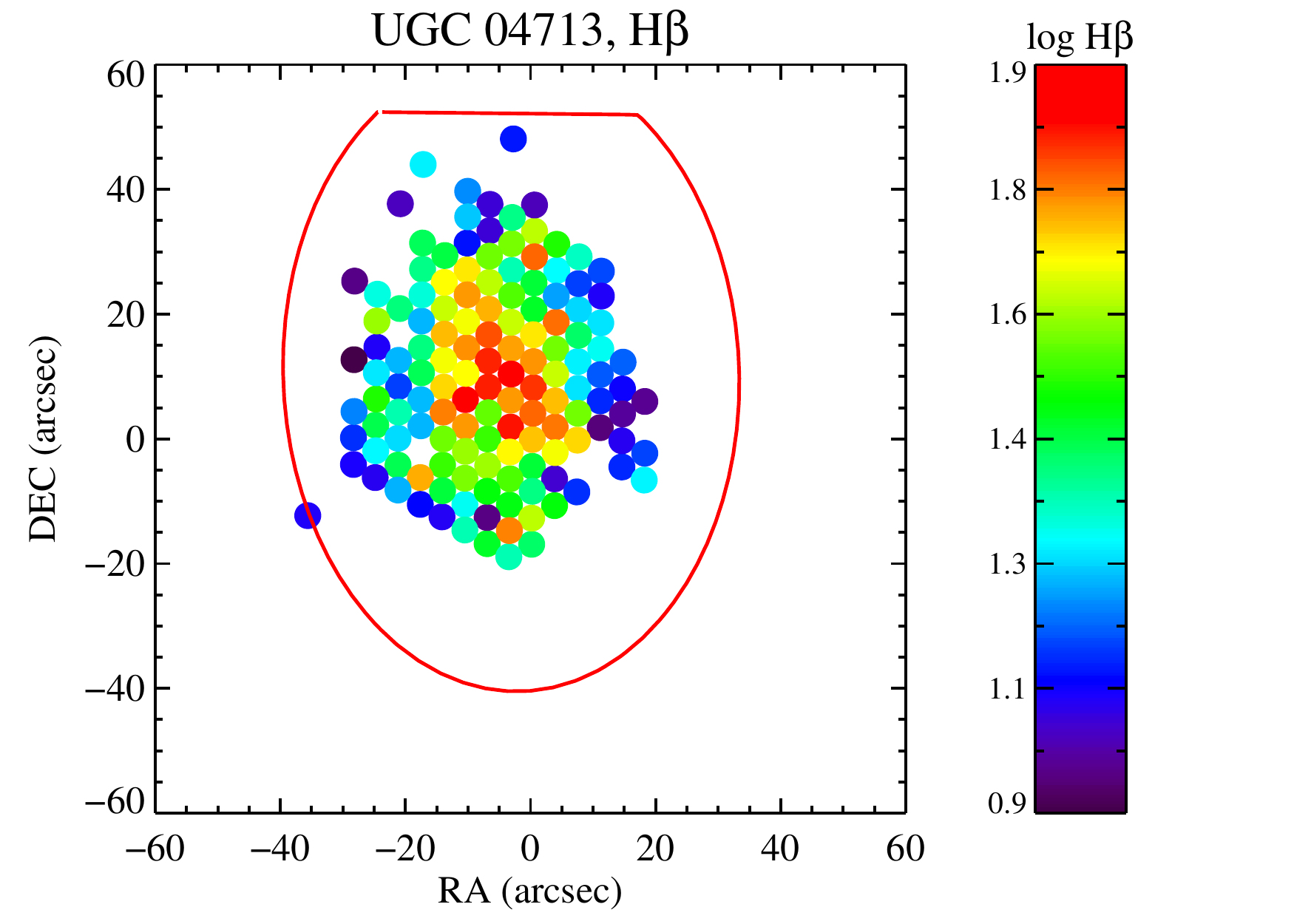}&
\includegraphics[scale=0.27]{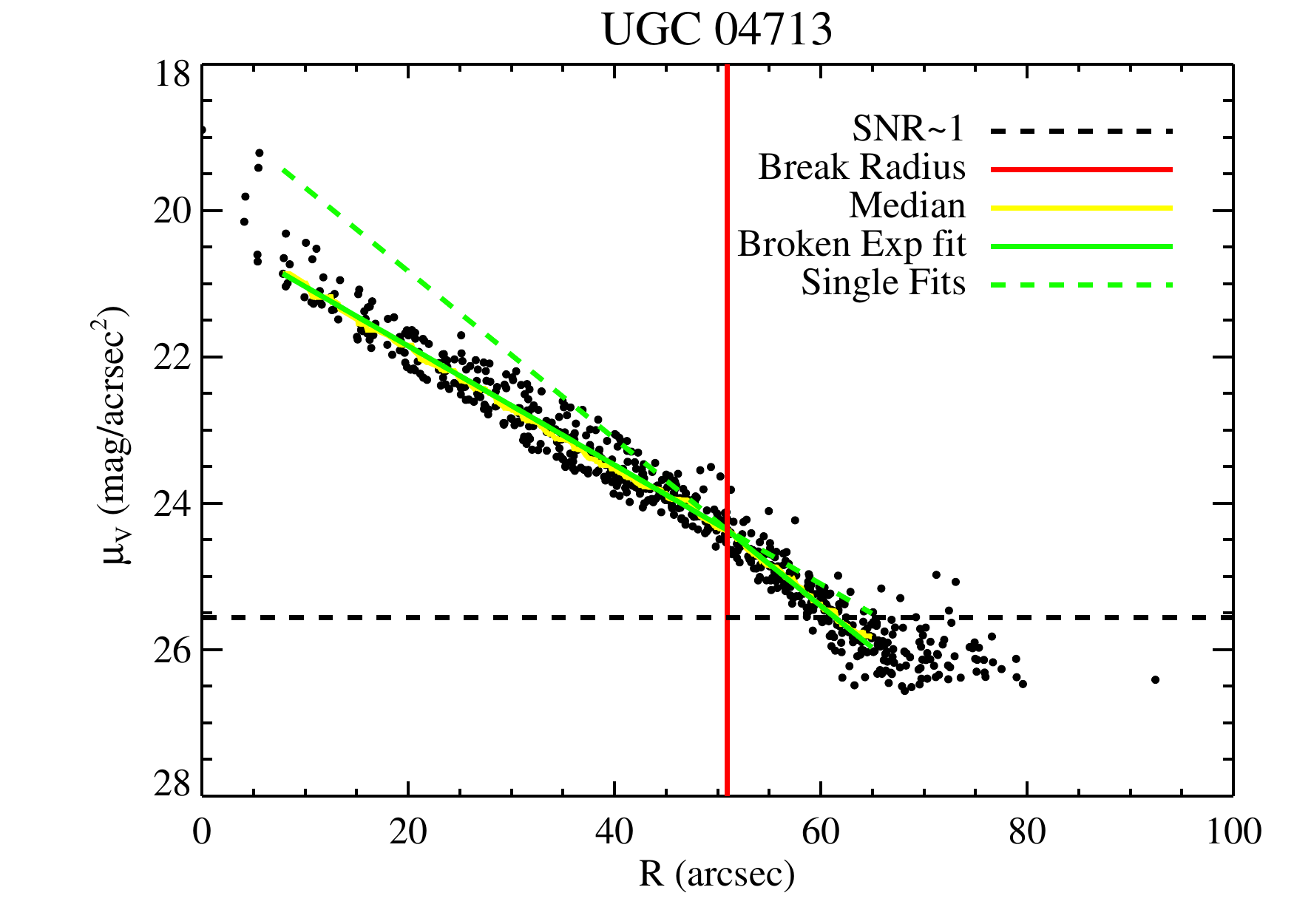}\\
\includegraphics[scale=0.27]{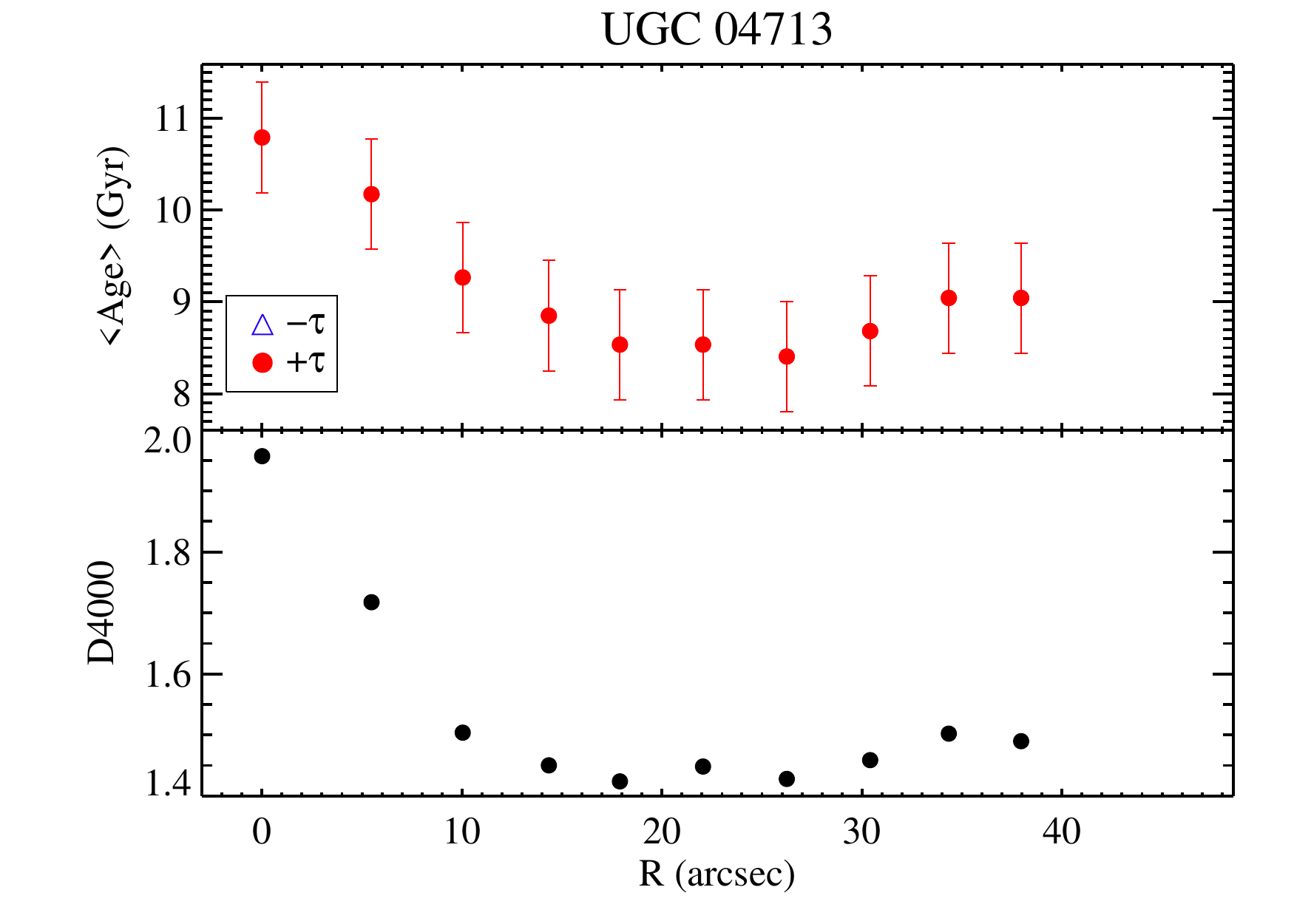}&
\includegraphics[scale=0.27]{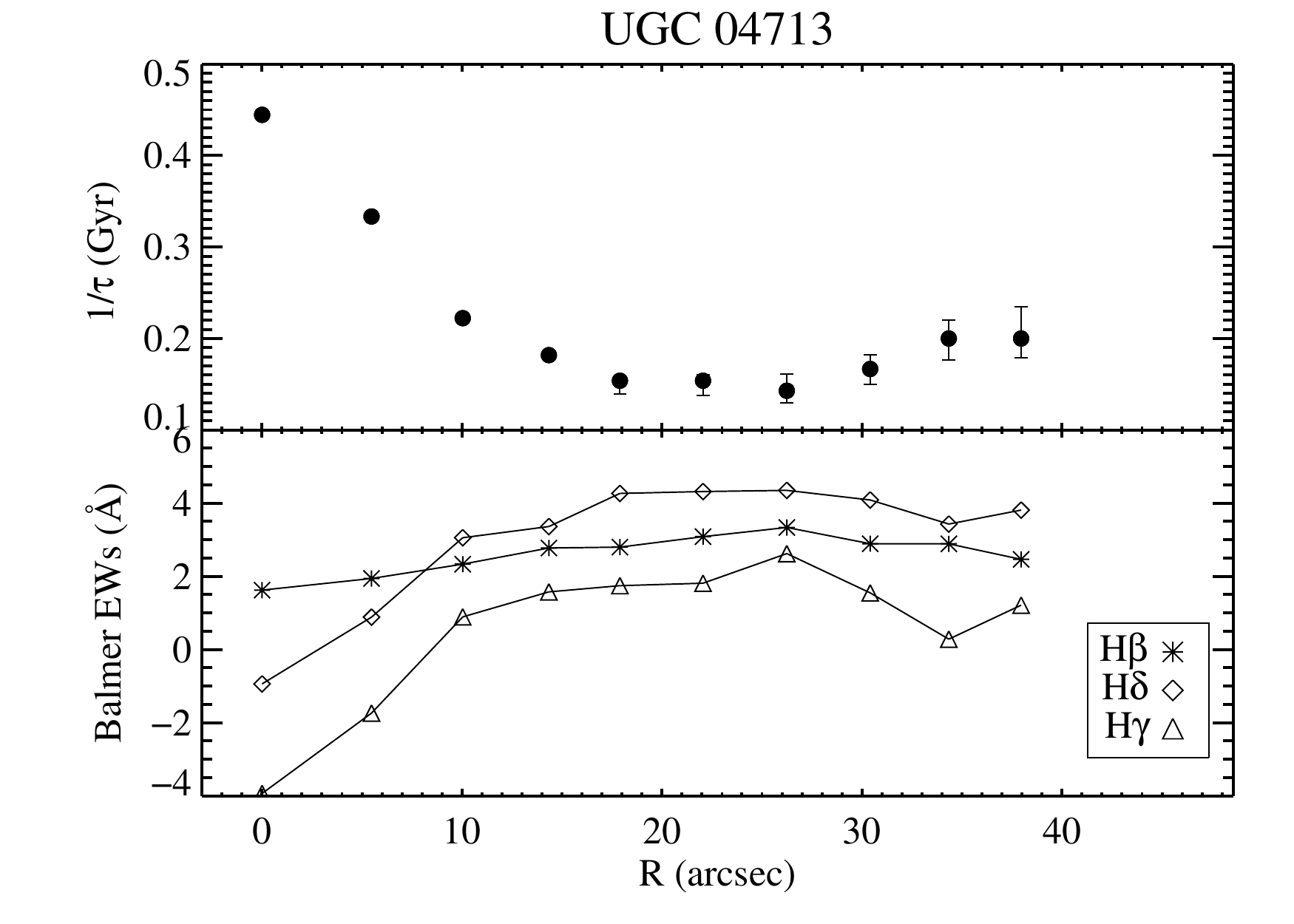}&
\includegraphics[scale=0.27]{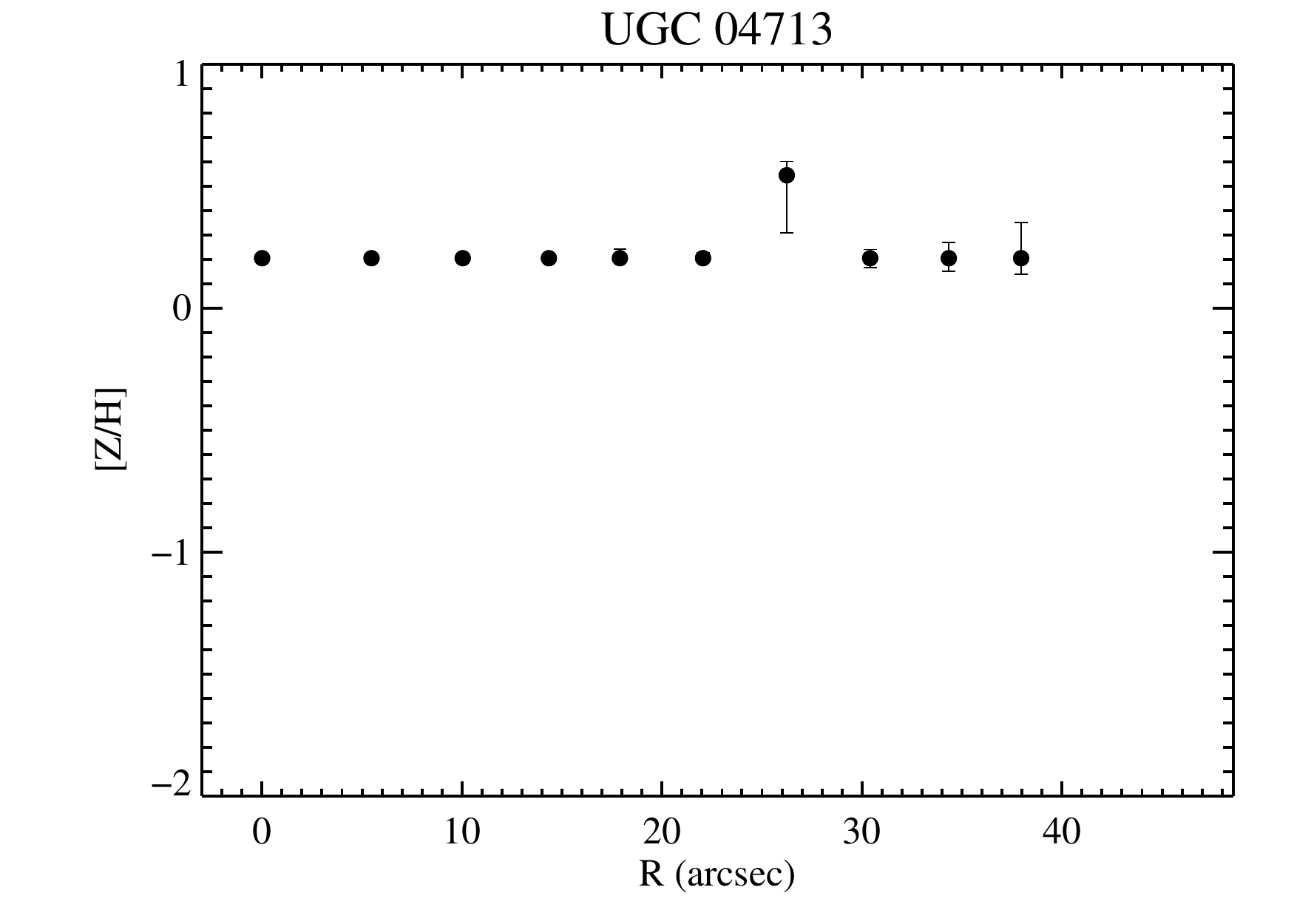}   
\end{array}$
    \epsscale{1}
 \end{center}
\caption{UGC 04713.  Same as Figure~\ref{NGC1058}. \label{UGC04713}}
\end{figure*}


\begin{deluxetable*}{l c c c c c c c c c }
\tablewidth{0pt}
\tablecaption{ Fitted Structural Parameters \label{Table_profilefits}}
\tablehead{\colhead{Name} & \colhead{$\mu_{0,V}$} &\colhead{$h_{inner}$} & \colhead{R$_{break}$} & \colhead{$h_{outer}$} \\
 & \colhead{$\frac{\rm{mag}}{\square^{\prime\prime}}$} & \colhead{\arcsec} & \colhead{\arcsec} & \colhead{\arcsec} 
}
\startdata
IC 1132  &  21.1 &  17.0 &   26.6   &   7.4      \\
NGC 1058 &  20.5 &  25.8 &  \nodata &  \nodata \\
NGC 2684 &  20.1 &   9.0 &   22.3   &   6.7 \\ 
NGC 2684$^{1}$ &  19.6 &   10.6&   24     &  5.6\\ 
NGC 2942 &  20.8 &  17.3 &   45.5   &  10.8\\
NGC 3888 &  19.0 &  11.0 &  \nodata &  \nodata\\
NGC 3888$^{1}$ &  18.7 &  10.1 & \nodata &  \nodata\\
NGC 4904 &  20.3 &  23.2 &   48.8   &  11.1\\
NGC 4904$^{1}$&  20.2 &  27.6 &   39     &  12.5 \\
NGC 5624 &  19.8 &   9.2 &   39.3   &  21.0\\
NGC 5624$^{1}$&  20.1 &   9.9 &   48     &  19.8 \\
NGC 6060 &  20.1 &  21.3 &   56.6   &  12.3\\
NGC 6155 &  19.7 &  11.6 &   35.6   &   7.8\\
NGC 6155$^{1}$&  19.4 &  12.2 &   34     &  8.1 \\
NGC 6691 &  20.6 &  16.1 &   36.5   &   8.6\\
NGC 7437 &  21.6 &  24.2 &   43.6   &   9.7\\
NGC 7437$^{1}$&  21.3 &  26.4 &   43     &  13.9 \\
UGC 04713& 20.2  & 13.3  &  51.1    &  9.5
\enddata
\tablenotetext{1}{$r$-band fits from \citet{Pohlen06}.}
\end{deluxetable*}

\begin{deluxetable}{l c c c c}
\tablewidth{0pt}
\tablecaption{Properties of Migration Signature Galaxies \label{table_mig}}
\tablehead{\colhead{Name} & \colhead{Break Strength} & \colhead{$\Delta$ age} & 
\colhead{$\Delta$ Z} & \colhead{ v$_{\rm{rot}}$ }  \\
 &   $h_{inner}/h_{outer}$   &       (Gyr)            &   (Dex) & (\kms) }
\startdata
NGC 2684 &       1.40   &     1.1  &      0.00 & 101\\
NGC 6155 &       1.46    &    2.4  &    -0.46 & 110\\
NGC 7437 &      2.46  &    1.1  &   -0.11  & 152
\enddata
\end{deluxetable}

\begin{deluxetable}{l c c c c}
\tablewidth{0pt}
\tablecaption{Properties of No Migration Signature Galaxies \label{table_nomig}}
\tablehead{\colhead{Name} & \colhead{Break Strength} & \colhead{$\Delta$ age} & 
\colhead{$\Delta$ Z} & \colhead{ v$_{\rm{rot}}$ }  \\
 &   $h_{inner}/h_{outer}$   &       (Gyr)            &   (Dex) & (\kms) }
\startdata
IC1132   &    2.31  &    0.2  &     0.0 & 106\\
NGC4904  &     2.10 &     0.5 &    -0.18 & 105\\
NGC6691  &     1.57 &   -0.1 &    -0.29 & 78 \\
\enddata
\end{deluxetable}

\subsection{Notes on Individual Galaxies}

\subsubsection{NGC 1058}

The galaxy nearly fills the field of view and is close to face-on, resulting in limited rotation curve information.  We do not detect any break in the surface brightness profile.  The stellar age profile shows a smooth transition from an old center to a young star forming outer region.  The age profile levels off and is constant in the outer region of the galaxy.  There is a surprising amount of spiral structure for such a low mass system.   

\subsubsection{NGC 5624}

The only Type III (up-bending) profile in our sample.  Our stellar population fits show this system has recently had a central burst of star formation, while the outer region is significantly older.  The \hb\ emission is very concentrated in the inner region of the galaxy.  Our observations were not deep enough to probe the region beyond the profile break.  Similar central star bursts have been found in other dwarf galaxies with up-bending profiles \citep{Herrmann11}.  Anti-truncation via a central burst is also in agreement with the simulations of \citet{Younger2007} who use minor-mergers to drive gas to the center of galaxies to create an anti-truncated surface brightness profile.

\subsubsection{NGC 6691}

The best-fit age does change slightly beyond the break radius, however, we do not consider this strong evidence for migration as the D4000 indicator shows little change across the break.  The stellar populations beyond the break are also still best fit by an increasing star formation rate, implying that the profile break does not correspond to a star formation threshold.  This could be a case where a recent burst of star formation in the outer disk is masking the dominant older stellar population.  Such bursty star formation histories will be poorly fit by our technique.

\subsubsection{NGC 2684}

\citet{Pohlen06} fit the break for this galaxy at 24\arcsec\ and classify it as a Type II-CT (classical truncation).  We find a similar break radius, although it is not particularly pronounced.  There is an increase in the stellar age around the break radius.  Interestingly, we find the age minimum is actually well inside the break radius at around 15\arcsec.  The few stellar velocity points we have beyond this radius still show rotation.  There is also a strong metallicity gradient, with the outskirts being metal-poor compared to the interior.  

\subsubsection{NGC 4904}

Our ellipse fitting for the photometry of this galaxy is not stable, with the center and position angle of the galaxy being difficult to pin down.  The kinematic position angle is also noticeably different from the photometric major axis.  Spectroscopically, it looks like active star formation across the disk, and also outside the break radius.  \citet{Pohlen06} find the break radius at 39\arcsec\ in $r$ and 104\arcsec\ in $g$ and suggest the break is due to the outer Lindblad resonance.  Our best-fit break radius is around 49\arcsec, between the \citet{Pohlen06} values.  The galaxy has a strong bar, and visually looks like the break is near the expected OLR radius.  

\subsubsection{IC 1132}

This galaxy shows a very strong profile break, but no significant change in the stellar population across the break.  Many fibers show \hb\ emission beyond the break radius as well, suggesting robust star formation beyond the break radius.

\subsubsection{NGC 6155}

A strong inner star forming region creates a minimum in the stellar age.  There is a clear increase in the stellar ages beyond the profile break as originally reported in \citet{Yoachim10c}.  We also detect a decrease in the stellar metallicity beyond the break radius, consistent with the migration hypothesis.  There is little \hb\ emission or UV-emission beyond the break radius.  \citet{Pohlen06} fit the break at 34\arcsec, while our fit is a comparable 36\arcsec.  Again, the stellar kinematics remain disk-like beyond the break radius.  

\subsubsection{NGC 7437}

The stellar ages increase beyond the profile break, but only one point is fit with a decreasing star formation history.  Similar to NGC 2684, the actual age minimum actually occurs inside the break radius.  Emission from \hb\ does get sparse beyond the break radius.  This galaxy is listed as a classical truncation in \citet{Pohlen06} where they find a similar break radius to our fit.  

\subsubsection{NGC 2942}
 
The break is not particularly well defined as bright spiral arms extend well beyond our best-fit break radius.  Spectroscopy failed to reach adequate signal-to-noise beyond the break.  The SDSS and GALEX imaging shows extensive star formation in the spiral arms extending beyond the break.  We also detect \hb\ emission beyond the break.  

\subsubsection{NGC 3888}
 
The surface brightness profile of NGC 3888 is well describes by a single exponential, as also found in \citet{Pohlen06}.  The stellar age profile shows a pronounced older population around 30\arcsec.

\subsubsection{NGC 6060}
  
We failed to reach adequate SNR to measure a spectroscopic age beyond the break.  The GALEX and \hb\ images do suggest that star formation truncates at the surface brightness profile break.  This is a high mass galaxy (v$_{\rm{rot}}$=252 \kms) with a very smoothly declining stellar age profile.  The best fit SFHs are all decreasing, implying the gas supply for this galaxy is being exhausted.  This galaxy was not in \citet{Pohlen06}.  While we detect a break radius, we did not have adequate signal to measure stellar populations beyond the break radius.  

\subsubsection{UGC 04713}

There is a slight upturn in stellar ages before the profile break, suggestive of possible stellar migration.  The \hb\ emission and UV flux are truncated well before the profile break radius.  Like NGC~6060, this is a high mass system (v$_{\rm{rot}}$=347 \kms) where all the SFHs are decreasing, implying very low levels of star formation.  Again, we did not reach adequate signal-to-noise to measure stellar populations beyond the break.

\section{Discussion}

Figures~\ref{NGC1058}-\ref{UGC04713} show our observations and best-fit stellar ages and metallicities.  We have ordered the galaxies by rotation velocity.  Three of the galaxies, listed in Table~\ref{table_mig}, show indications of stellar migration with the stars beyond the profile break being 1-2 Gyr older than those inside the break.  The star formation history for these outer disks are also best fit by declining rates, while the interiors show signs of active star formation and are best fit with increasing rates of star formation.  In all of these galaxies, both the D4000 index and the Balmer absorption lines show significant changes indicating the outskirts of the galaxies are older than the region inside the break radius.

Exploring low surface brightness levels, there is always the concern that the observations could become dominated by a stellar halo population.  For the few galaxies where we have adequate SNR to measure stellar kinematics beyond the break radius (NGC 2684 and NGC 6155), the velocity field remains disk-like, suggesting we have not become dominated by a halo or stream stellar population.  

While we find several examples where the surface brightness profile break corresponds with a change in stellar population, the actual minimum is rarely at the break radius.  Rather, the stellar age minimum is often located well inside the break radius.  In most cases, it appears the region inside the break radius is hosting the most active star formation, thus driving down the average age compared to the SFR near the profile break.  It should be emphasized that these results are in systems with v$_{\rm{rot}}\sim100$ \kms, and may not apply to higher mass systems.  Since the spectra can be dominated by bright current star formation, we could be biased to measuring younger ages by strong current star formation.  Alternatively, star formation may not proceed in a simple inside-out fashion, and the complex star formation histories of real galaxies can move the stellar age-minimum away from the break radius.

While three galaxies in our sample show signs of stellar migration, three of the galaxies (listed in Table~\ref{table_nomig}) show little or no sign of the outer disk being formed through stellar migration despite having some of the strongest profile breaks.  While there is no strong upturn in the age profiles of these galaxies, we also note that the ages do not seem to be decreasing either.  

Unfortunately, we failed to reach adequate signal-to-noise in higher mass systems, so we cannot comment on how efficient stellar migration is as a function of galaxy mass.  Interestingly, the galaxies with the weakest profile breaks show the strongest evidence for stellar migration, while galaxies with stronger profile breaks have outer regions dominated by younger stars.  This could be consistent with the migration hypothesis, as the galaxies that have experienced the most migration should have the weakest profile breaks.  

Our metallicity measurement is rather crude since we do not include a chemical enrichment model with the star formation histories.   It is notable that both NGC 2684 and NGC 6155 show both an age increase correlated with a metallicity decrease.   However, NGC 4904 and NGC 6691 also show a decrease in metallicity past the break radius but no change in stellar age.  \citet{Vlajic11} look at the outer disk of NGC 300 which is well described by a single exponential with resolved stars and find an upturn in the metallicity gradient.

Overall, we find a very wide range of stellar populations beyond profile breaks, with some outer disks dominated by old stars while others host vigorous star formation.  It would thus appear profile breaks in spiral galaxies will require multiple formation mechanisms.  For example, radial migration may be the primary driver of stellar population trends across the profile break for isolated galaxies, while recent infall of gas may drive star formation in the outskirts of a galaxy that generates a break.  Cosmological simulations should help determine which mechanisms dominate.

\section{Conclusions}

We spectroscopically confirm that stellar population ages increase near spiral galaxy surface brightness profile breaks.  However, we also find several examples where star formation is clearly active on both sides of the break radius and the outer disk is dominated by young stars.  

While \citet{Bakos08} found that, on average, the break radius corresponds to a change in stellar age, we see that in individual cases the stellar population can grow older or remain constant.  Our results suggest that the behavior of profile breaks is bi-modal, with some galaxies hosting a change in stellar populations, while others have similar stellar populations throughout the disk.  This is consistent with the \citet{Bakos08} result, but suggests that the population change they detect has been diluted by galaxies where there is no change in the stellar populations beyond the breaks.

In the future, we would like to extend our SFH fitting procedure to fit multiple epochs of star formation rather than being limited to continuous functions.  We would also like to expand to further constrain the star formation histories with broad-band colors, possibly combining optical spectra with IR and UV observations.  

There is obviously a wealth of data present in IFU observations.  While we have focused on stellar populations, our data also allow the construction of velocity maps and measurements of emission line strengths.  We plan to release our reduced data-cubes along with the larger VENGA survey \citep{Blanc10}.  The data are currently available on the author's web site\footnote{http://www.astro.washington.edu/users/yoachim/}.  

\acknowledgments
Thanks to David Radburn-Smith and Anne Sansom for commenting on early drafts of this paper.  Thanks to Julianne Dalcanton for incessant nagging to get this published.  Thanks to Josh Adams, Guillermo Blanc, and Jeremy Murphy for help with VIRUS-P.  Thanks to Dave Doss, Earl Green, and the rest of the McDonald Observatory support staff.   This work made use of Craig Markwardt's IDL curve fitting code \citep{Markwardt09}.

We thank the Cynthia and George Mitchell Foundation for funding the Mitchell Spectrograph, formerly known as VIRUS-P

We acknowledge the usage of the HyperLeda database (http://leda.univ-lyon1.fr).

    Funding for the SDSS and SDSS-II has been provided by the Alfred P. Sloan Foundation, the Participating Institutions, the National Science Foundation, the U.S. Department of Energy, the National Aeronautics and Space Administration, the Japanese Monbukagakusho, the Max Planck Society, and the Higher Education Funding Council for England. The SDSS Web Site is http://www.sdss.org/.

    The SDSS is managed by the Astrophysical Research Consortium for the Participating Institutions. The Participating Institutions are the American Museum of Natural History, Astrophysical Institute Potsdam, University of Basel, University of Cambridge, Case Western Reserve University, University of Chicago, Drexel University, Fermilab, the Institute for Advanced Study, the Japan Participation Group, Johns Hopkins University, the Joint Institute for Nuclear Astrophysics, the Kavli Institute for Particle Astrophysics and Cosmology, the Korean Scientist Group, the Chinese Academy of Sciences (LAMOST), Los Alamos National Laboratory, the Max-Planck-Institute for Astronomy (MPIA), the Max-Planck-Institute for Astrophysics (MPA), New Mexico State University, Ohio State University, University of Pittsburgh, University of Portsmouth, Princeton University, the United States Naval Observatory, and the University of Washington.

GALEX (Galaxy Evolution Explorer) is a NASA Small Explorer, launched in April 2003. We gratefully acknowledge NASA's support for construction, operation, and science analysis for the GALEX mission. 


\end{document}